\documentclass{article}
\usepackage{fancyheadings}
\usepackage{amssymb}
\usepackage{bm} 
\usepackage{cite}
\makeatletter
  \@addtoreset{equation}{subsection}
  
  \renewcommand\section{\@startsection {section}{1}{\z@}%
       {-3.5ex \@plus -1ex \@minus -.2ex}%
       {2.3ex \@plus.2ex}%
       {\normalfont\Large\bfseries\centering}}
  \renewcommand\subsection{\@startsection{subsection}{2}{\z@}%
       {-3.25ex\@plus -1ex \@minus -.2ex}%
       {1.5ex \@plus .2ex}%
       {\normalfont\large\it\centering}}
  \renewcommand{\sectionmark}[1]%
    {\markboth{\thesection\hspace*{10pt} #1}{}} 
  \renewcommand{\subsectionmark}[1]%
    {\markright{\thesubsection\hspace*{10pt} #1}} 
\renewcommand{\contentsname}{Contents} 
\renewcommand{\tableofcontents}{%
    \section*{\contentsname
        \@mkboth{\contentsname}{\contentsname}}%
    \@starttoc{toc}%
}
\renewcommand{\listfigurename}{List of figures} 
\renewcommand{\listoffigures}{%
    \section*{\listfigurename
      \@mkboth{\listfigurename}{\listfigurename}}%
    \@starttoc{lof}%
}
\makeatother 
\newcommand{\base}[2]{e^{#1}_{#2}} 

\begin{document}
\thispagestyle{empty} 

\begin{center} 
\LARGE \sc 
The motion of point particles in curved spacetime 
\end{center} 

\vspace*{.1in} 
\begin{center} 
\large
Eric Poisson \\ 
\small 
Department of Physics, University of Guelph, 
Guelph, Ontario, Canada N1G 2W1; and \\ 
Perimeter Institute for Theoretical Physics,  
35 King Street North, Waterloo, Ontario, Canada N2J 2W9 
\end{center} 

\begin{center} 
Final draft of review article, March 31, 2004 
\end{center} 

\vspace*{.1in} 
\begin{abstract} 
This review is concerned with the motion of a point scalar charge, a
point electric charge, and a point mass in a specified background
spacetime. In each of the three cases the particle produces a field 
that behaves as outgoing radiation in the wave zone, and therefore
removes energy from the particle. In the near zone the field acts on
the particle and gives rise to a self-force that prevents the
particle from moving on a geodesic of the background spacetime. The
self-force contains both conservative and dissipative terms, and the 
latter are responsible for the radiation reaction. The work done by
the self-force matches the energy radiated away by the particle. 

The field's action on the particle is difficult to calculate because
of its singular nature: the field diverges at the position of the
particle. But it is possible to isolate the field's singular part and
show that it exerts no force on the particle --- its only effect is to
contribute to the particle's inertia. What remains after subtraction
is a smooth field that is fully responsible for the
self-force. Because this field satisfies a homogeneous wave equation,  
it can be thought of as a free (radiative) field that interacts with   
the particle; it is this interaction that gives rise to the
self-force.  

The mathematical tools required to derive the equations of motion of a
point scalar charge, a point electric charge, and a point mass in a
specified background spacetime are developed here from scratch. The
review begins with a discussion of the basic theory of bitensors 
(part \ref{part1}). It then applies the theory to the construction of
convenient coordinate systems to chart a neighbourhood of the
particle's word line (part \ref{part2}). It continues with a thorough
discussion of Green's functions in curved spacetime (part
\ref{part3}). The review concludes with a detailed derivation of each
of the three equations of motion (part \ref{part4}).          
\end{abstract} 

\vspace*{.1in}
\tableofcontents

\newpage
%
\section{Introduction and summary} 
\label{1}

\subsection{Invitation}
\label{1.1} 

The motion of a point electric charge in flat spacetime was the 
subject of active investigation since the early work of Lorentz, 
Abrahams, and Poincar\'e, until Dirac \cite{dirac} produced a proper
relativistic derivation of the equations of motion in 1938. (The
field's early history is well related in Ref.~\cite{rohrlich}). In
1960 DeWitt and Brehme \cite{dewittbrehme} generalized
Dirac's result to curved spacetimes, and their calculation was
corrected by Hobbs \cite{hobbs} several years later. In 1997 the  
motion of a point mass in a curved background spacetime was
investigated by Mino, Sasaki, and Tanaka \cite{MST}, who derived an
expression for the particle's acceleration (which is not zero unless
the particle is a test mass); the same equations of motion were later
obtained by Quinn and Wald \cite{QW1} using an axiomatic approach. The 
case of a point scalar charge was finally considered by Quinn in 2000
\cite{quinn}, and this led to the realization that the mass of a
scalar particle is not necessarily a constant of the motion.  

This article reviews the achievements described in the preceding
paragraph; it is concerned with the motion of a point scalar charge 
$q$, a point electric charge $e$, and a point mass $m$ in a specified
background spacetime with metric $g_{\alpha\beta}$. These particles
carry with them fields that behave as outgoing radiation in the wave 
zone. The radiation removes energy and angular momentum from the
particle, which then undergoes a radiation reaction --- its world line
cannot be simply a geodesic of the background spacetime. The
particle's motion is affected by the near-zone field which acts
directly on the particle and produces a {\it self-force}. In curved
spacetime the self-force contains a radiation-reaction component that
is directly associated with dissipative effects, but it contains also
a conservative component that is not associated with energy or
angular-momentum transport. The self-force is proportional to $q^2$ in
the case of a scalar charge, proportional to $e^2$ in the case of an
electric charge, and proportional to $m^2$ in the case of a point mass.  

In this review I derive the equations that govern the motion of 
a point particle in a curved background spacetime. The presentation is 
entirely self-contained, and all relevant materials are developed  
{\it ab initio}. The reader, however, is assumed to have a solid grasp  
of differential geometry and a deep understanding of general
relativity. The reader is also assumed to have unlimited stamina, 
for the road to the equations of motion is a long one. One must first
assimilate the basic theory of bitensors (part \ref{part1}), then
apply the theory to construct convenient coordinate systems to chart a 
neighbourhood of the particle's world line (part \ref{part2}). One
must next formulate a theory of Green's functions in curved spacetimes
(part \ref{part3}), and finally calculate the scalar, electromagnetic,
and gravitational fields near the world line and figure out how
they should act on the particle (part \ref{part4}). The review is very
long, but the payoff, I hope, will be commensurate.   

In this introductory section I set the stage and present an
impressionistic survey of what the review contains. This should help
the reader get oriented and acquainted with some of the ideas and some
of the notation. Enjoy! 

\subsection{Radiation reaction in flat spacetime}  
\label{1.2}

Let us first consider the relatively simple and well-understood case
of a point electric charge $e$ moving in flat spacetime
\cite{rohrlich, jackson, TVW, poisson}. The charge produces an
electromagnetic vector potential $A^\alpha$ that satisfies the wave
equation    
\begin{equation} 
\Box A^\alpha = - 4\pi j^\alpha
\label{1.2.1}
\end{equation} 
together with the Lorenz gauge condition $\partial_\alpha A^\alpha 
= 0$. (On page 294 Jackson \cite{jackson} explains why the term
``Lorenz gauge'' is preferable to ``Lorentz gauge''.) 
The vector $j^\alpha$ is the charge's current density, which is 
formally written in terms of a four-dimensional Dirac functional
supported on the charge's world line: the density is zero everywhere, 
except at the particle's position where it is infinite. For
concreteness we will imagine that the particle moves around a
centre (perhaps another charge, which is taken to be fixed) and that
it emits outgoing radiation. We expect that the charge will undergo a
radiation reaction and that it will spiral down toward the
centre. This effect must be accounted for by the equations of motion,
and these must therefore include the action of the charge's own field,
which is the only available agent that could be responsible for the
radiation reaction. We seek to determine this self-force acting on the 
particle.   

An immediate difficulty presents itself: the vector potential, and 
also the electromagnetic field tensor, diverge on the particle's world
line, because the field of a point charge is necessarily infinite at
the charge's position. This behaviour makes it most difficult to
decide how the field is supposed to act on the particle.  

Difficult but not impossible. To find a way around this problem I  
note first that {\it the situation considered here, in which the
radiation is propagating outward and the charge is spiraling inward,
breaks the time-reversal invariance of Maxwell's theory}. A specific 
time direction was adopted when, among all possible solutions to the
wave equation, we chose $A^\alpha_{\rm ret}$, the {\it retarded
solution}, as the physically-relevant solution. Choosing instead 
the {\it advanced solution} $A^\alpha_{\rm adv}$ would produce a 
time-reversed picture in which the radiation is propagating inward 
and the charge is spiraling outward. Alternatively, choosing the
linear superposition  
\begin{equation}
A^\alpha_{\rm S} = 
\frac{1}{2} \bigl( A^\alpha_{\rm ret} + A^\alpha_{\rm adv} \bigr)
\label{1.2.2}
\end{equation}
would restore time-reversal invariance: outgoing and incoming
radiation would be present in equal amounts, there would be no net
loss nor gain of energy by the system, and the charge would not 
undergo any radiation reaction. In Eq.~(\ref{1.2.2}) the subscript `S'    
stands for `symmetric', as the vector potential depends symmetrically
upon future and past.  

My second key observation is that while the potential of
Eq.~(\ref{1.2.2}) does not exert a force on the charged
particle, {\it it is just as singular as the retarded potential in the
vicinity of the world line}. This follows from the fact that 
$A^\alpha_{\rm ret}$, $A^\alpha_{\rm adv}$, and $A^\alpha_{\rm S}$ all
satisfy Eq.~(\ref{1.2.1}), whose source term is infinite on the world
line. So while the wave-zone behaviours of these solutions are very
different (with the retarded solution describing outgoing waves, the
advanced solution describing incoming waves, and the symmetric
solution describing standing waves), the three vector potentials share
the same singular behaviour near the world line --- all three
electromagnetic fields are dominated by the particle's Coulomb field
and the different asymptotic conditions make no difference close to
the particle. This observation gives us an alternative interpretation 
for the subscript `S': it stands for `singular' as well as
`symmetric'. 

Because $A^\alpha_{\rm S}$ is just as singular as 
$A^\alpha_{\rm ret}$, removing it from the retarded solution gives
rise to a potential that is well behaved in a neighbourhood of the
world line. And because $A^\alpha_{\rm S}$ is known not to affect the
motion of the charged particle, {\it this new potential must be
entirely responsible for the radiation reaction}. We therefore
introduce the new potential  
\begin{equation} 
A^\alpha_{\rm R} = A^\alpha_{\rm ret} - A^\alpha_{\rm S} = 
\frac{1}{2} \bigl( A^\alpha_{\rm ret} - A^\alpha_{\rm adv} \bigr)
\label{1.2.3}
\end{equation}
and postulate that it, and it alone, exerts a force on the
particle. The subscript `R' stands for `regular', because 
$A^\alpha_{\rm R}$ is nonsingular on the world line. This property can
be directly inferred from the fact that the regular potential
satisfies the homogeneous version of Eq.~(\ref{1.2.1}), 
$\Box A^\alpha_{\rm R} = 0$; there is no singular source to produce a
singular behaviour on the world line. Since $A^\alpha_{\rm R}$
satisfies the homogeneous wave equation, it can be thought of as a
free radiation field, and the subscript `R' could also stand for
`radiative'. 

The self-action of the charge's own field is now clarified: a singular 
potential $A^\alpha_{\rm S}$ can be removed from the retarded
potential and shown not to affect the motion of the particle.  
(Establishing this last statement requires a careful analysis that is 
presented in the bulk of the paper; what really happens is that the
singular field contributes to the particle's inertia and renormalizes
its mass.) What remains is a well-behaved potential $A^\alpha_{\rm R}$ 
that must be solely responsible for the radiation reaction. From the
radiative potential we form an electromagnetic field tensor 
$F^{\rm R}_{\alpha\beta} = \partial_\alpha A^{\rm R}_\beta 
- \partial_\beta A^{\rm R}_\alpha$ and we take the particle's
equations of motion to be 
\begin{equation} 
m a_\mu = f_\mu^{\rm ext} + e F^{\rm R}_{\mu\nu} u^\nu, 
\label{1.2.4}
\end{equation} 
where $u^\mu = d z^\mu/d\tau$ is the charge's four-velocity 
[$z^\mu(\tau)$ gives the description of the world line and $\tau$ is
proper time], $a^\mu = du^\mu/d\tau$ its acceleration, $m$ its
(renormalized) mass, and $f^\mu_{\rm ext}$ an external force also
acting on the particle. Calculation of the radiative field yields the
more concrete expression   
\begin{equation} 
m a^\mu = f^\mu_{\rm ext} + \frac{2e^2}{3m} \bigl( \delta^\mu_{\ \nu}
+ u^\mu u_\nu \bigr) \frac{d f^\nu_{\rm ext}}{d\tau},  
\label{1.2.5}
\end{equation} 
in which the second-term is the self-force that is responsible for the
radiation reaction. We observe that the self-force is proportional to
$e^2$, it is orthogonal to the four-velocity, and it depends on the
rate of change of the external force. This is the result that was
first derived by Dirac \cite{dirac}. (Dirac's original expression
actually involved the rate of change of the acceleration vector on the
right-hand side. The resulting equation gives rise to the well-known
problem of runaway solutions. To avoid such unphysical behaviour I
have submitted Dirac's equation to a reduction-of-order procedure
whereby $d a^\nu/d\tau$ is replaced with 
$m^{-1} d f^\nu_{\rm ext}/d\tau$. This procedure is explained and
justified, for example, in Refs.~\cite{poisson, flanaganwald}.)     

\subsection{Green's functions in flat spacetime} 
\label{1.3} 

To see how Eq.~(\ref{1.2.5}) can eventually be generalized to curved 
spacetimes, I introduce a new layer of mathematical formalism and
show that the decomposition of the retarded potential into
symmetric-singular and regular-radiative pieces can be performed at
the level of the Green's functions associated with
Eq.~(\ref{1.2.1}). The retarded solution to the wave equation can be
expressed as  
\begin{equation} 
A^\alpha_{\rm ret}(x) = \int G^{\ \alpha}_{+\beta'}(x,x')   
  j^{\beta'}(x')\, dV', 
\label{1.3.1}
\end{equation}
in terms of the retarded Green's function 
$G^{\ \alpha}_{+\beta'}(x,x') = \delta^\alpha_{\beta'}
\delta(t-t'-|\bm{x}-\bm{x'}|)/|\bm{x}-\bm{x'}|$. Here $x = (t,\bm{x})$
is an arbitrary field point, $x' = (t',\bm{x'})$ is a source point,
and $dV' \equiv d^4 x'$; tensors at $x$ are identified with unprimed
indices, while primed indices refer to tensors at $x'$. Similarly, the
advanced solution can be expressed as 
\begin{equation} 
A^\alpha_{\rm adv}(x) = \int G^{\ \alpha}_{-\beta'}(x,x')   
  j^{\beta'}(x')\, dV',  
\label{1.3.2}
\end{equation}
in terms of the advanced Green's function 
$G^{\ \alpha}_{-\beta'}(x,x') = \delta^\alpha_{\beta'}
\delta(t-t'+|\bm{x}-\bm{x'}|)/|\bm{x}-\bm{x'}|$. The retarded Green's
function is zero whenever $x$ lies outside of the future light cone of  
$x'$, and $G^{\ \alpha}_{+\beta'}(x,x')$ is infinite at these
points. On the other hand, the advanced Green's function is zero 
whenever $x$ lies outside of the past light cone of $x'$, and  
$G^{\ \alpha}_{-\beta'}(x,x')$ is infinite at these points. The
retarded and advanced Green's functions satisfy the reciprocity
relation  
\begin{equation} 
G^-_{\beta'\alpha}(x',x) = G^+_{\alpha\beta'}(x,x'); 
\label{1.3.3}
\end{equation} 
this states that the retarded Green's function becomes the advanced
Green's function (and vice versa) when $x$ and $x'$ are interchanged.  

From the retarded and advanced Green's functions we can define a 
singular Green's function by 
\begin{equation} 
G^{\ \alpha}_{{\rm S}\,\beta'}(x,x') = \frac{1}{2} \Bigl[ 
G^{\ \alpha}_{+\beta'}(x,x') 
+ G^{\ \alpha}_{-\beta'}(x,x') \Bigr] 
\label{1.3.4}
\end{equation}
and a radiative Green's function by 
\begin{equation} 
G^{\ \alpha}_{{\rm R}\,\beta'}(x,x') = 
G^{\ \alpha}_{+\beta'}(x,x') - G^{\ \alpha}_{{\rm S}\,\beta'}(x,x') 
= \frac{1}{2} \Bigl[ 
G^{\ \alpha}_{+\beta'}(x,x') 
- G^{\ \alpha}_{-\beta'}(x,x') \Bigr]. 
\label{1.3.5}
\end{equation}
By virtue of Eq.~(\ref{1.3.3}) the singular Green's function is
symmetric in its indices and arguments: 
$G^{\rm S}_{\beta'\alpha}(x',x) 
= G^{\rm S}_{\alpha\beta'}(x,x')$. The radiative Green's function, 
on the other hand, is antisymmetric. The potential 
\begin{equation} 
A^\alpha_{\rm S}(x) = \int G^{\ \alpha}_{{\rm S}\,\beta'}(x,x')   
  j^{\beta'}(x')\, dV' 
\label{1.3.6}
\end{equation}
satisfies the wave equation of Eq.~(\ref{1.2.1}) and is singular on
the world line, while 
\begin{equation} 
A^\alpha_{\rm R}(x) = \int G^{\ \alpha}_{{\rm R}\,\beta'}(x,x')   
  j^{\beta'}(x')\, dV'  
\label{1.3.7}
\end{equation}
satisfies the homogeneous equation $\Box A^\alpha = 0$ and is 
well behaved on the world line.  

Equation (\ref{1.3.1}) implies that the retarded potential at $x$ is 
generated by a single event in spacetime: the intersection of the
world line and $x$'s past light cone (see Fig.~1). I shall call this
the {\it retarded point} associated with $x$ and denote it $z(u)$; $u$
is the {\it retarded time}, the value of the proper-time parameter at 
the retarded point. Similarly we find that the advanced potential of
Eq.~(\ref{1.3.2}) is generated by the intersection of the world 
line and the future light cone of the field point $x$. I shall call
this the {\it advanced point} associated with $x$ and denote it
$z(v)$; $v$ is the {\it advanced time}, the value of the proper-time
parameter at the advanced point. 

\begin{figure}[t]
\vspace*{2.0in}
\special{hscale=35 vscale=35 hoffset=100.0 voffset=180.0
         angle=-90.0 psfile=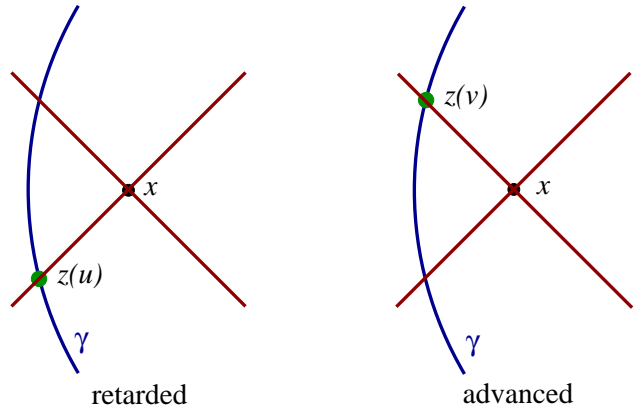}
\caption{In flat spacetime, the retarded potential at $x$ depends on
the particle's state of motion at the retarded point $z(u)$ on the
world line; the advanced potential depends on the state of motion at
the advanced point $z(v)$.} 
\end{figure} 

\subsection{Green's functions in curved spacetime} 
\label{1.4}

In a curved spacetime with metric $g_{\alpha\beta}$ the wave equation
for the vector potential becomes 
\begin{equation} 
\Box A^\alpha - R^\alpha_{\ \beta} A^\beta = -4\pi j^\alpha, 
\label{1.4.1}
\end{equation}
where $\Box = g^{\alpha\beta} \nabla_\alpha \nabla_\beta$ is the
covariant wave operator and $R_{\alpha\beta}$ is the spacetime's
Ricci tensor; the Lorenz gauge conditions becomes $\nabla_\alpha
A^\alpha = 0$, and $\nabla_\alpha$ denotes covariant
differentiation. Retarded and advanced Green's functions can be 
defined for this equation, and solutions to Eq.~(\ref{1.4.1}) take
the same form as in Eqs.~(\ref{1.3.1}) and (\ref{1.3.2}), except that
$dV'$ now stands for $\sqrt{-g(x')}\, d^4 x'$.  

The causal structure of the Green's functions is richer in curved 
spacetime: While in flat spacetime the retarded Green's function has
support only on the future light cone of $x'$, in curved spacetime its 
support extends {\it inside} the light cone as well; 
$G^{\ \alpha}_{+\beta'}(x,x')$ is therefore nonzero when $x \in
I^+(x')$, which denotes the chronological future of $x'$. This
property reflects the fact that in curved spacetime, electromagnetic
waves propagate not just at the speed of light, but at {\it all speeds
smaller than or equal to the speed of light}; the delay is caused by
an interaction between the radiation and the spacetime curvature. A
direct implication of this property is that the retarded potential at
$x$ is now generated by the point charge during its entire history
prior to the retarded time $u$ associated with $x$: the potential
depends on the particle's state of motion for all times $\tau \leq u$
(see Fig.~2).   

\begin{figure}[b]
\vspace*{2.0in}
\special{hscale=35 vscale=35 hoffset=100.0 voffset=180.0
         angle=-90.0 psfile=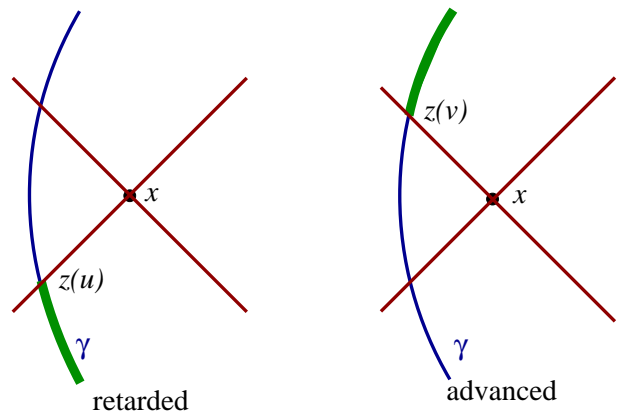}
\caption{In curved spacetime, the retarded potential at $x$ depends on 
the particle's history before the retarded time $u$; the advanced
potential depends on the particle's history after the advanced time
$v$.}  
\end{figure} 

Similar statements can be made about the advanced Green's function and
the advanced solution to the wave equation. While in flat spacetime
the advanced Green's function has support only on the past light cone
of $x'$, in curved spacetime its support extends inside the light
cone, and $G^{\ \alpha}_{-\beta'}(x,x')$ is nonzero when $x
\in I^-(x')$, which denotes the chronological past of $x'$. This
implies that the advanced potential at $x$ is generated by the point 
charge during its entire {\it future} history following the advanced 
time $v$ associated with $x$: the potential depends on the
particle's state of motion for all times $\tau \geq v$.  

The physically relevant solution to Eq.~(\ref{1.4.1}) is obviously the
retarded potential $A^\alpha_{\rm ret}(x)$, and as in flat spacetime,
this diverges on the world line. The cause of this singular
behaviour is still the pointlike nature of the source, and the
presence of spacetime curvature does not change the fact that the
potential diverges at the position of the particle. Once more this
behaviour makes it difficult to figure out how the retarded field is
supposed to act on the particle and determine its motion. As in flat
spacetime we shall attempt to decompose the retarded solution into a
singular part that exerts no force, and a smooth radiative part that
produces the entire self-force.  

To decompose the retarded Green's function into singular and radiative
parts is not a straightforward task in curved spacetime. The
flat-spacetime definition for the singular Green's function,
Eq.~(\ref{1.3.4}), cannot be adopted without modification: While the  
combination half-retarded plus half-advanced Green's functions does
have the property of being symmetric, and while the resulting vector
potential would be a solution to Eq.~(\ref{1.4.1}), this candidate for
the singular Green's function would produce a self-force with an
unacceptable dependence on the particle's future history. For suppose
that we made this choice. Then the radiative Green's function would be
given by the combination half-retarded minus half-advanced Green's
functions, just as in flat spacetime. The resulting radiative
potential would satisfy the homogeneous wave equation, and it would be
smooth on the world line, but it would also depend on the particle's
entire history, both past (through the retarded Green's function) and
future (through the advanced Green's function). More precisely stated,
we would find that the radiative potential at $x$ depends on the
particle's state of motion at all times $\tau$ outside the interval 
$u < \tau < v$; in the limit where $x$ approaches the world line, this
interval shrinks to nothing, and we would find that the radiative
potential is generated by the complete history of the particle. A
self-force constructed from this potential would be highly noncausal,
and we are compelled to reject these definitions for the singular and
radiative Green's functions. 

The proper definitions were identified by Detweiler and Whiting
\cite{detweilerwhiting}, who proposed the following generalization to
Eq.~(\ref{1.3.4}):  
\begin{equation} 
G^{\ \alpha}_{{\rm S}\,\beta'}(x,x') = \frac{1}{2} \Bigl[  
G^{\ \alpha}_{+\beta'}(x,x') 
+ G^{\ \alpha}_{-\beta'}(x,x')
- H^\alpha_{\ \beta'}(x,x') \Bigr]. 
\label{1.4.2} 
\end{equation}  
The two-point function $H^\alpha_{\ \beta'}(x,x')$ is introduced
specifically to cure the pathology described in the preceding
paragraph. It is symmetric in its indices and arguments, so that 
$G^{\rm S}_{\alpha\beta'}(x,x')$ will be also (since the retarded 
and advanced Green's functions are still linked by a reciprocity
relation); and it is a solution to the homogeneous wave equation,
$\Box H^\alpha_{\ \beta'}(x,x') 
- R^\alpha_{\ \gamma}(x) H^\gamma_{\ \beta'}(x,x') = 0$, so that the 
singular, retarded, and advanced Green's functions will all satisfy
the same wave equation. Furthermore, and this is its key property, the
two-point function is defined to agree with the advanced Green's
function when $x$ is in the chronological past of $x'$:  
$H^\alpha_{\ \beta'}(x,x') = G^{\ \alpha}_{-\beta'}(x,x')$ when 
$x \in I^-(x')$. This ensures that 
$G^{\ \alpha}_{{\rm S}\,\beta'}(x,x')$ vanishes when $x$ is in the 
chronological past of $x'$. In fact, reciprocity implies that 
$H^\alpha_{\ \beta'}(x,x')$ will also agree with the {\it retarded}
Green's function when $x$ is in the chronological future of $x'$, and
it follows that the symmetric Green's function vanishes also when $x$
is in the chronological future of $x'$.  

The potential $A^\alpha_{\rm S}(x)$ constructed from the singular 
Green's function can now be seen to depend on the particle's state of 
motion at times $\tau$ restricted to the interval $u \leq \tau \leq
v$ (see Fig.~3). Because this potential satisfies Eq.~(\ref{1.4.1}),
it is just as singular as the retarded potential in the vicinity of
the world line. And because the singular Green's function is symmetric
in its arguments, the singular potential can be shown to exert no
force on the charged particle. (This requires a lengthy analysis that
will be presented in the bulk of the paper.) 

\begin{figure}[t]
\vspace*{2.0in}
\special{hscale=35 vscale=35 hoffset=100.0 voffset=180.0
         angle=-90.0 psfile=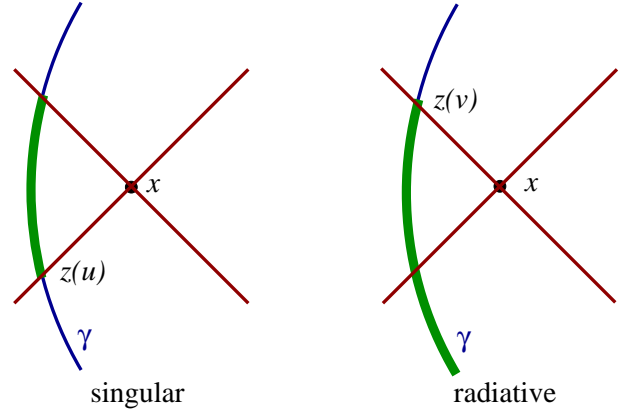}
\caption{In curved spacetime, the singular potential at $x$ depends on  
the particle's history during the interval $u \leq \tau \leq v$; for
the radiative potential the relevant interval is $-\infty < \tau 
\leq v$.}  
\end{figure} 

The Detweiler-Whiting \cite{detweilerwhiting} definition for the
radiative Green's function is then  
\begin{equation} 
G^{\ \alpha}_{{\rm R}\,\beta'}(x,x') = 
G^{\ \alpha}_{+\beta'}(x,x') - G^{\ \alpha}_{{\rm S}\,\beta'}(x,x') 
= \frac{1}{2} \Bigl[ 
G^{\ \alpha}_{+\beta'}(x,x') 
- G^{\ \alpha}_{-\beta'}(x,x') + H^\alpha_{\ \beta'}(x,x') \Bigr].  
\label{1.4.3}
\end{equation}
The potential $A^\alpha_{\rm R}(x)$ constructed from this depends on
the particle's state of motion at all times $\tau$ prior to the
advanced time $v$: $\tau \leq v$. Because this potential satisfies the 
homogeneous wave equation, it is well behaved on the world line and
its action on the point charge is well defined. And because the 
singular potential $A^\alpha_{\rm S}(x)$ can be shown to exert no
force on the particle, we conclude that $A^\alpha_{\rm R}(x)$ alone is   
responsible for the self-force.  

From the radiative potential we form an electromagnetic field tensor  
$F^{\rm R}_{\alpha\beta} = \nabla_\alpha A^{\rm R}_\beta 
- \nabla_\beta A^{\rm R}_\alpha$ and the curved-spacetime
generalization to Eq.~(\ref{1.2.4}) is 
\begin{equation} 
m a_\mu = f_\mu^{\rm ext} + e F^{\rm R}_{\mu\nu} u^\nu, 
\label{1.4.4}
\end{equation} 
where $u^\mu = d z^\mu/d\tau$ is again the charge's four-velocity, but    
$a^\mu = Du^\mu/d\tau$ is now its covariant acceleration. 

\subsection{World line and retarded coordinates} 
\label{1.5} 

To flesh out the ideas contained in the preceding subsection I 
add yet another layer of mathematical formalism and construct a
convenient coordinate system to chart a neighbourhood of 
the particle's world line. In the next subsection I will display
explicit expressions for the retarded, singular, and radiative fields
of a point electric charge.  

Let $\gamma$ be the world line of a point particle in a curved
spacetime. It is described by parametric relations 
$z^\mu(\tau)$ in which $\tau$ is proper time. Its tangent vector is
$u^\mu = dz^\mu/d\tau$ and its acceleration is 
$a^\mu = D u^\mu/d\tau$; we shall also encounter 
$\dot{a}^\mu \equiv D a^\mu/d\tau$. 

On $\gamma$ we erect an orthonormal basis that consists of the
four-velocity $u^\mu$ and three spatial vectors $\base{\mu}{a}$
labelled by a frame index $a = (1,2,3)$. These vectors satisfy the
relations $g_{\mu\nu} u^\mu u^\nu = -1$, $g_{\mu\nu} u^\mu
\base{\nu}{a} = 0$, and $g_{\mu\nu} \base{\mu}{a} \base{\nu}{b} =
\delta_{ab}$. We take the spatial vectors to be Fermi-Walker 
transported on the world line: $D \base{\mu}{a} / d\tau = a_a u^\mu$,
where  
\begin{equation}
a_a(\tau) = a_\mu \base{\mu}{a} 
\label{1.5.1}
\end{equation}
are frame components of the acceleration vector; it is easy to show
that Fermi-Walker transport preserves the orthonormality of the basis
vectors. We shall use the tetrad to decompose various tensors
evaluated on the world line. An example was already given in
Eq.~(\ref{1.5.1}) but we shall also encounter frame components of the
Riemann tensor,   
\begin{equation}
R_{a0b0}(\tau) = R_{\mu\lambda\nu\rho} \base{\mu}{a} u^\lambda 
\base{\nu}{b} u^\rho, \qquad 
R_{a0bc}(\tau) = R_{\mu\lambda\nu\rho} \base{\mu}{a} u^\lambda 
\base{\nu}{b} \base{\rho}{c}, \qquad 
R_{abcd}(\tau) = R_{\mu\lambda\nu\rho} \base{\mu}{a} \base{\lambda}{b}   
\base{\nu}{c} \base{\rho}{d},
\label{1.5.2}
\end{equation}
as well as frame components of the Ricci tensor, 
\begin{equation} 
R_{00}(\tau) = R_{\mu\nu} u^\mu u^\nu, \qquad 
R_{a0}(\tau) = R_{\mu\nu} \base{\mu}{a} u^\nu, \qquad 
R_{ab}(\tau) = R_{\mu\nu} \base{\mu}{a} \base{\nu}{b}. 
\label{1.5.3}
\end{equation} 
We shall use $\delta_{ab} = \mbox{diag}(1,1,1)$ and its inverse
$\delta^{ab} = \mbox{diag}(1,1,1)$ to lower and raise frame indices,
respectively. 

Consider a point $x$ in a neighbourhood of the world line $\gamma$. We
assume that $x$ is sufficiently close to the world line that a unique
geodesic links $x$ to any neighbouring point $z$ on $\gamma$. The
two-point function $\sigma(x,z)$, known as {\it Synge's world
function} \cite{synge}, is numerically equal to half the squared
geodesic distance between $z$ and $x$; it is positive if $x$ and $z$
are spacelike related, negative if they are timelike related, and
$\sigma(x,z)$ is zero if $x$ and $z$ are linked by a null geodesic. We
denote its gradient $\partial \sigma / \partial z^\mu$ by
$\sigma_\mu(x,z)$, and $-\sigma^\mu$ gives a meaningful notion of a
separation vector (pointing from $z$ to $x$).       

To construct a coordinate system in this neighbourhood we locate the 
unique point $x' \equiv z(u)$ on $\gamma$ which is linked to $x$ by a
future-directed null geodesic (this geodesic is directed from $x'$ to
$x$); I shall refer to $x'$ as the {\it retarded point} associated
with $x$, and $u$ will be called the {\it retarded time}. To tensors
at $x'$ we assign indices $\alpha'$, $\beta'$, \ldots; this will
distinguish them from tensors at a generic point $z(\tau)$ on the
world line, to which we have assigned indices $\mu$, $\nu$, \ldots. We
have $\sigma(x,x') = 0$ and $-\sigma^{\alpha'}(x,x')$ is a null vector
that can be interpreted as the separation between $x'$ and $x$.  

\begin{figure}[t]
\vspace*{2.2in}
\special{hscale=35 vscale=35 hoffset=115.0 voffset=-55.0
         psfile=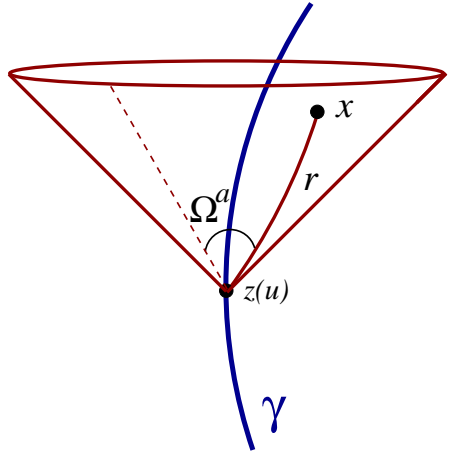}
\caption{Retarded coordinates of a point $x$ relative to a world
line $\gamma$. The retarded time $u$ selects a particular null cone,
the unit vector $\Omega^a \equiv \hat{x}^a/r$ selects a particular
generator of this null cone, and the retarded distance $r$ selects a
particular point on this generator.}  
\end{figure} 

The {\it retarded coordinates} of the point $x$ are $(u,\hat{x}^a)$,
where $\hat{x}^a = - \base{a}{\alpha'} \sigma^{\alpha'}$ are the frame
components of the separation vector. They come with a straightforward
interpretation (see Fig.~4). The invariant quantity    
\begin{equation}
r \equiv \sqrt{ \delta_{ab} \hat{x}^a \hat{x}^b } = u_{\alpha'}
\sigma^{\alpha'}  
\label{1.5.4}
\end{equation} 
is an affine parameter on the null geodesic that links $x$ to $x'$; it
can be loosely interpreted as the time delay between $x$ and $x'$ as
measured by an observer moving with the particle. This therefore gives
a meaningful notion of distance between $x$ and the retarded point,
and I shall call $r$ the {\it retarded distance} between $x$ and the
world line. The unit vector  
\begin{equation}
\Omega^a = \hat{x}^a/r  
\label{1.5.5}
\end{equation} 
is constant on the null geodesic that links $x$ to $x'$. Because
$\Omega^a$ is a different constant on each null geodesic that emanates
from $x'$, keeping $u$ fixed and varying $\Omega^a$ produces a
congruence of null geodesics that generate the future light cone of
the point $x'$ (the congruence is hypersurface orthogonal). Each light
cone can thus be labelled by its retarded time $u$, each generator on
a given light cone can be labelled by its direction vector $\Omega^a$,
and each point on a given generator can be labelled by its retarded
distance $r$. We therefore have a good coordinate system in a
neighbourhood of $\gamma$. 

To tensors at $x$ we assign indices $\alpha$, $\beta$, \ldots. These
tensors will be decomposed in a tetrad 
$(\base{\alpha}{0},\base{\alpha}{a})$ that is constructed as
follows: Given $x$ we locate its associated retarded point $x'$ on the
world line, as well as the null geodesic that links these two
points; we then take the tetrad $(u^{\alpha'},\base{\alpha'}{a})$ at   
$x'$ and parallel transport it to $x$ along the null geodesic to
obtain $(\base{\alpha}{0},\base{\alpha}{a})$.      

\subsection{Retarded, singular, and radiative electromagnetic fields
of a point electric charge} 
\label{1.6}
 
The retarded solution to Eq.~(\ref{1.4.1}) is 
\begin{equation}
A^\alpha(x) = e \int_\gamma G^{\ \alpha}_{+\mu}(x,z) u^\mu\, d\tau, 
\label{1.6.1}
\end{equation}
where the integration is over the world line of the point electric
charge. Because the retarded solution is the physically relevant
solution to the wave equation, it will not be necessary to put a
label `ret' on the vector potential. 

From the vector potential we form the electromagnetic field tensor
$F_{\alpha\beta}$, which we decompose in the tetrad 
$(\base{\alpha}{0},\base{\alpha}{a})$ introduced at the end of
Sec.~\ref{1.5}. We then express the frame components of the field  
tensor in retarded coordinates, in the form of an expansion in powers
of $r$. This gives 
\begin{eqnarray} 
F_{a0}(u,r,\Omega^a) &\equiv& F_{\alpha\beta}(x) 
\base{\alpha}{a}(x) \base{\beta}{0}(x) 
\nonumber \\  
&=& \frac{e}{r^2} \Omega_a 
- \frac{e}{r} \bigl( a_a - a_b \Omega^b \Omega_a \bigr) 
+ \frac{1}{3} e R_{b0c0} \Omega^b \Omega^c \Omega_a 
- \frac{1}{6} e \bigl( 5R_{a0b0} \Omega^b + R_{ab0c} \Omega^b \Omega^c
\bigr) 
\nonumber \\ & & \mbox{}
+ \frac{1}{12} e \bigl( 5 R_{00} + R_{bc} \Omega^b\Omega^c + R \bigr)
\Omega_a 
+ \frac{1}{3} e R_{a0} - \frac{1}{6} e R_{ab} \Omega^b 
+ F_{a0}^{\rm tail} + O(r), 
\label{1.6.2} \\ 
F_{ab}(u,r,\Omega^a) &\equiv& F_{\alpha\beta}(x) 
\base{\alpha}{a}(x) \base{\beta}{b}(x) 
\nonumber \\  
&=& \frac{e}{r} \bigl( a_a \Omega_b - \Omega_a a_b \bigr) 
+ \frac{1}{2} e \bigl( R_{a0bc} - R_{b0ac} + R_{a0c0} \Omega_b 
- \Omega_a R_{b0c0} \bigr) \Omega^c 
\nonumber \\ & & \mbox{}
- \frac{1}{2} e \bigl( R_{a0} \Omega_b - \Omega_a R_{b0} \bigr)
+ F_{ab}^{\rm tail} + O(r),  
\label{1.6.3}
\end{eqnarray} 
where 
\begin{equation} 
F_{a0}^{\rm tail} = F_{\alpha'\beta'}^{\rm tail}(x') \base{\alpha'}{a}
u^{\beta'}, \qquad 
F_{ab}^{\rm tail} = F_{\alpha'\beta'}^{\rm tail}(x') \base{\alpha'}{a}
\base{\beta'}{b} 
\label{1.6.4}
\end{equation}
are the frame components of the ``tail part'' of the field, which is
given by 
\begin{equation}
F_{\alpha'\beta'}^{\rm tail}(x') = 2 e \int_{-\infty}^{u^-}
\nabla_{[\alpha'} G_{+\beta']\mu}(x',z) u^\mu\, d\tau.  
\label{1.6.5}
\end{equation} 
In these expressions, all tensors (or their frame components) are 
evaluated at the retarded point $x' \equiv z(u)$ associated with $x$; 
for example, $a_a \equiv a_a(u) \equiv a_{\alpha'}
\base{\alpha'}{a}$. The tail part of the electromagnetic field tensor
is written as an integral over the portion of the world line that
corresponds to the interval $-\infty < \tau \leq u^- \equiv u - 0^+$;
this represents the past history of the particle. The integral is cut
short at $u^-$ to avoid the singular behaviour of the retarded Green's 
function when $z(\tau)$ coincides with $x'$; the portion of the
Green's function involved in the tail integral is smooth, and the
singularity at coincidence is completely accounted for by the other
terms in Eqs.~(\ref{1.6.2}) and (\ref{1.6.3}).  

The expansion of $F_{\alpha\beta}(x)$ near the world line does indeed
reveal many singular terms. We first recognize terms that diverge when 
$r \to 0$; for example the Coulomb field $F_{a0}$ diverges as $r^{-2}$ 
when we approach the world line. But there are also terms that, though  
they stay bounded in the limit, possess a directional ambiguity at
$r=0$; for example $F_{ab}$ contains a term proportional to
$R_{a0bc}\Omega^c$ whose limit depends on the direction of approach.  

This singularity structure is perfectly reproduced by the singular
field $F^{\rm S}_{\alpha\beta}$ obtained from the potential 
\begin{equation} 
A^\alpha_{\rm S}(x) = e \int_\gamma 
G^{\ \alpha}_{{\rm S}\,\mu}(x,z) u^\mu\, d\tau,  
\label{1.6.6}
\end{equation} 
where $G^{\ \alpha}_{{\rm S}\,\mu}(x,z)$ is the singular Green's
function of Eq.~(\ref{1.4.2}). Near the world line the singular field
is given by 
\begin{eqnarray} 
F^{\rm S}_{a0}(u,r,\Omega^a) &\equiv& F^{\rm S}_{\alpha\beta}(x)  
\base{\alpha}{a}(x) \base{\beta}{0}(x) 
\nonumber \\  
&=& \frac{e}{r^2} \Omega_a 
- \frac{e}{r} \bigl( a_a - a_b \Omega^b \Omega_a \bigr) 
- \frac{2}{3} e \dot{a}_a 
+ \frac{1}{3} e R_{b0c0} \Omega^b \Omega^c \Omega_a 
- \frac{1}{6} e \bigl( 5R_{a0b0} \Omega^b + R_{ab0c} \Omega^b \Omega^c
\bigr) 
\nonumber \\ & & \mbox{}
+ \frac{1}{12} e \bigl( 5 R_{00} + R_{bc} \Omega^b\Omega^c + R \bigr)
\Omega_a 
- \frac{1}{6} e R_{ab} \Omega^b 
+ O(r), 
\label{1.6.7} \\ 
F^{\rm S}_{ab}(u,r,\Omega^a) &\equiv& F^{\rm S}_{\alpha\beta}(x)  
\base{\alpha}{a}(x) \base{\beta}{b}(x) 
\nonumber \\  
&=& \frac{e}{r} \bigl( a_a \Omega_b - \Omega_a a_b \bigr) 
+ \frac{1}{2} e \bigl( R_{a0bc} - R_{b0ac} + R_{a0c0} \Omega_b 
- \Omega_a R_{b0c0} \bigr) \Omega^c 
\nonumber \\ & & \mbox{}
- \frac{1}{2} e \bigl( R_{a0} \Omega_b - \Omega_a R_{b0} \bigr)
+ O(r).   
\label{1.6.8}
\end{eqnarray} 
Comparison of these expressions with Eqs.~(\ref{1.6.2}) and
(\ref{1.6.3}) does indeed reveal that all singular terms are shared by
both fields. 

The difference between the retarded and singular fields defines the
radiative field $F^{\rm R}_{\alpha\beta}(x)$. Its frame components are  
\begin{eqnarray} 
F^{\rm R}_{a0} &=& \frac{2}{3} e \dot{a}_a + \frac{1}{3} e R_{a0}  
+ F_{a0}^{\rm tail} + O(r), 
\label{1.6.9} \\ 
F^{\rm R}_{ab} &=& F_{ab}^{\rm tail} + O(r), 
\label{1.6.10}
\end{eqnarray} 
and at $x'$ the radiative field becomes 
\begin{equation} 
F^{\rm R}_{\alpha'\beta'} = 
2e u_{[\alpha'} \bigl( g_{\beta']\gamma'}  
+ u_{\beta']} u_{\gamma'} \bigr) 
\biggl( \frac{2}{3} \dot{a}^{\gamma'} 
+ \frac{1}{3} R^{\gamma'}_{\ \delta'} u^{\delta'}
\biggr) + F_{\alpha'\beta'}^{\rm tail}, 
\label{1.6.11}
\end{equation}
where $\dot{a}^{\gamma'} = D a^{\gamma'}/d\tau$ is the rate of change
of the acceleration vector, and where the tail term was given by
Eq.~(\ref{1.6.5}). We see that $F^{\rm R}_{\alpha\beta}(x)$ is a
smooth tensor field, even on the world line.   

\subsection{Motion of an electric charge in curved spacetime} 
\label{1.7} 

I have argued in Sec.~\ref{1.4} that the self-force acting on a point
electric charge is produced by the radiative field, and that the
charge's equations of motion should take the form of $m a_\mu =
f_\mu^{\rm ext} + e F^{\rm R}_{\mu\nu} u^\nu$, where $f_\mu^{\rm ext}$
is an external force also acting on the particle. Substituting
Eq.~(\ref{1.6.11}) gives 
\begin{equation} 
m a^\mu = f_{\rm ext}^\mu 
+ e^2 \bigl( \delta^\mu_{\ \nu} + u^\mu u_\nu \bigr) 
\biggl( \frac{2}{3m} \frac{D f_{\rm ext}^\nu}{d \tau}   
+ \frac{1}{3} R^{\nu}_{\ \lambda} u^{\lambda} \biggr)  
+ 2 e^2 u_\nu \int_{-\infty}^{\tau^-}     
\nabla^{[\mu} G^{\ \nu]}_{+\,\lambda'}\bigl(z(\tau),z(\tau')\bigr)   
u^{\lambda'}\, d\tau',  
\label{1.7.1}  
\end{equation}   
in which all tensors are evaluated at $z(\tau)$, the current position
of the particle on the world line. The primed indices in the tail  
integral refer to a point $z(\tau')$ which represents a prior
position; the integration is cut short at $\tau' = \tau^- \equiv \tau
- 0^+$ to avoid the singular behaviour of the retarded Green's
function at coincidence. To get Eq.~(\ref{1.7.1}) I have reduced the
order of the differential equation by replacing $\dot{a}^{\nu}$
with $m^{-1} \dot{f}^{\nu}_{\rm ext}$ on the right-hand side; this
procedure was explained at the end of Sec.~\ref{1.2}. 

Equation (\ref{1.7.1}) is the result that was first derived by DeWitt
and Brehme \cite{dewittbrehme} and later corrected by Hobbs
\cite{hobbs}. (The original equation did not include the
Ricci-tensor term.) In flat spacetime the Ricci tensor is zero, the
tail integral disappears (because the Green's function vanishes
everywhere within the domain of integration), and Eq.~(\ref{1.7.1})
reduces to Dirac's result of Eq.~(\ref{1.2.5}). In curved spacetime
the self-force does not vanish even when the electric charge is moving
freely, in the absence of an external force: it is then given by the
tail integral, which represents radiation emitted earlier and coming
back to the particle after interacting with the spacetime
curvature. This delayed action implies that in general, the self-force
is nonlocal in time: it depends not only on the current state of
motion of the particle, but also on its past history. Lest this
behaviour should seem mysterious, it may help to keep in mind
that the physical process that leads to Eq.~(\ref{1.7.1}) is simply an
interaction between the charge and a free electromagnetic field
$F^{\rm R}_{\alpha\beta}$; it is this field that carries the
information about the charge's past.   

\subsection{Motion of a scalar charge in curved spacetime} 
\label{1.8}

The dynamics of a point scalar charge can be formulated in a way that 
stays fairly close to the electromagnetic theory. The particle's 
charge $q$ produces a scalar field $\Phi(x)$ which satisfies a wave 
equation 
\begin{equation} 
\bigl( \Box - \xi R \bigr) \Phi = -4\pi \mu 
\label{1.8.1}
\end{equation} 
that is very similar to Eq.~(\ref{1.4.1}). Here, $R$ is the
spacetime's Ricci scalar, and $\xi$ is an arbitrary coupling constant;
the scalar charge density $\mu(x)$ is given by a four-dimensional
Dirac functional supported on the particle's world line $\gamma$. The
retarded solution to the wave equation is 
\begin{equation}
\Phi(x) = q \int_\gamma G_+(x,z)\, d\tau, 
\label{1.8.2}
\end{equation}
where $G_+(x,z)$ is the retarded Green's function associated with
Eq.~(\ref{1.8.1}). The field exerts a force on the particle, whose
equations of motion are 
\begin{equation} 
m a^\mu = q \bigl( g^{\mu\nu} + u^\mu u^\nu \bigr) \nabla_\nu
\Phi,    
\label{1.8.3}
\end{equation} 
where $m$ is the particle's mass; this equation is very similar to the
Lorentz-force law. But the dynamics of a scalar charge comes with a
twist: If Eqs.~(\ref{1.8.1}) and (\ref{1.8.3}) are to follow from a 
variational principle, {\it the particle's mass should not be expected
to be a constant of the motion}. It is found instead to satisfy the
differential equation 
\begin{equation} 
\frac{d m}{d\tau} = -q u^\mu \nabla_\mu \Phi,  
\label{1.8.4}
\end{equation} 
and in general $m$ will vary with proper time. This phenomenon is
linked to the fact that a scalar field has zero spin: the particle can
radiate monopole waves and the radiated energy can come at the
expense of the rest mass. 

The scalar field of Eq.~(\ref{1.8.2}) diverges on the world line and
its singular part $\Phi_{\rm S}(x)$ must be removed before
Eqs.~(\ref{1.8.3}) and (\ref{1.8.4}) can be evaluated. This procedure
produces the radiative field $\Phi_{\rm R}(x)$, and it is this field
(which satisfies the homogeneous wave equation) that gives rise to a
self-force. The gradient of the radiative field takes the form of 
\begin{equation} 
\nabla_\mu \Phi_{\rm R} = - \frac{1}{12} (1-6\xi) q R u_{\mu}  
+ q \bigl( g_{\mu\nu} + u_{\mu} u_{\nu} \bigr) 
\biggl( \frac{1}{3} \dot{a}^{\nu} 
+ \frac{1}{6} R^{\nu}_{\ \lambda} u^{\lambda}
\biggr) + \Phi_{\mu}^{\rm tail}
\label{1.8.5}
\end{equation}
when it is evaluated of the world line. The last term is the tail
integral 
\begin{equation} 
\Phi_\mu^{\rm tail} = q \int_{-\infty}^{\tau^-}  
\nabla_{\mu} G_+\bigl(z(\tau),z(\tau')\bigr)\, d\tau',    
\label{1.8.6}
\end{equation}
and this brings the dependence on the particle's past.  

Substitution of Eq.~(\ref{1.8.5}) into Eqs.~(\ref{1.8.3}) and
(\ref{1.8.4}) gives the equations of motion of a point scalar
charge. (At this stage I introduce an external force 
$f_{\rm ext}^\mu$ and reduce the order of the differential equation.)
The acceleration is given by   
\begin{equation} 
m a^\mu = f_{\rm ext}^\mu 
+ q^2 \bigl( \delta^\mu_{\ \nu} + u^\mu u_\nu \bigr) 
\Biggl[ \frac{1}{3m} \frac{D f_{\rm ext}^\nu}{d \tau}   
+ \frac{1}{6} R^{\nu}_{\ \lambda} u^{\lambda} 
+ \int_{-\infty}^{\tau^-}   
\nabla^{\nu} G_+\bigl(z(\tau),z(\tau')\bigr)\, d\tau' \Biggr] 
\label{1.8.7} 
\end{equation}   
and the mass changes according to 
\begin{equation}
\frac{d m}{d\tau} = - \frac{1}{12} (1-6\xi) q^2 R 
- q^2 u^\mu \int_{-\infty}^{\tau^-} \nabla_{\mu}
G_+\bigl(z(\tau),z(\tau')\bigr)\, d\tau'. 
\label{1.8.8}
\end{equation} 
These equations were first derived by Quinn \cite{quinn}. (His
analysis was restricted to a minimally-coupled scalar field, so that
$\xi = 0$ in his expressions. The extension to an arbitrary coupling
constant was carried out by myself for this review.)  

In flat spacetime the Ricci-tensor term and the tail integral
disappear and Eq.~(\ref{1.8.7}) takes the form of Eq.~(\ref{1.2.5})
with $q^2/(3m)$ replacing the factor of $2e^2/(3m)$. In this simple
case Eq.~(\ref{1.8.8}) reduces to $dm/d\tau = 0$ and the mass is in
fact a constant. This property remains true in a conformally-flat 
spacetime when the wave equation is conformally invariant 
($\xi = 1/6$): in this case the Green's function possesses only a
light-cone part and the right-hand side of Eq.~(\ref{1.8.8})
vanishes. In generic situations the mass of a point scalar charge will
vary with proper time.   
         
\subsection{Motion of a point mass, or a black hole, in a background
spacetime} 
\label{1.9}

The case of a point mass moving in a specified background spacetime
presents itself with a serious conceptual challenge, as the
fundamental equations of the theory are nonlinear and the very notion
of a ``point mass'' is somewhat misguided. Nevertheless, to the extent
that the perturbation $h_{\alpha\beta}(x)$ created by the point mass
can be considered to be ``small'', the problem can be formulated in
close analogy with what was presented before. 

We take the metric $g_{\alpha\beta}$ of the background spacetime to be
a solution of the Einstein field equations in vacuum. (We impose this
condition globally.) We describe the gravitational perturbation
produced by a point particle of mass $m$ in terms of
trace-reversed potentials $\gamma_{\alpha\beta}$ defined by
\begin{equation}
\gamma_{\alpha\beta} = h_{\alpha\beta} - \frac{1}{2} \bigl(
g^{\gamma\delta} h_{\gamma\delta} \bigr) g_{\alpha\beta}, 
\label{1.9.1}
\end{equation} 
where $h_{\alpha\beta}$ is the difference between 
${\sf g}_{\alpha\beta}$, the actual metric of the perturbed spacetime,
and $g_{\alpha\beta}$. The potentials satisfy the wave equation 
\begin{equation} 
\Box \gamma^{\alpha\beta} + 2 R_{\gamma\ \delta}^{\ \alpha\ \beta} 
\gamma^{\gamma\delta} = -16\pi T^{\alpha\beta} 
\label{1.9.2}
\end{equation}
together with the Lorenz gauge condition 
$\gamma^{\alpha\beta}_{\ \ \ ;\beta} = 0$. Here and below, covariant
differentiation refers to a connection that is compatible with the
background metric,   
$\Box = g^{\alpha\beta} \nabla_\alpha \nabla_\beta$ is the wave 
operator for the background spacetime, and $T^{\alpha\beta}$ is the  
stress-energy tensor of the point mass; this is given by a Dirac
distribution supported on the particle's world line $\gamma$. The
retarded solution is  
\begin{equation} 
\gamma^{\alpha\beta}(x) = 4 m \int_\gamma 
G^{\ \alpha\beta}_{+\ \mu\nu}(x,z) u^\mu u^\nu\, d\tau, 
\label{1.9.3}
\end{equation}
where $G^{\ \alpha\beta}_{+\ \mu\nu}(x,z)$ is the retarded Green's
function associated with Eq.~(\ref{1.9.2}). The perturbation
$h_{\alpha\beta}(x)$ can be recovered by inverting
Eq.~(\ref{1.9.1}).   

Equations of motion for the point mass can be obtained by formally 
demanding that the motion be geodesic in the perturbed spacetime with
metric ${\sf g}_{\alpha\beta} = g_{\alpha\beta} 
+ h_{\alpha\beta}$. After a mapping to the background
spacetime, the equations of motion take the form of 
\begin{equation}
a^\mu = -\frac{1}{2} \bigl( g^{\mu\nu} + u^\mu
u^\nu \bigr) \bigl( 2 h_{\nu\lambda;\rho} - h_{\lambda\rho;\nu} \bigr)
u^\lambda u^\rho.  
\label{1.9.4} 
\end{equation}   
The acceleration is thus proportional to $m$; in the test-mass limit 
the world line of the particle is a geodesic of the background
spacetime.   

We now remove $h^{\rm S}_{\alpha\beta}(x)$ from the
retarded perturbation and postulate that it is the radiative field 
$h^{\rm S}_{\alpha\beta}(x)$ that should act on the particle. (Note
that $\gamma^{\rm S}_{\alpha\beta}$ satisfies the same wave equation
as the retarded potentials, but that $\gamma^{\rm R}_{\alpha\beta}$ is
a free gravitational field that satisfies the homogeneous wave
equation.) On the world line we have 
\begin{equation} 
h^{\rm R}_{\mu\nu;\lambda} = -4 m \Bigl( u_{(\mu}
R_{\nu)\rho\lambda\xi} + R_{\mu\rho\nu\xi} u_\lambda \Bigr) u^\rho
u^\xi + h^{\rm tail}_{\mu\nu\lambda}, 
\label{1.9.5}
\end{equation}
where the tail term is given by 
\begin{equation}
h^{\rm tail}_{\mu\nu\lambda} = 4 m \int_{-\infty}^{\tau^-}
\nabla_\lambda \biggl( G_{+\mu\nu\mu'\nu'}
- \frac{1}{2} g_{\mu\nu} G^{\ \ \rho}_{+\ \rho\mu'\nu'}
\biggr) \bigl( z(\tau), z(\tau')\bigr) u^{\mu'} u^{\nu'}\, d\tau'. 
\label{1.9.6} 
\end{equation}     
When Eq.~(\ref{1.9.5}) is substituted into Eq.~(\ref{1.9.4}) we find
that the terms that involve the Riemann tensor cancel out, and we
are left with  
\begin{equation}
a^\mu = -\frac{1}{2} \bigl( g^{\mu\nu} + u^\mu
u^\nu \bigr) \bigl( 2 h^{\rm tail}_{\nu\lambda\rho} 
- h^{\rm tail}_{\lambda\rho\nu} \bigr) u^\lambda u^\rho. 
\label{1.9.7} 
\end{equation}   
Only the tail integral appears in the final form of the equations
of motion. It involves the current position $z(\tau)$ of the particle,
at which all tensors with unprimed indices are evaluated, as well as
all prior positions $z(\tau')$, at which tensors with primed indices
are evaluated. As before the integral is cut short at $\tau' = \tau^-
\equiv \tau - 0^+$ to avoid the singular behaviour of the retarded
Green's function at coincidence.

The equations of motion of Eq.~(\ref{1.9.7}) were first derived by
Mino, Sasaki, and Tanaka \cite{MST}, and then reproduced with a
different analysis by Quinn and Wald \cite{QW1}. They are now known as
the MiSaTaQuWa equations of motion. Detweiler and Whiting
\cite{detweilerwhiting} have contributed the compelling interpretation
that the motion is actually geodesic in a spacetime with metric 
$g_{\alpha\beta} + h^{\rm R}_{\alpha\beta}$. This metric satisfies the 
Einstein field equations {\it in vacuum} and is perfectly smooth on
the world line. This spacetime can thus be viewed as the
background spacetime perturbed by a free gravitational wave produced
by the particle at an earlier stage of its history.       

While Eq.~(\ref{1.9.7}) does indeed give the correct equations of
motion for a small mass $m$ moving in a background spacetime with
metric $g_{\alpha\beta}$, the derivation outlined here leaves much to
be desired --- to what extent should we trust an analysis based on the  
existence of a point mass? Fortunately, Mino, Sasaki, and Tanaka
\cite{MST} gave two different derivations of their result, and the
second derivation was concerned not with the motion of a point mass,
but with the motion of a small nonrotating black hole. In this
alternative derivation of the MiSaTaQuWa equations, the metric of the
black hole perturbed by the tidal gravitational field of the external
universe is matched to the metric of the background spacetime
perturbed by the moving black hole. Demanding that this metric be a
solution to the vacuum field equations determines the motion of the 
black hole: it must move according to Eq.~(\ref{1.9.7}). This
alternative derivation is entirely free of conceptual and technical
pitfalls, and we conclude that the MiSaTaQuWa equations can be trusted
to describe the motion of any gravitating body in a curved background 
spacetime (so long as the body's internal structure can be ignored).   

It is important to understand that unlike Eqs.~(\ref{1.7.1}) and
(\ref{1.8.7}), which are true tensorial equations, Eq.~(\ref{1.9.7}) 
reflects a specific choice of coordinate system and its form would not
be preserved under a coordinate transformation. In other words, 
{\it the MiSaTaQuWa equations are not gauge invariant}, and they
depend upon the Lorenz gauge condition 
$\gamma^{\alpha\beta}_{\ \ \ ;\beta} = 0$. Barack and Ori \cite{BO1} 
have shown that under a coordinate transformation of the form
$x^\alpha \to x^\alpha + \xi^\alpha$, where $x^\alpha$ are the
coordinates of the background spacetime and $\xi^\alpha$ is a smooth
vector field of order $m$, the particle's acceleration changes
according to $a^\mu \to a^\mu + a[\xi]^\mu$, where 
\begin{equation}
a[\xi]^\mu = \bigl( \delta^\mu_{\ \nu} + u^\mu u_\nu \bigr) \biggl( 
\frac{D^2 \xi^\nu}{d\tau^2} + R^\nu_{\ \rho\omega\lambda} u^\rho
\xi^\omega u^\lambda \biggr) 
\label{1.9.8}
\end{equation} 
is the ``gauge acceleration''; $D^2 \xi^\nu/d\tau^2 
= (\xi^\nu_{\ ;\mu} u^\mu)_{;\rho} u^\rho$ is the second covariant
derivative of $\xi^\nu$ in the direction of the world line. This 
implies that the particle's acceleration can be altered at will by a
gauge transformation; $\xi^\alpha$ could even be chosen so as to
produce $a^\mu = 0$, making the motion geodesic after all. This
observation provides a dramatic illustration of the following point: 
{\it The MiSaTaQuWa equations of motion are not gauge invariant and
they cannot by themselves produce a meaningful answer to a well-posed
physical question; to obtain such answers it shall always be necessary
to combine the equations of motion with the metric perturbation 
$h_{\alpha\beta}$ so as to form gauge-invariant quantities that will
correspond to direct observables.} This point is very important and
cannot be over-emphasized.      

\subsection{Evaluation of the self-force} 
\label{1.10} 

To concretely evaluate the self-force, whether it be for a scalar 
charge, an electric charge, or a point mass, is a difficult
undertaking. The difficulty resides mostly with the computation of the
retarded Green's function for the spacetime under
consideration. Because Green's functions are known for a very limited
number of spacetimes, the self-force has so far been evaluated in a
rather limited number of situations.     

The first evaluation of the electromagnetic self-force was carried out
by DeWitt and DeWitt \cite{deWdeW} for a charge moving freely in a
weakly-curved spacetime characterized by a Newtonian potential
$\Phi \ll 1$. (This condition must be imposed globally, and
requires the spacetime to contain a matter distribution.) In this
context the right-hand side of Eq.~(\ref{1.7.1}) reduces to the tail
integral, since there is no external force acting on the charge. They 
found the spatial components of the self-force to be given by  
\begin{equation} 
\bm{f}_{\rm em} = e^2 \frac{M}{r^3}\, \bm{\hat{r}} + \frac{2}{3} e^2
\frac{d \bm{g}}{dt}, 
\label{1.10.1}
\end{equation} 
where $M$ is the total mass contained in the spacetime, $r =
|\bm{x}|$ is the distance from the centre of mass, $\bm{\hat{r}} =
\bm{x}/r$, and $\bm{g} = -\bm{\nabla} \Phi$ is the Newtonian
gravitational field. (In these expressions the bold-faced symbols
represent vectors in three-dimensional flat space.) 
The first term on the right-hand side of Eq.~(\ref{1.10.1}) is a
conservative correction to the Newtonian force $m \bm{g}$. The
second term is the standard radiation-reaction force; although it
comes from the tail integral, this is the same result that would be  
obtained in flat spacetime if an external force $m \bm{g}$ were acting  
on the particle. This agreement is necessary, but remarkable!  

A similar expression was obtained by Pfenning and Poisson
\cite{pfenningpoisson} for the case of a scalar charge. Here  
\begin{equation} 
\bm{f}_{\rm scalar} = 2 \xi q^2 \frac{M}{r^3}\, \bm{\hat{r}} 
+ \frac{1}{3} q^2 \frac{d \bm{g}}{dt}, 
\label{1.10.2}
\end{equation} 
where $\xi$ is the coupling of the scalar field to the spacetime
curvature; the conservative term disappears when the field is
minimally coupled. Pfenning and Poisson also computed the
gravitational self-force acting on a massive particle moving in a
weakly curved spacetime. The expression they obtained is in complete 
agreement (within its domain of validity) with the standard
post-Newtonian equations of motion.   

The force required to hold an electric charge in place in a
Schwarzschild spacetime was computed, without approximations, by Smith
and Will \cite{smithwill}. As measured by a free-falling observer
momentarily at rest at the position of the charge, the total force is  
\begin{equation} 
f = \frac{Mm}{r^2} \biggl( 1 - \frac{2M}{r} \biggr)^{-1/2} 
- e^2 \frac{M}{r^3}    
\label{1.10.3}
\end{equation}
and it is directed in the radial direction. Here, $m$ is the mass of
the charge, $M$ the mass of the black hole, and $r$ is the charge's
radial coordinate (the expression is valid in Schwarzschild
coordinates). The first term on the right-hand side of
Eq.~(\ref{1.10.3}) is the force required to keep a neutral test
particle stationary in a Schwarzschild spacetime; the second term is
the negative of the electromagnetic self-force, and its expression
agrees with the weak-field result of Eq.~(\ref{1.10.1}). Wiseman
\cite{wiseman} performed a similar calculation for a scalar charge. He
found that in this case the self-force vanishes. This result is not
incompatible with Eq.~(\ref{1.10.2}), even for nonminimal coupling,
because the computation of the weak-field self-force requires the
presence of matter, while Wiseman's scalar charge lives in a purely
vacuum spacetime.  

The intriguing phenomenon of mass loss by a scalar charge was studied
by Burko, Harte, and Poisson \cite{BHP} in the simple context of a
particle at rest in an expanding universe. For the special cases of a
de Sitter cosmology, or a spatially-flat matter-dominated universe,
the retarded Green's function could be computed, and the action of the 
scalar field on the particle determined, without approximations. In de
Sitter spacetime the particle is found to radiate all of its rest mass
into monopole scalar waves. In the matter-dominated cosmology this
happens only if the charge of the particle is sufficiently large; for
smaller charges the particle first loses a fraction of its mass, but
then regains it eventually.  

In recent years a large effort has been devoted to the elaboration of 
a practical method to compute the (scalar, electromagnetic, and
gravitational) self-force in the Schwarzschild spacetime. This work
originated with Barack and Ori \cite{BO2} and was pursued by Barack
\cite{barack1, barack2} until it was put in its definitive form by
Barack, Mino, Nakano, Ori, and Sasaki \cite{BMNOS, BO3, BO4, MNS}. The
idea is take advantage of the spherical symmetry of the Schwarzschild
solution by decomposing the retarded Green's function $G_+(x,x')$ into
spherical-harmonic modes which can be computed individually. (To be
concrete I refer here to the scalar case, but the method works just as
well for the electromagnetic and gravitational cases.) From the
mode-decomposition of the Green's function one obtains a
mode-decomposition of the field gradient $\nabla_\alpha \Phi$, and
from this subtracts a mode-decomposition of the singular field 
$\nabla_\alpha \Phi_{\rm S}$, for which a local expression is
known. This results in the radiative field 
$\nabla_\alpha \Phi_{\rm R}$ decomposed into modes, and since this 
field is well behaved on the world line, it can be directly evaluated 
at the position of the particle by summing over all modes. (This sum
converges because the radiative field is smooth; the mode sums for
the retarded or singular fields, on the other hand, do not converge.)
An extension of this method to the Kerr spacetime has recently been
presented \cite{ori:03, loustowhiting, BO5}, and Mino \cite{mino} has 
devised a surprisingly simple prescription to calculate the
time-averaged evolution of a generic orbit around a Kerr black hole.       
 
The mode-sum method was applied to a number of different
situations. Burko computed the self-force acting on an electric charge
in circular motion in flat spacetime \cite{burko1}, as well as on a
scalar and electric charge kept stationary in a Schwarzschild
spacetime \cite{burko2}, in a spacetime that contains a spherical
matter shell (Burko, Liu, and Soren \cite{BLS}), and in a Kerr
spacetime (Burko and Liu \cite{burkoliu}). Burko also computed the
scalar self-force acting on a particle in circular motion around a 
Schwarzschild black hole \cite{burko3}, a calculation that was
recently revisited by Detweiler, Messaritaki, and Whiting
\cite{DMW}. Barack and Burko considered the case of a particle falling
radially into a Schwarzschild black hole, and evaluated the scalar
self-force acting on such a particle \cite{barackburko}; Lousto
\cite{lousto} and Barack and Lousto \cite{baracklousto}, on the other
hand, calculated the gravitational self-force.   

\subsection{Organization of this review} 
\label{1.11} 

The main body of the review begins in Part \ref{part1} (Secs.~\ref{2}
to \ref{6}) with a description of the general theory of bitensors,
the name designating tensorial functions of two points in
spacetime. I introduce Synge's world function $\sigma(x,x')$ and
its derivatives in Sec.~\ref{2}, the parallel propagator 
$g^\alpha_{\ \alpha'}(x,x')$ in Sec.~\ref{4}, and the van Vleck
determinant $\Delta(x,x')$ in Sec.~\ref{6}. An important portion of
the theory (covered in Secs.~\ref{3} and \ref{5}) is concerned with
the expansion of bitensors when $x$ is very close to $x'$; expansions
such as those displayed in Eqs.~(\ref{1.6.2}) and (\ref{1.6.3}) are
based on these techniques. The presentation in Part \ref{part1}
borrows heavily from Synge's book \cite{synge} and the article by
DeWitt and Brehme \cite{dewittbrehme}. These two sources use different  
conventions for the Riemann tensor, and I have adopted Synge's
conventions (which agree with those of Misner, Thorne, and Wheeler
\cite{MTW}). The reader is therefore warned that formulae derived in
Part \ref{part1} may look superficially different from what can be
found in DeWitt and Brehme. 

In Part \ref{part2} (Secs.~\ref{7} to \ref{10}) I introduce a number
of coordinate systems that play an important role in later parts of
the review. As a warmup exercise I first construct (in Sec.~\ref{7}) 
Riemann normal coordinates in a neighbourhood of a reference point
$x'$. I then move on (in Sec.~\ref{8}) to Fermi normal coordinates
\cite{manassemisner}, which are defined in a neighbourhood of a world
line $\gamma$. The retarded coordinates, which are also based at a
world line and which were briefly introduced in Sec.~\ref{1.5}, are
covered systematically in Sec.~\ref{9}. The relationship between 
Fermi and retarded coordinates is worked out in Sec.~\ref{10}, which
also locates the advanced point $z(v)$ associated with a field point
$x$. The presentation in Part \ref{part2} borrows heavily from Synge's
book \cite{synge}. In fact, I am much indebted to Synge for initiating
the construction of retarded coordinates in a neighbourhood of a world 
line. I have implemented his program quite differently (Synge was
interested in a large neighbourhood of the world line in a weakly
curved spacetime, while I am interested in a small neighbourhood in a 
strongly curved spacetime), but the idea is originally his.    

In Part \ref{part3} (Secs.~\ref{11} to \ref{15}) I review the theory
of Green's functions for (scalar, vectorial, and tensorial) wave
equations in curved spacetime. I begin in Sec.~\ref{11} with a
pedagogical introduction to the retarded and advanced Green's
functions for a massive scalar field in flat spacetime; in this simple
context the all-important Hadamard decomposition \cite{hadamard} of
the Green's function into ``light-cone'' and ``tail'' parts can be
displayed explicitly. The invariant Dirac functional is defined in
Sec.~\ref{12} along with its restrictions on the past and future null
cones of a reference point $x'$. The retarded, advanced, singular, and
radiative Green's functions for the scalar wave equation are
introduced in Sec.~\ref{13}. In Secs.~\ref{14} and \ref{15} I cover
the vectorial and tensorial wave equations, respectively. The
presentation in Part \ref{part3} is based partly on the paper by
DeWitt and Brehme \cite{dewittbrehme}, but it is inspired mostly by
Friedlander's book \cite{friedlander}. The reader should be warned
that in one important aspect, my notation differs from the notation of
DeWitt and Brehme: While they denote the tail part of the Green's
function by $-v(x,x')$, I have taken the liberty of eliminating the
silly minus sign and I call it instead $+V(x,x')$. The reader should
also note that all my Green's functions are normalized in the same
way, with a factor of $-4\pi$ multiplying a four-dimensional Dirac
functional of the right-hand side of the wave equation. (The
gravitational Green's function is sometimes normalized with a $-16\pi$
on the right-hand side.)  

In Part \ref{part4} (Secs.~\ref{16} to \ref{19}) I compute the
retarded, singular, and radiative fields associated with a point
scalar charge (Sec.~\ref{16}), a point electric charge
(Sec.~\ref{17}), and a point mass (Sec.~\ref{18}). I provide two
different derivations for each of the equations of motion. The first
type of derivation was outlined previously: I follow Detweiler and
Whiting \cite{detweilerwhiting} and postulate that only the radiative
field exerts a force on the particle. In the second type of derivation
I take guidance from Quinn and Wald \cite{QW1} and postulate that the
net force exerted on a point particle is given by an average of the
retarded field over a surface of constant proper distance orthogonal
to the world line --- this rest-frame average is easily carried out in
Fermi normal coordinates. The averaged field is still infinite on  
the world line, but the divergence points in the direction of the
acceleration vector and it can thus be removed by mass
renormalization. Such calculations show that while the singular field
does not affect the motion of the particle, it nonetheless contributes
to its inertia. In Sec.~\ref{19} I present an alternative derivation
of the MiSaTaQuWa equations of motion based on the method of matched
asymptotic expansions \cite{manasse, kates, thornehartle, death, alvi,
detweiler}; the derivation applies to a small nonrotating black hole
instead of a point mass. The ideas behind this derivation were
contained in the original paper by Mino, Sasaki, and Tanaka
\cite{MST}, but the implementation given here, which involves the
retarded coordinates of Sec.~\ref{9} and displays explicitly the
transformation between external and internal coordinates, is original
work.      

Concluding remarks are presented in Sec.~\ref{20}. Throughout this
review I use geometrized units and adopt the notations and conventions
of Misner, Thorne, and Wheeler \cite{MTW}.   

\subsection*{Acknowledgments} 

My understanding of the work presented in this review was shaped by a
series of annual meetings named after the movie director Frank
Capra. The first of these meetings took place in 1998 and was held at
Capra's ranch in Southern California; the ranch now belongs to
Caltech, Capra's alma mater. Subsequent meetings were held in Dublin,
Pasadena, Potsdam, State College PA, and Kyoto. At these meetings and 
elsewhere I have enjoyed many instructive conversations with Warren
Anderson, Patrick Brady, Claude Barrab\`es, Leor Barack, Lior Burko,
Manuella Campanelli, Steve Detweiler, Eanna Flanagan, Scott Hughes,
Werner Israel, Carlos Lousto, Yasushi Mino, Hiroyuki Nakano, Amos Ori,
Misao Sasaki, Takahiro Tanaka, Bill Unruh, Bob Wald, Alan Wiseman, and
Bernard Whiting. This work was supported by the Natural Sciences and
Engineering Research Council of Canada.

\newpage
\hrule
\hrule
\part{General theory of bitensors}
\label{part1}
\hrule
\hrule 
\vspace*{.25in} 
%
\section{Synge's world function}
\label{2}

\subsection{Definition} 
\label{2.1}

In this and the following sections we will construct a number of 
{\it bitensors}, tensorial functions of two points in spacetime. The
first is $x'$, to which we refer as the ``base point'', and to which
we assign indices $\alpha'$, $\beta'$, etc. The second is $x$, to
which we refer as the ``field point'', and to which we assign indices
$\alpha$, $\beta$, etc. We assume that $x$ belongs to ${\cal N}(x')$,
the {\it normal convex neighbourhood} of $x'$; this is the set of
points that are linked to $x'$ by a {\it unique} geodesic. The
geodesic $\beta$ that links $x$ to $x'$ is described by relations
$z^\mu(\lambda)$ in which $\lambda$ is an affine parameter that ranges 
from $\lambda_0$ to $\lambda_1$; we have $z(\lambda_0) \equiv x'$ and
$z(\lambda_1) \equiv x$. To an arbitrary point $z$ on the geodesic we
assign indices $\mu$, $\nu$, etc. The vector $t^\mu = dz^\mu/d\lambda$
is tangent to the geodesic, and it obeys the geodesic equation 
$D t^\mu/d\lambda = 0$. The situation is illustrated in Fig.~5. 

\begin{figure}[b]
\vspace*{1.8in}
\special{hscale=40 vscale=40 hoffset=115.0 voffset=-85.0
         psfile=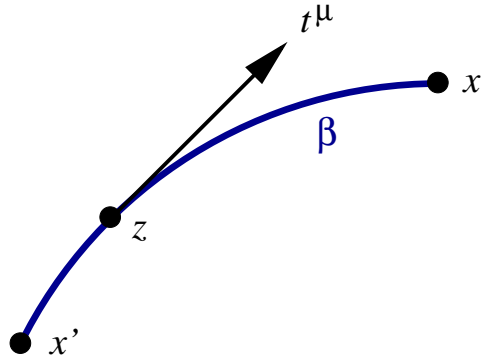}
\caption{The base point $x'$, the field point $x$, and the geodesic
$\beta$ that links them. The geodesic is described by parametric
relations $z^\mu(\lambda)$ and $t^\mu = dz^\mu/d\lambda$ is its
tangent vector.}
\end{figure} 

Synge's world function is a scalar function of the base point $x'$ and
the field point $x$. It is defined by 
\begin{equation}
\sigma(x,x') = \frac{1}{2} (\lambda_1 - \lambda_0)
\int_{\lambda_0}^{\lambda_1} g_{\mu\nu}(z) 
t^\mu t^\nu\, d\lambda, 
\label{2.1.1}
\end{equation}       
and the integral is evaluated on the geodesic $\beta$ that links $x$
to $x'$. You may notice that $\sigma$ is invariant under a constant
rescaling of the affine parameter, $\lambda \to \bar{\lambda} = a
\lambda + b$, where $a$ and $b$ are constants. 

By virtue of the geodesic equation, the quantity $\varepsilon \equiv
g_{\mu\nu} t^\mu t^\nu$ is constant on the geodesic. The world
function is therefore numerically equal to $\frac{1}{2} \varepsilon
(\lambda_1-\lambda_0)^2$. If the geodesic is timelike, then $\lambda$
can be set equal to the proper time $\tau$, which implies that
$\varepsilon = -1$ and $\sigma = -\frac{1}{2} (\Delta \tau)^2$. If the
geodesic is spacelike, then $\lambda$ can be set equal to the proper
distance $s$, which implies that $\varepsilon = 1$ and $\sigma =
\frac{1}{2} (\Delta s)^2$. If the geodesic is null, then $\sigma =
0$. Quite generally, therefore, the world function is half the squared
geodesic distance between the points $x'$ and $x$.  

In flat spacetime, the geodesic linking $x$ to $x'$ is a straight
line, and $\sigma = \frac{1}{2} \eta_{\alpha\beta} (x-x')^\alpha
(x-x')^\beta$ in Lorentzian coordinates.  

\subsection{Differentiation of the world function} 
\label{2.2}

The world function $\sigma(x,x')$ can be differentiated with respect
to either argument. We let $\sigma_\alpha = \partial \sigma / \partial
x^\alpha$ be its partial derivative with respect to $x$, and
$\sigma_{\alpha'} = \partial \sigma / \partial x^{\alpha'}$ its
partial derivative with respect to $x'$. It is clear that
$\sigma_\alpha$ behaves as a dual vector with respect to tensorial 
operations carried out at $x$, but as a scalar with respect to
operations carried out $x'$. Similarly, $\sigma_{\alpha'}$ is a
scalar at $x$ but a dual vector at $x'$. 

We let $\sigma_{\alpha\beta} \equiv \nabla_\beta \sigma_{\alpha}$ be
the covariant derivative of $\sigma_{\alpha}$ with respect to $x$;
this is a rank-2 tensor at $x$ and a scalar at $x'$. Because $\sigma$
is a scalar at $x$, we have that this tensor is symmetric:
$\sigma_{\beta\alpha} = \sigma_{\alpha\beta}$. Similarly, we let
$\sigma_{\alpha\beta'} \equiv \partial_{\beta'} \sigma_{\alpha} = 
\partial^2 \sigma / \partial x^{\beta'} \partial x^\alpha$ be the
partial derivative of $\sigma_{\alpha}$ with respect to $x'$; this
is a dual vector both at $x$ and $x'$. We can also define 
$\sigma_{\alpha'\beta} \equiv \partial_\beta \sigma_{\alpha'} =
\partial^2 \sigma / \partial x^{\beta} \partial x^{\alpha'}$ to be the
partial derivative of $\sigma_{\alpha'}$ with respect to $x$. Because
partial derivatives commute, these bitensors are equal:
$\sigma_{\beta'\alpha} = \sigma_{\alpha\beta'}$. Finally, we let
$\sigma_{\alpha'\beta'} \equiv \nabla_{\beta'} \sigma_{\alpha'}$ be
the covariant derivative of $\sigma_{\alpha'}$ with respect to $x'$;
this is a symmetric rank-2 tensor at $x'$ and a scalar at $x$. 

The notation is easily extended to any number of derivatives. For
example, we let $\sigma_{\alpha\beta\gamma\delta'} \equiv 
\nabla_{\delta'} \nabla_\gamma \nabla_\beta \nabla_\alpha \sigma$, 
which is a rank-3 tensor at $x$ and a dual vector at $x'$. This
bitensor is symmetric in the pair of indices $\alpha$ and $\beta$, but
not in the pairs $\alpha$ and $\gamma$, nor $\beta$ and
$\gamma$. Because $\nabla_{\delta'}$ is here an ordinary partial
derivative with respect to $x'$, the bitensor is symmetric in any pair
of indices involving $\delta'$. The ordering of the primed index
relative to the unprimed indices is therefore irrelevant: the same
bitensor can be written as $\sigma_{\delta'\alpha\beta\gamma}$ or
$\sigma_{\alpha\delta'\beta\gamma}$ or
$\sigma_{\alpha\beta\delta'\gamma}$, making sure that the ordering of
the unprimed indices is not altered.    
       
More generally, we can show that derivatives of any bitensor
$\Omega_{\cdots}(x,x')$ satisfy the property  
\begin{equation}
\Omega_{\cdots ; \beta \alpha' \cdots} 
= \Omega_{\cdots ; \alpha' \beta \cdots}, 
\label{2.2.1} 
\end{equation}
in which ``$\cdots$'' stands for any combination of primed and
unprimed indices. We start by establishing the symmetry of
$\Omega_{\cdots ;\alpha \beta'}$ with respect to the pair $\alpha$ and  
$\beta'$. This is most easily done by adopting Fermi normal
coordinates (see Sec.~\ref{8}) adapted to the geodesic $\beta$ and 
setting the connection to zero both at $x$ and $x'$. In these
coordinates, the bitensor $\Omega_{\cdots ;\alpha}$ is the partial
derivative of $\Omega_{\cdots}$ with respect to $x^\alpha$, and 
$\Omega_{\cdots ;\alpha\beta'}$ is obtained by taking an additional
partial derivative with respect to $x^{\beta'}$. These two operations
commute, and $\Omega_{\cdots ;\beta' \alpha} = 
\Omega_{\cdots ;\alpha \beta'}$ follows as a bitensorial
identity. Equation (\ref{2.2.1}) then follows by further
differentiation with respect to either $x$ or $x'$.     

The message of Eq.~(\ref{2.2.1}), when applied to derivatives of the 
world function, is that while the ordering of the primed and
unprimed indices relative to themselves is important, their 
ordering with respect to each other is arbitrary. For example, 
$\sigma_{\alpha'\beta'\gamma\delta'\epsilon} =
\sigma_{\alpha'\beta'\delta'\gamma\epsilon} =
\sigma_{\gamma\epsilon\alpha'\beta'\delta'}$. 
 
\subsection{Evaluation of first derivatives} 
\label{2.3}

We can compute $\sigma_\alpha$ by examining how $\sigma$ varies when  
the field point $x$ moves. We let the new field point be $x +
\delta x$, and $\delta \sigma \equiv \sigma(x+\delta x,x') -
\sigma(x,x')$ is the corresponding variation of the world function. We
let $\beta + \delta \beta$ be the unique geodesic that links $x + 
\delta x$ to $x'$; it is described by relations $z^\mu(\lambda) 
+ \delta z^\mu(\lambda)$, in which the affine parameter is scaled in
such a way that it runs from $\lambda_0$ to $\lambda_1$ also on the
new geodesic. We note that $\delta z (\lambda_0) = \delta x' \equiv 0$
and $\delta z (\lambda_1) = \delta x$.  

Working to first order in the variations, Eq.~(\ref{2.1.1}) implies   
\[ 
\delta \sigma = \Delta \lambda \int_{\lambda_0}^{\lambda_1} 
\biggl( g_{\mu\nu} \dot{z}^\mu\, \delta\dot{z}^\nu + \frac{1}{2}\, 
g_{\mu\nu,\lambda} \dot{z}^\mu \dot{z}^\nu\, \delta z^\lambda
\biggr)\, d\lambda, 
\]
where $\Delta \lambda = \lambda_1 - \lambda_0$, an overdot
indicates differentiation with respect to $\lambda$, and the metric
and its derivatives are evaluated on $\beta$. Integrating the 
first term by parts gives 
\[ 
\delta \sigma = \Delta \lambda \Bigl[ g_{\mu\nu}
\dot{z}^\mu\, \delta z^\nu \Bigr]^{\lambda_1}_{\lambda_0} \\
- \Delta \lambda \int_{\lambda_0}^{\lambda_1} 
\Bigl( g_{\mu\nu} \ddot{z}^\nu + \Gamma_{\mu\nu\lambda}  
\dot{z}^\nu \dot{z}^\lambda \Bigr)\, \delta z^\mu\, d\lambda. 
\] 
The integral vanishes because $z^\mu(\lambda)$ satisfies the geodesic 
equation. The boundary term at $\lambda_0$ is zero because the 
variation $\delta z^\mu$ vanishes there. We are left with 
$\delta \sigma = \Delta \lambda g_{\alpha\beta} t^\alpha 
\delta x^\beta$, or
\begin{equation} 
\sigma_\alpha(x,x') = (\lambda_1 - \lambda_0)\, 
g_{\alpha\beta} t^\beta,  
\label{2.3.1}
\end{equation} 
in which the metric and the tangent vector are both evaluated at
$x$. Apart from a factor $\Delta \lambda$, we see that
$\sigma^\alpha(x,x')$ is equal to the geodesic's tangent vector at
$x$. If in Eq.~(\ref{2.3.1}) we replace $x$ by a generic point
$z(\lambda)$ on $\beta$, and if we correspondingly replace
$\lambda_1$ by $\lambda$, we obtain $\sigma^\mu(z,x') = (\lambda -
\lambda_0) t^\mu$; we therefore see that $\sigma^\mu(z,x')$ is a
rescaled tangent vector on the geodesic.    

A virtually identical calculation reveals how $\sigma$ varies under a 
change of base point $x'$. Here the variation of the geodesic is such
that $\delta z (\lambda_0) = \delta x'$ and $\delta z (\lambda_1) =
\delta x = 0$, and we obtain $\delta \sigma = - \Delta \lambda 
g_{\alpha'\beta'} t^{\alpha'} \delta x^{\beta'}$. This shows that 
\begin{equation}  
\sigma_{\alpha'}(x,x') = -(\lambda_1 - \lambda_0)\,
g_{\alpha'\beta'} t^{\beta'},   
\label{2.3.2}
\end{equation} 
in which the metric and the tangent vector are both evaluated at
$x'$. Apart from a factor $\Delta \lambda$, we see that
$\sigma^{\alpha'}(x,x')$ is minus the geodesic's tangent 
vector at $x'$. 

It is interesting to compute the norm of $\sigma_{\alpha}$. According
to Eq.~(\ref{2.3.1}) we have $g_{\alpha\beta} \sigma^{\alpha}
\sigma^{\beta} = (\Delta \lambda)^2 g_{\alpha\beta} t^\alpha t^\beta =  
(\Delta \lambda)^2 \varepsilon$. According to Eq.~(\ref{2.1.1}),
this is equal to $2 \sigma$. We have obtained 
\begin{equation}
g^{\alpha\beta} \sigma_\alpha \sigma_\beta = 2 \sigma,  
\label{2.3.3}
\end{equation} 
and similarly,  
\begin{equation}
g^{\alpha'\beta'} \sigma_{\alpha'} \sigma_{\beta'} = 2 \sigma.  
\label{2.3.4} 
\end{equation} 
These important relations will be the starting point of many
computations to be described below.  

We note that in flat spacetime, $\sigma_\alpha = \eta_{\alpha\beta}
(x-x')^\beta$ and $\sigma_{\alpha'} = - \eta_{\alpha\beta} 
(x-x')^\beta$ in Lorentzian coordinates. From this it follows that
$\sigma_{\alpha\beta} = \sigma_{\alpha'\beta'} =
-\sigma_{\alpha\beta'} = -\sigma_{\alpha'\beta} = \eta_{\alpha\beta}$,
and finally, $g^{\alpha\beta} \sigma_{\alpha\beta} = 4 =
g^{\alpha'\beta'} \sigma_{\alpha'\beta'}$.   

\subsection{Congruence of geodesics emanating from $x'$} 
\label{2.4}

If the base point $x'$ is kept fixed, $\sigma$ can be considered to be
an ordinary scalar function of $x$. According to Eq.~(\ref{2.3.3}),
this function is a solution to the nonlinear
differential equation $\frac{1}{2} g^{\alpha\beta} \sigma_\alpha 
\sigma_\beta = \sigma$. Suppose that we are presented with such a
scalar field. What can we say about it? 

An additional differentiation of the defining equation reveals that
the vector $\sigma^\alpha \equiv \sigma^{;\alpha}$ satisfies  
\begin{equation}
\sigma^\alpha_{\ ;\beta} \sigma^\beta = \sigma^\alpha, 
\label{2.4.1}
\end{equation}
which is the geodesic equation in a non-affine parameterization. The  
vector field is therefore tangent to a congruence of geodesics. The
geodesics are timelike where $\sigma < 0$, they are spacelike where
$\sigma > 0$, and they are null where $\sigma = 0$. Here, for
concreteness, we shall consider only the timelike subset of the
congruence.    

The vector 
\begin{equation}
u^\alpha = \frac{\sigma^\alpha}{|2\sigma|^{1/2}} 
\label{2.4.2}
\end{equation}
is a normalized tangent vector that satisfies the geodesic equation 
in affine-parameter form: $u^\alpha_{\ ;\beta} u^\beta = 0$. The 
parameter $\lambda$ is then proper time $\tau$. If $\lambda^*$
denotes the original parameterization of the geodesics, we have that
$d\lambda^*/d\tau = |2\sigma|^{-1/2}$, and we see that the original 
parameterization is singular at $\sigma = 0$. 

In the affine parameterization, the expansion of the congruence is
calculated to be 
\begin{equation}
\theta = \frac{\theta^*}{|2\sigma|^{1/2}}, \qquad
\theta^* = \sigma^{\alpha}_{\ ;\alpha} - 1, 
\label{2.4.3}
\end{equation} 
where $\theta^* = (\delta V)^{-1} (d/d\lambda^*) (\delta V)$ is the
expansion in the original parameterization ($\delta V$ is the
congruence's cross-sectional volume). While $\theta^*$ is well 
behaved in the limit $\sigma \to 0$ (we shall see below that $\theta^* 
\to 3$), we have that $\theta \to \infty$. This means that the point
$x'$ at which $\sigma = 0$ is a caustic of the congruence: all
geodesics emanate from this point.  

These considerations, which all follow from a postulated relation
$\frac{1}{2} g^{\alpha\beta} \sigma_\alpha \sigma_\beta = \sigma$,
are clearly compatible with our preceding explicit construction of the  
world function.            

\section{Coincidence limits} 
\label{3}

It is useful to determine the limiting behaviour of the bitensors 
$\sigma_{\cdots}$ as $x$ approaches $x'$. We introduce the notation 
\[
\bigl[ \Omega_{\cdots} \bigr] 
= \lim_{x \to x'} \Omega_{\cdots}(x,x') 
= \mbox{a tensor at $x'$}   
\]
to designate the limit of any bitensor $\Omega_{\cdots}(x,x')$ as $x$
approaches $x'$; this is called the {\it coincidence limit} of the
bitensor. We assume that the coincidence limit is a unique tensorial
function of the base point $x'$, independent of the direction in which
the limit is taken. In other words, if the limit is computed by
letting $\lambda \to \lambda_0$ after evaluating
$\Omega_{\cdots}(z,x')$ as a function of $\lambda$ on a specified
geodesic $\beta$, it is assumed that the answer does not depend on the
choice of geodesic.    

\subsection{Computation of coincidence limits} 
\label{3.1}

From Eqs.~(\ref{2.1.1}), (\ref{2.3.1}), and (\ref{2.3.2}) we already
have    
\begin{equation} 
\bigl[ \sigma \bigr] = 0, \qquad
\bigl[ \sigma_\alpha \bigr] = \bigl[ \sigma_{\alpha'} \bigr] = 0. 
\label{3.1.1}
\end{equation} 
Additional results are obtained by repeated differentiation of the 
relations (\ref{2.3.3}) and (\ref{2.3.4}). For example,
Eq.~(\ref{2.3.3}) implies $\sigma_\gamma = g^{\alpha\beta}
\sigma_\alpha \sigma_{\beta\gamma} = \sigma^\beta
\sigma_{\beta\gamma}$, or $(g_{\beta\gamma} - \sigma_{\beta\gamma})
t^\beta = 0$ after using Eq.~(\ref{2.3.1}). From the assumption stated
in the preceding paragraph, $\sigma_{\beta\gamma}$ becomes independent
of $t^\beta$ in the limit $x \to x'$, and we arrive at
$[\sigma_{\alpha\beta}] = g_{\alpha'\beta'}$. By very similar
calculations we obtain all other coincidence limits for the second
derivatives of the world function. The results are   
\begin{equation} 
\bigl[ \sigma_{\alpha\beta} \bigr] =  
\bigl[ \sigma_{\alpha'\beta'} \bigr] = 
g_{\alpha'\beta'}, \qquad
\bigl[ \sigma_{\alpha\beta'} \bigr] = 
\bigl[ \sigma_{\alpha'\beta} \bigr] = - g_{\alpha'\beta'}. 
\label{3.1.2}
\end{equation}
From these relations we infer that
$[\sigma^\alpha_{\ \alpha}] = 4$, so that $[\theta^*] = 3$, where
$\theta^*$ was defined in Eq.~(\ref{2.4.3}). 

To generate coincidence limits of bitensors involving primed 
indices, it is efficient to invoke Synge's rule, 
\begin{equation} 
\bigl[ \sigma_{\cdots \alpha'} \bigr] = \bigl[ \sigma_{\cdots}
\bigr]_{;\alpha'} - \bigl[ \sigma_{\cdots \alpha} \bigr], 
\label{3.1.3}
\end{equation} 
in which ``$\cdots$'' designates any combination of primed and
unprimed indices; this rule will be established below. For example,
according to Synge's rule we have $[\sigma_{\alpha\beta'}] =
[\sigma_\alpha]_{;\beta'} - [\sigma_{\alpha\beta}]$, and since the 
coincidence limit of $\sigma_\alpha$ is zero, this gives us  
$[\sigma_{\alpha\beta'}] = - [\sigma_{\alpha\beta}] =
-g_{\alpha'\beta'}$, as was stated in Eq.~(\ref{3.1.2}). Similarly,  
$[\sigma_{\alpha'\beta'}] = [\sigma_{\alpha'}]_{;\beta'} -
[\sigma_{\alpha'\beta}] = - [\sigma_{\beta\alpha'}] =
g_{\alpha'\beta'}$. The results of Eq.~(\ref{3.1.2}) can thus all be 
generated from the known result for $[\sigma_{\alpha\beta}]$. 

The coincidence limits of Eq.~(\ref{3.1.2}) were derived from the
relation $\sigma_\alpha = \sigma^\delta_{\ \alpha} \sigma_\delta$. We
now differentiate this twice more and obtain
$\sigma_{\alpha\beta\gamma} = 
\sigma^\delta_{\ \alpha\beta\gamma} \sigma_\delta 
+ \sigma^\delta_{\ \alpha\beta} \sigma_{\delta \gamma}
+ \sigma^\delta_{\ \alpha\gamma} \sigma_{\delta \beta} 
+ \sigma^\delta_{\ \alpha} \sigma_{\delta\beta\gamma}$. 
At coincidence we have  
\[
\bigl[ \sigma_{\alpha\beta\gamma} \bigr] = \bigl[ \sigma^\delta_{\
\alpha\beta} \bigr] g_{\delta'\gamma'} + \bigl[ \sigma^\delta_{\
\alpha\gamma} \bigr] g_{\delta'\beta'} + \delta^{\delta'}_{\ \alpha'}
\bigl[ \sigma_{\delta\beta\gamma} \bigr], 
\]
or $[\sigma_{\gamma\alpha\beta}] + [\sigma_{\beta\alpha\gamma}] = 0$
if we recognize that the operations of raising or lowering indices and 
taking the limit $x \to x'$ commute. Noting the symmetries of
$\sigma_{\alpha\beta}$, this gives us
$[\sigma_{\alpha\gamma\beta}] + [\sigma_{\alpha\beta\gamma}] = 0$, 
or $2[\sigma_{\alpha\beta\gamma}] - [R^\delta_{\ \alpha\beta\gamma}
\sigma_\delta] = 0$, or $2[\sigma_{\alpha\beta\gamma}] =
R^{\delta'}_{\ \alpha'\beta'\gamma'}[\sigma_{\delta'}]$. Since the
last factor is zero, we arrive at  
\begin{equation}
\bigl[\sigma_{\alpha\beta\gamma} \bigr] = 
\bigl[\sigma_{\alpha\beta\gamma'} \bigr] = 
\bigl[\sigma_{\alpha\beta'\gamma'} \bigr] = 
\bigl[\sigma_{\alpha'\beta'\gamma'} \bigr] = 0. 
\label{3.1.4}
\end{equation}
The last three results were derived from $[\sigma_{\alpha\beta\gamma}]  
= 0$ by employing Synge's rule. 

We now differentiate the relation $\sigma_\alpha = \sigma^\delta_{\
\alpha} \sigma_\delta$ three times and obtain 
\[ 
\sigma_{\alpha\beta\gamma\delta} = 
\sigma^\epsilon_{\ \alpha\beta\gamma\delta} \sigma_\epsilon 
+ \sigma^\epsilon_{\ \alpha\beta\gamma} \sigma_{\epsilon\delta} 
+ \sigma^\epsilon_{\ \alpha\beta\delta} \sigma_{\epsilon\gamma}  
+ \sigma^\epsilon_{\ \alpha\gamma\delta} \sigma_{\epsilon\beta}  
+ \sigma^\epsilon_{\ \alpha\beta} \sigma_{\epsilon\gamma\delta} 
+ \sigma^\epsilon_{\ \alpha\gamma} \sigma_{\epsilon\beta\delta} 
+ \sigma^\epsilon_{\ \alpha\delta} \sigma_{\epsilon\beta\gamma}
+ \sigma^\epsilon_{\ \alpha} \sigma_{\epsilon\beta\gamma\delta}. 
\]
At coincidence this reduces to $[\sigma_{\alpha\beta\gamma\delta}] 
+ [\sigma_{\alpha\delta\beta\gamma}] 
+ [\sigma_{\alpha\gamma\beta\delta}] = 0$. To simplify the third term
we differentiate Ricci's identity $\sigma_{\alpha\gamma\beta} =
\sigma_{\alpha\beta\gamma} - R^\epsilon_{\ \alpha\beta\gamma}
\sigma_\epsilon$ with respect to $x^\delta$ and then take the
coincidence limit. This gives us $[\sigma_{\alpha\gamma\beta\delta}] 
= [\sigma_{\alpha\beta\gamma\delta}] +
R_{\alpha'\delta'\beta'\gamma'}$. The same manipulations on the second
term give $[\sigma_{\alpha\delta\beta\gamma}] =
[\sigma_{\alpha\beta\delta\gamma}] +
R_{\alpha'\gamma'\beta'\delta'}$. Using the identity
$\sigma_{\alpha\beta\delta\gamma} = \sigma_{\alpha\beta\gamma\delta} 
- R^\epsilon_{\ \alpha\gamma\delta} \sigma_{\epsilon\beta}   
- R^\epsilon_{\ \beta\gamma\delta} \sigma_{\alpha\epsilon}$ and the
symmetries of the Riemann tensor, it is then easy to show that
$[\sigma_{\alpha\beta\delta\gamma}] =
[\sigma_{\alpha\beta\gamma\delta}]$. Gathering the results, we obtain 
$3 [\sigma_{\alpha\beta\gamma\delta}] +
R_{\alpha'\gamma'\beta'\delta'} + R_{\alpha'\delta'\beta'\gamma'} = 
0$, and Synge's rule allows us to generalize this to any combination
of primed and unprimed indices. Our final results are 
\begin{eqnarray} 
\bigl[ \sigma_{\alpha\beta\gamma\delta} \bigr] &=& 
- \frac{1}{3} \bigl( R_{\alpha'\gamma'\beta'\delta'} 
+ R_{\alpha'\delta'\beta'\gamma'} \bigr), 
\qquad 
\bigl[ \sigma_{\alpha\beta\gamma\delta'} \bigr] = 
\frac{1}{3} \bigl( R_{\alpha'\gamma'\beta'\delta'} 
+ R_{\alpha'\delta'\beta'\gamma'} \bigr),
\nonumber \\
\bigl[ \sigma_{\alpha\beta\gamma'\delta'} \bigr] &=&  
- \frac{1}{3} \bigl( R_{\alpha'\gamma'\beta'\delta'} 
+ R_{\alpha'\delta'\beta'\gamma'} \bigr), 
\qquad 
\bigl[ \sigma_{\alpha\beta'\gamma'\delta'} \bigr] =  
- \frac{1}{3} \bigl( R_{\alpha'\beta'\gamma'\delta'} 
+ R_{\alpha'\gamma'\beta'\delta'} \bigr),
\nonumber \\ 
\bigl[ \sigma_{\alpha'\beta'\gamma'\delta'} \bigr] &=&  
- \frac{1}{3} \bigl( R_{\alpha'\gamma'\beta'\delta'} 
+ R_{\alpha'\delta'\beta'\gamma'} \bigr). 
\label{3.1.5}
\end{eqnarray} 

\subsection{Derivation of Synge's rule} 
\label{3.2}

We begin with {\it any} bitensor $\Omega_{AB'}(x,x')$ in which $A =
\alpha \cdots \beta$ is a multi-index that represents any number of
unprimed indices, and $B' = \gamma' \cdots \delta'$ a multi-index
that represents any number of primed indices. (It does not matter
whether the primed and unprimed indices are segregated or mixed.) On
the geodesic $\beta$ that links $x$ to $x'$ we introduce an ordinary
tensor $P^M(z)$ where $M$ is a multi-index that contains the same
number of indices as $A$. This tensor is arbitrary, but we assume that
it is parallel transported on $\beta$; this means that it satisfies 
$P^{A}_{\ \ ;\alpha} t^\alpha = 0$ at $x$. Similarly, we
introduce an ordinary tensor $Q^{N}(z)$ in which $N$ contains the same
number of indices as $B'$. This tensor is arbitrary, but we assume
that it is parallel transported on $\beta$; at $x'$ it satisfies 
$Q^{B'}_{\ \ ;\alpha'} t^{\alpha'} = 0$. With $\Omega$, 
$P$, and $Q$ we form a biscalar $H(x,x')$ defined by   
\[
H(x,x') = \Omega_{AB'}(x,x') P^A(x) Q^{B'}(x').   
\]
Having specified the geodesic that links $x$ to $x'$, we can consider  
$H$ to be a function of $\lambda_0$ and $\lambda_1$. If $\lambda_1$ is 
not much larger than $\lambda_0$ (so that $x$ is not far from $x'$),
we can express $H(\lambda_1,\lambda_0)$ as 
\[
H(\lambda_1,\lambda_0) = H(\lambda_0,\lambda_0) + (\lambda_1 -
\lambda_0)\, \frac{\partial H}{\partial \lambda_1} \biggr|_{\lambda_1
= \lambda_0} + \cdots.
\]
Alternatively, 
\[
H(\lambda_1,\lambda_0) = H(\lambda_1,\lambda_1) - (\lambda_1 -
\lambda_0)\, \frac{\partial H}{\partial \lambda_0} \biggr|_{\lambda_0
= \lambda_1} + \cdots, 
\]
and these two expressions give 
\[
\frac{d}{d\lambda_0} H(\lambda_0,\lambda_0) = 
\frac{\partial H}{\partial \lambda_0} \biggr|_{\lambda_0 =
\lambda_1} + \frac{\partial H}{\partial \lambda_1} \biggr|_{\lambda_1
= \lambda_0},
\]
because the left-hand side is the limit of $[H(\lambda_1,\lambda_1) - 
H(\lambda_0,\lambda_0)]/(\lambda_1-\lambda_0)$ when $\lambda_1 \to
\lambda_0$. The partial derivative of $H$ with respect to $\lambda_0$
is equal to $\Omega_{AB';\alpha'} t^{\alpha'} P^A Q^{B'}$, and in
the limit this becomes
$[\Omega_{AB';\alpha'}] t^{\alpha'} P^{A'} Q^{B'}$. Similarly, the 
partial derivative of $H$ with respect to $\lambda_1$ is  
$\Omega_{AB';\alpha} t^{\alpha} P^A Q^{B'}$, and in the limit
$\lambda_1 \to \lambda_0$ this becomes $[\Omega_{AB';\alpha}]
t^{\alpha'} P^{A'} Q^{B'}$. Finally, $H(\lambda_0,\lambda_0) =
[\Omega_{AB'}] P^{A'} Q^{B'}$, and its derivative with respect to 
$\lambda_0$ is $[\Omega_{AB'}]_{;\alpha'} t^{\alpha'} P^{A'} Q^{B'}$. 
Gathering the results we find that  
\[
\Bigl\{ \bigl[ \Omega_{AB'} \bigr]_{;\alpha'} 
- \bigl[ \Omega_{AB';\alpha'} \bigr]  
- \bigl[ \Omega_{AB';\alpha} \bigr] \Bigr\} 
t^{\alpha'} P^{A'} Q^{B'} = 0,  
\]
and the final statement of Synge's rule, 
\begin{equation} 
\bigl[ \Omega_{AB'} \bigr]_{;\alpha'} = 
\bigl[ \Omega_{AB';\alpha'} \bigr] 
+ \bigl[ \Omega_{AB';\alpha} \bigr],  
\label{3.2.1}
\end{equation}
follows from the fact that the tensors $P^M$ and $Q^N$, and the
direction of the selected geodesic $\beta$, are all
arbitrary. Equation (\ref{3.2.1}) reduces to Eq.~(\ref{3.1.3}) when
$\sigma_{\cdots}$ is substituted in place of $\Omega_{AB'}$.   

\section{Parallel propagator} 
\label{4}

\subsection{Tetrad on $\beta$} 
\label{4.1}

On the geodesic $\beta$ that links $x$ to $x'$ we introduce an
orthonormal basis $\base{\mu}{\sf a}(z)$ that is parallel transported
on the geodesic. The frame indices $\sf a$, $\sf b$, \ldots, run from
0 to 3 and the frame vectors satisfy 
\begin{equation}
g_{\mu\nu}\, \base{\mu}{\sf a} \base{\nu}{\sf b} = \eta_{\sf ab},
\qquad   
\frac{D \base{\mu}{\sf a}}{d\lambda} = 0,  
\label{4.1.1}
\end{equation}    
where $\eta_{\sf ab} = \mbox{diag}(-1,1,1,1)$ is the Minkowski metric 
(which we shall use to raise and lower frame indices). We have the  
completeness relations 
\begin{equation}
g^{\mu\nu} = \eta^{\sf ab}\, \base{\mu}{\sf a} \base{\nu}{\sf b}, 
\label{4.1.2}
\end{equation} 
and we define a dual tetrad $\base{\sf a}{\mu}(z)$ by  
\begin{equation}           
\base{\sf a}{\mu} \equiv \eta^{\sf ab} g_{\mu\nu}\, \base{\nu}{\sf b};  
\label{4.1.3}
\end{equation}
this is also parallel transported on $\beta$. In terms of the dual
tetrad the completeness relations take the form 
\begin{equation}
g_{\mu\nu} = \eta_{\sf ab}\, \base{\sf a}{\mu} \base{\sf b}{\nu},  
\label{4.1.4}
\end{equation} 
and it is easy to show that the tetrad and its dual satisfy
$\base{\sf a}{\mu} \base{\mu}{\sf b} = \delta^{\sf a}_{\ \sf b}$ and
$\base{\sf a}{\nu} \base{\mu}{\sf a} = \delta^\mu_{\ \nu}$. Equations 
(\ref{4.1.1})--(\ref{4.1.4}) hold everywhere on $\beta$. In 
particular, with an appropriate change of notation they hold at  
$x'$ and $x$; for example, $g_{\alpha\beta} = \eta_{\sf ab}\,
\base{\sf a}{\alpha} \base{\sf b}{\beta}$ is the metric at $x$. 

(You will have noticed that I use {\sf sans-serif} symbols for the
frame indices. This is to distinguish them from another set of frame
indices that will appear below. The frame indices introduced here run
from 0 to 3; those to be introduced later will run from 1 to 3.)      

\subsection{Definition and properties of the parallel propagator}  
\label{4.2}

Any vector field $A^\mu(z)$ on $\beta$ can be decomposed in the basis 
$\base{\mu}{\sf a}$: $A^\mu = A^{\sf a}\, \base{\mu}{\sf a}$, and
the vector's frame components are given by $A^{\sf a} 
= A^\mu\, \base{\sf a}{\mu}$. If $A^\mu$ is parallel transported on
the geodesic, then the coefficients $A^{\sf a}$ are constants. The
vector at $x$ can then be expressed as $A^\alpha 
= (A^{\alpha'}\, \base{\sf a}{\alpha'}) \base{\alpha}{\sf a}$, or   
\begin{equation}
A^{\alpha}(x) = g^\alpha_{\ \alpha'}(x,x')\, A^{\alpha'}(x'), 
\qquad
g^\alpha_{\ \alpha'}(x,x') \equiv \base{\alpha}{\sf a}(x)\, 
\base{\sf a}{\alpha'}(x').
\label{4.2.1}
\end{equation} 
The object $g^\alpha_{\ \alpha'} = \base{\alpha}{\sf a} 
\base{\sf a}{\alpha'}$ is the {\it parallel propagator}: it takes a
vector at $x'$ and parallel-transports it to $x$ along the unique
geodesic that links these points.  

Similarly, we find that  
\begin{equation}
A^{\alpha'}(x') = g^{\alpha'}_{\ \alpha}(x',x)\, A^{\alpha}(x),
\qquad 
g^{\alpha'}_{\ \alpha}(x',x) \equiv \base{\alpha'}{\sf a}(x')\,
\base{\sf a}{\alpha}(x),
\label{4.2.2} 
\end{equation} 
and we see that $g^{\alpha'}_{\ \alpha} = \base{\alpha'}{\sf a} 
\base{\sf a}{\alpha}$ performs the inverse operation: it takes a
vector at $x$ and parallel-transports it back to $x'$. Clearly, 
\begin{equation} 
g^\alpha_{\ \alpha'} g^{\alpha'}_{\ \beta} = \delta^\alpha_{\ \beta},
\qquad 
g^{\alpha'}_{\ \alpha} g^{\alpha}_{\ \beta'} = 
\delta^{\alpha'}_{\ \beta'},
\label{4.2.3}
\end{equation} 
and these relations formally express the fact that 
$g^{\alpha'}_{\ \alpha}$ is the inverse of 
$g^\alpha_{\ \alpha'}$.   

The relation $g^\alpha_{\ \alpha'} = \base{\alpha}{\sf a} 
\base{\sf a}{\alpha'}$ can also be expressed as 
$g_\alpha^{\ \alpha'} = \base{\sf a}{\alpha} \base{\alpha'}{\sf a}$,
and this reveals that 
\begin{equation}
g_\alpha^{\ \alpha'}(x,x') = g^{\alpha'}_{\ \alpha}(x',x), 
\qquad
g_{\alpha'}^{\ \alpha}(x',x) = g^\alpha_{\ \alpha'}(x,x').   
\label{4.2.4}
\end{equation}
The ordering of the indices, and the ordering of the arguments, are
therefore arbitrary.  

The action of the parallel propagator on tensors of arbitrary ranks is 
easy to figure out. For example, suppose that the dual vector $p_\mu =
p_a\, \base{a}{\mu}$ is parallel transported on $\beta$. Then the
frame components $p_{\sf a} = p_\mu\, \base{\mu}{\sf a}$ are
constants, and the dual vector at $x$ can be expressed as $p_\alpha =
(p_{\alpha'} \base{\alpha'}{\sf a}) \base{\alpha}{\sf a}$, or 
\begin{equation}
p_\alpha(x) = g^{\alpha'}_{\ \alpha}(x',x)\, p_{\alpha'}(x').
\label{4.2.5}
\end{equation}
It is therefore the inverse propagator $g^{\alpha'}_{\ \alpha}$ that
takes a dual vector at $x'$ and parallel-transports it to $x$. As
another example, it is easy to show that a tensor $A^{\alpha\beta}$ at
$x$ obtained by parallel transport from $x'$ must be given by 
\begin{equation}
A^{\alpha\beta}(x) = g^\alpha_{\ \alpha'}(x,x') 
g^\beta_{\ \beta'}(x,x')\, A^{\alpha'\beta'}(x'). 
\label{4.2.6}
\end{equation}       
Here we need two occurrences of the parallel propagator, one for each 
tensorial index. Because the metric tensor is covariantly constant, it
is automatically parallel transported on $\beta$, and a special case
of Eq.~(\ref{4.2.6}) is therefore $g_{\alpha\beta} = 
g^{\alpha'}_{\ \alpha} g^{\beta'}_{\ \beta}\, g_{\alpha'\beta'}$.   

Because the basis vectors are parallel transported on $\beta$,
they satisfy $e^\alpha_{{\sf a};\beta} \sigma^\beta = 0$ at $x$ and 
$e^{\alpha'}_{{\sf a};\beta'} \sigma^{\beta'} = 0$ at $x'$. This
immediately implies that the parallel propagators must satisfy 
\begin{equation}
g^{\alpha}_{\ \alpha';\beta} \sigma^{\beta} =  
g^{\alpha}_{\ \alpha';\beta'} \sigma^{\beta'} = 0,
\qquad
g^{\alpha'}_{\ \alpha;\beta} \sigma^{\beta} = 
g^{\alpha'}_{\ \alpha;\beta'} \sigma^{\beta'} = 0. 
\label{4.2.7}
\end{equation} 
Another useful property of the parallel propagator follows from the
fact that if $t^\mu = dz^\mu/d\lambda$ is tangent to the geodesic
connecting $x$ to $x'$, then $t^\alpha = g^\alpha_{\ \alpha'}
t^{\alpha'}$. Using Eqs.~(\ref{2.3.1}) and (\ref{2.3.2}), this
observation gives us the relations 
\begin{equation}
\sigma_{\alpha} = -g^{\alpha'}_{\ \alpha} \sigma_{\alpha'}, \qquad 
\sigma_{\alpha'} = -g^{\alpha}_{\ \alpha'} \sigma_{\alpha}. 
\label{4.2.8}
\end{equation} 
 
\subsection{Coincidence limits} 
\label{4.3}

Equation (\ref{4.2.1}) and the completeness relations of
Eqs.~(\ref{4.1.2}) or (\ref{4.1.4}) imply that 
\begin{equation} 
\bigl[ g^\alpha_{\ \beta'} \bigr] = \delta^{\alpha'}_{\ \beta'}. 
\label{4.3.1}
\end{equation} 
Other coincidence limits are obtained by differentiation of 
Eqs.~(\ref{4.2.7}). For example, the relation 
$g^{\alpha}_{\ \beta';\gamma} \sigma^\gamma = 0$ implies 
$g^{\alpha}_{\ \beta';\gamma\delta} \sigma^\gamma + 
g^{\alpha}_{\ \beta';\gamma} \sigma^\gamma_{\ \delta} = 0$,  
and at coincidence we have 
\begin{equation} 
\bigl[ g^\alpha_{\ \beta';\gamma} \bigr] = 
\bigl[ g^\alpha_{\ \beta';\gamma'} \bigr] = 0; 
\label{4.3.2}
\end{equation}
the second result was obtained by applying Synge's rule on the first
result. Further differentiation gives 
\[
g^{\alpha}_{\ \beta';\gamma\delta\epsilon} \sigma^\gamma +  
g^{\alpha}_{\ \beta';\gamma\delta} \sigma^\gamma_{\ \epsilon} +  
g^{\alpha}_{\ \beta';\gamma\epsilon} \sigma^\gamma_{\ \delta} + 
g^{\alpha}_{\ \beta';\gamma} \sigma^\gamma_{\ \delta\epsilon} = 0, 
\]
and at coincidence we have $[g^{\alpha}_{\ \beta';\gamma\delta}] +
[g^{\alpha}_{\ \beta';\delta\gamma}] = 0$, or 
$2[g^{\alpha}_{\ \beta';\gamma\delta}] 
+ R^{\alpha'}_{\ \beta'\gamma'\delta'} = 0$. The coincidence limit for 
$g^{\alpha}_{\ \beta';\gamma\delta'} 
= g^{\alpha}_{\ \beta';\delta'\gamma}$ can then be obtained from
Synge's rule, and an additional application of the rule gives
$[g^{\alpha}_{\ \beta';\gamma'\delta'}]$. Our results are 
\begin{eqnarray} 
\bigl[ g^{\alpha}_{\ \beta';\gamma\delta} \bigr] &=& 
-\frac{1}{2}\, R^{\alpha'}_{\ \beta'\gamma'\delta'}, 
\qquad
\bigl[ g^{\alpha}_{\ \beta';\gamma\delta'} \bigr] = 
\frac{1}{2}\, R^{\alpha'}_{\ \beta'\gamma'\delta'}, 
\nonumber \\ 
& & \label{4.3.3} \\
\bigl[ g^{\alpha}_{\ \beta';\gamma'\delta} \bigr] &=& 
-\frac{1}{2}\, R^{\alpha'}_{\ \beta'\gamma'\delta'}, 
\qquad
\bigl[ g^{\alpha}_{\ \beta';\gamma'\delta'} \bigr] = 
\frac{1}{2}\, R^{\alpha'}_{\ \beta'\gamma'\delta'}. 
\nonumber
\end{eqnarray} 

\section{Expansion of bitensors near coincidence} 
\label{5}

\subsection{General method} 
\label{5.1}

We would like to express a bitensor $\Omega_{\alpha'\beta'}(x,x')$
near coincidence as an expansion in powers of
$-\sigma^{\alpha'}(x,x')$, the closest analogue in curved spacetime to
the flat-spacetime quantity $(x-x')^\alpha$. For concreteness we shall
consider the case of rank-2 bitensor, and for the moment we will
assume that the bitensor's indices all refer to the base point $x'$.   

The expansion we seek is of the form 
\begin{equation}
\Omega_{\alpha'\beta'}(x,x') = A_{\alpha'\beta'}
+ A_{\alpha'\beta'\gamma'}\, \sigma^{\gamma'}  
+ \frac{1}{2}\, A_{\alpha'\beta'\gamma'\delta'}\, 
  \sigma^{\gamma'} \sigma^{\delta'} 
+ O(\epsilon^3), 
\label{5.1.1}
\end{equation} 
in which the ``expansion coefficients'' $A_{\alpha'\beta'}$,
$A_{\alpha'\beta'\gamma'}$, and $A_{\alpha'\beta'\gamma'\delta'}$ are
all ordinary tensors at $x'$; this last tensor is symmetric in the
pair of indices $\gamma'$ and $\delta'$, and $\epsilon$ measures the 
size of a typical component of $\sigma^{\alpha'}$. 

To find the expansion coefficients we differentiate Eq.~(\ref{5.1.1}) 
repeatedly and take coincidence limits. Equation (\ref{5.1.1})   
immediately implies $[\Omega_{\rm \alpha'\beta'}] =
A_{\alpha'\beta'}$. After one differentiation we obtain
$\Omega_{\alpha'\beta';\gamma'} = A_{\alpha'\beta';\gamma'} +
A_{\alpha'\beta'\epsilon';\gamma'} \sigma^{\epsilon'} +
A_{\alpha'\beta'\epsilon'} \sigma^{\epsilon'}_{\ \gamma'} +
\frac{1}{2}\, A_{\alpha'\beta'\epsilon'\iota';\gamma'}
\sigma^{\epsilon'} \sigma^{\iota'} + A_{\alpha'\beta'\epsilon'\iota'}
\sigma^{\epsilon'} \sigma^{\iota'}_{\ \gamma'} + O(\epsilon^2)$, and
at coincidence this reduces to $[\Omega_{\alpha'\beta';\gamma'}] = 
A_{\alpha'\beta';\gamma'} + A_{\alpha'\beta'\gamma'}$. Taking the
coincidence limit after two differentiations yields
$[\Omega_{\alpha'\beta';\gamma'\delta'}] =
A_{\alpha'\beta';\gamma'\delta'} + A_{\alpha'\beta'\gamma';\delta'} +
A_{\alpha'\beta'\delta';\gamma'} + A_{\alpha'\beta'\gamma'\delta'}$. 
The expansion coefficients are therefore  
\begin{eqnarray}
A_{\alpha'\beta'} &=& \bigl[ \Omega_{\alpha'\beta'} \bigr], 
\nonumber \\
A_{\alpha'\beta'\gamma'} &=& \bigl[ \Omega_{\alpha'\beta';\gamma'}
\bigr] - A_{\alpha'\beta';\gamma'}, 
\nonumber \\
A_{\alpha'\beta'\gamma'\delta'} &=& 
\bigl[ \Omega_{\alpha'\beta';\gamma'\delta'} \bigr] 
- A_{\alpha'\beta';\gamma'\delta'}
- A_{\alpha'\beta'\gamma';\delta'} 
- A_{\alpha'\beta'\delta';\gamma'}. 
\label{5.1.2}
\end{eqnarray}  
These results are to be substituted into Eq.~(\ref{5.1.1}), and this
gives us $\Omega_{\alpha'\beta'}(x,x')$ to second order in $\epsilon$.    

Suppose now that the bitensor is $\Omega_{\alpha'\beta}$, with one
index referring to $x'$ and the other to $x$. The previous procedure 
can be applied directly if we introduce an auxiliary bitensor 
$\tilde{\Omega}_{\alpha'\beta'} \equiv g^\beta_{\ \beta'}
\Omega_{\alpha'\beta}$ whose indices all refer to the point $x'$. Then
$\tilde{\Omega}_{\alpha'\beta'}$ can be expanded as in
Eq.~(\ref{5.1.1}), and the original bitensor is reconstructed as 
$\Omega_{\alpha'\beta} = g^{\beta'}_{\ \beta} 
\tilde{\Omega}_{\alpha'\beta'}$, or 
\begin{equation}
\Omega_{\alpha'\beta}(x,x') = g^{\beta'}_{\ \beta}
\biggl( B_{\alpha'\beta'}
+ B_{\alpha'\beta'\gamma'}\, \sigma^{\gamma'}  
+ \frac{1}{2}\, B_{\alpha'\beta'\gamma'\delta'}\, 
  \sigma^{\gamma'} \sigma^{\delta'} \biggr) 
+ O(\epsilon^3). 
\label{5.1.3}
\end{equation} 
The expansion coefficients can be obtained from the coincidence limits
of $\tilde{\Omega}_{\alpha'\beta'}$ and its derivatives. It is 
convenient, however, to express them directly in terms of the original
bitensor $\Omega_{\alpha'\beta}$ by substituting the relation
$\tilde{\Omega}_{\alpha'\beta'} = g^\beta_{\ \beta'}
\Omega_{\alpha'\beta}$ and its derivatives. After using the results of 
Eq.~(\ref{4.3.1})--(\ref{4.3.3}) we find 
\begin{eqnarray}
B_{\alpha'\beta'} &=& \bigl[ \Omega_{\alpha'\beta} \bigr], 
\nonumber \\
B_{\alpha'\beta'\gamma'} &=& \bigl[ \Omega_{\alpha'\beta;\gamma'}
\bigr] - B_{\alpha'\beta';\gamma'}, 
\nonumber \\
B_{\alpha'\beta'\gamma'\delta'} &=& 
\bigl[ \Omega_{\alpha'\beta;\gamma'\delta'} \bigr] 
+ \frac{1}{2}\, B_{\alpha'\epsilon'} 
  R^{\epsilon'}_{\ \beta'\gamma'\delta'}
- B_{\alpha'\beta';\gamma'\delta'} 
- B_{\alpha'\beta'\gamma';\delta'} 
- B_{\alpha'\beta'\delta';\gamma'}. 
\label{5.1.4}
\end{eqnarray}  
The only difference with respect to Eq.~(\ref{5.1.3}) is the presence
of a Riemann-tensor term in $B_{\alpha'\beta'\gamma'\delta'}$.  

Suppose finally that the bitensor to be expanded is
$\Omega_{\alpha\beta}$, whose indices all refer to $x$. Much as we did 
before, we introduce an auxiliary bitensor
$\tilde{\Omega}_{\alpha'\beta'} =  
g^\alpha_{\ \alpha'} g^\beta_{\ \beta'} \Omega_{\alpha\beta}$ whose
indices all refer to $x'$, we expand
$\tilde{\Omega}_{\alpha'\beta'}$ as in Eq.~(\ref{5.1.1}), and we then 
reconstruct the original bitensor. This gives us 
\begin{equation}
\Omega_{\alpha\beta}(x,x') = g^{\alpha'}_{\ \alpha}
g^{\beta'}_{\ \beta} \biggl( C_{\alpha'\beta'}    
+ C_{\alpha'\beta'\gamma'}\, \sigma^{\gamma'}  
+ \frac{1}{2}\, C_{\alpha'\beta'\gamma'\delta'}\, 
  \sigma^{\gamma'} \sigma^{\delta'} \biggr) 
+ O(\epsilon^3), 
\label{5.1.5}
\end{equation} 
and the expansion coefficients are now 
\begin{eqnarray}
\hspace*{-10pt} 
C_{\alpha'\beta'} &=& \bigl[ \Omega_{\alpha\beta} \bigr], 
\nonumber \\
\hspace*{-10pt} 
C_{\alpha'\beta'\gamma'} &=& \bigl[ \Omega_{\alpha\beta;\gamma'} 
\bigr] - C_{\alpha'\beta';\gamma'}, 
\nonumber \\
\hspace*{-10pt} 
C_{\alpha'\beta'\gamma'\delta'} &=& 
\bigl[ \Omega_{\alpha\beta;\gamma'\delta'} \bigr] 
+ \frac{1}{2}\, C_{\alpha'\epsilon'} 
  R^{\epsilon'}_{\ \beta'\gamma'\delta'}
+ \frac{1}{2}\, C_{\epsilon'\beta'} 
  R^{\epsilon'}_{\ \alpha'\gamma'\delta'}
- C_{\alpha'\beta';\gamma'\delta'} 
- C_{\alpha'\beta'\gamma';\delta'} 
- C_{\alpha'\beta'\delta';\gamma'}. 
\label{5.1.6}
\end{eqnarray}  
This differs from Eq.~(\ref{5.1.4}) by the presence of an additional  
Riemann-tensor term in $C_{\alpha'\beta'\gamma'\delta'}$. 

\subsection{Special cases} 
\label{5.2}

We now apply the general expansion method developed in the preceding
subsection to the bitensors $\sigma_{\alpha'\beta'}$,
$\sigma_{\alpha'\beta}$, and $\sigma_{\alpha\beta}$. In the first
instance we have $A_{\alpha'\beta'} = g_{\alpha'\beta'}$,
$A_{\alpha'\beta'\gamma'} = 0$, and $A_{\alpha'\beta'\gamma'\delta'} =
-\frac{1}{3}(R_{\alpha'\gamma'\beta'\delta'} 
+ R_{\alpha'\delta'\beta'\gamma'})$. In the second instance we have   
$B_{\alpha'\beta'} = -g_{\alpha'\beta'}$,
$B_{\alpha'\beta'\gamma'} = 0$, and $B_{\alpha'\beta'\gamma'\delta'} = 
-\frac{1}{3}(R_{\beta'\alpha'\gamma'\delta'} 
+ R_{\beta'\gamma'\alpha'\delta'}) 
- \frac{1}{2} R_{\alpha'\beta'\gamma'\delta'} = 
-\frac{1}{3} R_{\alpha'\delta'\beta'\gamma'} 
-\frac{1}{6} R_{\alpha'\beta'\gamma'\delta'}$. In the third instance
we have $C_{\alpha'\beta'} = g_{\alpha'\beta'}$,
$C_{\alpha'\beta'\gamma'} = 0$, and $C_{\alpha'\beta'\gamma'\delta'} =  
-\frac{1}{3} (R_{\alpha'\gamma'\beta'\delta'} 
+ R_{\alpha'\delta'\beta'\gamma'})$. This gives us the expansions 
\begin{eqnarray} 
\sigma_{\alpha'\beta'} &=& g_{\alpha'\beta'} - \frac{1}{3}\,
R_{\alpha'\gamma'\beta'\delta'}\, \sigma^{\gamma'} \sigma^{\delta'}
+ O(\epsilon^3), 
\label{5.2.1} \\ 
\sigma_{\alpha'\beta} &=& -g^{\beta'}_{\ \beta} 
\Bigl( g_{\alpha'\beta'} 
+ \frac{1}{6}\, R_{\alpha'\gamma'\beta'\delta'}\, \sigma^{\gamma'}
\sigma^{\delta'} \Bigr) + O(\epsilon^3),
\label{5.2.2} \\ 
\sigma_{\alpha\beta} &=& g^{\alpha'}_{\ \alpha} g^{\beta'}_{\ \beta'} 
\Bigl( g_{\alpha'\beta'} - \frac{1}{3}\,
R_{\alpha'\gamma'\beta'\delta'}\, \sigma^{\gamma'} 
\sigma^{\delta'} \Bigr) + O(\epsilon^3).
\label{5.2.3} 
\end{eqnarray}
Taking the trace of the last equation returns 
$\sigma^\alpha_{\ \alpha} = 4 
- \frac{1}{3} R_{\gamma'\delta'}\, \sigma^{\gamma'}
\sigma^{\delta'} + O(\epsilon^3)$, or 
\begin{equation}
\theta^* = 3 - \frac{1}{3}\, R_{\alpha'\beta'}\, \sigma^{\alpha'} 
\sigma^{\beta'} + O(\epsilon^3), 
\label{5.2.4}
\end{equation}
where $\theta^* \equiv \sigma^{\alpha}_{\ \alpha} - 1$ was shown in
Sec.~\ref{2.4} to describe the expansion of the congruence of
geodesics that emanate from $x'$. Equation (\ref{5.2.4}) reveals that
timelike geodesics are focused if the Ricci tensor is nonzero and
the strong energy condition holds: when $R_{\alpha'\beta'}\,
\sigma^{\alpha'} \sigma^{\beta'} > 0$ we see that $\theta^*$ is
smaller than 3, the value it would take in flat spacetime. 

The expansion method can easily be extended to bitensors of other
tensorial ranks. In particular, it can be adapted to give expansions
of the first derivatives of the parallel propagator. The expansions
\begin{equation}
g^\alpha_{\ \beta';\gamma'} = \frac{1}{2}\, g^\alpha_{\ \alpha'}
R^{\alpha'}_{\ \beta'\gamma'\delta'}\, \sigma^{\delta'} 
+ O(\epsilon^2), \qquad
g^\alpha_{\ \beta';\gamma} = \frac{1}{2}\, 
g^\alpha_{\ \alpha'} g^{\gamma'}_{\ \gamma} 
R^{\alpha'}_{\ \beta'\gamma'\delta'}\, \sigma^{\delta'} 
+ O(\epsilon^2) 
\label{5.2.5}
\end{equation}
and thus easy to establish, and they will be needed in part
\ref{part3} of this review.   

\subsection{Expansion of tensors} 
\label{5.3}

The expansion method can also be applied to ordinary tensor
fields. For concreteness, suppose that we wish to express a rank-2
tensor $A_{\alpha\beta}$ at a point $x$ in terms of its values (and
that of its covariant derivatives) at a neighbouring point $x'$. The
tensor can be written as an expansion in powers
of $-\sigma^{\alpha'}(x,x')$ and in this case we have 
\begin{equation}
A_{\alpha\beta}(x) = g^{\alpha'}_{\ \alpha} g^{\beta'}_{\ \beta} 
\biggl( A_{\alpha'\beta'} 
- A_{\alpha'\beta';\gamma'}\, \sigma^{\gamma'}   
+ \frac{1}{2}\, A_{\alpha'\beta';\gamma'\delta'}\, 
  \sigma^{\gamma'} \sigma^{\delta'} \biggr) 
+ O(\epsilon^3).  
\label{5.3.1}
\end{equation} 
If the tensor field is parallel transported on the geodesic $\beta$
that links $x$ to $x'$, then Eq.~(\ref{5.3.1}) reduces to
Eq.~(\ref{4.2.6}). The extension of this formula to tensors of other
ranks is obvious.   

To derive this result we express $A_{\mu\nu}(z)$, the restriction of 
the tensor field on $\beta$, in terms of its tetrad components
$A_{\sf ab}(\lambda) = A_{\mu\nu} \base{\mu}{\sf a} 
\base{\nu}{\sf b}$. Recall from Sec.~\ref{4.1} that 
$\base{\mu}{\sf a}$ is an orthonormal basis that
is parallel transported on $\beta$; recall also that the affine
parameter $\lambda$ ranges from $\lambda_0$ (its value at $x'$) to
$\lambda_1$ (its value at $x$). We have $A_{\alpha'\beta'}(x') = 
A_{\sf ab}(\lambda_0) \base{\sf a}{\alpha'} \base{\sf b}{\beta'}$,  
$A_{\alpha\beta}(x) = A_{\sf ab}(\lambda_1)
\base{\sf a}{\alpha} \base{\sf b}{\beta}$, and $A_{\sf ab}(\lambda_1)$
can be expressed in terms of quantities at $\lambda = \lambda_0$ by
straightforward Taylor expansion. Since, for example, 
\[
(\lambda_1 - \lambda_0) \frac{d A_{\sf ab}}{d\lambda}
\biggr|_{\lambda_0} = (\lambda_1 - \lambda_0) \bigl(A_{\mu\nu}
\base{\mu}{\sf a} \base{\nu}{\sf b} \bigr)_{;\lambda} t^\lambda
\Bigr|_{\lambda_0} = (\lambda_1 - \lambda_0) A_{\mu\nu;\lambda}
\base{\mu}{\sf a} \base{\nu}{\sf b} t^\lambda \Bigr|_{\lambda_0} 
= - A_{\alpha'\beta';\gamma'} \base{\alpha'}{\sf a} 
\base{\beta'}{\sf b} \sigma^{\gamma'}, 
\]
where we have used Eq.~(\ref{2.3.2}), we arrive at Eq.~(\ref{5.3.1}) 
after involving Eq.~(\ref{4.2.2}). 

\section{van Vleck determinant} 
\label{6}

\subsection{Definition and properties} 
\label{6.1}

The van Vleck biscalar $\Delta(x,x')$ is defined by 
\begin{equation}
\Delta(x,x') \equiv \mbox{det}\bigl[ \Delta^{\alpha'}_{\ \beta'}(x,x')
\bigr], \qquad 
\Delta^{\alpha'}_{\ \beta'}(x,x') \equiv -g^{\alpha'}_{\ \alpha}(x',x) 
\sigma^\alpha_{\ \beta'}(x,x').    
\label{6.1.1}
\end{equation}
As we shall show below, it can also be expressed as  
\begin{equation} 
\Delta(x,x') = - \frac{ \mbox{det}\bigl[ -\sigma_{\alpha\beta'}(x,x')
\bigr]} {\sqrt{-g}\sqrt{-g'}}, 
\label{6.1.2}
\end{equation}
where $g$ is the metric determinant at $x$ and $g'$ the metric
determinant at $x'$. 

Equations (\ref{3.1.2}) and (\ref{4.3.1}) imply that at coincidence, 
$[\Delta^{\alpha'}_{\ \beta'}] = \delta^{\alpha'}_{\ \beta'}$ and
$[\Delta] = 1$. Equation (\ref{5.2.2}), on the other hand, implies
that near coincidence, 
\begin{equation} 
\Delta^{\alpha'}_{\ \beta'} = \delta^{\alpha'}_{\ \beta'} +
\frac{1}{6}\, R^{\alpha'}_{\ \gamma'\beta'\delta'}\, \sigma^{\gamma'}
\sigma^{\delta'} + O(\epsilon^3), 
\label{6.1.3}
\end{equation}
so that 
\begin{equation}
\Delta = 1 + \frac{1}{6}\, R_{\alpha'\beta'}\, \sigma^{\alpha'}
\sigma^{\beta'} + O(\epsilon^3). 
\label{6.1.4}
\end{equation}  
This last result follows from the fact that for a ``small'' matrix
$\bm{a}$, $\mbox{det}(\bm{1} + \bm{a}) = 1 + \mbox{tr}(\bm{a}) +
O(\bm{a}^2)$.      

We shall prove below that the van Vleck determinant satisfies the
differential equation 
\begin{equation}
\frac{1}{\Delta} \bigl( \Delta \sigma^\alpha \bigr)_{;\alpha} = 4, 
\label{6.1.5}
\end{equation}
which can also be written as $(\ln \Delta)_{,\alpha} \sigma^\alpha = 4
- \sigma^\alpha_{\ \alpha}$, or 
\begin{equation}
\frac{d}{d\lambda^*}(\ln \Delta) = 3 - \theta^*  
\label{6.1.6}
\end{equation}
in the notation introduced in Sec.~\ref{2.4}. Equation (\ref{6.1.6})
reveals that the behaviour of the van Vleck determinant is
governed by the expansion of the congruence of geodesics that emanate
from $x'$. If $\theta^* < 3$, then the congruence expands less rapidly
than it would in flat spacetime, and $\Delta$ {\it increases} along
the geodesics. If, on the other hand, $\theta^* > 3$, then the
congruence expands more rapidly than it would in flat spacetime, and
$\Delta$ {\it decreases} along the geodesics. Thus, $\Delta > 1$
indicates that the geodesics are undergoing focusing, while $\Delta <
1$ indicates that the geodesics are undergoing defocusing. The
connection between the van Vleck determinant and the strong energy
condition is well illustrated by Eq.~(\ref{6.1.4}): the sign of
$\Delta - 1$ near $x'$ is determined by the sign of
$R_{\alpha'\beta'}\, \sigma^{\alpha'} \sigma^{\beta'}$.   

\subsection{Derivations} 
\label{6.2}

To show that Eq.~(\ref{6.1.2}) follows from Eq.~(\ref{6.1.1}) we
rewrite the completeness relations at $x$, $g^{\alpha\beta} =
\eta^{\sf ab} \base{\alpha}{\sf a} \base{\beta}{\sf b}$, in the matrix
form $\bm{g}^{-1} = \bm{E} \bm{\eta} \bm{E}^T$, where $\bm{E}$ denotes
the $4 \times 4$ matrix whose entries correspond to
$\base{\alpha}{\sf a}$. (In this translation we put tensor and frame
indices on equal footing.) With $e$ denoting the determinant of this
matrix, we have $1/g = -e^2$, or $e = 1/\sqrt{-g}$. Similarly, we 
rewrite the completeness relations at $x'$, $g^{\alpha'\beta'} =
\eta^{\sf ab} \base{\alpha'}{\sf a} \base{\beta'}{\sf b}$, in the
matrix form $\bm{g'}^{-1} = \bm{E'} \bm{\eta} \bm{E'}^T$, where
$\bm{E'}$ is the matrix corresponding to $\base{\alpha'}{\sf a}$. With
$e'$ denoting its determinant, we have $1/g' = -e^{\prime 2}$, or $e'
= 1/\sqrt{-g'}$. Now, the parallel propagator is defined by
$g^{\alpha}_{\ \alpha'} = \eta^{\sf ab} g_{\alpha'\beta'}
\base{\alpha}{\sf a} \base{\beta'}{\sf b}$, and the matrix form of
this equation is $\bm{\hat{g}} = \bm{E} \bm{\eta} \bm{E'}^T
\bm{g'}^T$. The determinant of the parallel propagator is therefore
$\hat{g} = -ee'g' = \sqrt{-g'}/\sqrt{-g}$. So we have 
\begin{equation}
\mbox{det}\bigl[ g^{\alpha}_{\ \alpha'} \bigr] =
\frac{\sqrt{-g'}}{\sqrt{-g}}, \qquad 
\mbox{det}\bigl[ g^{\alpha'}_{\ \alpha} \bigr] =
\frac{\sqrt{-g}}{\sqrt{-g'}},
\label{6.2.1}
\end{equation}
and Eq.~(\ref{6.1.2}) follows from the fact that the matrix form of 
Eq.~(\ref{6.1.1}) is $\bm{\Delta} = - \bm{\hat{g}}^{-1} \bm{g}^{-1}  
\bm{\sigma}$, where $\bm{\sigma}$ is the matrix corresponding to
$\sigma_{\alpha\beta'}$. 

To establish Eq.~(\ref{6.1.5}) we differentiate the relation 
$\sigma = \frac{1}{2} \sigma^\gamma \sigma_\gamma$ twice and obtain 
$\sigma_{\alpha\beta'} = \sigma^\gamma_{\ \alpha}
\sigma_{\gamma\beta'} + \sigma^\gamma \sigma_{\gamma\alpha\beta'}$. If 
we replace the last factor by $\sigma_{\alpha\beta'\gamma}$ and
multiply both sides by $-g^{\alpha'\alpha}$ we find
\[
\Delta^{\alpha'}_{\ \beta'} = -g^{\alpha'\alpha} \bigl(   
\sigma^\gamma_{\ \alpha} \sigma_{\gamma\beta'} 
+ \sigma^\gamma \sigma_{\alpha\beta'\gamma} \bigr).  
\]
In this expression we make the substitution $\sigma_{\alpha\beta'} =
-g_{\alpha\alpha'} \Delta^{\alpha'}_{\ \beta'}$, which follows
directly from Eq.~(\ref{6.1.1}). This gives us  
\begin{equation} 
\Delta^{\alpha'}_{\ \beta'} = g^{\alpha'}_{\ \alpha} 
g^{\gamma}_{\ \gamma'} \sigma^\alpha_{\ \gamma} 
\Delta^{\gamma'}_{\ \beta'} 
+ \Delta^{\alpha'}_{\ \beta';\gamma} \sigma^\gamma, 
\label{6.2.2}
\end{equation}  
where we have used Eq.~(\ref{4.2.7}). At this stage we introduce
an inverse $(\Delta^{-1})^{\alpha'}_{\ \beta'}$ to the van Vleck
bitensor, defined by $\Delta^{\alpha'}_{\ \beta'} 
(\Delta^{-1})^{\beta'}_{\ \gamma'} = 
\delta^{\alpha'}_{\ \gamma'}$. After multiplying both sides of
Eq.~(\ref{6.2.2}) by $(\Delta^{-1})^{\beta'}_{\ \gamma'}$ we find  
\[
\delta^{\alpha'}_{\ \beta'} = g^{\alpha'}_{\ \alpha} 
g^{\beta}_{\ \beta'} \sigma^\alpha_{\ \beta} 
+ (\Delta^{-1})^{\gamma'}_{\ \beta'}
\Delta^{\alpha'}_{\ \gamma';\gamma} \sigma^\gamma, 
\]
and taking the trace of this equation yields 
\[
4 = \sigma^{\alpha}_{\ \alpha} + (\Delta^{-1})^{\beta'}_{\ \alpha'} 
\Delta^{\alpha'}_{\ \beta';\gamma} \sigma^\gamma.  
\]
We now recall the identity $\delta \ln \mbox{det}\bm{M} =
\mbox{Tr}(\bm{M}^{-1} \delta \bm{M})$, which relates the variation of
a determinant to the variation of the matrix elements. It implies, in
particular, that $(\Delta^{-1})^{\beta'}_{\ \alpha'}  
\Delta^{\alpha'}_{\ \beta';\gamma} = (\ln \Delta)_{,\gamma}$, and we
finally obtain  
\begin{equation}
4 = \sigma^\alpha_{\ \alpha} + (\ln \Delta)_{,\alpha} \sigma^\alpha, 
\label{6.2.3}
\end{equation} 
which is equivalent to Eq.~(\ref{6.1.5}) or Eq.~(\ref{6.1.6}).

\newpage
\hrule
\hrule
\part{Coordinate systems} 
\label{part2}
\hrule
\hrule
\vspace*{.25in} 
%
\section{Riemann normal coordinates} 
\label{7}

\subsection{Definition and coordinate transformation} 
\label{7.1}

Given a fixed base point $x'$ and a tetrad 
$\base{\alpha'}{\sf a}(x')$, we assign to a neighbouring point $x$ the
four coordinates  
\begin{equation}
\hat{x}^{\sf a} = - \base{\sf a}{\alpha'}(x')\,
\sigma^{\alpha'}(x,x'),   
\label{7.1.1}
\end{equation}
where $\base{\sf a}{\alpha'} = \eta^{\sf ab} g_{\alpha'\beta'}
\base{\beta'}{\sf b}$ is the dual tetrad attached to $x'$. The new  
coordinates $\hat{x}^{\sf a}$ are called {\it Riemann normal
coordinates} (RNC), and they are such that $\eta_{\sf ab} 
\hat{x}^{\sf a} \hat{x}^{\sf b} = \eta_{\sf ab} 
\base{\sf a}{\alpha'} \base{\sf b}{\beta'} \sigma^{\alpha'}
\sigma^{\beta'} = g_{\alpha'\beta'} \sigma^{\alpha'} \sigma^{\beta'}$, 
or  
\begin{equation}
\eta_{\sf ab} \hat{x}^{\sf a} \hat{x}^{\sf b} = 2\sigma(x,x'). 
\label{7.1.2}
\end{equation}
Thus, $\eta_{\sf ab} \hat{x}^{\sf a} \hat{x}^{\sf b}$ is the squared
geodesic distance between $x$ and the base point $x'$. It is obvious
that $x'$ is at the origin of the RNC, where $\hat{x}^{\sf a} = 0$.       

If we move the point $x$ to $x + \delta x$, the new coordinates change
to $\hat{x}^{\sf a} + \delta \hat{x}^{\sf a} = - \base{\sf a}{\alpha'} 
\sigma^{\alpha'}(x+\delta x,x') = \hat{x}^{\sf a} 
- \base{\sf a}{\alpha'} \sigma^{\alpha'}_{\ \beta}\, \delta x^\beta$,
so that   
\begin{equation}
d \hat{x}^{\sf a} = - \base{\sf a}{\alpha'} 
\sigma^{\alpha'}_{\ \beta}\, d x^\beta. 
\label{7.1.3}
\end{equation}
The coordinate transformation is therefore determined by $\partial 
\hat{x}^{\sf a}/\partial x^\beta = -\base{\sf a}{\alpha'} 
\sigma^{\alpha'}_{\ \beta}$, and at coincidence we have 
\begin{equation} 
\biggl[ \frac{\partial \hat{x}^{\sf a}}{\partial x^\alpha} \biggr] = 
\base{\sf a}{\alpha'}, 
\qquad 
\biggl[ \frac{\partial x^\alpha}{\partial \hat{x}^{\sf a}} \biggr] = 
\base{\alpha'}{\sf a}; 
\label{7.1.4}
\end{equation} 
the second result follows from the identities
$\base{\sf a}{\alpha'} \base{\alpha'}{\sf b} = 
\delta^{\sf a}_{\ \sf b}$ and $\base{\alpha'}{\sf a} 
\base{\sf a}{\beta'}  = \delta^{\alpha'}_{\ \beta'}$.  

It is interesting to note that the Jacobian of the
transformation of Eq.~(\ref{7.1.3}), $J \equiv 
\mbox{det}(\partial \hat{x}^{\sf a}/\partial x^\beta)$, is given by $J
= \sqrt{-g} \Delta(x,x')$, where $g$ is the determinant of the metric
in the original coordinates, and $\Delta(x,x')$ is the Van Vleck
determinant of Eq.~(\ref{6.1.2}). This result follows simply by
writing the coordinate transformation in the form $\partial
\hat{x}^{\sf a}/\partial x^\beta = -\eta^{\sf ab} 
\base{\alpha'}{\sf b} \sigma_{\alpha'\beta}$ and computing the product
of the determinants. It allows us to deduce that in the RNC, the
determinant of the metric is given by
\begin{equation} 
\sqrt{-g(\mbox{RNC})} = \frac{1}{\Delta(x,x')}.  
\label{7.1.5}
\end{equation} 
It is easy to show that the geodesics emanating from $x'$ are straight
lines in the RNC. The proper volume of a small comoving region is then
equal to $d V = \Delta^{-1}\, d^4 \hat{x}$, and this is smaller than
the flat-spacetime value of $d^4 \hat{x}$ if $\Delta > 1$, that is, if
the geodesics are focused by the spacetime curvature.       

\subsection{Metric near $x'$} 
\label{7.2}

We now would like to invert Eq.~(\ref{7.1.3}) in order to express the 
line element $ds^2 = g_{\alpha\beta}\, dx^\alpha dx^\beta$ in terms of
the displacements $d\hat{x}^{\sf a}$. We shall do this approximately,
by working in a small neighbourhood of $x'$. We recall the expansion
of Eq.~(\ref{5.2.2}),    
\[
\sigma^{\alpha'}_{\ \beta} = - g^{\beta'}_{\ \beta} 
\biggl( \delta^{\alpha'}_{\ \beta'} + \frac{1}{6}\, 
R^{\alpha'}_{\ \gamma'\beta'\delta'} \sigma^{\gamma'} \sigma^{\delta'}
\biggr) + O(\epsilon^3),
\]
and in this we substitute the frame decomposition of the Riemann
tensor,  
$R^{\alpha'}_{\ \gamma'\beta'\delta'} = R^{\sf a}_{\ \sf cbd}\, 
\base{\alpha'}{\sf a} \base{\sf c}{\gamma'} \base{\sf b}{\beta'}
\base{\sf d}{\delta'}$, and the tetrad decomposition of the parallel  
propagator, $g^{\beta'}_{\ \beta} = \base{\beta'}{\sf b} 
\base{\sf b}{\beta}$, where $\base{\sf b}{\beta}(x)$ is the dual
tetrad at $x$ obtained by parallel transport of 
$\base{\sf b}{\beta'}(x')$. After some algebra we obtain   
\[
\sigma^{\alpha'}_{\ \beta} = 
- \base{\alpha'}{\sf a} \base{\sf a}{\beta} 
- \frac{1}{6}\, R^{\sf a}_{\ \sf cbd}\, \base{\alpha'}{\sf a} 
\base{\sf b}{\beta} \hat{x}^{\sf c} \hat{x}^{\sf d} 
+ O(\epsilon^3),  
\]
where we have used Eq.~(\ref{7.1.1}). Substituting this into
Eq.~(\ref{7.1.3}) yields
\begin{equation}
d \hat{x}^{\sf a} = \biggl[ \delta^{\sf a}_{\ \sf b} 
+ \frac{1}{6}\, R^{\sf a}_{\ \sf cbd} \hat{x}^{\sf c} \hat{x}^{\sf d} 
+ O(x^3) \biggr] \base{\sf b}{\beta}\, d x^\beta,  
\label{7.2.1}
\end{equation} 
and this is easily inverted to give 
\begin{equation}
\base{\sf a}{\alpha}\, d x^\alpha = \biggl[ \delta^{\sf a}_{\ \sf b} - 
\frac{1}{6}\, R^{\sf a}_{\ \sf cbd} \hat{x}^{\sf c} \hat{x}^{\sf d} 
+ O(x^3) \biggr] d \hat{x}^{\sf b}.   
\label{7.2.2}
\end{equation} 
This is the desired approximate inversion of
Eq.~(\ref{7.1.3}). It is useful to note that Eq.~(\ref{7.2.2}), when
specialized from the arbitrary coordinates $x^\alpha$ to 
$\hat{x}^{\sf a}$, gives us the components of the dual tetrad at $x$
in the RNC. 

We are now in a position to calculate the metric in the new
coordinates. We have $ds^2 = g_{\alpha\beta}\, dx^\alpha dx^\beta =
(\eta_{\sf ab} \base{\sf a}{\alpha} \base{\sf b}{\beta}) dx^\alpha
dx^\beta = \eta_{\sf ab} (\base{\sf a}{\alpha}\, dx^\alpha) 
(\base{\sf b}{\beta}\, dx^\beta)$, and in this we substitute
Eq.~(\ref{7.2.2}). The final result is $ds^2 = g_{\sf ab}\, 
d\hat{x}^{\sf a} d\hat{x}^{\sf b}$, with  
\begin{equation}
g_{\sf ab} = \eta_{\sf ab} - \frac{1}{3} R_{\sf acbd} 
\hat{x}^{\sf c} \hat{x}^{\sf d} + O(x^3).  
\label{7.2.3}
\end{equation} 
The quantities $R_{\sf acbd}$ appearing in Eq.~(\ref{7.2.3}) are the 
frame components of the Riemann tensor evaluated at the base point
$x'$,  
\begin{equation}
R_{\sf acbd} = R_{\alpha'\gamma'\beta'\delta'}\, \base{\alpha'}{\sf a} 
\base{\gamma'}{\sf c} \base{\beta'}{\sf b} \base{\delta'}{\sf d}, 
\label{7.2.4}
\end{equation}
and these are independent of $\hat{x}^{\sf a}$. They are also, by
virtue of Eq.~(\ref{7.1.4}), the components of the (base-point)
Riemann tensor in the RNC, because Eq.~(\ref{7.2.4}) can also
be expressed as 
\[
R_{\sf acdb} = R_{\alpha'\gamma'\beta'\delta'} 
\biggl[ \frac{\partial x^\alpha}{\partial \hat{x}^{\sf a}} \biggr] 
\biggl[ \frac{\partial x^\gamma}{\partial \hat{x}^{\sf c}} \biggr] 
\biggl[ \frac{\partial x^\beta}{\partial \hat{x}^{\sf b}} \biggr] 
\biggl[ \frac{\partial x^\delta}{\partial \hat{x}^{\sf d}} \biggr],  
\]
which is the standard transformation law for tensor components.    

It is obvious from Eq.~(\ref{7.2.3}) that $g_{\sf ab}(x') = 
\eta_{\sf ab}$ and $\Gamma^{\sf a}_{\ \sf bc}(x') = 0$, where
$\Gamma^{\sf a}_{\ \sf bc} = -\frac{1}{3} 
(R^{\sf a}_{\ \sf bcd} + R^{\sf a}_{\ \sf cbd}) \hat{x}^{\sf d} 
+ O(x^2)$ is the connection compatible with the metric 
$g_{\sf ab}$. The Riemann normal coordinates therefore provide a
constructive proof of the local flatness theorem.   

\section{Fermi normal coordinates} 
\label{8}

\subsection{Fermi-Walker transport} 
\label{8.1}

Let $\gamma$ be a timelike curve described by parametric relations
$z^\mu(\tau)$ in which $\tau$ is proper time. Let $u^\mu =
dz^\mu/d\tau$ be the curve's normalized tangent vector, and let $a^\mu
= D u^\mu/d\tau$ be its acceleration vector.  

A vector field $v^{\mu}$ is said to be {\it Fermi-Walker transported}
on $\gamma$ if it is a solution to the differential equation 
\begin{equation}
\frac{D v^{\mu}}{d\tau} = \bigl( v_{\nu} a^{\nu} \bigr)
u^{\mu} - \bigl( v_{\nu} u^{\nu} \bigr) a^{\mu}. 
\label{8.1.1}
\end{equation} 
Notice that this reduces to parallel transport if $a^{\mu} = 0$
and $\gamma$ is a geodesic. 

The operation of Fermi-Walker (FW) transport satisfies two important  
properties. The first is that $u^{\mu}$ is automatically FW
transported along $\gamma$; this follows at once from
Eq.~(\ref{8.1.1}) and the fact that $u^{\mu}$ is orthogonal to
$a^{\mu}$. The second is that if the vectors $v^{\mu}$ and
$w^{\mu}$ are both FW transported along $\gamma$, then their inner
product $v_{\mu} w^{\mu}$ is constant on $\gamma$:
$D(v_{\mu} w^{\mu})/d\tau = 0$; this also follows immediately
from Eq.~(\ref{8.1.1}).  

\subsection{Tetrad and dual tetrad on $\gamma$} 
\label{8.2}

Let $\bar{z}$ be an arbitrary reference point on $\gamma$. At this
point we erect an orthonormal tetrad $(u^{\bar{\mu}},
\base{\bar{\mu}}{a})$ where, contrary to former usage, the frame
index $a$ runs from 1 to 3. We then propagate each frame vector on
$\gamma$ by FW transport; this guarantees that the tetrad remains
orthonormal everywhere on $\gamma$. At a generic point $z(\tau)$ we
have  
\begin{equation}
\frac{D \base{\mu}{a}}{d \tau} = \bigl( a_{\nu} \base{\nu}{a}
\bigr) u^{\mu}, \qquad
g_{\mu\nu} u^{\mu} u^{\nu} = -1, \qquad
g_{\mu\nu} \base{\mu}{a} u^{\nu} = 0, \qquad
g_{\mu\nu} \base{\mu}{a} \base{\nu}{b} = \delta_{ab}. 
\label{8.2.1}
\end{equation} 
From the tetrad on $\gamma$ we define a dual tetrad $(\base{0}{\mu}, 
\base{a}{\mu})$ by the relations 
\begin{equation} 
\base{0}{\mu} = - u_{\mu}, \qquad
\base{a}{\mu} = \delta^{ab} g_{\mu\nu} \base{\nu}{b}; 
\label{8.2.2}
\end{equation}
this is also FW transported on $\gamma$. The tetrad and its dual give 
rise to the completeness relations 
\begin{equation}
g^{\mu\nu} = -u^{\mu} u^{\nu} 
+ \delta^{ab} \base{\mu}{a} \base{\nu}{b}, \qquad
g_{\mu\nu} = -\base{0}{\mu} \base{0}{\nu} 
+ \delta_{ab}\, \base{a}{\mu} \base{b}{\nu}.
\label{8.2.3}
\end{equation} 

\subsection{Fermi normal coordinates} 
\label{8.3}

To construct the Fermi normal coordinates (FNC) of a point $x$ in 
the normal convex neighbourhood of $\gamma$ we locate the unique
{\it spacelike geodesic} $\beta$ that passes through $x$ and
intersects $\gamma$ {\it orthogonally}. We denote the intersection
point by $\bar{x} \equiv z(t)$, with $t$ denoting the value of the 
proper-time parameter at this point. To tensors at $\bar{x}$ we
assign indices $\bar{\alpha}$, $\bar{\beta}$, and so on. The FNC of
$x$ are defined by  
\begin{equation}
\hat{x}^0 = t, \qquad
\hat{x}^a = -\base{a}{\bar{\alpha}}(\bar{x}) 
\sigma^{\bar{\alpha}}(x,\bar{x}), \qquad
\sigma_{\bar{\alpha}}(x,\bar{x}) u^{\bar{\alpha}}(\bar{x}) = 0; 
\label{8.3.1}
\end{equation}
the last statement determines $\bar{x}$ from the requirement that 
$-\sigma^{\bar{\alpha}}$, the vector tangent to $\beta$ at $\bar{x}$,
be orthogonal to $u^{\bar{\alpha}}$, the vector tangent to
$\gamma$. From the definition of the FNC and the completeness
relations of Eq.~(\ref{8.2.3}) it follows that 
\begin{equation}
s^2 \equiv \delta_{a b} \hat{x}^a \hat{x}^b = 2 \sigma(x,\bar{x}),  
\label{8.3.2}
\end{equation} 
so that $s$ is the spatial distance between $\bar{x}$ and $x$ along
the geodesic $\beta$. This statement gives an immediate meaning to
$\hat{x}^a$, the spatial Fermi normal coordinates; and the time
coordinate $\hat{x}^0$ is simply proper time at the intersection point
$\bar{x}$. The situation is illustrated in Fig.~6.  

\begin{figure}[b]
\vspace*{2.2in}
\special{hscale=35 vscale=35 hoffset=115.0 voffset=-55.0
         psfile=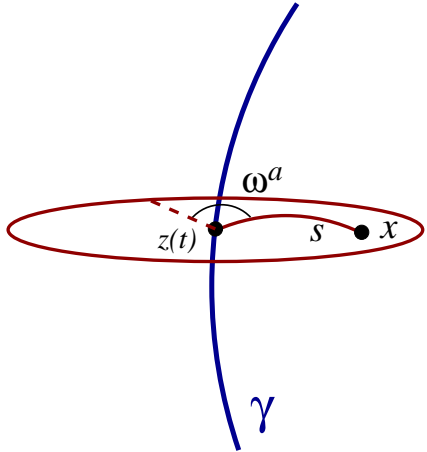}
\caption{Fermi normal coordinates of a point $x$ relative to a world 
line $\gamma$. The time coordinate $t$ selects a particular point on
the word line, and the disk represents the set of spacelike
geodesics that intersect $\gamma$ orthogonally at $z(t)$. The   
unit vector $\omega^a \equiv \hat{x}^a/s$ selects a particular
geodesic among this set, and the spatial distance $s$ selects a
particular point on this geodesic.}    
\end{figure} 

Suppose that $x$ is moved to $x + \delta x$. This typically induces a 
change in the spacelike geodesic $\beta$, which moves to $\beta +
\delta \beta$, and a corresponding change in the intersection point
$\bar{x}$, which moves to $x'' \equiv \bar{x} + \delta \bar{x}$, with
$\delta x^{\bar{\alpha}} = u^{\bar{\alpha}} \delta t$. The FNC of the
new point are then $\hat{x}^0(x+\delta x) = t + \delta t$ and 
$\hat{x}^a(x + \delta x) = -\base{a}{\alpha''}(x'') 
\sigma^{\alpha''}(x + \delta x, x'')$, with $x''$ determined by 
$\sigma_{\alpha''}(x + \delta x,x'') u^{\alpha''}(x'') = 0$. Expanding
these relations to first order in the displacements, and simplifying
using Eqs.~(\ref{8.2.1}), yields      
\begin{equation}
dt = \mu\, \sigma_{\bar{\alpha}\beta} u^{\bar{\alpha}}\, 
d x^{\beta}, \qquad 
d \hat{x}^a = -\base{a}{\bar{\alpha}} \bigl( 
\sigma^{\bar{\alpha}}_{\ \beta} 
+ \mu\, \sigma^{\bar{\alpha}}_{\ \bar{\beta}} u^{\bar{\beta}} 
\sigma_{\beta \bar{\gamma}} u^{\bar{\gamma}} \bigr) d x^{\beta},  
\label{8.3.3}
\end{equation} 
where $\mu$ is determined by $\mu^{-1} = - 
( \sigma_{\bar{\alpha}\bar{\beta}} u^{\bar{\alpha}} u^{\bar{\beta}} 
+ \sigma_{\bar{\alpha}} a^{\bar{\alpha}})$.    

\subsection{Coordinate displacements near $\gamma$}  
\label{8.4}

The relations of Eq.~(\ref{8.3.3}) can be expressed as expansions in
powers of $s$, the spatial distance from $\bar{x}$ to $x$. For this we
use the expansions of Eqs.~(\ref{5.2.1}) and (\ref{5.2.2}), in which
we substitute $\sigma^{\bar{\alpha}} = -\base{\bar{\alpha}}{a}
\hat{x}^a$ and $g^{\bar{\alpha}}_{\ \alpha} = u^{\bar{\alpha}}
\bar{e}^{0}_{\alpha} + \base{\bar{\alpha}}{a} \bar{e}^{a}_{\alpha}$, 
where $(\bar{e}^{0}_{\alpha},\bar{e}^{a}_{\alpha})$ is a dual tetrad
at $x$ obtained by parallel transport of
$(-u_{\bar{\alpha}},\base{a}{\bar{\alpha}})$ on the 
spacelike geodesic $\beta$. After some algebra we obtain  
\[
\mu^{-1} = 1 + a_a \hat{x}^a + \frac{1}{3} R_{0c0d} \hat{x}^c
\hat{x}^d + O(s^3), 
\]
where $a_a(t) \equiv a_{\bar{\alpha}} \base{\bar{\alpha}}{a}$ are
frame components of the acceleration vector, and $R_{0c0d}(t) \equiv 
R_{\bar{\alpha}\bar{\gamma}\bar{\beta}\bar{\delta}} u^{\bar{\alpha}}
\base{\bar{\gamma}}{c} u^{\bar{\beta}} \base{\bar{\delta}}{d}$ are
frame components of the Riemann tensor evaluated on $\gamma$. This
last result is easily inverted to give  
\[
\mu = 1 - a_a \hat{x}^a + \bigl( a_a \hat{x}^a \bigr)^2      
- \frac{1}{3} R_{0c0d} \hat{x}^c \hat{x}^d + O(s^3).
\]
Proceeding similarly for the other relations of Eq.~(\ref{8.3.3}), we 
obtain 
\begin{equation}
dt = \biggl[ 1 - a_a \hat{x}^a 
+ \bigl( a_a \hat{x}^a \bigr)^2       
- \frac{1}{2} R_{0c0d} \hat{x}^c \hat{x}^d + O(s^3) \biggr] 
\bigl( \bar{e}^{0}_{\beta} d x^\beta \bigr) 
+ \biggl[ -\frac{1}{6} R_{0cbd} \hat{x}^c \hat{x}^d + O(s^3) \biggr]  
\bigl( \bar{e}^{b}_{\beta} d x^\beta \bigr)
\label{8.4.1}
\end{equation} 
and
\begin{equation}
d \hat{x}^a = \biggl[ \frac{1}{2} R^a_{\ c0d} \hat{x}^c \hat{x}^d  
+ O(s^3) \biggr] 
\bigl( \bar{e}^{0}_{\beta} d x^\beta \bigr) 
+ \biggl[ \delta^a_{\ b} +\frac{1}{6} R^a_{\ cbd} \hat{x}^c \hat{x}^d  
+ O(s^3) \biggr] 
\bigl( \bar{e}^{b}_{\beta} d x^\beta \bigr),
\label{8.4.2}
\end{equation} 
where $R_{ac0d}(t) \equiv
R_{\bar{\alpha}\bar{\gamma}\bar{\beta}\bar{\delta}}  
\base{\bar{\alpha}}{a} \base{\bar{\gamma}}{c} u^{\bar{\beta}}
\base{\bar{\delta}}{d}$ and $R_{acbd}(t) \equiv 
R_{\bar{\alpha}\bar{\gamma}\bar{\beta}\bar{\delta}}  
\base{\bar{\alpha}}{a} \base{\bar{\gamma}}{c} \base{\bar{\beta}}{b} 
\base{\bar{\delta}}{d}$ are additional frame components of the Riemann 
tensor evaluated on $\gamma$. (Note that frame indices are raised with 
$\delta^{ab}$.) 

As a special case of Eqs.~(\ref{8.4.1}) and (\ref{8.4.2}) we find that  
\begin{equation}
\frac{\partial t}{\partial x^\alpha} \biggr|_\gamma =
-u_{\bar{\alpha}}, \qquad 
\frac{\partial \hat{x}^a}{\partial x^\alpha} \biggr|_\gamma =
\base{a}{\bar{\alpha}},
\label{8.4.3}
\end{equation}
because in the limit $x \to \bar{x}$, the dual tetrad
$(\bar{e}^{0}_{\alpha},\bar{e}^{a}_{\alpha})$ at $x$ coincides with
the dual tetrad $(-u_{\bar{\alpha}},\base{a}{\bar{\alpha}})$ at
$\bar{x}$. It follows that on $\gamma$, the transformation matrix
between the original coordinates $x^\alpha$ and the FNC
$(t,\hat{x}^a)$ is formed by the Fermi-Walker transported tetrad:  
\begin{equation} 
\frac{\partial x^\alpha}{\partial t} \biggr|_\gamma =
u^{\bar{\alpha}}, \qquad
\frac{\partial x^\alpha}{\partial \hat{x}^a} \biggr|_\gamma =
\base{\bar{\alpha}}{a}. 
\label{8.4.4}
\end{equation} 
This implies that the frame components of the acceleration vector,
$a_a(t)$, are also the {\it components} of the acceleration
vector in the FNC; orthogonality between $u^{\bar{\alpha}}$ and
$a^{\bar{\alpha}}$ means that $a_0 = 0$. Similarly, $R_{0c0d}(t)$, 
$R_{0cbd}(t)$, and $R_{acbd}(t)$ are the {\it components} of the
Riemann tensor (evaluated on $\gamma$) in the Fermi normal
coordinates.   

\subsection{Metric near $\gamma$} 
\label{8.5}

Inversion of Eqs.~(\ref{8.4.1}) and (\ref{8.4.2}) gives 
\begin{equation} 
\bar{e}^{0}_{\alpha} dx^\alpha = \biggl[ 1 + a_a \hat{x}^a 
+ \frac{1}{2} R_{0c0d} \hat{x}^c \hat{x}^d 
+ O(s^3) \biggr]\, dt  
+ \biggl[ \frac{1}{6} R_{0cbd} \hat{x}^c \hat{x}^d 
+ O(s^3) \biggr]\, d\hat{x}^b 
\label{8.5.1}
\end{equation}
and 
\begin{equation} 
\bar{e}^{a}_{\alpha} dx^\alpha = \biggl[ \delta^a_{\ b} 
- \frac{1}{6} R^a_{\ cbd} \hat{x}^c \hat{x}^d  
+ O(s^3) \biggr]\, d\hat{x}^b 
+ \biggl[ - \frac{1}{2} R^a_{\ c0d} \hat{x}^c \hat{x}^d  
+ O(s^3) \biggr]\, dt. 
\label{8.5.2}
\end{equation} 
These relations, when specialized to the FNC, give the components of
the dual tetrad at $x$. They can also be used to compute the metric at
$x$, after invoking the completeness relations $g_{\alpha\beta} =
-\bar{e}^{0}_{\alpha} \bar{e}^{0}_{\beta} + \delta_{ab}
\bar{e}^{a}_{\alpha} \bar{e}^{b}_{\beta}$. This gives 
\[
ds^2 = g_{tt}\, dt^2 + 2 g_{ta}\, dt d\hat{x}^a + g_{ab}\,
d\hat{x}^a d\hat{x}^b,   
\]
with 
\begin{eqnarray}
g_{tt} &=& - \Bigl[ 1 + 2 a_a \hat{x}^a 
+ \bigl( a_a \hat{x}^a \bigr)^2 
+ R_{0c0d} \hat{x}^c \hat{x}^d + O(s^3) \Bigr], 
\label{8.5.3} \\ 
g_{ta} &=& - \frac{2}{3} R_{0cad} \hat{x}^c \hat{x}^d + O(s^3),  
\label{8.5.4} \\ 
g_{ab} &=& \delta_{ab} - \frac{1}{3} R_{acbd} \hat{x}^c \hat{x}^d  
+ O(s^3). 
\label{8.5.5}
\end{eqnarray} 
This is the metric near $\gamma$ in the Fermi normal
coordinates. Recall that $a_a(t)$ are the components of the
acceleration vector of $\gamma$ --- the timelike curve described by 
$\hat{x}^a = 0$ --- while $R_{0c0d}(t)$, $R_{0cbd}(t)$,
and $R_{acbd}(t)$ are the components of the Riemann tensor evaluated
on $\gamma$.  

Notice that on $\gamma$, the metric of
Eqs.~(\ref{8.5.3})--(\ref{8.5.5}) reduces to $g_{tt} = -1$ and $g_{ab}
= \delta_{ab}$. On the other hand, the nonvanishing Christoffel
symbols (on $\gamma$) are $\Gamma^t_{\ ta} = \Gamma^a_{\ tt} = a_a$;
these are zero (and the FNC enforce local flatness on the entire
curve) when $\gamma$ is a geodesic.     

\subsection{Thorne-Hartle coordinates}
\label{8.6}

The form of the metric can be simplified if the Ricci tensor
vanishes on the world line:  
\begin{equation}
R_{\mu\nu}(z) = 0. 
\label{8.6.1}
\end{equation}
In such circumstances, a transformation from the Fermi normal
coordinates $(t,\hat{x}^a)$ to the {\it Thorne-Hartle coordinates}
$(t,\hat{y}^a)$ brings the metric to the form 
\begin{eqnarray}
g_{tt} &=& - \Bigl[ 1 + 2 a_a \hat{y}^a 
+ \bigl( a_a \hat{y}^a \bigr)^2 
+ R_{0c0d} \hat{y}^c \hat{y}^d + O(s^3) \Bigr], 
\label{8.6.2} \\ 
g_{ta} &=& -\frac{2}{3} R_{0cad} \hat{y}^c \hat{y}^d + O(s^3),  
\label{8.6.3} \\ 
g_{ab} &=& \delta_{ab} \bigl( 1 
- R_{0c0d} \hat{y}^c \hat{y}^d  \bigr) + O(s^3). 
\label{8.6.4}
\end{eqnarray} 
We see that the transformation leaves $g_{tt}$ and $g_{ta}$ unchanged,
but that it diagonalizes $g_{ab}$. This metric was first displayed in 
Ref.~\cite{thornehartle} and the coordinate transformation was later
produced by Zhang \cite{zhang}. 

The key to the simplification comes from Eq.~(\ref{8.6.1}), which 
dramatically reduces the number of independent components of the 
Riemann tensor. In particular, Eq.~(\ref{8.6.1}) implies that the
frame components $R_{acbd}$ of the Riemann tensor are completely
determined by ${\cal E}_{ab} \equiv R_{0a0b}$, which in this special
case is a symmetric-tracefree tensor. To prove this we invoke the
completeness relations of Eq.~(\ref{8.2.3}) and take frame components
of Eq.~(\ref{8.6.1}). This produces the three independent equations 
\[
\delta^{cd} R_{acbd} = {\cal E}_{ab}, \qquad
\delta^{cd} R_{0cad} = 0, \qquad
\delta^{cd} {\cal E}_{cd} = 0, 
\]
the last of which stating that ${\cal E}_{ab}$ has a vanishing
trace. Taking the trace of the first equation gives 
$\delta^{ab} \delta^{cd} R_{acbd} = 0$, and this implies that
$R_{acbd}$ has five independent components. Since this is also the
number of independent components of ${\cal E}_{ab}$, we see that the
first equation can be inverted --- $R_{acbd}$ can be expressed in
terms of ${\cal E}_{ab}$. A complete listing of the relevant relations
is $R_{1212} = {\cal E}_{11} + {\cal E}_{22} = -{\cal E}_{33}$,
$R_{1213} = {\cal E}_{23}$, $R_{1223} = -{\cal E}_{13}$, $R_{1313} =  
{\cal E}_{11} + {\cal E}_{33} = -{\cal E}_{22}$, $R_{1323} = 
{\cal E}_{12}$,  and $R_{2323} = {\cal E}_{22} + {\cal E}_{33} = -
{\cal E}_{11}$. These are summarized by 
\begin{equation} 
R_{acbd} = \delta_{ab} {\cal E}_{cd} + \delta_{cd} {\cal E}_{ab} -
\delta_{ad} {\cal E}_{bc} - \delta_{bc} {\cal E}_{ad}, 
\label{8.6.5}
\end{equation}
and ${\cal E}_{ab} \equiv R_{0a0b}$ satisfies
$\delta^{ab} {\cal E}_{ab} = 0$. 

We may also note that the relation $\delta^{cd} R_{0cad} = 0$,
together with the usual symmetries of the Riemann tensor, imply that
$R_{0cad}$ too possesses five independent components. These may thus
be related to another symmetric-tracefree tensor ${\cal B}_{ab}$. We 
take the independent components to be $R_{0112} \equiv 
-{\cal B}_{13}$, $R_{0113} \equiv {\cal B}_{12}$, $R_{0123} \equiv  
-{\cal B}_{11}$, $R_{0212} \equiv -{\cal B}_{23}$, and $R_{0213}
\equiv {\cal B}_{22}$, and it is easy to see that all other
components can be expressed in terms of these. For example, $R_{0223}
= -R_{0113} = -{\cal B}_{12}$, $R_{0312} = -R_{0123} + R_{0213} =
{\cal B}_{11} + {\cal B}_{22} = -{\cal B}_{33}$, $R_{0313}
= -R_{0212} = {\cal B}_{23}$, and $R_{0323} = R_{0112} = 
-{\cal B}_{13}$. These relations are summarized by 
\begin{equation}
R_{0abc} = -\varepsilon_{bcd} {\cal B}^d_{\ a}, 
\label{8.6.6}
\end{equation}
where $\varepsilon_{abc}$ is the three-dimensional permutation
symbol. The inverse relation is ${\cal B}^a_{\ b} = \frac{1}{2}
\varepsilon^{acd} R_{0bcd}$.   

Substitution of Eq.~(\ref{8.6.5}) into Eq.~(\ref{8.5.5}) gives  
\[
g_{ab} = \delta_{ab} \Bigl( 1 - \frac{1}{3} {\cal E}_{cd} \hat{x}^c
\hat{x}^d \Bigr) - \frac{1}{3} \bigl( \hat{x}_c \hat{x}^c \bigr)
{\cal E}_{ab} + \frac{1}{3} \hat{x}_a {\cal E}_{bc} \hat{x}^c +  
\frac{1}{3} \hat{x}_b {\cal E}_{ac} \hat{x}^c + O(s^3), 
\]
and we have not yet achieved the simple form of Eq.~(\ref{8.6.4}). The  
missing step is the transformation from the FNC $\hat{x}^a$ to the
Thorne-Hartle coordinates $\hat{y}^a$. This is given by 
\begin{equation}
\hat{y}^a = \hat{x}^a + \xi^a, \qquad
\xi^a = - \frac{1}{6} \bigl( \hat{x}_c \hat{x}^c \bigr)  
{\cal E}_{ab} \hat{x}^b + \frac{1}{3} \hat{x}_a {\cal E}_{bc}
\hat{x}^b \hat{x}^c + O(s^4). 
\label{8.6.7}
\end{equation} 
It is easy to see that this transformation does not affect $g_{tt}$
nor $g_{ta}$ at orders $s$ and $s^2$. The remaining components of the
metric, however, transform according to $g_{ab}(\mbox{THC}) =
g_{ab}(\mbox{FNC}) - \xi_{a;b} - \xi_{b;a}$, where  
\[
\xi_{a;b} = \frac{1}{3} \delta_{ab} {\cal E}_{cd} \hat{x}^c \hat{x}^d
- \frac{1}{6} \bigl( \hat{x}_c \hat{x}^c \bigr) {\cal E}_{ab}  
- \frac{1}{3} {\cal E}_{ac} \hat{x}^c \hat{x}_b 
+ \frac{2}{3} \hat{x}_a {\cal E}_{bc} \hat{x}^c + O(s^3).
\]
It follows that $g_{ab}^{\rm THC} = \delta_{ab}(1 - {\cal E}_{cd}
\hat{y}^c \hat{y}^d) + O(\hat{y}^3)$, which is just the same statement
as in Eq.~(\ref{8.6.4}).   

Alternative expressions for the components of the Thorne-Hartle metric
are  
\begin{eqnarray}
g_{tt} &=& - \Bigl[ 1 + 2 a_a \hat{y}^a 
+ \bigl( a_a \hat{y}^a \bigr)^2 
+ {\cal E}_{ab} \hat{y}^a \hat{y}^b + O(s^3) \Bigr],  
\label{8.6.8} \\ 
g_{ta} &=& - \frac{2}{3} \varepsilon_{abc} {\cal B}^b_{\ d} \hat{y}^c
\hat{y}^d + O(s^3), 
\label{8.6.9} \\ 
g_{ab} &=& \delta_{ab} \bigl( 1 
- {\cal E}_{cd} \hat{y}^c \hat{y}^d  \bigr)  + O(s^3). 
\label{8.6.10}
\end{eqnarray} 

\section{Retarded coordinates}
\label{9}

\subsection{Geometrical elements}
\label{9.1}

We introduce the same geometrical elements as in Sec.~\ref{8}: we have
a timelike curve $\gamma$ described by relations $z^\mu(\tau)$, its
normalized tangent vector $u^\mu = dz^\mu/d\tau$, and its acceleration
vector $a^\mu = D u^\mu/d\tau$. We also have an orthonormal triad 
$\base{\mu}{a}$ that is transported on the world line according to   
\begin{equation}
\frac{D \base{\mu}{a}}{d \tau} = a_a u^{\mu} 
+ \omega_a^{\ b} \base{\mu}{b},  
\label{9.1.1}
\end{equation}  
where $a_a(\tau) = a_{\mu} \base{\mu}{a}$ are the frame components of
the acceleration vector and $\omega_{ab}(\tau) = -\omega_{ba}(\tau)$
is a prescribed rotation tensor. Here the triad is {\it not}
Fermi-Walker transported: for added generality we allow the spatial
vectors to rotate as they are transported on the world line. While
$\omega_{ab}$ will be set to zero in most sections of this paper, the
freedom to perform such a rotation can be useful and will be exploited 
in Sec.~\ref{19}. It is easy to check that Eq.~(\ref{9.1.1}) is
compatible with the requirement that the tetrad
$(u^\mu,\base{\mu}{a})$ be orthonormal everywhere on
$\gamma$. Finally, we have a dual tetrad
$(\base{0}{\mu},\base{a}{\mu})$, with $\base{0}{\mu} = -u_\mu$ and
$\base{a}{\mu} = \delta^{ab} g_{\mu\nu} \base{\nu}{b}$. The tetrad and
its dual give rise to the completeness relations  
\begin{equation}
g^{\mu\nu} = -u^{\mu} u^{\nu} 
+ \delta^{ab} \base{\mu}{a} \base{\nu}{b}, \qquad
g_{\mu\nu} = -\base{0}{\mu} \base{0}{\nu} 
+ \delta_{ab}\, \base{a}{\mu} \base{b}{\nu}, 
\label{9.1.2}
\end{equation} 
which are the same as in Eq.~(\ref{8.2.3}). 

The Fermi normal coordinates of Sec.~\ref{8} were constructed on the
basis of a spacelike geodesic connecting a field point $x$ to the
world line. The retarded coordinates are based
instead on a {\it null geodesic} going from the world line to the
field point. We thus let $x$ be within the normal convex neighbourhood
of $\gamma$, $\beta$ be the unique future-directed null geodesic that
goes from the world line to $x$, and $x' \equiv z(u)$ be the point at
which $\beta$ intersects the world line, with $u$ denoting the value
of the proper-time parameter at this point.   

From the tetrad at $x'$ we obtain another tetrad
$(\base{\alpha}{0},\base{\alpha}{a})$ at $x$ by parallel transport 
on $\beta$. By raising the frame index and lowering the vectorial
index we also obtain a dual tetrad at $x$: $\base{0}{\alpha} =
-g_{\alpha\beta} \base{\beta}{0}$ and $\base{a}{\alpha} = \delta^{ab}
g_{\alpha\beta} \base{\beta}{b}$. The metric at $x$ can be then be
expressed as 
\begin{equation}
g_{\alpha\beta} = -\base{0}{\alpha} \base{0}{\beta} + \delta_{ab} 
\base{a}{\alpha} \base{b}{\beta},   
\label{9.1.3}
\end{equation} 
and the parallel propagator from $x'$ to $x$ is given by 
\begin{equation} 
g^{\alpha}_{\ \alpha'}(x,x') = -\base{\alpha}{0} u_{\alpha'} + 
\base{\alpha}{a} \base{a}{\alpha'}, \qquad 
g^{\alpha'}_{\ \alpha}(x',x) = u^{\alpha'} \base{0}{\alpha} + 
\base{\alpha'}{a} \base{a}{\alpha}. 
\label{9.1.4}  
\end{equation} 
 
\subsection{Definition of the retarded coordinates} 
\label{9.2}

The quasi-Cartesian version of the retarded coordinates are defined by  
\begin{equation}
\hat{x}^0 = u, \qquad
\hat{x}^a = -\base{a}{\alpha'}(x') \sigma^{\alpha'}(x,x'), \qquad 
\sigma(x,x') = 0; 
\label{9.2.1}
\end{equation}
the last statement indicates that $x'$ and $x$ are linked by a null
geodesic. From the fact that $\sigma^{\alpha'}$ is a null vector we
obtain  
\begin{equation} 
r \equiv (\delta_{ab} \hat{x}^a \hat{x}^b)^{1/2} 
= u_{\alpha'} \sigma^{\alpha'}, 
\label{9.2.2}
\end{equation} 
and $r$ is a positive quantity by virtue of the fact that $\beta$ is a 
future-directed null geodesic --- this makes $\sigma^{\alpha'}$
past-directed. In flat spacetime, $\sigma^{\alpha'} =
-(x-x')^{\alpha}$, and in a Lorentz frame that is momentarily comoving
with the world line, $r = t-t' > 0$; with the speed of light set equal
to unity, $r$ is also the spatial distance between $x'$ and $x$ as
measured in this frame. In curved spacetime, the quantity $r =
u_{\alpha'} \sigma^{\alpha'}$ can still be called the {\it retarded 
distance} between the point $x$ and the world line. Another
consequence of Eq.~(\ref{9.2.1}) is that 
\begin{equation}
\sigma^{\alpha'} = -r \bigl( u^{\alpha'} + \Omega^a \base{\alpha'}{a}
\bigr), 
\label{9.2.3}
\end{equation}
where $\Omega^a \equiv \hat{x}^a/r$ is a spatial vector that satisfies  
$\delta_{ab} \Omega^a \Omega^b = 1$. 

A straightforward calculation reveals that under a displacement of the
point $x$, the retarded coordinates change according to  
\begin{equation} 
d u = -k_\alpha\, d x^\alpha, \qquad 
d \hat{x}^a = - \bigl( r a^a - \omega^a_{\ b} \hat{x}^b 
+ \base{a}{\alpha'} \sigma^{\alpha'}_{\ \beta'} u^{\beta'} 
\bigr)\, du 
- \base{a}{\alpha'} \sigma^{\alpha'}_{\ \beta}\, d x^\beta, 
\label{9.2.4}
\end{equation}
where $k_\alpha = \sigma_{\alpha}/r$ is a future-directed null vector
at $x$ that is tangent to the geodesic $\beta$. To obtain these
results we must keep in mind that a displacement of $x$ typically
induces a simultaneous displacement of $x'$ because the new points 
$x + \delta x$ and $x' + \delta x'$ must also be linked by a null 
geodesic. We therefore have $0 = \sigma(x+\delta x, x' + \delta x') 
= \sigma_{\alpha}\, \delta x^\alpha + \sigma_{\alpha'}\, 
\delta x^{\alpha'}$, and the first relation of Eq.~(\ref{9.2.4}) 
follows from the fact that a displacement along the world line is
described by $\delta x^{\alpha'} = u^{\alpha'}\, \delta u$.  
 
\subsection{The scalar field $r(x)$ and the vector field
$k^\alpha(x)$}  
\label{9.3}

If we keep $x'$ linked to $x$ by the relation $\sigma(x,x') = 0$, then 
the quantity 
\begin{equation}
r(x) = \sigma_{\alpha'}(x,x') u^{\alpha'}(x') 
\label{9.3.1}
\end{equation}
can be viewed as an ordinary scalar field defined in a neighbourhood
of $\gamma$. We can compute the gradient of $r$ by finding how $r$
changes under a displacement of $x$ (which again induces a
displacement of $x'$). The result is 
\begin{equation} 
\partial_\beta r = - \bigl( \sigma_{\alpha'} a^{\alpha'} +
\sigma_{\alpha'\beta'} u^{\alpha'} u^{\beta'} \bigr) k_\beta +
\sigma_{\alpha'\beta} u^{\alpha'}.
\label{9.3.2}
\end{equation} 

Similarly, we can view 
\begin{equation} 
k^{\alpha}(x) = \frac{\sigma^{\alpha}(x,x')}{r(x)} 
\label{9.3.3}
\end{equation}
as an ordinary vector field, which is tangent to the congruence of
null geodesics that emanate from $x'$. It is easy to check that this 
vector satisfies the identities 
\begin{equation} 
\sigma_{\alpha\beta} k^\beta = k_\alpha, \qquad 
\sigma_{\alpha'\beta} k^\beta = \frac{\sigma_{\alpha'}}{r}, 
\label{9.3.4}
\end{equation}
from which we also obtain $\sigma_{\alpha'\beta} u^{\alpha'} k^\beta 
= 1$. From this last result and Eq.~(\ref{9.3.2}) we deduce the
important relation  
\begin{equation}
k^\alpha \partial_\alpha r = 1. 
\label{9.3.5}
\end{equation} 
In addition, combining the general statement $\sigma^{\alpha} = 
-g^{\alpha}_{\ \alpha'} \sigma^{\alpha'}$ --- cf.\ Eq.~(\ref{4.2.8})
--- with Eq.~(\ref{9.2.3}) gives
\begin{equation} 
k^\alpha = g^{\alpha}_{\ \alpha'} \bigl( u^{\alpha'} + \Omega^a
\base{\alpha'}{a} \bigr); 
\label{9.3.6}
\end{equation} 
the vector at $x$ is therefore obtained by parallel transport of
$u^{\alpha'} + \Omega^a \base{\alpha'}{a}$ on $\beta$. From this and
Eq.~(\ref{9.1.4}) we get the alternative expression 
\begin{equation} 
k^\alpha = \base{\alpha}{0} + \Omega^a \base{\alpha}{a},  
\label{9.3.7}
\end{equation} 
which confirms that $k^\alpha$ is a future-directed null vector field
(recall that $\Omega^a = \hat{x}^a/r$ is a unit vector).    

The covariant derivative of $k_\alpha$ can be computed by finding 
how the vector changes under a displacement of $x$. (It is in fact
easier to first calculate how $r k_\alpha$ changes, and then
substitute our previous expression for $\partial_\beta r$.) The result
is  
\begin{equation} 
r k_{\alpha;\beta} = \sigma_{\alpha\beta} - k_{\alpha} \sigma_{\beta
\gamma'} u^{\gamma'} - k_{\beta} \sigma_{\alpha \gamma'} u^{\gamma'} +
\bigl( \sigma_{\alpha'} a^{\alpha'} + \sigma_{\alpha'\beta'}
u^{\alpha'} u^{\beta'} \bigr) k_{\alpha} k_{\beta}. 
\label{9.3.8}
\end{equation} 
From this we infer that $k^\alpha$ satisfies the geodesic equation in
affine-parameter form, $k^\alpha_{\ ;\beta} k^\beta = 0$, and
Eq.~(\ref{9.3.5}) informs us that the affine parameter is in fact
$r$. A displacement along a member of the congruence is therefore
given by $dx^\alpha = k^\alpha\, dr$. Specializing to retarded
coordinates, and using Eqs.~(\ref{9.2.4}) and (\ref{9.3.4}), we find
that this statement becomes $du = 0$ and $d\hat{x}^a = (\hat{x}^a/r)\,
dr$, which integrate to $u = \mbox{constant}$ and $\hat{x}^a = r
\Omega^a$, respectively, with $\Omega^a$ still denoting a constant
unit vector. We have found that the congruence of null geodesics
emanating from $x'$ is described by   
\begin{equation}
u = \mbox{constant}, \qquad
\hat{x}^a = r \Omega^a(\theta^A)
\label{9.3.9}
\end{equation} 
in the retarded coordinates. Here, the two angles $\theta^A$ ($A = 1,
2$) serve to parameterize the unit vector $\Omega^a$, which is
independent of $r$.       

Equation (\ref{9.3.8}) also implies that the expansion of the
congruence is given by 
\begin{equation} 
\theta = k^\alpha_{\ ;\alpha} = \frac{\sigma^\alpha_{\ \alpha} 
- 2}{r}.
\label{9.3.10}
\end{equation} 
Using the expansion for $\sigma^\alpha_{\ \alpha}$ given by
Eq.~(\ref{5.2.4}), we find that this becomes $r\theta = 2 
- \frac{1}{3} R_{\alpha'\beta'} \sigma^{\alpha'} \sigma^{\beta'} 
+ O(r^3)$, or 
\begin{equation}
r \theta =  
2 - \frac{1}{3} r^2 \bigl( R_{00} + 2 R_{0a} \Omega^a + R_{ab}
\Omega^a \Omega^b \bigr) + O(r^3)
\label{9.3.11} 
\end{equation}
after using Eq.~(\ref{9.2.3}). Here, 
$R_{00} = R_{\alpha'\beta'} u^{\alpha'} u^{\beta'}$,
$R_{0a} = R_{\alpha'\beta'} u^{\alpha'} \base{\beta'}{a}$, and  
$R_{ab} = R_{\alpha'\beta'} \base{\alpha'}{a} \base{\beta'}{b}$
are the frame components of the Ricci tensor evaluated at $x'$. This
result confirms that the congruence is singular at $r=0$, because
$\theta$ diverges as $2/r$ in this limit; the caustic coincides with
the point $x'$.     

Finally, we infer from Eq.~(\ref{9.3.8}) that $k^{\alpha}$ is 
hypersurface orthogonal. This, together with the property that
$k^\alpha$ satisfies the geodesic equation in affine-parameter form, 
implies that there exists a scalar field $u(x)$ such that 
\begin{equation}
k_\alpha = -\partial_\alpha u. 
\label{9.3.12}
\end{equation} 
This scalar field was already identified in Eq.~(\ref{9.2.4}): it is 
numerically equal to the proper-time parameter of the world line at
$x'$. We conclude that the geodesics to which $k^\alpha$ is
tangent are the generators of the null cone $u = \mbox{constant}$. As 
Eq.~(\ref{9.3.9}) indicates, a specific generator is selected by
choosing a direction $\Omega^a$ (which can be parameterized by two
angles $\theta^A$), and $r$ is an affine parameter on each
generator. The geometrical meaning of the retarded coordinates is now
completely clear; it is illustrated in Fig.~7. 

\begin{figure}[t]
\vspace*{2.2in}
\special{hscale=35 vscale=35 hoffset=115.0 voffset=-55.0
         psfile=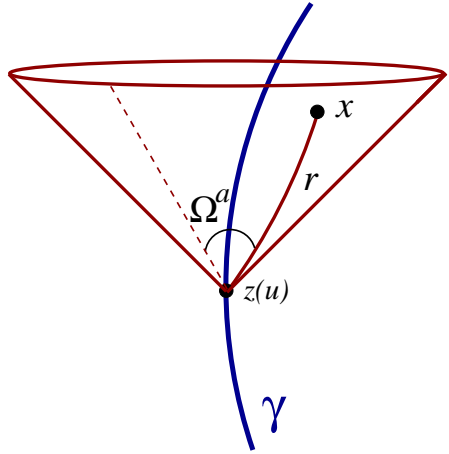}
\caption{Retarded coordinates of a point $x$ relative to a world
line $\gamma$. The retarded time $u$ selects a particular null cone,
the unit vector $\Omega^a \equiv \hat{x}^a/r$ selects a particular
generator of this null cone, and the retarded distance $r$ selects a
particular point on this generator. This figure is identical to
Fig.~4.} 
\end{figure} 

\subsection{Frame components of tensor fields on the world line}   
\label{9.4}

The metric at $x$ in the retarded coordinates will be expressed in
terms of frame components of vectors and tensors evaluated on the
world line $\gamma$. For example, if $a^{\alpha'}$ is the acceleration
vector at $x'$, then as we have seen,  
\begin{equation}
a_a(u) = a_{\alpha'}\, \base{\alpha'}{a}
\label{9.4.1}
\end{equation}
are the frame components of the acceleration at proper time $u$.  

Similarly,   
\begin{equation}
R_{a0b0}(u) = R_{\alpha'\gamma'\beta'\delta'}\, \base{\alpha'}{a} 
u^{\gamma'} \base{\beta'}{b} u^{\delta'}, \quad
R_{a0bd}(u) = R_{\alpha'\gamma'\beta'\delta'}\, \base{\alpha'}{a}
u^{\gamma'} \base{\beta'}{b} \base{\delta'}{d}, \quad
R_{acbd}(u) = R_{\alpha'\gamma'\beta'\delta'}\, \base{\alpha'}{a}
\base{\gamma'}{c} \base{\beta'}{b} \base{\delta'}{d} 
\label{9.4.2}
\end{equation}
are the frame components of the Riemann tensor evaluated on
$\gamma$. From these we form the useful combinations  
\begin{eqnarray}
S_{ab}(u,\theta^A) &=& R_{a0b0} + R_{a0bc} \Omega^c 
+ R_{b0ac} \Omega^c + R_{acbd} \Omega^c \Omega^d = S_{ba}, 
\label{9.4.3} \\
S_{a}(u,\theta^A) &=& S_{ab}\Omega^b = R_{a0b0} \Omega^b - R_{ab0c} 
\Omega^b \Omega^c, 
\label{9.4.4} \\
S(u,\theta^A) &=& S_{a} \Omega^a = R_{a0b0} \Omega^a \Omega^b, 
\label{9.4.5}
\end{eqnarray}
in which the quantities $\Omega^a \equiv \hat{x}^a / r$ depend on the 
angles $\theta^A$ only --- they are independent of $u$ and $r$. 

We have previously introduced the frame components of the Ricci tensor
in Eq.~(\ref{9.3.11}). The identity  
\begin{equation}    
R_{00} + 2 R_{0a} \Omega^a + R_{ab} \Omega^a \Omega^b 
= \delta^{ab} S_{ab} - S
\label{9.4.6}
\end{equation}
follows easily from Eqs.~(\ref{9.4.3})--(\ref{9.4.5}) and the
definition of the Ricci tensor. 

In Sec.~\ref{8} we saw that the frame components of a given tensor
were also the components of this tensor (evaluated on the world line)
in the Fermi normal coordinates. We should not expect this
property to be true also in the case of the retarded coordinates: 
{\it the frame components  of a tensor are not to be
identified with the components of this tensor in the retarded
coordinates}. The reason is that the retarded coordinates are in fact 
{\it singular} on the world line. As we shall see, they give rise to a 
metric that possesses a directional ambiguity at $r = 0$. (This can 
easily be seen in Minkowski spacetime by performing the coordinate
transformation $u = t - \sqrt{x^2+y^2+z^2}$.) Components of tensors
are therefore not defined on the world line, although they are
perfectly well defined for $r \neq 0$. Frame components, on the other
hand, are well defined both off and on the world line, and working
with them will eliminate any difficulty associated with the singular
nature of the retarded coordinates.      

\subsection{Coordinate displacements near $\gamma$} 
\label{9.5}

The changes in the quasi-Cartesian retarded coordinates under a
displacement of $x$ are given by Eq.~(\ref{9.2.4}). In these we
substitute the standard expansions for $\sigma_{\alpha'\beta'}$ and
$\sigma_{\alpha'\beta}$, as given by Eqs.~(\ref{5.2.1}) and
(\ref{5.2.2}), as well as Eqs.~(\ref{9.2.3}) and
(\ref{9.3.6}). After a straightforward (but fairly lengthy)
calculation, we obtain the following expressions for the coordinate
displacements:  
\begin{eqnarray}
du &=& \bigl( \base{0}{\alpha}\, dx^\alpha \bigr) 
- \Omega_a \bigl( \base{b}{\alpha}\, dx^\alpha \bigr),  
\label{9.5.1} \\ 
d\hat{x}^a &=& - \Bigl[ r a^a - r \omega^a_{\ b} \Omega^b 
+ \frac{1}{2} r^2 S^a + O(r^3) \Bigr]
\bigl( \base{0}{\alpha}\, dx^\alpha \bigr) 
\nonumber \\ 
& & \mbox{} + \Bigl[ \delta^a_{\ b} + \Bigl( r a^a - r \omega^a_{\ c}
\Omega^c + \frac{1}{3} r^2 S^a \Bigr) \Omega_b + \frac{1}{6} r^2
S^a_{\ b} + O(r^3) \Bigr] \bigl( \base{b}{\alpha}\, dx^\alpha \bigr).    
\label{9.5.2}
\end{eqnarray} 
Notice that the result for $du$ is exact, but that $d\hat{x}^a$ is
expressed as an expansion in powers of $r$. 

These results can also be expressed in the form of gradients of the
retarded coordinates: 
\begin{eqnarray} 
\partial_\alpha u &=& \base{0}{\alpha} - \Omega_a \base{a}{\alpha},   
\label{9.5.3} \\ 
\partial_\alpha \hat{x}^a &=& - \Bigl[ r a^a - r \omega^a_{\ b}
\Omega^b + \frac{1}{2} r^2 S^a + O(r^3) \Bigr] \base{0}{\alpha} 
\nonumber \\ & & \mbox{} 
+ \Bigl[ \delta^a_{\ b} + \Bigl( r a^a - r \omega^a_{\ c}
\Omega^c + \frac{1}{3} r^2 S^a \Bigr) \Omega_b 
+ \frac{1}{6} r^2 S^a_{\ b} + O(r^3) \Bigr] \base{b}{\alpha}. 
\label{9.5.4}
\end{eqnarray} 
Notice that Eq.~(\ref{9.5.3}) follows immediately from
Eqs.~(\ref{9.3.7}) and (\ref{9.3.12}). From Eq.~(\ref{9.5.4}) and the 
identity $\partial_\alpha r = \Omega_a \partial_\alpha \hat{x}^a$ we
also infer 
\begin{equation} 
\partial_\alpha r = - \Bigl[ r a_a \Omega^a + \frac{1}{2} r^2 S 
+ O(r^3) \Bigr] \base{0}{\alpha} + \Bigl[ \Bigl(1 + r a_b \Omega^b 
+ \frac{1}{3} r^2 S \Bigr) \Omega_a + \frac{1}{6} r^2 S_a + O(r^3)
\Bigr] \base{a}{\alpha}, 
\label{9.5.5}
\end{equation} 
where we have used the facts that $S_a = S_{ab} \Omega^b$ and $S = S_a 
\Omega^a$; these last results were derived in Eqs.~(\ref{9.4.4}) and
(\ref{9.4.5}). It may be checked that Eq.~(\ref{9.5.5}) agrees with
Eq.~(\ref{9.3.2}).  
    
\subsection{Metric near $\gamma$}   
\label{9.6}

It is straightforward (but fairly tedious) to invert the relations of
Eqs.~(\ref{9.5.1}) and (\ref{9.5.2}) and solve for $\base{0}{\alpha}\, 
dx^\alpha$ and $\base{a}{\alpha}\, dx^\alpha$. The results are  
\begin{eqnarray} 
\base{0}{\alpha}\, dx^\alpha &=& \Bigl[ 1 + r a_a \Omega^a + \frac{1}{2} 
r^2 S + O(r^3) \Bigr]\, du + \Bigl[ \Bigl( 1 + \frac{1}{6} r^2 S
\Bigr) \Omega_a - \frac{1}{6} r^2 S_a + O(r^3) \Bigr]\, d\hat{x}^a, 
\label{9.6.1} \\ 
\base{a}{\alpha}\, dx^\alpha &=& \Bigl[ r \bigl(a^a - \omega^a_{\ b}
\Omega^b \bigr) + \frac{1}{2} r^2 S^a + O(r^3) \Bigr]\, du 
+ \Bigl[ \delta^a_{\ b} - \frac{1}{6} r^2 S^a_{\ b} 
+ \frac{1}{6} r^2 S^a \Omega_b + O(r^3) \Bigr]\, d\hat{x}^b.
\label{9.6.2}
\end{eqnarray}  
These relations, when specialized to the retarded coordinates, give us 
the components of the dual tetrad $(\base{0}{\alpha},
\base{a}{\alpha})$ at $x$. The metric is then computed by using the 
completeness relations of Eq.~(\ref{9.1.3}). We find 
\[
ds^2 = g_{uu}\, du^2 + 2 g_{ua}\, du d\hat{x}^a + g_{ab}\, d\hat{x}^a 
d\hat{x}^b,
\]
with    
\begin{eqnarray}
g_{uu} &=& - \bigl( 1 + r a_a \Omega^a \bigr)^2 + r^2 
\bigl(a_a - \omega_{ab} \Omega^b \bigr) \bigl(a^a - \omega^a_{\ c}
\Omega^c \bigr) - r^2 S + O(r^3),  
\label{9.6.3} \\ 
g_{ua} &=& -\Bigl( 1 + r a_b \Omega^b + \frac{2}{3} r^2 S \Bigr) 
\Omega_a + r \bigl(a_a - \omega_{ab} \Omega^b \bigr) 
+ \frac{2}{3} r^2 S_a + O(r^3), 
\label{9.6.4} \\
g_{ab} &=& \delta_{ab} - \Bigl( 1 + \frac{1}{3} r^2 S \Bigr) \Omega_a 
\Omega_b - \frac{1}{3} r^2 S_{ab} + \frac{1}{3} r^2 \bigl( S_a
\Omega_b + \Omega_a S_b \bigr) + O(r^3).  
\label{9.6.5}
\end{eqnarray} 
We see (as was pointed out in Sec.~\ref{9.4}) that the metric
possesses a directional ambiguity on the world line: the metric at
$r=0$ still depends on the vector $\Omega^a = \hat{x}^a/r$ that
specifies the direction to the point $x$. The retarded coordinates are
therefore singular on the world line, and tensor components cannot be
defined on $\gamma$. 

By setting $S_{ab} = S_a = S = 0$ in Eqs.~(\ref{9.6.3})--(\ref{9.6.5})
we obtain the metric of flat spacetime in the retarded
coordinates. This we express as  
\begin{eqnarray} 
\eta_{uu} &=& - \bigl( 1 + r a_a \Omega^a \bigr)^2 
+ r^2 \bigl(a_a - \omega_{ab} \Omega^b \bigr) 
\bigl(a^a - \omega^a_{\ c} \Omega^c \bigr), \nonumber \\   
\eta_{ua} &=& - \bigl( 1 + r a_b \Omega^b \bigr) \Omega_a 
+ r \bigl(a_a - \omega_{ab} \Omega^b \bigr),
\label{9.6.6} \\ 
\eta_{ab} &=& \delta_{ab} - \Omega_a \Omega_b. \nonumber 
\end{eqnarray}
In spite of the directional ambiguity, the metric of flat spacetime
has a unit determinant everywhere, and it is easily inverted:   
\begin{equation}
\eta^{uu} = 0, \qquad
\eta^{ua} = -\Omega^a, \qquad
\eta^{ab} = \delta^{ab} + r \bigl(a^a - \omega^a_{\ c}
\Omega^c \bigr) \Omega^b + r \Omega^a \bigl(a^b - \omega^b_{\ c}
\Omega^c \bigr).  
\label{9.6.7}
\end{equation}
The inverse metric also is ambiguous on the world line.  

To invert the curved-spacetime metric of
Eqs.~(\ref{9.6.3})--(\ref{9.6.5}) we express it as $g_{\alpha\beta} =
\eta_{\alpha\beta} + h_{\alpha\beta} + O(r^3)$ and treat
$h_{\alpha\beta} = O(r^2)$ as a perturbation. The inverse metric is
then $g^{\alpha\beta} = \eta^{\alpha\beta} - \eta^{\alpha\gamma}
\eta^{\beta\delta} h_{\gamma \delta} + O(r^3)$, or  
\begin{eqnarray} 
\hspace*{-15pt} g^{uu} &=& 0, 
\label{9.6.8} \\
\hspace*{-15pt} g^{ua} &=& -\Omega^a, 
\label{9.6.9} \\
\hspace*{-15pt} g^{ab} &=& \delta^{ab} + r \bigl(a^a - \omega^a_{\ c}
\Omega^c \bigr) \Omega^b + r \Omega^a \bigl(a^b - \omega^b_{\ c}
\Omega^c \bigr) + \frac{1}{3} r^2 S^{ab} 
+ \frac{1}{3} r^2 \bigl( S^a \Omega^b + \Omega^a S^b \bigr) + O(r^3).  
\label{9.6.10}
\end{eqnarray} 
The results for $g^{uu}$ and $g^{ua}$ are exact, and they follow from 
the general relations $g^{\alpha\beta} (\partial_\alpha u)
(\partial_\beta u) = 0$ and $g^{\alpha\beta} (\partial_\alpha u)
(\partial_\beta r) = -1$ that are derived from Eqs.~(\ref{9.3.5}) and 
(\ref{9.3.12}).  

The metric determinant is computed from $\sqrt{-g} = 1 + \frac{1}{2} 
\eta^{\alpha\beta} h_{\alpha\beta} + O(r^3)$, which gives 
\begin{equation}
\sqrt{-g} = 1 - \frac{1}{6} r^2 \bigl( \delta^{ab} S_{ab} - S \bigr) +
O(r^3) = 1 - \frac{1}{6} r^2 \bigl( R_{00} + 2 R_{0a} \Omega^a +
R_{ab} \Omega^a \Omega^b \bigr) + O(r^3), 
\label{9.6.11}
\end{equation}
where we have substituted the identity of
Eq.~(\ref{9.4.6}). Comparison with Eq.~(\ref{9.3.11}) then gives us
the interesting relation $\sqrt{-g} = \frac{1}{2} r \theta + O(r^3)$,
where $\theta$ is the expansion of the generators of the null cones 
$u = \mbox{constant}$.  

\subsection{Transformation to angular coordinates} 
\label{9.7}

Because the vector $\Omega^a = \hat{x}^a/r$ satisfies $\delta_{ab}
\Omega^a \Omega^b = 1$, it can be parameterized by two angles
$\theta^A$. A canonical choice for the parameterization is $\Omega^a =
(\sin\theta \cos\phi, \sin\theta \sin\phi, \cos\theta)$. It is then
convenient to perform a coordinate transformation from $\hat{x}^a$ to 
$(r,\theta^A)$, using the relations $\hat{x}^a = r
\Omega^a(\theta^A)$. (Recall from Sec.~\ref{9.3} that the angles  
$\theta^A$ are constant on the generators of the null cones $u =
\mbox{constant}$, and that $r$ is an affine parameter on these
generators. The relations $\hat{x}^a = r \Omega^a$ therefore
describe the behaviour of the generators.) The differential form of
the coordinate transformation is     
\begin{equation}
d\hat{x}^a = \Omega^a\, dr + r \Omega^a_A\, d\theta^A, 
\label{9.7.1}
\end{equation}
where the transformation matrix 
\begin{equation} 
\Omega^a_A \equiv \frac{\partial \Omega^a}{\partial \theta^A}
\label{9.7.2}
\end{equation} 
satisfies the identity $\Omega_a \Omega^a_A = 0$. 

We introduce the quantities 
\begin{equation}
\Omega_{AB} = \delta_{ab} \Omega^a_A \Omega^b_B, 
\label{9.7.3}
\end{equation}
which act as a (nonphysical) metric in the subspace spanned by the
angular coordinates. In the canonical parameterization, $\Omega_{AB} = 
\mbox{diag}(1,\sin^2\theta)$. We use the inverse of $\Omega_{AB}$, 
denoted $\Omega^{AB}$, to raise upper-case latin indices. We then
define the new object 
\begin{equation}
\Omega^A_a = \delta_{ab} \Omega^{AB} \Omega^b_B 
\label{9.7.4}
\end{equation} 
which satisfies the identities 
\begin{equation}
\Omega^A_a \Omega^a_B = \delta^A_B, \qquad
\Omega^a_A \Omega^A_b = \delta^a_{\ b} - \Omega^a \Omega_b. 
\label{9.7.5}
\end{equation} 
The second result follows from the fact that both sides are
simultaneously symmetric in $a$ and $b$, orthogonal to $\Omega_a$ and 
$\Omega^b$, and have the same trace.  

From the preceding results we establish that the transformation from
$\hat{x}^a$ to $(r,\theta^A)$ is accomplished by 
\begin{equation}
\frac{\partial \hat{x}^a}{\partial r} = \Omega^a, \qquad
\frac{\partial \hat{x}^a}{\partial \theta^A} = r \Omega^a_A, 
\label{9.7.6}
\end{equation}
while the transformation from $(r,\theta^A)$ to $\hat{x}^a$ is
accomplished by 
\begin{equation}
\frac{\partial r}{\partial \hat{x}^a} = \Omega_a, \qquad 
\frac{\partial \theta^A}{\partial \hat{x}^a} = \frac{1}{r}\Omega^A_a.  
\label{9.7.7}
\end{equation}
With these transformation rules it is easy to show that in the
angular coordinates, the metric takes the form of 
\[
ds^2 = g_{uu}\, du^2 + 2 g_{ur}\, dudr + 2 g_{uA}\, du d\theta^A 
+ g_{AB}\, d\theta^A d\theta^B,
\]
with  
\begin{eqnarray}
g_{uu} &=& - \bigl( 1 + r a_a \Omega^a \bigr)^2 
+ r^2 \bigl(a_a - \omega_{ab} \Omega^b \bigr) 
\bigl(a^a - \omega^a_{\ c} \Omega^c \bigr) - r^2 S + O(r^3),  
\label{9.7.8} \\
g_{ur} &=& -1, 
\label{9.7.9} \\ 
g_{uA} &=& r \Bigl[ r \bigl(a_a - \omega_{ab} \Omega^b \bigr) 
+ \frac{2}{3} r^2 S_a + O(r^3) \Bigr] \Omega^a_A, 
\label{9.7.10} \\ 
g_{AB} &=& r^2 \Bigl[ \Omega_{AB} - \frac{1}{3} r^2 S_{ab} \Omega^a_A
\Omega^b_B + O(r^3) \Bigr]. 
\label{9.7.11}
\end{eqnarray} 
The results $g_{ru} = -1$, $g_{rr} = 0$, and $g_{rA} = 0$ are exact,
and they follow from the fact that in the retarded coordinates,
$k_\alpha\, dx^\alpha = - du$ and $k^\alpha \partial_\alpha =
\partial_r$.   

The nonvanishing components of the inverse metric are  
\begin{eqnarray}
g^{ur} &=& -1, 
\label{9.7.12} \\ 
g^{rr} &=& 1 + 2r a_a \Omega^a + r^2 S + O(r^3),   
\label{9.7.13} \\ 
g^{rA} &=& \frac{1}{r} \Bigl[ r \bigl(a^a - \omega^a_{\ b} \Omega^b
\bigr) + \frac{2}{3}r^2 S^a + O(r^3) \Bigr] \Omega^A_a, 
\label{9.7.14} \\
g^{AB} &=& \frac{1}{r^2} \Bigl[ \Omega^{AB} + \frac{1}{3} r^2 S^{ab}
\Omega^A_a \Omega^B_b + O(r^3) \Bigr].  
\label{9.7.15}
\end{eqnarray} 
The results $g^{uu} = 0$, $g^{ur} = -1$, and $g^{uA} = 0$ are exact,
and they follow from the same reasoning as before.  

Finally, we note that in the angular coordinates, the metric
determinant is given by  
\begin{equation}
\sqrt{-g} = r^2 \sqrt{\Omega} \Bigl[ 1 - \frac{1}{6} r^2 \bigl( R_{00}
+ 2 R_{0a} \Omega^a + R_{ab} \Omega^a \Omega^b \bigr) + O(r^3) \Bigr], 
\label{9.7.16}
\end{equation}
where $\Omega$ is the determinant of $\Omega_{AB}$; in the canonical 
parameterization, $\sqrt{\Omega} = \sin\theta$. 
 
\subsection{Specialization to $a^\mu = 0 = R_{\mu\nu}$}  
\label{9.8} 

In this subsection we specialize our previous results to a situation 
where $\gamma$ is a geodesic on which the Ricci tensor vanishes. We
therefore set $a^\mu = 0 = R_{\mu\nu}$ everywhere on $\gamma$, and for
simplicity we also set $\omega_{ab}$ to zero.        

We have seen in Sec.~\ref{8.6} that when the Ricci tensor vanishes on
$\gamma$, all frame components of the Riemann tensor can be expressed
in terms of the symmetric-tracefree tensors 
${\cal E}_{ab}(u)$ and ${\cal B}_{ab}(u)$. The relations are 
$R_{a0b0} = {\cal E}_{ab}$, $R_{a0bc} = \varepsilon_{bcd} 
{\cal B}^d_{\ a}$, and $R_{acbd} = \delta_{ab} {\cal E}_{cd} 
+ \delta_{cd} {\cal E}_{ab} - \delta_{ad} {\cal E}_{bc} 
- \delta_{bc} {\cal E}_{ad}$. These can be substituted into 
Eqs.~(\ref{9.4.3})--(\ref{9.4.5}) to give 
\begin{eqnarray}
S_{ab}(u,\theta^A) &=& 2 {\cal E}_{ab} 
- \Omega_a {\cal E}_{bc} \Omega^c 
- \Omega_b {\cal E}_{ac} \Omega^c 
+ \delta_{ab} {\cal E}_{bc} \Omega^c \Omega^d 
+ \varepsilon_{acd} \Omega^c {\cal B}^d_{\ b} 
+ \varepsilon_{bcd} \Omega^c {\cal B}^d_{\ a}, 
\label{9.8.1} \\ 
S_{a}(u,\theta^A) &=& {\cal E}_{ab} \Omega^b 
+ \varepsilon_{abc} \Omega^b {\cal B}^c_{\ d} \Omega^d, 
\label{9.8.2} \\ 
S(u,\theta^A) &=& {\cal E}_{ab} \Omega^a \Omega^b.  
\label{9.8.3}
\end{eqnarray} 
In these expressions the dependence on retarded time $u$ is contained
in ${\cal E}_{ab}$ and ${\cal B}_{ab}$, while the angular dependence
is encoded in the unit vector $\Omega^a$. 

It is convenient to introduce the irreducible quantities 
\begin{eqnarray}  
{\cal E}^* &=& {\cal E}_{ab} \Omega^a \Omega^b, 
\label{9.8.4} \\ 
{\cal E}^*_a &=& \bigl(\delta_a^{\ b} - \Omega_a \Omega^b \bigr) 
{\cal E}_{bc} \Omega^c,
\label{9.8.5} \\ 
{\cal E}^*_{ab} &=& 2 {\cal E}_{ab} 
- 2\Omega_a {\cal E}_{bc} \Omega^c  
- 2\Omega_b {\cal E}_{ac} \Omega^c
+ (\delta_{ab} + \Omega_a \Omega_b) {\cal E}^*, 
\label{9.8.6} \\ 
{\cal B}^*_a &=& \varepsilon_{abc} \Omega^b {\cal B}^c_{\ d} \Omega^d, 
\label{9.8.7} \\ 
{\cal B}^*_{ab} &=& \varepsilon_{acd} \Omega^c {\cal B}^d_{\ e} 
\bigl(\delta^e_{\ b} - \Omega^e \Omega_b \bigr) 
+ \varepsilon_{bcd} \Omega^c {\cal B}^d_{\ e} 
\bigl(\delta^e_{\ a} - \Omega^e \Omega_a \bigr).  
\label{9.8.8}
\end{eqnarray}    
These are all orthogonal to $\Omega^a$: ${\cal E}^*_a \Omega^a = 
{\cal B}^*_a \Omega^a = 0$ and ${\cal E}^*_{ab} \Omega^b = 
{\cal B}^*_{ab} \Omega^b = 0$. In terms of these
Eqs.~(\ref{9.8.1})--(\ref{9.8.3}) become 
\begin{eqnarray} 
S_{ab} &=& {\cal E}_{ab}^* + \Omega_a {\cal E}^*_b + {\cal E}^*_a 
\Omega_b + \Omega_a \Omega_b {\cal E}^* + {\cal B}^*_{ab} 
+ \Omega_a {\cal B}^*_b + {\cal B}^*_a \Omega_b, 
\label{9.8.9} \\ 
S_{a} &=& {\cal E}^*_a + \Omega_a {\cal E}^* + {\cal B}^*_a, 
\label{9.8.10} \\ 
S &=& {\cal E}^*. 
\label{9.8.11}
\end{eqnarray} 

When Eqs.~(\ref{9.8.9})--(\ref{9.8.11}) are substituted into the
metric tensor of Eqs.~(\ref{9.6.3})--(\ref{9.6.5}) --- in which $a_a$
and $\omega_{ab}$ are both set equal to zero --- we obtain the compact 
expressions  
\begin{eqnarray}  
g_{uu} &=& - 1 - r^2 {\cal E}^* + O(r^3),
\label{9.8.12} \\ 
g_{ua} &=& -\Omega_a + \frac{2}{3} r^2 \bigl( {\cal E}^*_a 
+ {\cal B}^*_a \bigr) + O(r^3), 
\label{9.8.13} \\ 
g_{ab} &=& \delta_{ab} - \Omega_a \Omega_b - \frac{1}{3} r^2  
\bigl( {\cal E}^*_{ab} + {\cal B}^*_{ab} \bigr) + O(r^3). 
\label{9.8.14}
\end{eqnarray}   
The metric becomes 
\begin{eqnarray} 
g_{uu} &=& - 1 - r^2 {\cal E}^* + O(r^3),
\label{9.8.15} \\
g_{ur} &=& - 1, 
\label{9.8.16} \\
g_{uA} &=& \frac{2}{3} r^3 \bigl( {\cal E}^*_A 
+ {\cal B}^*_A \bigr) + O(r^4), 
\label{9.8.17} \\ 
g_{AB} &=& r^2 \Omega_{AB} - \frac{1}{3} r^4  
\bigl( {\cal E}^*_{AB} + {\cal B}^*_{AB} \bigr) + O(r^5)  
\label{9.8.18}
\end{eqnarray}   
after transforming to angular coordinates using the rules of 
Eq.~(\ref{9.7.6}). Here we have introduced the projections 
\begin{eqnarray} 
{\cal E}^*_A &\equiv& {\cal E}^*_{a} \Omega^a_A 
= {\cal E}_{ab} \Omega^a_A \Omega^b, 
\label{9.8.19} \\
{\cal E}^*_{AB} &\equiv& {\cal E}^*_{ab} \Omega^a_A \Omega^b_B 
= 2 {\cal E}_{ab} \Omega^a_A \Omega^b_B + {\cal E}^* \Omega_{AB}, 
\label{9.8.20} \\
{\cal B}^*_A &\equiv& {\cal B}^*_{a} \Omega^a_A 
= \varepsilon_{abc} \Omega^a_A \Omega^b {\cal B}^c_{\ d} \Omega^d, 
\label{9.8.21} \\
{\cal B}^*_{AB} &\equiv& {\cal B}^*_{ab} \Omega^a_A \Omega^b_B 
= 2 \varepsilon_{acd} \Omega^c {\cal B}^d_{\ b} \Omega^{a}_{(A}  
\Omega^{b}_{B)}.  
\label{9.8.22}
\end{eqnarray} 
It may be noted that the inverse relations are ${\cal E}^*_a    
= {\cal E}^*_{A} \Omega^A_a$, ${\cal B}^*_a 
= {\cal B}^*_{A} \Omega^A_a$, ${\cal E}^*_{ab} 
= {\cal E}^*_{AB} \Omega^A_a \Omega^B_b$, and ${\cal B}^*_{ab}  
= {\cal B}^*_{AB} \Omega^A_a \Omega^B_b$, where $\Omega^A_a$ was
introduced in Eq.~(\ref{9.7.4}). 
  
\section{Transformation between Fermi and retarded coordinates;
advanced point}  
\label{10}

A point $x$ in the normal convex neighbourhood of a world line
$\gamma$ can be assigned a set of Fermi normal coordinates
(as in Sec.~\ref{8}), or it can be assigned a set of retarded
coordinates (Sec.~\ref{9}). These coordinate systems can obviously be 
related to one another, and our first task in this section (which will
occupy us in Secs.~\ref{10.1}--\ref{10.3}) will be to derive the
transformation rules. We begin by refining our notation so as to
eliminate any danger of ambiguity.    

\begin{figure}[t]
\vspace*{2.2in}
\special{hscale=35 vscale=35 hoffset=115.0 voffset=-55.0
         psfile=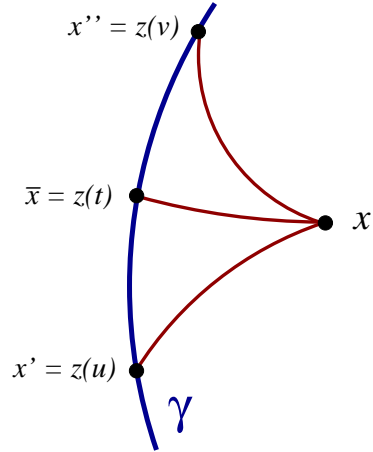}
\caption{The retarded, simultaneous, and advanced points on a world
line $\gamma$. The retarded point $x' \equiv z(u)$ is linked to $x$ by
a future-directed null geodesic. The simultaneous point $\bar{x}
\equiv z(t)$ is linked to $x$ by a spacelike geodesic that intersects
$\gamma$ orthogonally. The advanced point $x'' \equiv z(v)$ is linked
to $x$ by a past-directed null geodesic.}  
\end{figure} 

The Fermi normal coordinates of $x$ refer to a point $\bar{x} \equiv
z(t)$ on $\gamma$ that is related to $x$ by a spacelike geodesic that
intersects $\gamma$ orthogonally; see Fig.~8. We refer to this point
as $x$'s {\it simultaneous point}, and to tensors at $\bar{x}$ we
assign indices $\bar{\alpha}$, $\bar{\beta}$, etc. We let
$(t,s\omega^a)$ be the Fermi normal coordinates of $x$, with $t$
denoting the value of $\gamma$'s proper-time parameter at $\bar{x}$,
$s = \sqrt{2\sigma(x,\bar{x})}$ representing the proper distance from 
$\bar{x}$ to $x$ along the spacelike geodesic, and $\omega^a$ denoting
a unit vector ($\delta_{a b} \omega^a \omega^b = 1$) that determines
the direction of the geodesic. The Fermi normal coordinates are
defined by $s \omega^a = -\base{a}{\bar{\alpha}}
\sigma^{\bar{\alpha}}$ and $\sigma_{\bar{\alpha}} u^{\bar{\alpha}} =
0$. Finally, we denote by $(\bar{e}^{\alpha}_0, \bar{e}^\alpha_a)$ the
tetrad at $x$ that is obtained by parallel transport of
$(u^{\bar{\alpha}}, \base{\bar{\alpha}}{a})$ on the spacelike
geodesic.  

The retarded coordinates of $x$ refer to a point $x' \equiv z(u)$ on
$\gamma$ that is linked to $x$ by a future-directed null geodesic; see
Fig.~8. We refer to this point as $x$'s {\it retarded point}, and to
tensors at $x'$ we assign indices $\alpha'$, $\beta'$, etc. We let 
$(u,r\Omega^a)$ be the retarded coordinates of $x$, with $u$ denoting
the value of $\gamma$'s proper-time parameter at $x'$, $r =
\sigma_{\alpha'} u^{\alpha'}$ representing the affine-parameter
distance from $x'$ to $x$ along the null geodesic, and $\Omega^a$
denoting a unit vector ($\delta_{a b} \Omega^a \Omega^b = 1$) that
determines the direction of the geodesic. The retarded coordinates are
defined by $r \Omega^a = -\base{a}{\alpha'} \sigma^{\alpha'}$ and
$\sigma(x,x') = 0$. Finally, we denote by
$(\base{\alpha}{0},\base{\alpha}{a})$ the tetrad at $x$ that is
obtained by parallel transport of $(u^{\alpha'}, \base{\alpha'}{a})$
on the null geodesic.     

The reader not interested in following the details of this discussion
can be informed that: (i) our results concerning the transformation 
from the retarded coordinates $(u,r,\Omega^a)$ to the Fermi normal
coordinates $(t,s,\omega^a)$ are contained in
Eqs.~(\ref{10.1.1})--(\ref{10.1.3}) below; (ii) our results  
concerning the transformation from the Fermi normal coordinates
$(t,s,\omega^a)$ to the retarded coordinates $(u,r,\Omega^a)$ are
contained in Eqs.~(\ref{10.2.1})--(\ref{10.2.3}); (iii) the
decomposition of each member of $(\bar{e}^{\alpha}_0,
\bar{e}^\alpha_a)$ in the tetrad
$(\base{\alpha}{0},\base{\alpha}{a})$ is given in retarded coordinates
by Eqs.~(\ref{10.3.1}) and (\ref{10.3.2}); and (iv) the
decomposition of each member of $(\base{\alpha}{0},\base{\alpha}{a})$
in the tetrad $(\bar{e}^{\alpha}_0, \bar{e}^\alpha_a)$ is given in
Fermi normal coordinates by Eqs.~(\ref{10.3.3}) and (\ref{10.3.4}). 

Our final task will be to define, along with the retarded and
simultaneous points, an {\it advanced point} $x''$ on the world line 
$\gamma$; see Fig.~8. This is taken on in Sec.~\ref{10.4}. Throughout 
this section we shall set $\omega_{ab} = 0$, where $\omega_{ab}$ is
the rotation tensor defined by Eq.~(\ref{9.1.1}) --- the tetrad
vectors $\base{\mu}{a}$ will be assumed to be Fermi-Walker transported
on $\gamma$.
  
\subsection{From retarded to Fermi coordinates} 
\label{10.1}

Quantities at $\bar{x} \equiv z(t)$ can be related to quantities at
$x' \equiv z(u)$ by Taylor expansion along the world line $\gamma$. To
implement this strategy we must first find an expression for $\Delta
\equiv t - u$. (Although we use the same notation, this should not be
confused with the van Vleck determinant introduced in Sec.~\ref{6}.) 

Consider the function $p(\tau)$ of the proper-time parameter $\tau$
defined by 
\[
p(\tau) = \sigma_\mu\bigl(x,z(\tau)\bigr) u^\mu(\tau),  
\]
in which $x$ is kept fixed and in which $z(\tau)$ is an arbitrary
point on the world line. We have that $p(u) = r$ and $p(t) = 0$, and 
$\Delta$ can ultimately be obtained by expressing $p(t)$ as $p(u +
\Delta)$ and expanding in powers of $\Delta$. Formally, 
\[
p(t) = p(u) + \dot{p}(u) \Delta + \frac{1}{2} \ddot{p}(u) \Delta^2 +
\frac{1}{6} p^{(3)}(u) \Delta^3 + O(\Delta^4), 
\]
where overdots (or a number within brackets) indicate repeated 
differentiation with respect to $\tau$. We have  
\begin{eqnarray*} 
\dot{p}(u) &=& \sigma_{\alpha'\beta'} u^{\alpha'} u^{\beta'} +
\sigma_{\alpha'} a^{\alpha'}, \\
\ddot{p}(u) &=& \sigma_{\alpha'\beta'\gamma'} u^{\alpha'} u^{\beta'} 
u^{\gamma'} + 3 \sigma_{\alpha'\beta'} u^{\alpha'} a^{\beta'} 
+ \sigma_{\alpha'} \dot{a}^{\alpha'}, \\ 
p^{(3)}(u) &=& \sigma_{\alpha'\beta'\gamma'\delta'} u^{\alpha'}
u^{\beta'} u^{\gamma'} u^{\delta'} + \sigma_{\alpha'\beta'\gamma'}
\bigl( 5 a^{\alpha'} u^{\beta'} u^{\gamma'} + u^{\alpha'} u^{\beta'}
a^{\gamma'} \bigr) + \sigma_{\alpha'\beta'} \bigl( 3 a^{\alpha'}
a^{\beta'} + 4 u^{\alpha'} \dot{a}^{\beta'} \bigr) 
+ \sigma_{\alpha'} \ddot{a}^{\alpha'},
\end{eqnarray*}
where $a^{\mu} = D u^{\mu}/d\tau$, $\dot{a}^{\mu} = D a^{\mu}/d\tau$,
and $\ddot{a}^{\mu} = D \dot{a}^{\mu}/d\tau$. 

We now express all of this in retarded coordinates by invoking the 
expansion of Eq.~(\ref{5.2.1}) for $\sigma_{\alpha'\beta'}$ (as well
as additional expansions for the higher derivatives of the world 
function, obtained by further differentiation of this result) and the
relation $\sigma^{\alpha'} = - r(u^{\alpha'} + \Omega^a
\base{\alpha'}{a})$ first derived in Eq.~(\ref{9.2.3}). With a degree
of accuracy sufficient for our purposes we obtain   
\begin{eqnarray*} 
\dot{p}(u) &=& -\Bigl[ 1 + r a_a \Omega^a + \frac{1}{3} r^2 S + O(r^3)
\Bigr], \\ 
\ddot{p}(u) &=& - r\bigl( \dot{a}_0 + \dot{a}_a \Omega^a \bigr) +
O(r^2), \\ 
p^{(3)}(u) &=& \dot{a}_0 + O(r), 
\end{eqnarray*}
where $S = R_{a0b0} \Omega^a \Omega^b$ was first introduced in
Eq.~(\ref{9.4.5}), and where $\dot{a}_0 \equiv \dot{a}_{\alpha'} 
u^{\alpha'}$, $\dot{a}_a \equiv \dot{a}_{\alpha'} \base{\alpha'}{a}$
are the frame components of the covariant derivative of the
acceleration vector. To arrive at these results we made use of the 
identity $a_{\alpha'} a^{\alpha'} + \dot{a}_{\alpha'} u^{\alpha'} = 0$
that follows from the fact that $a^\mu$ is orthogonal to
$u^\mu$. Notice that there is no distinction between the two possible 
interpretations $\dot{a}_a \equiv d a_a / d\tau$ and $\dot{a}_a \equiv
\dot{a}_\mu \base{\mu}{a}$ for the quantity $\dot{a}_a(\tau)$; their
equality follows at once from the substitution of 
$D \base{\mu}{a}/d\tau = a_a u^\mu$ (which states that the basis
vectors are Fermi-Walker transported on the world line) into the
identity $d a_a / d\tau = D(a_\nu \base{\nu}{a})/d\tau$.       

Collecting our results we obtain
\[
r = \Bigl[ 1 + r a_a \Omega^a + \frac{1}{3} r^2 S + O(r^3)
\Bigr] \Delta + \frac{1}{2} r \Bigl[ \dot{a}_0 + \dot{a}_a \Omega^a +
O(r) \Bigr] \Delta^2 - \frac{1}{6} \Bigl[ \dot{a}_0 + O(r) \Bigr]
\Delta^3 + O(\Delta^4), 
\] 
which can readily be solved for $\Delta \equiv t - u$ expressed as an
expansion in powers of $r$. The final result is  
\begin{equation} 
t = u + r \biggl\{ 1 - r a_a(u) \Omega^a + r^2 \bigl[ a_a(u)
\Omega^a \bigr]^2 - \frac{1}{3} r^2 \dot{a}_0(u) - \frac{1}{2} r^2 
\dot{a}_a(u) \Omega^a - \frac{1}{3} r^2 R_{a0b0}(u) \Omega^a \Omega^b
+ O(r^3) \biggr\},  
\label{10.1.1}
\end{equation}
where we show explicitly that all frame components are evaluated at
the retarded point $z(u)$.  

To obtain relations between the spatial coordinates we consider the
functions 
\[
p_a(\tau) = -\sigma_\mu\bigl(x,z(\tau)\bigr) \base{\mu}{a}(\tau),  
\]
in which $x$ is fixed and $z(\tau)$ is an arbitrary point on
$\gamma$. We have that the retarded coordinates are given by $r
\Omega^a = p^a(u)$, while the Fermi coordinates are given instead by
$s \omega^a = p^a(t) = p^a(u + \Delta)$. This last expression can be
expanded in powers of $\Delta$, producing 
\[
s\omega^a = p^a(u) + \dot{p}^a(u) \Delta + \frac{1}{2} \ddot{p}^a(u)
\Delta^2 + \frac{1}{6} p^{a(3)}(u) \Delta^3 + O(\Delta^4) 
\]
with 
\begin{eqnarray*}
\dot{p}_a(u) &=& -\sigma_{\alpha'\beta'} \base{\alpha'}{a} u^{\beta'}
- \bigl(\sigma_{\alpha'} u^{\alpha'}\bigr) \bigl( a_{\beta'}
  \base{\beta'}{a} \bigr) \\ 
&=& -r a_a - \frac{1}{3} r^2 S_a + O(r^3), \\ 
\ddot{p}_a(u) &=& -\sigma_{\alpha'\beta'\gamma'} \base{\alpha'}{a}
  u^{\beta'} u^{\gamma'} - \bigl( 2 \sigma_{\alpha'\beta'} u^{\alpha'}
u^{\beta'} + \sigma_{\alpha'} a^{\alpha'} \bigr) \bigl(a_{\gamma'}
\base{\gamma'}{a} \bigr) - \sigma_{\alpha'\beta'} \base{\alpha'}{a}
a^{\beta'} - \bigl(\sigma_{\alpha'} u^{\alpha'}\bigr)
\bigl(\dot{a}_{\beta'} \base{\beta'}{a} \bigr) \\ 
&=& \bigl(1 + r a_b \Omega^b \bigr) a_a - r \dot{a}_a 
+ \frac{1}{3} r R_{a0b0}\Omega^b + O(r^2), \\ 
p^{(3)}_a(u) &=& -\sigma_{\alpha'\beta'\gamma'\delta'}
\base{\alpha'}{a} u^{\beta'} u^{\gamma'} u^{\delta'} 
- \bigl( 3 \sigma_{\alpha'\beta'\gamma'} u^{\alpha'}
u^{\beta'} u^{\gamma'} + 6 \sigma_{\alpha'\beta'} u^{\alpha'}
a^{\beta'} + \sigma_{\alpha'} \dot{a}^{\alpha'} + \sigma_{\alpha'}
u^{\alpha'} \dot{a}_{\beta'} u^{\beta'} \bigr) \bigl(a_{\delta'}
\base{\delta'}{a} \bigr) \\ 
& & \mbox{} - \sigma_{\alpha'\beta'\gamma'}
\base{\alpha'}{a} \bigl( 2 a^{\beta'} u^{\gamma'} + u^{\beta'}
a^{\gamma'} \bigr) - \bigl(3 \sigma_{\alpha'\beta'} u^{\alpha'}
u^{\beta'} + 2 \sigma_{\alpha'} a^{\alpha'} \bigr)
\bigl(\dot{a}_{\gamma'} \base{\gamma'}{a} \bigr) -
\sigma_{\alpha'\beta'} \base{\alpha'}{a} \dot{a}^{\beta'} \\ 
& & \mbox{} - \bigl(\sigma_{\alpha'} u^{\alpha'} \bigr) \bigl(
\ddot{a}_{\beta'} \base{\beta'}{a} \bigr) \\ 
&=& 2 \dot{a}_a + O(r).  
\end{eqnarray*} 
To arrive at these results we have used the same expansions as before
and re-introduced $S_a = R_{a0b0} \Omega^b - R_{ab0c} \Omega^b
\Omega^c$, as it was first defined in Eq.~(\ref{9.4.4}).  

Collecting our results we obtain 
\begin{eqnarray*} 
s\omega^a &=& r\Omega^a - r \Bigl[ a^a + \frac{1}{3} r S^a + O(r^2) 
\Bigr] \Delta + \frac{1}{2} \Bigl[ \bigl(1 + r a_b \Omega^b\bigr) a^a
- r \dot{a}^a + \frac{1}{3} r R^a_{\ 0b0}\Omega^b + O(r^2) \Bigr]
\Delta^2 \\ & & \mbox{} 
+ \frac{1}{3} \Bigl[ \dot{a}^a + O(r) \Bigr] \Delta^3 
+ O(\Delta^4),   
\end{eqnarray*} 
which becomes 
\begin{equation}
s\omega^a = r \biggl\{ \Omega^a - \frac{1}{2} r \bigl[ 1 - r a_b(u)
\Omega^b \bigr] a^a(u) - \frac{1}{6} r^2 \dot{a}^a(u) - \frac{1}{6}
r^2 R^a_{\ 0b0}(u)\Omega^b + \frac{1}{3} r^2 R^a_{\ b0c}(u)
\Omega^b\Omega^c + O(r^3) \biggr\} 
\label{10.1.2}
\end{equation} 
after substituting Eq.~(\ref{10.1.1}) for $\Delta \equiv t-u$. From
squaring Eq.~(\ref{10.1.2}) and using the identity $\delta_{ab} 
\omega^a \omega^b = 1$ we can also deduce 
\begin{equation} 
s = r \biggl\{ 1 - \frac{1}{2} r a_a(u) \Omega^a + \frac{3}{8} r^2
\bigl[ a_a(u) \Omega^a \bigr]^2 - \frac{1}{8} r^2 \dot{a}_0(u) -
\frac{1}{6} r^2 \dot{a}_a(u) \Omega^a - \frac{1}{6} r^2 R_{a0b0}(u)
\Omega^a \Omega^b + O(r^3) \biggr\}
\label{10.1.3} 
\end{equation}  
for the spatial distance between $x$ and $z(t)$. 

\subsection{From Fermi to retarded coordinates} 
\label{10.2}

The techniques developed in the preceding subsection can easily be
adapted to the task of relating the retarded coordinates of $x$ to its
Fermi normal coordinates. Here we use $\bar{x} \equiv z(t)$ as
the reference point and express all quantities at $x' \equiv z(u)$ as
Taylor expansions about $\tau = t$. 

We begin by considering the function  
\[
\sigma(\tau) = \sigma\bigl(x,z(\tau)\bigr) 
\]
of the proper-time parameter $\tau$ on $\gamma$. We have that
$\sigma(t) = \frac{1}{2} s^2$ and $\sigma(u) = 0$, and $\Delta \equiv
t-u$ is now obtained by expressing $\sigma(u)$ as $\sigma(t-\Delta)$
and expanding in powers of $\Delta$. Using the fact that
$\dot{\sigma}(\tau) = p(\tau)$, we have 
\[
\sigma(u) = \sigma(t) - p(t) \Delta + \frac{1}{2} \dot{p}(t) \Delta^2
- \frac{1}{6} \ddot{p}(t) \Delta^3 + \frac{1}{24} p^{(3)}(t)
  \Delta^4 + O(\Delta^5). 
\]
Expressions for the derivatives of $p(\tau)$ evaluated at $\tau = t$
can be constructed from results derived previously in Sec.~\ref{10.1}:
it suffices to replace all primed indices by barred indices and then
substitute the relation $\sigma^{\bar{\alpha}} = -s \omega^a
\base{\bar{\alpha}}{a}$ that follows immediately from
Eq.~(\ref{8.3.1}). This gives  
\begin{eqnarray*} 
\dot{p}(t) &=& -\Bigl[ 1 + s a_a \omega^a + \frac{1}{3} s^2 R_{a0b0} 
\omega^a \omega^b + O(s^3) \Bigr], \\ 
\ddot{p}(t) &=& - s \dot{a}_a \omega^a + O(s^2), \\ 
p^{(3)}(t) &=& \dot{a}_0 + O(s), 
\end{eqnarray*} 
and then 
\[
s^2 = \Bigl[ 1 + s a_a \omega^a + \frac{1}{3} s^2 R_{a0b0} 
\omega^a \omega^b + O(s^3) \Bigr] \Delta^2 - \frac{1}{3} s \Bigl[
\dot{a}_a \omega^a + O(s) \Bigr] \Delta^3 - \frac{1}{12} \Bigl[
\dot{a}_0 + O(s) \Bigr] \Delta^4 + O(\Delta^5) 
\]
after recalling that $p(t) = 0$. Solving for $\Delta$ as an expansion
in powers of $s$ returns  
\begin{equation} 
u = t - s \biggl\{ 1 - \frac{1}{2} s a_a(t) \omega^a + \frac{3}{8} s^2
\bigl[ a_a(t) \omega^a \bigr]^2 + \frac{1}{24} s^2 \dot{a}_0(t) 
+ \frac{1}{6} s^2 \dot{a}_a(t) \omega^a  
- \frac{1}{6} s^2 R_{a0b0}(t) \omega^a \omega^b + O(s^3) \biggr\}, 
\label{10.2.1} 
\end{equation}
in which we emphasize that all frame components are evaluated at the 
simultaneous point $z(t)$.  

An expression for $r = p(u)$ can be obtained by expanding
$p(t-\Delta)$ in powers of $\Delta$. We have  
\[
r = - \dot{p}(t) \Delta + \frac{1}{2} \ddot{p}(t) \Delta^2 -
\frac{1}{6} p^{(3)}(t) \Delta^3 + O(\Delta^4), 
\]
and substitution of our previous results gives 
\begin{equation} 
r = s \biggl\{ 1 + \frac{1}{2} s a_a(t) \omega^a - \frac{1}{8} s^2
\bigl[ a_a(t) \omega^a \bigr]^2 - \frac{1}{8} s^2 \dot{a}_0(t) 
- \frac{1}{3} s^2 \dot{a}_a(t) \omega^a 
+ \frac{1}{6} s^2 R_{a0b0}(t) \omega^a \omega^b + O(s^3) \biggr\}  
\label{10.2.2} 
\end{equation}     
for the retarded distance between $x$ and $z(u)$. 

Finally, the retarded coordinates $r\Omega^a = p^a(u)$ can be related
to the Fermi coordinates by expanding $p^a(t-\Delta)$ in powers of
$\Delta$, so that  
\[
r\Omega^a = p^a(t) - \dot{p}^a(t) \Delta + \frac{1}{2} \ddot{p}^a(t)
\Delta^2 - \frac{1}{6} p^{a(3)}(t) \Delta^3 + O(\Delta^4). 
\]
Results from the preceding subsection can again be imported with mild
alterations, and we find 
\begin{eqnarray*} 
\dot{p}_a(t) &=& \frac{1}{3} s^2 R_{ab0c} \omega^b \omega^c + O(s^3), 
\\ 
\ddot{p}_a(t) &=& \bigl( 1 + s a_b \omega^b \bigr) a_a + \frac{1}{3} s
R_{a0b0} \omega^b + O(s^2), \\ 
p^{(3)}_a(t) &=& 2 \dot{a}_a(t) + O(s).  
\end{eqnarray*} 
This, together with Eq.~(\ref{10.2.1}), gives 
\begin{equation}
r\Omega^a = s \biggl\{ \omega^a + \frac{1}{2} s a^a(t) - \frac{1}{3}
s^2 \dot{a}^a(t) - \frac{1}{3} s^2 R^a_{\ b0c}(t) \omega^b \omega^c +
\frac{1}{6} s^2 R^a_{\ 0b0}(t) \omega^b + O(s^3) \biggr\}.  
\label{10.2.3}
\end{equation}   
It may be checked that squaring this equation and using the identity
$\delta_{ab} \Omega^a \Omega^b = 1$ returns the same result as
Eq.~(\ref{10.2.2}).   

\subsection{Transformation of the tetrads at $x$} 
\label{10.3} 

Recall that we have constructed two sets of basis vectors at $x$. The
first set is the tetrad $(\bar{e}^{\alpha}_0, \bar{e}^\alpha_a)$ that
is obtained by parallel transport of $(u^{\bar{\alpha}},
\base{\bar{\alpha}}{a})$ on the spacelike geodesic that links $x$ to
the simultaneous point $\bar{x} \equiv z(t)$. The second set is the
tetrad $(\base{\alpha}{0},\base{\alpha}{a})$ that is obtained by
parallel transport of $(u^{\alpha'}, \base{\alpha'}{a})$ on the null
geodesic that links $x$ to the retarded point $x' \equiv z(u)$. Since
each tetrad forms a complete set of basis vectors, each member of
$(\bar{e}^{\alpha}_0, \bar{e}^\alpha_a)$ can be decomposed in the
tetrad $(\base{\alpha}{0},\base{\alpha}{a})$, and correspondingly,
each member of $(\base{\alpha}{0},\base{\alpha}{a})$ can be decomposed
in the tetrad $(\bar{e}^{\alpha}_0, \bar{e}^\alpha_a)$. These
decompositions are worked out in this subsection. For this purpose we 
shall consider the functions 
\[
p^\alpha(\tau) = g^\alpha_{\ \mu}\bigl(x,z(\tau)\bigr) u^\mu(\tau),
\qquad 
p^\alpha_a(\tau) = g^\alpha_{\ \mu}\bigl(x,z(\tau)\bigr)
\base{\mu}{a}(\tau), 
\]
in which $x$ is a fixed point in a neighbourhood of $\gamma$,
$z(\tau)$ is an arbitrary point on the world line, and 
$g^\alpha_{\ \mu}(x,z)$ is the parallel propagator on the unique
geodesic that links $x$ to $z$. We have $\bar{e}^\alpha_0 =
p^\alpha(t)$, $\bar{e}^\alpha_a = p^\alpha_a(t)$, $\base{\alpha}{0} =
p^\alpha(u)$, and $\base{\alpha}{a} = p^\alpha_a(u)$. 

We begin with the decomposition of $(\bar{e}^{\alpha}_0,
\bar{e}^\alpha_a)$ in the tetrad $(\base{\alpha}{0},\base{\alpha}{a})$
associated with the retarded point $z(u)$. This decomposition will be
expressed in the retarded coordinates as an expansion in powers of
$r$. As in Sec.~\ref{8.1} we express quantities at $z(t)$ in terms 
of quantities at $z(u)$ by expanding in powers of $\Delta \equiv t-u$.  
We have
\[
\bar{e}^\alpha_0 = p^\alpha(u) + \dot{p}^\alpha(u) \Delta +
\frac{1}{2} \ddot{p}^\alpha(u) \Delta^2 + O(\Delta^3), 
\]
with 
\begin{eqnarray*}
\dot{p}^\alpha(u) &=& g^\alpha_{\ \alpha';\beta'} u^{\alpha'}
u^{\beta'} + g^{\alpha}_{\ \alpha'} a^{\alpha'} \\ 
&=& \Bigl[ a^a + \frac{1}{2} r R^a_{\ 0b0} \Omega^b + O(r^2) \Bigr]
\base{\alpha}{a}, \\ 
\ddot{p}^\alpha(u) &=& g^\alpha_{\ \alpha';\beta'\gamma'} u^{\alpha'} 
u^{\beta'} u^{\gamma'} + g^{\alpha}_{\ \alpha';\beta'} \bigl( 2
a^{\alpha'} u^{\beta'} + u^{\alpha'} a^{\beta'} \bigr) 
+ g^\alpha_{\ \alpha'} \dot{a}^{\alpha'} \\ 
&=& \Bigl[ -\dot{a}_0 + O(r) \Bigr] \base{\alpha}{0} + \Bigl[
\dot{a}^a + O(r) \Bigr] \base{\alpha}{a}, 
\end{eqnarray*} 
where we have used the expansions of Eq.~(\ref{5.2.5}) as well as the
decompositions of Eq.~(\ref{9.1.4}). Collecting these results and
substituting Eq.~(\ref{10.1.1}) for $\Delta$ yields  
\begin{equation} 
\bar{e}^\alpha_0 = \Bigl[ 1 - \frac{1}{2} r^2 \dot{a}_0(u) + O(r^3) 
\Bigr]\, \base{\alpha}{0} + \Bigl[ r \bigl( 1 - a_b \Omega^b \bigr)
a^a(u) + \frac{1}{2} r^2 \dot{a}^a(u) + \frac{1}{2} r^2 
R^a_{\ 0b0}(u) \Omega^b + O(r^3) \Bigr]\, \base{\alpha}{a}.  
\label{10.3.1}       
\end{equation} 
Similarly, we have  
\[
\bar{e}^\alpha_a = p^\alpha_a(u) + \dot{p}^\alpha_a(u) \Delta + 
\frac{1}{2} \ddot{p}^\alpha_a(u) \Delta^2 + O(\Delta^3), 
\]
with 
\begin{eqnarray*} 
\dot{p}^\alpha_a(u) &=& g^\alpha_{\ \alpha';\beta'} \base{\alpha'}{a}
u^{\beta'} + \bigl( g^\alpha_{\ \alpha'} u^{\alpha'} \bigr) \bigl(
a_{\beta'} \base{\beta'}{a} \bigr) \\ 
&=& \Bigl[a_a + \frac{1}{2} r R_{a0b0} \Omega^b + O(r^2) \Bigr]
\base{\alpha}{0} + \Bigl[ - \frac{1}{2} r R^b_{\ a0c} \Omega^c +
O(r^2) \Bigr] \base{\alpha}{b}, \\ 
\ddot{p}^\alpha_a(u) &=& g^\alpha_{\ \alpha';\beta'\gamma'}
\base{\alpha'}{a} u^{\beta'} u^{\gamma'} + g^\alpha_{\ \alpha';\beta'}
\bigl( 2 u^{\alpha'} u^{\beta'} a_{\gamma'} \base{\gamma'}{a} +
\base{\alpha'}{a} a^{\beta'} \bigr) + \bigl( g^\alpha_{\ \alpha'}
a^{\alpha'} \bigr) \bigl( a_{\beta'} \base{\beta'}{a} \bigr) + \bigl(
g^\alpha_{\ \alpha'} u^{\alpha'} \bigr) \bigl( \dot{a}_{\beta'}
\base{\beta'}{a} \bigr) \\ 
&=& \Bigl[ \dot{a}_a + O(r) \Bigr] \base{\alpha}{0} + \Bigl[ a_a a^b +
O(r) \Bigr] \base{\alpha}{b},
\end{eqnarray*} 
and all this gives 
\begin{eqnarray} 
\bar{e}^\alpha_a &=& \Bigl[ \delta^b_{\ a} + \frac{1}{2} r^2 a^b(u)
a_a(u) - \frac{1}{2} r^2 R^b_{\ a0c}(u) \Omega^c + O(r^3) \Bigr]\,
\base{\alpha}{b} 
\nonumber \\ & & \mbox{} 
+ \Bigl[ r \bigl( 1 - r a_b \Omega^b \bigr) a_a(u) +
\frac{1}{2} r^2 \dot{a}_a(u) + \frac{1}{2} r^2 R_{a0b0}(u) \Omega^b +
O(r^3) \Bigr]\, \base{\alpha}{0}. 
\label{10.3.2} 
\end{eqnarray}   

We now turn to the decomposition of $(\base{\alpha}{0},
\base{\alpha}{a})$ in the tetrad $(\bar{e}^{\alpha}_0,
\bar{e}^\alpha_a)$ associated with the simultaneous point $z(t)$. This
decomposition will be expressed in the Fermi normal coordinates as an
expansion in powers of $s$. Here, as in Sec.~\ref{8.2}, we shall
express quantities at $z(u)$ in terms of quantities at $z(t)$. We
begin with  
\[
\base{\alpha}{0} = p^\alpha(t) - \dot{p}^\alpha(t) \Delta +
\frac{1}{2} \ddot{p}^\alpha(t) \Delta^2 + O(\Delta^3) 
\]
and we evaluate the derivatives of $p^\alpha(\tau)$ at $\tau = t$. To
accomplish this we rely on our previous results (replacing primed
indices with barred indices), on the expansions of Eq.~(\ref{5.2.5}),
and on the decomposition of $g^\alpha_{\ \bar{\alpha}}(x,\bar{x})$ in
the tetrads at $x$ and $\bar{x}$. This gives 
\begin{eqnarray*} 
\dot{p}^\alpha(t) &=& \Bigl[ a^a + \frac{1}{2} s R^a_{\ 0b0} \omega^b +
O(s^2) \Bigr] \bar{e}^\alpha_a, \\ 
\ddot{p}^\alpha(t) &=& \Bigl[ -\dot{a}_0 + O(s) \Bigr]
\bar{e}^\alpha_0 + \Bigl[ \dot{a}^a + O(s) \Bigr] \bar{e}^\alpha_a,  
\end{eqnarray*} 
and we finally obtain 
\begin{equation} 
\base{\alpha}{0} = \Bigl[ 1 - \frac{1}{2} s^2 \dot{a}_0(t) + O(s^3) 
\Bigr]\, \bar{e}^\alpha_0 + \Bigl[ -s \Bigl( 1 - \frac{1}{2} s a_b
\omega^b \Bigr) a^a(t) + \frac{1}{2} s^2 \dot{a}^a(t) 
- \frac{1}{2} s^2 R^a_{\ 0b0}(t) \omega^b + O(s^3) \Bigr]\,
\bar{e}^\alpha_a.  
\label{10.3.3}
\end{equation}    
Similarly, we write 
\[
\base{\alpha}{a} = p^\alpha_a(t) - \dot{p}^\alpha_a(t) \Delta +
\frac{1}{2} \ddot{p}^\alpha_a(t) \Delta^2 + O(\Delta^3), 
\]
in which we substitute 
\begin{eqnarray*} 
\dot{p}^\alpha_a(t) &=& \Bigl[ a_a + \frac{1}{2} s R_{a0b0} \omega^b +
O(s^2) \Bigr] \bar{e}^\alpha_0 + \Bigl[ -\frac{1}{2} s R^b_{\ a0c}
\omega^c + O(s^2) \Bigr] \bar{e}^\alpha_b, \\ 
\ddot{p}^\alpha_a(t) &=& \Bigl[ \dot{a}_a + O(s) \Bigr]
\bar{e}^\alpha_0 + \Bigl[ a_a a^b + O(s) \Bigr] \bar{e}^\alpha_b, 
\end{eqnarray*} 
as well as Eq.~(\ref{10.2.1}) for $\Delta \equiv t - u$. Our final
result is  
\begin{eqnarray} 
\base{\alpha}{a} &=& \Bigl[ \delta^b_{\ a} + \frac{1}{2} s^2 a^b(t) 
a_a(t) + \frac{1}{2} s^2 R^b_{\ a0c}(t) \omega^c + O(s^3) \Bigr]\,
\bar{e}^\alpha_b  
\nonumber \\ & & \mbox{} 
+ \Bigl[ -s \Bigl( 1 - \frac{1}{2} s a_b \omega^b \Bigr) a_a(t) +
\frac{1}{2} s^2 \dot{a}_a(t) - \frac{1}{2} s^2 R_{a0b0}(u) \omega^b + 
O(s^3) \Bigr]\, \bar{e}^\alpha_0.   
\label{10.3.4} 
\end{eqnarray}   

\subsection{Advanced point}  
\label{10.4}

It will prove convenient to introduce on the world line, along with
the retarded and simultaneous points, an {\it advanced point}
associated with the field point $x$. The advanced point will be
denoted $x'' \equiv z(v)$, with $v$ denoting the value of the
proper-time parameter at $x''$; to tensors at this point we assign
indices $\alpha''$, $\beta''$, etc. The advanced point is linked to
$x$ by a {\it past-directed null geodesic} (refer back to Fig.~8), and
it can be located by solving $\sigma(x,x'') = 0$ together with the
requirement that $\sigma^{\alpha''}(x,x'')$ be a future-directed null
vector. The affine-parameter distance between $x$ and $x''$ along the
null geodesic is given by  
\begin{equation} 
r_{\rm adv} = - \sigma_{\alpha''} u^{\alpha''}, 
\label{10.4.1}
\end{equation} 
and we shall call this the {\it advanced distance} between $x$ and the
world line. Notice that $r_{\rm adv}$ is a positive quantity. 

We wish first to find an expression for $v$ in terms of the retarded
coordinates of $x$. For this purpose we define $\Delta^{\!\prime}
\equiv v - u$ and re-introduce the function $\sigma(\tau) \equiv
\sigma(x,z(\tau))$ first considered in Sec.~\ref{10.2}. We have that
$\sigma(v) = \sigma(u) = 0$, and $\Delta^{\!\prime}$ can ultimately be
obtained by expressing $\sigma(v)$ as $\sigma(u+\Delta^{\!\prime})$
and expanding in powers of $\Delta^{\!\prime}$. Recalling that
$\dot{\sigma}(\tau) = p(\tau)$, we have  
\[
\sigma(v) = \sigma(u) + p(u) \Delta^{\!\prime} + \frac{1}{2}
\dot{p}(u) \Delta^{\!\prime 2} + \frac{1}{6} \ddot{p}(u)
\Delta^{\!\prime 3} + \frac{1}{24} p^{(3)}(u) \Delta^{\!\prime 4}
+ O(\Delta^{\!\prime 5}).  
\]
Using the expressions for the derivatives of $p(\tau)$ that were first 
obtained in Sec.~\ref{10.1}, we write this as 
\[ 
r = \frac{1}{2} \Bigl[ 1 + r a_a \Omega^a + \frac{1}{3} r^2 S + O(r^3)  
\Bigr] \Delta^{\!\prime} + \frac{1}{6} r \Bigl[ \dot{a}_0 +
\dot{a}_a \Omega^a + O(r) \Bigr] \Delta^{\!\prime 2} - \frac{1}{24}
\Bigl[ \dot{a}_0 + O(r) \Bigr] \Delta^{\!\prime 3} +
O(\Delta^{\!\prime 4}).  
\]
Solving for $\Delta^{\!\prime}$ as an expansion in powers of $r$, we
obtain  
\begin{equation} 
v = u + 2 r \biggl\{ 1 - r a_a(u) \Omega^a + r^2 \bigl[ a_a(u)
\Omega^a \bigr]^2 - \frac{1}{3} r^2 \dot{a}_0(u) - \frac{2}{3} r^2  
\dot{a}_a(u) \Omega^a - \frac{1}{3} r^2 R_{a0b0}(u) \Omega^a \Omega^b 
+ O(r^3) \biggr\}, 
\label{10.4.2} 
\end{equation} 
in which all frame components are evaluated at the retarded point
$z(u)$. 

Our next task is to derive an expression for the advanced distance 
$r_{\rm adv}$. For this purpose we observe that $r_{\rm adv} = -p(v) = 
-p(u+\Delta^{\!\prime})$, which we can expand in powers of
$\Delta^{\!\prime} \equiv v - u$. This gives 
\[
r_{\rm adv} = -p(u) - \dot{p}(u) \Delta^{\!\prime} - \frac{1}{2}
\ddot{p}(u) \Delta^{\!\prime 2} - \frac{1}{6} p^{(3)}(u)
\Delta^{\!\prime 3} + O(\Delta^{\!\prime 4}),  
\]
which then becomes 
\[
r_{\rm adv} = -r + \Bigl[ 1 + r a_a \Omega^a + \frac{1}{3} r^2 S + 
O(r^3) \Bigr] \Delta^{\!\prime} + \frac{1}{2} r \Bigl[ \dot{a}_0 +
\dot{a}_a \Omega^a + O(r) \Bigr] \Delta^{\!\prime 2} - \frac{1}{6}
\Bigl[ \dot{a}_0 + O(r) \Bigr] \Delta^{\!\prime 3} +
O(\Delta^{\!\prime 4}).  
\] 
After substituting Eq.~(\ref{10.4.2}) for $\Delta^{\!\prime}$ and 
witnessing a number of cancellations, we arrive at the simple
expression    
\begin{equation} 
r_{\rm adv} = r \biggl[ 1 + \frac{2}{3} r^2 \dot{a}_a(u) \Omega^a +
O(r^3) \biggr].  
\label{10.4.3} 
\end{equation} 

From Eqs.~(\ref{9.5.3}), (\ref{9.5.4}), and (\ref{10.4.2}) we deduce
that the gradient of the advanced time $v$ is given by   
\begin{equation} 
\partial_\alpha v = \Bigl[ 1 - 2 r a_a \Omega^a + O(r^2) \Bigr]\,
\base{0}{\alpha} + \Bigr[ \Omega_a - 2 r a_a + O(r^2) \Bigr]\,
\base{a}{\alpha}, 
\label{10.4.4} 
\end{equation}
where the expansion in powers of $r$ was truncated to a sufficient
number of terms. Similarly, Eqs.~(\ref{9.5.4}), (\ref{9.5.5}), and  
(\ref{10.4.3}) imply that the gradient of the advanced distance is
given by  
\begin{eqnarray} 
\partial_\alpha r_{\rm adv} &=& \Bigl[ \Bigl( 1 + r a_b \Omega^b +  
\frac{4}{3} r^2 \dot{a}_b \Omega^b + \frac{1}{3} r^2 S \Bigr) \Omega_a
+ \frac{2}{3} r^2 \dot{a}_a + \frac{1}{6} r^2 S_a + O(r^3) \Bigl]\,
\base{a}{\alpha} \nonumber \\ & & \mbox{} 
+ \Bigl[ -r a_a \Omega^a - \frac{1}{2} r^2 S +
O(r^3) \Bigr]\, \base{0}{\alpha}, 
\label{10.4.5} 
\end{eqnarray} 
where $S_a$ and $S$ were first introduced in Eqs.~(\ref{9.4.4}) and 
(\ref{9.4.5}), respectively. We emphasize that in Eqs.~(\ref{10.4.4})
and (\ref{10.4.5}), all frame components are evaluated at the retarded
point $z(u)$.  

\newpage
\hrule
\hrule
\part{Green's functions} 
\label{part3}
\hrule
\hrule
\vspace*{.25in} 
%
\section{Scalar Green's functions in flat spacetime} 
\label{11}  

\subsection{Green's equation for a massive scalar field} 
\label{11.1} 

To prepare the way for our discussion of Green's functions in curved  
spacetime, we consider first the slightly nontrivial case of a massive
scalar field $\Phi(x)$ in flat spacetime. This field satisfies the
wave equation 
\begin{equation} 
(\Box - k^2) \Phi(x) = -4\pi \mu(x), 
\label{11.1.1}
\end{equation} 
where $\Box = \eta^{\alpha\beta} \partial_\alpha \partial_\beta$ is   
the wave operator, $\mu(x)$ a prescribed source, and where the mass
parameter $k$ has a dimension of inverse length. We seek a Green's
function $G(x,x')$ such that a solution to Eq.~(\ref{11.1.1}) can be
expressed as 
\begin{equation} 
\Phi(x) = \int G(x,x') \mu(x')\, d^4 x', 
\label{11.1.2} 
\end{equation} 
where the integration is over all of Minkowski spacetime. The relevant
wave equation for the Green's function is
\begin{equation} 
(\Box - k^2) G(x,x') = -4\pi \delta_4(x-x'), 
\label{11.1.3}
\end{equation}
where $\delta_4(x-x') = \delta(t-t') \delta(x-x') \delta(y-y')
\delta(z-z')$ is a four-dimensional Dirac distribution in flat
spacetime. Two types of Green's functions will be of particular 
interest: the retarded Green's function, a solution to 
Eq.~(\ref{11.1.3}) with the property that it vanishes if $x$ is in the 
past of $x'$, and the advanced Green's function, which vanishes when
$x$ is in the future of $x'$.     

To solve Eq.~(\ref{11.1.3}) we use Lorentz invariance and the fact 
that the spacetime is homogeneous to argue that the retarded and
advanced Green's functions must be given by expressions of the form   
\begin{equation}
G_{\rm ret}(x,x') = \theta(t-t') g(\sigma), \qquad 
G_{\rm adv}(x,x') = \theta(t'-t) g(\sigma),
\label{11.1.4}
\end{equation}
where $\sigma = \frac{1}{2} \eta_{\alpha\beta} (x-x')^\alpha
(x-x')^\beta$ is Synge's world function in flat spacetime, and where
$g(\sigma)$ is a function to be determined. For the remainder of this
section we set $x'=0$ without loss of generality.   

\subsection{Integration over the source} 
\label{11.2} 

The Dirac functional on the right-hand side of Eq.~(\ref{11.1.3}) is a
highly singular quantity, and we can avoid dealing with it by
integrating the equation over a small four-volume $V$ that contains 
$x' \equiv 0$. This volume is bounded by a closed hypersurface
$\partial V$. After using Gauss' theorem on the first term of
Eq.~(\ref{11.1.3}), we obtain $\oint_{\partial V} G^{;\alpha}
d\Sigma_\alpha - k^2 \int_V G\, dV = -4\pi$, where $d\Sigma_{\alpha}$ 
is a surface element on $\partial V$. Assuming that the integral of 
$G$ over $V$ goes to zero in the limit $V \to 0$, we have    
\begin{equation}
\lim_{V \to 0} \oint_{\partial V} G^{;\alpha}
d\Sigma_{\alpha} = -4\pi. 
\label{11.2.1}
\end{equation} 
It should be emphasized that the four-volume $V$ must contain the
point $x'$.  

To examine Eq.~(\ref{11.2.1}) we introduce coordinates 
$(w,\chi,\theta,\phi)$ defined by 
\[
t = w \cos\chi, \qquad
x = w \sin\chi \sin\theta \cos\phi, \qquad
y = w \sin\chi \sin\theta \sin\phi, \qquad
z = w \sin\chi \cos\theta, 
\]
and we let $\partial V$ be a surface of constant $w$. The metric of
flat spacetime is given by 
\[
ds^2 = -\cos2\chi\, dw^2 + 2w \sin2\chi\, dwd\chi + w^2\cos2\chi\,
d\chi^2 + w^2 \sin^2\chi\, d\Omega^2
\]
in the new coordinates, where $d\Omega^2 = d\theta^2 + \sin^2\theta\,
d\phi^2$. Notice that $w$ is a timelike coordinate when $\cos 2\chi >
0$, and that $\chi$ is then a spacelike coordinate; the roles are
reversed when $\cos 2\chi < 0$. Straightforward computations reveal
that in these coordinates, 
$\sigma = -\frac{1}{2} w^2 \cos 2\chi$, $\sqrt{-g} = w^3 \sin^2\chi
\sin\theta$, $g^{ww} = -\cos 2\chi$, $g^{w\chi} = w^{-1} \sin 2\chi$,
$g^{\chi\chi} = w^{-2} \cos2\chi$, and the only nonvanishing component
of the surface element is $d\Sigma_w = w^3 \sin^2\chi\, d\chi
d\Omega$, where $d\Omega = \sin\theta\, d\theta d\phi$. To calculate
the gradient of the Green's function we express it as $G = \theta(\pm
t) g(\sigma) = \theta(\pm w\cos\chi) g(-\frac{1}{2} w^2 \cos 2\chi)$,
with the upper (lower) sign belonging to the retarded (advanced)
Green's function. Calculation gives $G^{;\alpha} d\Sigma_{\alpha} = 
\theta(\pm \cos\chi) w^4 \sin^2 \chi g'(\sigma)\, d\chi d\Omega$, with
a prime indicating differentiation with respect to $\sigma$; it should
be noted that derivatives of the step function do not appear in this
expression. 

Integration of $G^{;\alpha} d\Sigma_{\alpha}$ with respect to
$d\Omega$ is immediate, and we find that Eq.~(\ref{11.2.1}) reduces to   
\begin{equation} 
\lim_{w \to 0} \int_0^\pi \theta(\pm \cos\chi) w^4 \sin^2\chi
g'(\sigma)\, d\chi = -1. 
\label{11.2.2}
\end{equation} 
For the retarded Green's function, the step function restricts the
domain of integration to $0 < \chi < \pi/2$, in which $\sigma$
increases from $-\frac{1}{2} w^2$ to $\frac{1}{2} w^2$. Changing the  
variable of integration from $\chi$ to $\sigma$ transforms
Eq.~(\ref{11.2.2}) into 
\begin{equation}
\lim_{\epsilon \to 0} \epsilon \int_{-\epsilon}^\epsilon
w(\sigma/\epsilon)\, g'(\sigma)\, d\sigma
= - 1, \qquad
w(\xi) \equiv \sqrt{ \frac{1+\xi}{1-\xi} }, 
\label{11.2.3}
\end{equation}
where $\epsilon \equiv \frac{1}{2} w^2$. For the advanced Green's 
function, the domain of integration is $\pi/2 < \chi < \pi$, in which 
$\sigma$ decreases from $\frac{1}{2} w^2$ to $-\frac{1}{2}
w^2$. Changing the variable of integration from $\chi$ to $\sigma$
also produces Eq.~(\ref{11.2.3}).    

\subsection{Singular part of $g(\sigma)$} 
\label{11.3} 

We have seen that Eq.~(\ref{11.2.3}) properly encodes the influence of 
the singular source term on both the retarded and advanced Green's
function. The function $g(\sigma)$ that enters into the expressions of
Eq.~(\ref{11.1.4}) must therefore be such that Eq.~(\ref{11.2.3}) is 
satisfied. It follows immediately that $g(\sigma)$ must be a singular
function, because for a smooth function the integral of
Eq.~(\ref{11.2.3}) would be of order $\epsilon$ and the left-hand side
of Eq.~(\ref{11.2.3}) could never be made equal to $-1$. The
singularity, however, must be integrable, and this leads us to assume
that $g'(\sigma)$ must be made out of Dirac $\delta$-functions and  
derivatives.  

We make the ansatz
\begin{equation}
g(\sigma) = V(\sigma) \theta(-\sigma) + A \delta(\sigma) + B
\delta'(\sigma) + C \delta''(\sigma) + \cdots, 
\label{11.3.1}
\end{equation} 
where $V(\sigma)$ is a smooth function, and $A$, $B$, $C$, \ldots are 
constants. The first term represents a function supported within the
past and future light cones of $x' \equiv 0$; we exclude a term
proportional to $\theta(\sigma)$ for reasons of causality. The other
terms are supported on the past and future light cones. It is
sufficient to take the coefficients in front of the $\delta$-functions
to be constants. To see this we invoke the distributional identities   
\begin{equation} 
\sigma \delta(\sigma) = 0 
\quad \rightarrow \quad 
\sigma \delta'(\sigma) + \delta(\sigma) = 0
\quad \rightarrow \quad 
\sigma \delta''(\sigma) + 2 \delta'(\sigma) = 0 
\quad \rightarrow \quad \cdots
\label{11.3.2}
\end{equation}
from which it follows that $\sigma^2 \delta'(\sigma) = \sigma^3
\delta''(\sigma) = \cdots = 0$. A term like $f(\sigma)
\delta(\sigma)$ is then distributionally equal to $f(0)
\delta(\sigma)$, while a term like $f(\sigma)
\delta'(\sigma)$ is distributionally equal to $f(0)
\delta'(\sigma) - f'(0)\delta(\sigma)$, and a term like $f(\sigma) 
\delta''(\sigma)$ is distributionally equal to $f(0)
\delta''(\sigma) - 2f'(0)\delta'(\sigma) + 2f''(0)
\delta(\sigma)$; here $f(\sigma)$ is an arbitrary test
function. Summing over such terms, we recover an expression of
the form of Eq.~(\ref{11.3.2}), and there is no need to make $A$, $B$,
$C$, \ldots functions of $\sigma$.  

Differentiation of Eq.~(\ref{11.3.1}) and substitution into
Eq.~(\ref{11.2.3}) yields 
\[
\epsilon \int_{-\epsilon}^\epsilon
w(\sigma/\epsilon)\, g'(\sigma)\, d\sigma        
= \epsilon \biggl[ \int_{-\epsilon}^\epsilon V'(\sigma)
w(\sigma/\epsilon)\, d\sigma - V(0) w(0) - \frac{A}{\epsilon}
\dot{w}(0) + \frac{B}{\epsilon^2} \ddot{w}(0) - \frac{C}{\epsilon^3}
w^{(3)}(0) + \cdots \biggr],  
\]
where overdots (or a number within brackets) indicate repeated
differentiation with respect to $\xi \equiv \sigma/\epsilon$. The
limit $\epsilon \to 0$ exists if and only if $B = C =\cdots = 0$. In
the limit we must then have $A \dot{w}(0) = 1$, which implies
$A=1$. We conclude that $g(\sigma)$ must have the form of 
\begin{equation}
g(\sigma) = \delta(\sigma) + V(\sigma) \theta(-\sigma), 
\label{11.3.3} 
\end{equation}
with $V(\sigma)$ a smooth function that cannot be determined from
Eq.~(\ref{11.2.3}) alone.  

\subsection{Smooth part of $g(\sigma)$}
\label{11.4}

To determine $V(\sigma)$ we must go back to the differential equation
of Eq.~(\ref{11.1.3}). Because the singular structure of the Green's 
function is now under control, we can safely set $x \neq x'\equiv 0$
in the forthcoming operations. This means that the equation to solve
is in fact $(\Box - k^2) g(\sigma) = 0$, the homogeneous version of
Eq.~(\ref{11.1.3}). We have $\nabla_\alpha g = g' \sigma_\alpha$,
$\nabla_\alpha \nabla_\beta g = g'' \sigma_\alpha \sigma_\beta + g'
\sigma_{\alpha\beta}$, $\Box g = 2\sigma g'' + 4 g'$, so that Green's
equation reduces to the ordinary differential equation  
\begin{equation}
2\sigma g'' + 4 g' - k^2 g = 0. 
\label{11.4.1}
\end{equation} 
If we substitute Eq.~(\ref{11.3.3}) into this we get 
\[
-(2V + k^2) \delta(\sigma) + (2\sigma V'' + 4 V' - k^2 V)
 \theta(-\sigma) = 0, 
\]
where we have used the identities of Eq.~(\ref{11.3.2}). The left-hand 
side will vanish as a distribution if we set 
\begin{equation}
2\sigma V'' + 4 V' - k^2 V = 0, \qquad
V(0) = -\frac{1}{2} k^2. 
\label{11.4.2}
\end{equation}
These equations determine $V(\sigma)$ uniquely, even in the absence of
a second boundary condition at $\sigma = 0$, because the differential
equation is singular at $\sigma = 0$ and $V$ is known to be smooth.        

To solve Eq.~(\ref{11.4.2}) we let $V = F(z)/z$, with $z \equiv  
k\sqrt{-2\sigma}$. This gives rise to Bessel's equation for the new 
function $F$:
\[
z^2 F_{zz} + z F_z + (z^2 - 1) F = 0. 
\]
The solution that is well behaved near $z=0$ is $F = aJ_1(z)$, where 
$a$ is a constant to be determined. We have that $J_1(z) \sim
\frac{1}{2} z$ for small values of $z$, and it follows that $V \sim
a/2$. From Eq.~(\ref{11.4.2}) we see that $a = -k^2$. So we have found 
that the only acceptable solution to Eq.~(\ref{11.4.2}) is  
\begin{equation} 
V(\sigma) = -\frac{k}{\sqrt{-2\sigma}}\, 
J_1\bigl( k\sqrt{-2\sigma} \bigr).   
\label{11.4.3} 
\end{equation}

To summarize, the retarded and advanced solutions to
Eq.~(\ref{11.1.3}) are given by Eq.~(\ref{11.1.4}) with $g(\sigma)$
given by Eq.~(\ref{11.3.3}) and $V(\sigma)$ given by
Eq.~(\ref{11.4.3}).  

\subsection{Advanced distributional methods}
\label{11.5}

The techniques developed previously to find Green's functions for the
scalar wave equation are limited to flat spacetime, and they would not
be very useful for curved spacetimes. To pursue this  
generalization we must introduce more powerful distributional
methods. We do so in this subsection, and in the next we shall use
them to recover our previous results.    

Let $\theta_+(x,\Sigma)$ be a generalized step function, defined to be 
one if $x$ is in the future of the spacelike hypersurface $\Sigma$, 
and defined to be zero otherwise. Similarly, define
$\theta_-(x,\Sigma) \equiv 1 - \theta_+(x,\Sigma)$ to be one if $x$ is 
in the past of the spacelike hypersurface $\Sigma$, and zero
otherwise. Then define the light-cone step functions   
\begin{equation} 
\theta_\pm(-\sigma) = \theta_\pm(x,\Sigma) \theta(-\sigma), \qquad 
x' \in \Sigma, 
\label{11.5.1}
\end{equation}
so that $\theta_+(-\sigma)$ is one if $x$ is an element of $I^+(x')$,
the chronological future of $x'$, and zero otherwise, and
$\theta_-(-\sigma)$ is one if $x$ is an element of $I^-(x')$, the
chronological past of $x'$, and zero otherwise; the choice of
hypersurface is immaterial so long as $\Sigma$ is spacelike and
contains the reference point $x'$. Notice that $\theta_+(-\sigma) +
\theta_-(-\sigma) = \theta(-\sigma)$. Define also the light-cone Dirac
functionals     
\begin{equation} 
\delta_\pm(\sigma) = \theta_\pm(x,\Sigma) \delta(\sigma), \qquad 
x' \in \Sigma,  
\label{11.5.2}
\end{equation}
so that $\delta_+(\sigma)$, when viewed as a function of $x$, is 
supported on the future light cone of $x'$, while $\delta_-(\sigma)$
is supported on its past light cone. Notice that $\delta_+(\sigma) +
\delta_-(\sigma) = \delta(\sigma)$. In Eqs.~(\ref{11.5.1}) and
(\ref{11.5.2}), $\sigma$ is the world function for flat spacetime; it
is negative if $x$ and $x'$ are timelike related, and positive if they
are spacelike related.    

The distributions $\theta_\pm(-\sigma)$ and $\delta_\pm(\sigma)$ are
not defined at $x=x'$ and they cannot be differentiated there. This
pathology can be avoided if we shift $\sigma$ by a small positive
quantity $\epsilon$. We can therefore use the distributions 
$\theta_\pm(-\sigma - \epsilon)$ and $\delta_\pm(\sigma + \epsilon)$  
in some sensitive computations, and then take the limit $\epsilon \to 
0^+$. Notice that the equation $\sigma + \epsilon = 0$ describes a
two-branch hyperboloid that is located just {\it within} the light 
cone of the reference point $x'$. The hyperboloid does not include
$x'$, and $\theta_+(x,\Sigma)$ is one everywhere on its future branch,
while $\theta_-(x,\Sigma)$ is one everywhere on its past branch. These
factors, therefore, become invisible to differential operators. For
example, $\theta_+'(-\sigma-\epsilon) = \theta_+(x,\Sigma)
\theta'(-\sigma-\epsilon) = -\theta_+(x,\Sigma)
\delta(\sigma+\epsilon) = -\delta_+(\sigma + \epsilon)$. This
manipulation shows that after the shift from $\sigma$ to $\sigma +
\epsilon$, the distributions of Eqs.~(\ref{11.5.1}) and (\ref{11.5.2})
can be straightforwardly differentiated with respect to $\sigma$.  

In the next paragraphs we shall establish the distributional
identities  
\begin{eqnarray} 
\lim_{\epsilon \to 0^+} \epsilon \delta_\pm(\sigma + \epsilon) &=& 0, 
\label{11.5.3} \\
\lim_{\epsilon \to 0^+} \epsilon \delta'_\pm(\sigma + \epsilon) &=& 0, 
\label{11.5.4} \\
\lim_{\epsilon \to 0^+} \epsilon \delta''_\pm(\sigma + \epsilon) &=&
2\pi \delta_4(x - x')  
\label{11.5.5} 
\end{eqnarray}
in four-dimensional flat spacetime. These will be used in the next
subsection to recover the Green's functions for the scalar wave
equation, and they will be generalized to curved spacetime in 
Sec.~\ref{12}.   

The derivation of Eqs.~(\ref{11.5.3})--(\ref{11.5.5}) relies on
a ``master'' distributional identity, formulated in three-dimensional
flat space:   
\begin{equation}
\lim_{\epsilon \to 0^+} \frac{\epsilon}{R^5} = \frac{2\pi}{3}
\delta_3(\bm{x}), \qquad 
R \equiv \sqrt{r^2 + 2\epsilon}, 
\label{11.5.6}
\end{equation}
with $r \equiv |\bm{x}| \equiv \sqrt{x^2+y^2+z^2}$. This follows from
yet another identity, $\nabla^2 r^{-1} = -4\pi \delta_3(\bm{x})$, in
which we write the left-hand side as $\lim_{\epsilon = 0^+}\nabla^2
R^{-1}$; since $R^{-1}$ is nonsingular at $\bm{x} = 0$ it can be
straightforwardly differentiated, and the result is $\nabla^2 R^{-1} =
-6\epsilon/R^5$, from which Eq.~(\ref{11.5.6}) follows.  

To prove Eq.~(\ref{11.5.3}) we must show that $\epsilon 
\delta_\pm(\sigma+\epsilon)$ vanishes as a distribution in the limit
$\epsilon \to 0^+$. For this we must prove that a functional of the 
form  
\[
A_\pm[f] = \lim_{\epsilon \to 0^+}  
\int \epsilon \delta_\pm(\sigma + \epsilon) f(x)\, d^4 x, 
\]
where $f(x) = f(t,\bm{x})$ is a smooth test function, vanishes for all
such functions $f$. Our first task will be to find a more convenient
expression for $\delta_\pm(\sigma+\epsilon)$. Once more we set $x' =
0$ (without loss of generality) and we note that $2(\sigma+\epsilon) =
-t^2 + r^2 + 2 \epsilon = -(t-R)(t+R)$, where we have used
Eq.~(\ref{11.5.6}). It follows that      
\begin{equation}
\delta_\pm(\sigma + \epsilon) = \frac{\delta(t \mp R)}{R},         
\label{11.5.7} 
\end{equation} 
and from this we find 
\[
A_\pm[f] = \lim_{\epsilon \to 0^+} \int \epsilon 
\frac{f(\pm R, \bm{x})}{R}\, d^3 x 
= \lim_{\epsilon \to 0^+}  
\int \frac{\epsilon}{R^5} R^4 f(\pm R, \bm{x})\, d^3 x 
= \frac{2\pi}{3} \int \delta_3(\bm{x}) r^4 f(\pm r, \bm{x})\, d^3 x
= 0, 
\] 
which establishes Eq.~(\ref{11.5.3}). 

The validity of Eq.~(\ref{11.5.4}) is established by a similar
computation. Here we must show that a functional of the form   
\[
B_\pm[f] = \lim_{\epsilon \to 0^+}  
\int \epsilon \delta'_\pm(\sigma + \epsilon) f(x)\, d^4 x 
\]
vanishes for all test functions $f$. We have  
\begin{eqnarray*}
B_\pm[f] &=& \lim_{\epsilon \to 0^+} \epsilon \frac{d}{d\epsilon}   
\int \delta_\pm(\sigma+\epsilon) f(x)\, d^4 x 
= \lim_{\epsilon \to 0^+} \epsilon \frac{d}{d\epsilon}
  \int \frac{f(\pm R, \bm{x})}{R}\, d^3 x 
= \lim_{\epsilon \to 0^+} \epsilon 
  \int \biggl( \pm \frac{\dot{f}}{R^2} - \frac{f}{R^3} \biggr)\, d^3 x 
\\
&=& \lim_{\epsilon \to 0^+} 
  \int \frac{\epsilon}{R^5} \bigl( \pm R^3 \dot{f} 
  - R^2 f \bigr)\, d^3 x 
= \frac{2\pi}{3} \int \delta_3(\bm{x}) \bigl( \pm r^3 \dot{f} 
  - r^2 f \bigr)\, d^3 x
= 0,
\end{eqnarray*}
and the identity of Eq.~(\ref{11.5.4}) is proved. In these
manipulations we have let an overdot indicate partial differentiation
with respect to $t$, and we have used $\partial R/\partial \epsilon =
1/R$.         

To establish Eq.~(\ref{11.5.5}) we consider the functional
\[
C_\pm[f] = \lim_{\epsilon \to 0^+}  
\int \epsilon \delta''_\pm(\sigma + \epsilon) f(x)\, d^4 x 
\]
and show that it evaluates to $2\pi f(0,\bm{0})$. We have 
\begin{eqnarray*}
C_\pm[f] &=& \lim_{\epsilon \to 0^+} \epsilon \frac{d^2}{d\epsilon^2}    
\int \delta_\pm(\sigma+\epsilon) f(x)\, d^4 x 
= \lim_{\epsilon \to 0^+} \epsilon \frac{d^2}{d\epsilon^2} 
  \int \frac{f(\pm R, \bm{x})}{R}\, d^3 x \\ 
&=& \lim_{\epsilon \to 0^+} \epsilon \int \biggl( \frac{\ddot{f}}{R^3} 
\mp 3 \frac{\dot{f}}{R^4} + 3 \frac{f}{R^5} \biggr)\, d^3 x 
= 2\pi \int \delta_3(\bm{x}) \biggl( \frac{1}{3} r^2 \ddot{f}  
    \pm r \dot{f} + f \biggr)\, d^3 x \\ 
&=& 2\pi f(0,\bm{0}), 
\end{eqnarray*} 
as required. This proves that Eq.~(\ref{11.5.5}) holds as a 
distributional identity in four-dimensional flat spacetime. 

\subsection{Alternative computation of the Green's functions} 
\label{11.6}

The retarded and advanced Green's functions for the scalar wave
equation are now defined as the limit of the functions 
$G^\epsilon_\pm(x,x')$ as $\epsilon \to 0^+$. For these we make  
the ansatz 
\begin{equation}
G^\epsilon_\pm(x,x') = \delta_\pm(\sigma + \epsilon) + V(\sigma)
\theta_\pm(-\sigma -\epsilon),
\label{11.6.1} 
\end{equation}
and we shall prove that $G^\epsilon_\pm(x,x')$ satisfies
Eq.~(\ref{11.1.3}) in the limit. We recall that the distributions
$\theta_\pm$ and $\delta_\pm$ were defined in the preceding
subsection, and we assume that $V(\sigma)$ is a smooth function of
$\sigma(x,x') = \frac{1}{2} \eta_{\alpha\beta} (x-x')^\alpha
(x-x')^\beta$; because this function is smooth, it is not necessary to
evaluate $V$ at $\sigma + \epsilon$ in Eq.~(\ref{11.6.1}). We recall
also that $\theta_+$ and $\delta_+$ are nonzero when $x$ is in the
future of $x'$, while $\theta_-$ and $\delta_-$ are nonzero when $x$
is in the past of $x'$. We will therefore prove that the retarded and
advanced Green's functions are of the form   
\begin{equation}
G_{\rm ret}(x,x') = \lim_{\epsilon \to 0^+} G_+^\epsilon(x,x')
= \theta_+(x,\Sigma) \bigl[ \delta(\sigma) + V(\sigma) \theta(-\sigma) 
\bigr] 
\label{11.6.2}
\end{equation}
and 
\begin{equation}
G_{\rm adv}(x,x') = \lim_{\epsilon \to 0^+} G_-^\epsilon(x,x')
= \theta_-(x,\Sigma) \bigl[ \delta(\sigma) + V(\sigma) \theta(-\sigma) 
\bigr],    
\label{11.6.3}
\end{equation}
where $\Sigma$ is a spacelike hypersurface that contains $x'$. We will 
also determine the form of the function $V(\sigma)$.    

The functions that appear in Eq.~(\ref{11.6.1}) can be
straightforwardly differentiated. The manipulations are similar to
what was done in Sec.~\ref{11.4}, and dropping all labels, we obtain
$(\Box - k^2) G = 2\sigma G'' + 4 G' - k^2 G$, with a prime indicating
differentiation with respect to $\sigma$. From Eq.~(\ref{11.6.1}) we
obtain $G' = \delta' - V \delta + V' \theta$ and $G'' = \delta'' - V
\delta' - 2V' \delta + V'' \theta$. The identities of
Eq.~(\ref{11.3.2}) can be expressed as $(\sigma +
\epsilon)\delta'(\sigma + \epsilon) = - \delta(\sigma + \epsilon)$ and
$(\sigma + \epsilon)\delta''(\sigma + \epsilon) = - 2\delta'(\sigma +
\epsilon)$, and combining this with our previous results gives  
\begin{eqnarray*} 
(\Box - k^2) G^\epsilon_\pm(x,x') &=& 
(-2V - k^2) \delta_\pm(\sigma + \epsilon)   
+ (2\sigma V'' + 4 V' - k^2 V) \theta_\pm(-\sigma - \epsilon) 
\\ & & \mbox{} 
- 2 \epsilon \delta''_\pm(\sigma+\epsilon) 
+ 2 V \epsilon \delta'_\pm(\sigma + \epsilon) 
+ 4 V' \epsilon \delta_\pm(\sigma + \epsilon). 
\end{eqnarray*}
According to Eq.~(\ref{11.5.3})--(\ref{11.5.5}), the last two terms on
the right-hand side disappear in the limit $\epsilon \to 0^+$, and the 
third term becomes $-4\pi \delta_4(x-x')$. Provided that the first two
terms vanish also, we recover $(\Box - k^2)G(x,x') = -4\pi
\delta_4(x-x')$ in the limit, as required. Thus, the limit of
$G^\epsilon_\pm(x,x')$ as $\epsilon \to 0^+$ will indeed satisfy
Green's equation provided that $V(\sigma)$ is a solution to  
\begin{equation}
2\sigma V'' + 4 V' - k^2 V = 0, \qquad
V(0) = - \frac{1}{2} k^2; 
\label{11.6.4}
\end{equation} 
these are the same statements as in Eq.~(\ref{11.4.2}). The solution
to these equations was produced in Eq.~(\ref{11.4.3}): 
\begin{equation} 
V(\sigma) = -\frac{k}{\sqrt{-2\sigma}}\, 
J_1 \bigl( k\sqrt{-2\sigma} \bigr), 
\label{11.6.5}
\end{equation}
and this completely determines the Green's functions of
Eqs.~(\ref{11.6.2}) and (\ref{11.6.3}). 
 
\section{Distributions in curved spacetime} 
\label{12} 

The distributions introduced in Sec.~\ref{11.5} can also be defined in
a four-dimensional spacetime with metric $g_{\alpha\beta}$. Here we 
produce the relevant generalizations of the results derived in that
section. 

\subsection{Invariant Dirac distribution} 
\label{12.1} 

We first introduce $\delta_4(x,x')$, an {\it invariant} Dirac
functional in a four-dimensional curved spacetime. This is
defined by the relations 
\begin{equation} 
\int_V f(x) \delta_4(x,x') \sqrt{-g}\, d^4 x = f(x'), \qquad 
\int_{V'} f(x') \delta_4(x,x') \sqrt{-g'}\, d^4 x' = f(x),
\label{12.1.1}
\end{equation}
where $f(x)$ is a smooth test function, $V$ any four-dimensional
region that contains $x'$, and $V'$ any four-dimensional region that
contains $x$. These relations imply that $\delta_4(x,x')$ is symmetric
in its arguments, and it is easy to see that  
\begin{equation}
\delta_4(x,x') = \frac{\delta_4(x-x')}{\sqrt{-g}} 
= \frac{\delta_4(x-x')}{\sqrt{-g'}}
= (gg')^{-1/4} \delta_4(x-x'), 
\label{12.1.2} 
\end{equation}
where $\delta_4(x-x') = \delta(x^0 - x^{\prime 0}) \delta(x^1 - 
x^{\prime 1}) \delta(x^2 - x^{\prime 2}) \delta(x^3 - x^{\prime 3})$
is the ordinary (coordinate) four-dimensional Dirac functional. The 
relations of Eq.~(\ref{12.1.2}) are all equivalent because $f(x)
\delta_4(x,x') = f(x') \delta_4(x,x')$ is a distributional identity;
the last form is manifestly symmetric in $x$ and $x'$. 

The invariant Dirac distribution satisfies the identities 
\begin{eqnarray}
\Omega_{\cdots}(x,x') \delta_4(x,x') &=& \bigl[ \Omega_{\cdots} 
\bigr] \delta_4(x,x'), 
\nonumber \\ 
& & \label{12.1.3} \\ 
\bigl( g^\alpha_{\ \alpha'}(x,x') \delta_4(x,x') \bigr)_{;\alpha} =
-\partial_{\alpha'} \delta_4(x,x'), &  &
\bigl( g^{\alpha'}_{\ \alpha}(x',x) \delta_4(x,x') \bigr)_{;\alpha'} = 
-\partial_{\alpha} \delta_4(x,x'), 
\nonumber
\end{eqnarray}
where $\Omega_{\cdots}(x,x')$ is any bitensor and 
$g^\alpha_{\ \alpha'}(x,x')$, $g^{\alpha'}_{\ \alpha}(x,x')$ are 
parallel propagators. The first identity follows immediately from the
definition of the $\delta$-function. The second and third identities
are established by showing that integration against a test function
$f(x)$ gives the same result from both sides. For example, the first
of the Eqs.~(\ref{12.1.1}) implies   
\[
\int_V f(x) \partial_{\alpha'}\delta_4(x,x') 
\sqrt{-g}\, d^4 x = \partial_{\alpha'} f(x'),  
\]
and on the other hand, 
\[
-\int_V f(x) \bigl( g^\alpha_{\ \alpha'} \delta_4(x,x')
\bigr)_{;\alpha} \sqrt{-g}\, d^4 x = - \oint_{\partial V} f(x)
g^\alpha_{\ \alpha'} \delta_4(x,x') d\Sigma_\alpha 
+ \bigl[ f_{,\alpha} g^\alpha_{\ \alpha'} \bigr] 
= \partial_{\alpha'} f(x'), 
\]
which establishes the second identity of Eq.~(\ref{12.1.3}). Notice
that in these manipulations, the integrations involve {\it scalar} 
functions of the coordinates $x$; the fact that these functions are
also vectors with respect to $x'$ does not invalidate the
procedure. The third identity of Eq.~(\ref{12.1.3}) is proved in a 
similar way.  

\subsection{Light-cone distributions} 
\label{12.2}

For the remainder of Sec.~\ref{12} we assume that $x \in 
{\cal N}(x')$, so that a unique geodesic $\beta$ links these two
points. We then let $\sigma(x,x')$ be the curved spacetime world
function, and we define light-cone step functions by   
\begin{equation} 
\theta_\pm(-\sigma) = \theta_\pm(x,\Sigma) \theta(-\sigma), \qquad  
x' \in \Sigma,
\label{12.2.1}
\end{equation}
where $\theta_+(x,\Sigma)$ is one if $x$ is in the future of the
spacelike hypersurface $\Sigma$ and zero otherwise, and
$\theta_-(x,\Sigma) = 1 - \theta_+(x,\Sigma)$. These are immediate
generalizations to curved spacetime of the objects defined in flat
spacetime by Eq.~(\ref{11.5.1}). We have that $\theta_+(-\sigma)$ is
one if $x$ is an element of $I^+(x')$, the chronological future of
$x'$, and zero otherwise, and $\theta_-(-\sigma)$ is one if $x$ is an
element of $I^-(x')$, the chronological past of $x'$, and zero
otherwise. We also have $\theta_+(-\sigma) + \theta_-(-\sigma) = 
\theta(-\sigma)$.  

We define the curved-spacetime version of the light-cone Dirac
functionals by     
\begin{equation} 
\delta_\pm(\sigma) = \theta_\pm(x,\Sigma) \delta(\sigma), \qquad
x' \in \Sigma,
\label{12.2.2}
\end{equation}
an immediate generalization of Eq.~(\ref{11.5.2}). We have that 
$\delta_+(\sigma)$, when viewed as a function of $x$, is supported on 
the future light cone of $x'$, while $\delta_-(\sigma)$ is supported
on its past light cone. We also have $\delta_+(\sigma) +
\delta_-(\sigma) = \delta(\sigma)$, and we recall that $\sigma$ is
negative if $x$ and $x'$ are timelike related, and positive if they
are spacelike related. 

For the same reasons as those mentioned in Sec.~\ref{11.5}, it is
sometimes convenient to shift the argument of the step and
$\delta$-functions from $\sigma$ to $\sigma + \epsilon$, where
$\epsilon$ is a small positive quantity. With this shift, the
light-cone distributions can be straightforwardly differentiated with
respect to $\sigma$. For example, $\delta_\pm(\sigma + \epsilon) = 
-\theta'_\pm(-\sigma-\epsilon)$, with a prime indicating 
differentiation with respect to $\sigma$.  

We now prove that the identities of Eq.~(\ref{11.5.3})--(\ref{11.5.5}) 
generalize to 
\begin{eqnarray} 
\lim_{\epsilon \to 0^+} \epsilon \delta_\pm(\sigma + \epsilon) &=& 0, 
\label{12.2.3} \\
\lim_{\epsilon \to 0^+} \epsilon \delta'_\pm(\sigma + \epsilon) &=& 0, 
\label{12.2.4} \\
\lim_{\epsilon \to 0^+} \epsilon \delta''_\pm(\sigma + \epsilon) &=&
2\pi \delta_4(x,x')  
\label{12.2.5} 
\end{eqnarray}
in a four-dimensional curved spacetime; the only differences lie with  
the definition of the world function and the fact that it is the
invariant Dirac functional that appears in Eq.~(\ref{12.2.5}). To 
establish these identities in curved spacetime we use the fact that
they hold in flat spacetime --- as was shown in Sec.~\ref{11.5} ---
and that they are scalar relations that must be valid in any
coordinate system if they are found to hold in one. Let us then
examine Eqs.~(\ref{12.2.3})--(\ref{12.2.4}) in the Riemann normal
coordinates of Sec.~\ref{7}; these are denoted
$\hat{x}^\alpha$ and are based at $x'$. We have that
$\sigma(x,x') = \frac{1}{2} \eta_{\alpha\beta} \hat{x}^\alpha
\hat{x}^\beta$ and $\delta_4(x,x') = \Delta(x,x') \delta_4(x-x') =
\delta_4(x-x')$, where $\Delta(x,x')$ is the van Vleck determinant,
whose coincidence limit is unity. In Riemann normal coordinates,
therefore, Eqs.~(\ref{12.2.3})--(\ref{12.2.5}) take exactly the same
form as Eqs.~(\ref{11.5.3})--(\ref{11.5.5}). Because the identities
are true in flat spacetime, they must be true also in curved spacetime
(in Riemann normal coordinates based at $x'$); and because these are
scalar relations, they must be valid in any coordinate system.     

\section{Scalar Green's functions in curved spacetime} 
\label{13}

\subsection{Green's equation for a massless scalar field in curved
spacetime} 
\label{13.1} 

We consider a massless scalar field $\Phi(x)$ in a curved spacetime 
with metric $g_{\alpha\beta}$. The field satisfies the wave equation 
\begin{equation}
( \Box - \xi R ) \Phi(x) = -4\pi \mu(x),   
\label{13.1.1}
\end{equation}
where $\Box = g^{\alpha\beta} \nabla_\alpha \nabla_\beta$ is the wave
operator, $R$ the Ricci scalar, $\xi$ an arbitrary coupling constant,
and $\mu(x)$ is a prescribed source. We seek a Green's function
$G(x,x')$ such that a solution to Eq.~(\ref{13.1.1}) can be expressed
as 
\begin{equation}
\Phi(x) = \int G(x,x') \mu(x') \sqrt{-g'}\, d^4 x', 
\label{13.1.2}
\end{equation} 
where the integration is over the entire spacetime. The wave equation
for the Green's function is 
\begin{equation}
( \Box - \xi R ) G(x,x') = -4\pi \delta_4(x,x'),  
\label{13.1.3}
\end{equation}
where $\delta_4(x,x')$ is the invariant Dirac functional introduced in
Sec.~\ref{12.1}. It is easy to verify that the field defined by
Eq.~(\ref{13.1.2}) is truly a solution to Eq.~(\ref{13.1.1}). 

We let $G_+(x,x')$ be the retarded solution to
Eq.~(\ref{13.1.3}), and $G_-(x,x')$ is the advanced solution;
when viewed as functions of $x$, $G_+(x,x')$ is nonzero in the causal 
future of $x'$, while $G_-(x,x')$ is nonzero in its causal past. We
assume that the retarded and advanced Green's functions exist as
distributions and can be defined globally in the entire spacetime.    

\subsection{Hadamard construction of the Green's functions} 
\label{13.2} 

Assuming throughout this subsection that $x$ is restricted to the
normal convex neighbourhood of $x'$, we make the ansatz     
\begin{equation}
G_\pm(x,x') = U(x,x') \delta_\pm(\sigma) 
+ V(x,x') \theta_\pm(-\sigma), 
\label{13.2.1}
\end{equation}
where $U(x,x')$ and $V(x,x')$ are smooth biscalars; the fact that the
spacetime is no longer homogeneous means that these functions cannot
depend on $\sigma$ alone.    

Before we substitute the Green's functions of Eq.~(\ref{13.2.1}) into
the differential equation of Eq.~(\ref{13.1.3}) we proceed as in
Sec.~\ref{11.6} and shift $\sigma$ by the small positive quantity 
$\epsilon$. We shall therefore consider the distributions
\[
G^\epsilon_\pm(x,x') = U(x,x') \delta_\pm(\sigma+\epsilon)  
+ V(x,x') \theta_\pm(-\theta-\epsilon),
\]
and later recover the Green's functions by taking the limit  
$\epsilon \to 0^+$. Differentiation of these objects is
straightforward, and in the following manipulations we will repeatedly 
use the relation $\sigma^\alpha \sigma_\alpha = 2 \sigma$
satisfied by the world function. We will also use the distributional
identities $\sigma \delta_\pm(\sigma + \epsilon) = -\epsilon
\delta_\pm(\sigma + \epsilon)$, $\sigma \delta'_\pm(\sigma + \epsilon)
= -\delta_\pm(\sigma + \epsilon) - \epsilon \delta'_\pm(\sigma +
\epsilon)$, and $\sigma \delta''_\pm(\sigma + \epsilon) = - 2
\delta'(\sigma+\epsilon) - \epsilon \delta''(\sigma +
\epsilon)$. After a routine calculation we obtain 
\begin{eqnarray*}
( \Box - \xi R ) G^\epsilon_\pm &=& 
- 2\epsilon \delta''_\pm(\sigma + \epsilon) U 
+ 2\epsilon \delta'_\pm(\sigma + \epsilon) V 
+ \delta'_\pm(\sigma+\epsilon) \Bigl\{ 2 U_{,\alpha} \sigma^\alpha 
  + (\sigma^\alpha_{\ \alpha} - 4) U \Bigr\} 
\\ & & \mbox{} 
+ \delta_\pm(\sigma+\epsilon) \Bigl\{ - 2 V_{,\alpha} \sigma^\alpha 
  + (2-\sigma^\alpha_{\ \alpha}) V + (\Box - \xi R) U \Bigr\} 
+ \theta_\pm(-\sigma-\epsilon) \Bigl\{ (\Box - \xi R) V \Bigr\}, 
\end{eqnarray*} 
which becomes 
\begin{eqnarray}
( \Box - \xi R ) G_\pm &=&  
- 4\pi \delta_4(x,x') U  
+ \delta'_\pm(\sigma) \Bigl\{ 2 U_{,\alpha} \sigma^\alpha  
  + (\sigma^\alpha_{\ \alpha} - 4) U \Bigr\} 
\nonumber \\ & & \mbox{} 
+ \delta_\pm(\sigma) \Bigl\{ - 2 V_{,\alpha} \sigma^\alpha 
  + (2-\sigma^\alpha_{\ \alpha}) V + (\Box - \xi R) U \Bigr\} 
+ \theta_\pm(-\sigma) \Bigl\{ (\Box - \xi R) V \Bigr\} \qquad
\label{13.2.2} 
\end{eqnarray} 
in the limit $\epsilon \to 0^+$,  after using the identities of
Eqs.~(\ref{12.2.3})--(\ref{12.2.5}).   
 
According to Eq.~(\ref{13.1.3}), the right-hand side of
Eq.~(\ref{13.2.2}) should be equal to $-4\pi \delta_4(x,x')$. This
immediately gives us the coincidence condition   
\begin{equation}
\bigl[ U \bigr] = 1
\label{13.2.3} 
\end{equation} 
for the biscalar $U(x,x')$. To eliminate the $\delta'_\pm$ term we 
make its coefficient vanish:
\begin{equation} 
2 U_{,\alpha} \sigma^\alpha + (\sigma^\alpha_{\ \alpha} - 4) U = 0. 
\label{13.2.4} 
\end{equation}
As we shall now prove, these two equations determine $U(x,x')$ 
uniquely. 

Recall from Sec.~\ref{2.3} that $\sigma^\alpha$ is a vector at $x$
that is tangent to the unique geodesic $\beta$ that connects $x$ to
$x'$. This geodesic is affinely parameterized by $\lambda$ and a
displacement along $\beta$ is described by $dx^\alpha =
(\sigma^\alpha/\lambda) d\lambda$. The first term of
Eq.~(\ref{13.2.4}) therefore represents the rate of change of 
$U(x,x')$ along $\beta$, and this can be expressed as $2 \lambda 
dU/d\lambda$. For the second term we recall from Sec.~\ref{6.1} the
differential equation $\Delta^{-1} (\Delta \sigma^\alpha)_{;\alpha} 
= 4$ satisfied by $\Delta(x,x')$, the van Vleck determinant. This
gives us $\sigma^\alpha_{\ \alpha} - 4 = \Delta^{-1} \Delta_{,\alpha} 
\sigma^\alpha = \Delta^{-1} \lambda d \Delta/d\lambda$, and  
Eq.~(\ref{13.2.4}) becomes   
\[
\lambda \frac{d}{d\lambda} \bigl( 2 \ln U - \ln \Delta \bigr) = 0.
\]
It follows that $U^2/\Delta$ is constant on $\beta$, and it must
therefore be equal to its value at the starting point $x'$:
$U^2/\Delta = [U^2/\Delta] = 1$, by virtue of Eq.~(\ref{13.2.3}) and
the property $[\Delta] = 1$ of the van Vleck determinant. Since this 
statement must be true for all geodesics $\beta$ that emanate from
$x'$, we have found that the unique solution to Eqs.~(\ref{13.2.3})
and (\ref{13.2.4}) is  
\begin{equation}
U(x,x') = \Delta^{1/2}(x,x').  
\label{13.2.5}
\end{equation}   

We must still consider the remaining terms in Eq.~(\ref{13.2.2}). The 
$\delta_\pm$ term can be eliminated by demanding that its coefficient
vanish when $\sigma = 0$. This, however, does not constrain its value
away from the light cone, and we thus obtain information about
$V|_{\sigma = 0}$ only. Denoting this by $\check{V}(x,x')$ --- the
restriction of $V(x,x')$ on the light cone $\sigma(x,x') = 0$ --- we
have  
\begin{equation} 
\check{V}_{,\alpha} \sigma^\alpha 
+ \frac{1}{2} \bigl( \sigma^\alpha_{\ \alpha} - 2 \bigr) \check{V} 
= \frac{1}{2} \bigl( \Box - \xi R \bigr) U \Bigr|_{\sigma=0}, 
\label{13.2.6}
\end{equation}
where we indicate that the right-hand side also must be
restricted to the light cone. The first term of Eq.~(\ref{13.2.6}) can 
be expressed as $\lambda d\check{V}/d\lambda$ and this equation can be  
integrated along any null geodesic that generates the null cone
$\sigma(x,x') = 0$. For these integrations to be well posed, however, 
we must provide initial values at $x = x'$. As we shall now see, these
can be inferred from Eq.~(\ref{13.2.6}) and the fact that $V(x,x')$
must be smooth at coincidence. 

Equations (\ref{6.1.4}) and (\ref{13.2.5}) imply that near
coincidence, $U(x,x')$ admits the expansion    
\begin{equation} 
U = 1 + \frac{1}{12} R_{\alpha'\beta'} \sigma^{\alpha'}
\sigma^{\beta'} + O(\lambda^3), 
\label{13.2.7}
\end{equation} 
where $R_{\alpha'\beta'}$ is the Ricci tensor at $x'$ and $\lambda$ 
is the affine-parameter distance to $x$ (which can be either on or off   
the light cone). Differentiation of this relation gives  
\begin{equation}
U_{,\alpha} = -\frac{1}{6} g^{\alpha'}_{\ \alpha} R_{\alpha'\beta'}
\sigma^{\beta'} + O(\lambda^2), \qquad 
U_{,\alpha'} = \frac{1}{6} R_{\alpha'\beta'} \sigma^{\beta'} +
O(\lambda^2), 
\label{13.2.8}
\end{equation} 
and eventually, 
\begin{equation}
\bigl[ \Box U \bigr] = \frac{1}{6} R(x'). 
\label{13.2.9} 
\end{equation} 
Using also $[\sigma^\alpha_{\ \alpha}] = 4$, we find that the
coincidence limit of Eq.~(\ref{13.2.6}) gives  
\begin{equation}
\bigl[ V \bigr] = \frac{1}{12} \Bigl( 1 - 6\xi \Bigr) R(x'), 
\label{13.2.10} 
\end{equation}
and this provides the initial values required for the integration of 
Eq.~(\ref{13.2.6}) on the null cone.   

Equations (\ref{13.2.6}) and (\ref{13.2.10}) give us a means to
construct $\check{V}(x,x')$, the restriction of $V(x,x')$ on the null
cone $\sigma(x,x') = 0$. These values can then be used as
characteristic data for the wave equation   
\begin{equation}
(\Box - \xi R) V(x,x') = 0, 
\label{13.2.11} 
\end{equation} 
which is obtained by elimination of the $\theta_\pm$ term in
Eq.~(\ref{13.2.2}). While this certainly does not constitute a
practical method to compute the biscalar $V(x,x')$, these
considerations show that $V(x,x')$ exists and is unique.   

To summarize: We have shown that with $U(x,x')$ given by 
Eq.~(\ref{13.2.5}) and $V(x,x')$ determined uniquely by the wave 
equation of Eq.~(\ref{13.2.11}) and the characteristic data
constructed with Eqs.~(\ref{13.2.6}) and (\ref{13.2.10}), the retarded
and advanced Green's functions of Eq.~(\ref{13.2.1}) do indeed satisfy 
Eq.~(\ref{13.1.3}). It should be emphasized that the construction
provided in this subsection is restricted to ${\cal N}(x')$, the
normal convex neighbourhood of the reference point $x'$.  

\subsection{Reciprocity} 
\label{13.3} 

We shall now establish the following reciprocity relation between the
(globally defined) retarded and advanced Green's functions:   
\begin{equation}
G_-(x',x) = G_+(x,x').
\label{13.3.1}   
\end{equation}
Before we get to the proof we observe that by virtue of
Eq.~(\ref{13.3.1}), the biscalar $V(x,x')$ must be symmetric in its 
arguments:   
\begin{equation}
V(x',x) = V(x,x'). 
\label{13.3.2}  
\end{equation}
To go from Eq.~(\ref{13.3.1}) to Eq.~(\ref{13.3.2}) we need simply
note that if $x \in {\cal N}(x')$ and belongs to $I^+(x')$, then
$G_+(x,x') = V(x,x')$ and $G_-(x',x) = V(x',x)$.  

To prove the reciprocity relation we invoke the identities 
\[
G_+(x,x') (\Box - \xi R) G_-(x,x'') = -4\pi G_+(x,x') \delta_4(x,x'') 
\]
and 
\[
G_-(x,x'') (\Box - \xi R) G_+(x,x') = -4\pi G_-(x,x'')
\delta_4(x,x')
\]
and take their difference. On the left-hand side we have
\[
G_+(x,x') \Box G_-(x,x'') - G_-(x,x'') \Box G_+(x,x') = \nabla_\alpha 
\Bigl( G_+(x,x') \nabla^\alpha G_-(x,x'') - G_-(x,x'') \nabla^\alpha
G_+(x,x') \Bigr), 
\]
while the right-hand side gives 
\[
-4\pi \Bigl( G_+(x,x') \delta_4(x,x'') - G_-(x,x'') \delta_4(x,x')
\Bigr). 
\]
Integrating both sides over a large four-dimensional region $V$ that
contains both $x'$ and $x''$, we obtain 
\[
\oint_{\partial V} \Bigl( G_+(x,x') \nabla^\alpha G_-(x,x'') -
G_-(x,x'') \nabla^\alpha G_+(x,x') \Bigr)\, d\Sigma_\alpha = 
-4\pi \Bigl( G_+(x'',x') - G_-(x',x'') \Bigr), 
\]
where $\partial V$ is the boundary of $V$. Assuming that the Green's
functions fall off sufficiently rapidly at infinity (in the limit
$\partial V \to \infty$; this statement imposes some
restriction on the spacetime's asymptotic structure), we have that the
left-hand side of the equation evaluates to zero in the limit. This
gives us the statement $G_+(x'',x') = G_-(x',x'')$, which is just
Eq.~(\ref{13.3.1}) with $x''$ replacing $x$.  

\subsection{Kirchhoff representation}  
\label{13.4} 

Suppose that the values for a scalar field $\Phi(x')$ and its normal
derivative $n^{\alpha'} \nabla_{\alpha'} \Phi(x')$ are known on a
spacelike hypersurface $\Sigma$. Suppose also that the scalar field 
satisfies the homogeneous wave equation  
\begin{equation}
(\Box - \xi R) \Phi(x) = 0. 
\label{13.4.1}
\end{equation} 
Then the value of the field at a point $x$ in the future of $\Sigma$ 
is given by Kirchhoff's formula,  
\begin{equation} 
\Phi(x) = -\frac{1}{4\pi} \int_\Sigma \Bigl( G_+(x,x')
\nabla^{\alpha'} \Phi(x') - \Phi(x') \nabla^{\alpha'} G_+(x,x')
\Bigr)\, d\Sigma_{\alpha'}, 
\label{13.4.2} 
\end{equation}
where $d\Sigma_{\alpha'}$ is the surface element on $\Sigma$. If
$n_{\alpha'}$ is the future-directed unit normal, then
$d\Sigma_{\alpha'} = - n_{\alpha'} dV$, with $dV$ denoting the
invariant volume element on $\Sigma$; notice that $d\Sigma_{\alpha'}$
is past directed.  

To establish this result we start with the equations 
\[
G_-(x',x) (\Box' - \xi R') \Phi(x') = 0, \qquad 
\Phi(x') (\Box' - \xi R') G_-(x',x) = -4\pi \delta_4(x',x) \Phi(x'),  
\]
in which $x$ and $x'$ refer to arbitrary points in spacetime. Taking
their difference gives 
\[
\nabla_{\alpha'} \Bigl( G_-(x',x) \nabla^{\alpha'} \Phi(x') - \Phi(x')
\nabla^{\alpha'} G_-(x',x) \Bigr) = 4\pi \delta_4(x',x) \Phi(x'), 
\]
and this we integrate over a four-dimensional region $V$ that is
bounded in the past by the hypersurface $\Sigma$. We suppose that $V$ 
contains $x$ and we obtain  
\[
\oint_{\partial V} \Bigl( G_-(x',x) \nabla^{\alpha'} \Phi(x') -
\Phi(x') \nabla^{\alpha'} G_-(x',x) \Bigr)\, d\Sigma_{\alpha'} 
= 4\pi \Phi(x), 
\]
where $d\Sigma_{\alpha'}$ is the outward-directed surface element on
the boundary $\partial V$. Assuming that the Green's function falls
off sufficiently rapidly into the future, we have that the only
contribution to the hypersurface integral is the one that comes from
$\Sigma$. Since the surface element on $\Sigma$ points in the
direction opposite to the outward-directed surface element on
$\partial V$, we must change the sign of the left-hand side to be
consistent with the convention adopted previously. With this change 
we have 
\[
\Phi(x) = -\frac{1}{4\pi} \oint_{\partial V} \Bigl( G_-(x',x)
\nabla^{\alpha'} \Phi(x') - \Phi(x') \nabla^{\alpha'} G_-(x',x)
\Bigr)\, d\Sigma_{\alpha'},
\]
which is the same as Eq.~(\ref{13.4.2}) if we take into account the 
reciprocity relation of Eq.~(\ref{13.3.1}). 

\subsection{Singular and radiative Green's functions}  
\label{13.5} 

In part \ref{part4} of this review we will compute the retarded field
of a moving scalar charge, and we will analyze its singularity
structure near the world line; this will be part of our effort to
understand the effect of the field on the particle's motion. The
retarded solution to the scalar wave equation is the physically
relevant solution because it properly incorporates outgoing-wave
boundary conditions at infinity --- the advanced solution would come
instead with incoming-wave boundary conditions. The retarded field is
singular on the world line because a point particle produces a Coulomb
field that diverges at the particle's position. In view of this
singular behaviour, it is a subtle matter to describe the field's
action on the particle, and to formulate meaningful equations of
motion.       

When facing this problem in flat spacetime (recall the discussion of
Sec.~\ref{1.3}) it is convenient to decompose the retarded Green's
function $G_+(x,x')$ into a {\it singular} Green's function 
$G_{\rm S}(x,x') \equiv \frac{1}{2}
[G_+(x,x') + G_-(x,x')]$ and a {\it radiative} Green's
function $G_{\rm R}(x,x') \equiv \frac{1}{2} [G_+(x,x') 
- G_-(x,x')]$. The singular Green's function takes its name from
the fact that it produces a field with the same singularity
structure as the retarded solution: the diverging field near the
particle is insensitive to the boundary conditions imposed at 
infinity. We note also that $G_{\rm S}(x,x')$ satisfies the same wave 
equation as the retarded Green's function (with a Dirac functional as
a source), and that by virtue of the reciprocity relations, it is
symmetric in its arguments. The radiative Green's function, on the
other hand, takes its name from the fact that it satisfies the 
{\it homogeneous} wave equation, without the Dirac functional on the
right-hand side; it produces a field that is smooth on the world line
of the moving scalar charge.  

Because the singular Green's function is symmetric in its argument, it
does not distinguish between past and future, and it produces a field
that contains equal amounts of outgoing and incoming radiation --- the
singular solution describes standing waves at infinity.  
Removing $G_{\rm S}(x,x')$ from the retarded Green's
function will therefore have the effect of removing the singular
behaviour of the field {\it without affecting the motion of the
particle}. The motion is not affected because it is intimately
tied to the boundary conditions: If the waves are outgoing, the  
particle loses energy to the radiation and its motion is affected; if
the waves are incoming, the particle gains energy from the radiation
and its motion is affected differently. With equal amounts of outgoing
and incoming radiation, the particle neither loses nor gains energy
and its interaction with the scalar field cannot affect its
motion. Thus, subtracting $G_{\rm S}(x,x')$ from the retarded
Green's function eliminates the singular part of the field without
affecting the motion of the scalar charge. The subtraction leaves 
behind the radiative Green's function, which produces a field that is
smooth on the world line; it is this field that will govern the  
motion of the particle. The action of this field is well defined, 
and it properly encodes the outgoing-wave boundary conditions: the 
particle will lose energy to the radiation.       

In this subsection we attempt a decomposition of the curved-spacetime
retarded Green's function into singular and radiative Green's
functions. The flat-spacetime relations will have to be amended,
however, because of the fact that in a curved spacetime, the advanced
Green's function is generally nonzero when $x'$ is in the
chronological future of $x$. This implies that the value of the
advanced field at $x$ depends on events $x'$ that will unfold {\it in
the future}; this dependence would be inherited by the radiative field
(which acts on the particle and determines its motion) if the naive
definition $G_{\rm R}(x,x') \equiv \frac{1}{2} [G_{+}(x,x') 
- G_{-}(x,x')]$ were to be adopted.   

We shall not adopt this definition. Instead, we shall follow Detweiler
and Whiting \cite{detweilerwhiting} and introduce a singular Green's
function with the properties    
\begin{description} 
\item[\qquad {\sf S1}:] $G_{\rm S}(x,x')$ satisfies the inhomogeneous
scalar wave equation,     
\begin{equation} 
(\Box - \xi R) G_{\rm S}(x,x') = -4\pi \delta_4(x,x'); 
\label{13.5.1} 
\end{equation} 
\item[\qquad {\sf S2}:] $G_{\rm S}(x,x')$ is symmetric in its
arguments,  
\begin{equation} 
G_{\rm S}(x',x) = G_{\rm S}(x,x');
\label{13.5.2}
\end{equation} 
\item[\qquad {\sf S3}:] $G_{\rm S}(x,x')$ vanishes if $x$ is in the 
chronological past or future of $x'$, 
\begin{equation}
G_{\rm S}(x,x') = 0 \qquad \mbox{when $x \in I^\pm(x')$}. 
\label{13.5.3}
\end{equation} 
\end{description} 
Properties {\sf S1} and {\sf S2} ensure that the singular Green's
function will properly reproduce the singular behaviour of the
retarded solution without distinguishing between past and future; and  
as we shall see, property {\sf S3} ensures that the support of the
radiative Green's function will not include the chronological future
of $x$. 

The radiative Green's function is then defined by 
\begin{equation}
G_{\rm R}(x,x') = G_+(x,x') - G_{\rm S}(x,x'), 
\label{13.5.4}
\end{equation}
where $G_+(x,x')$ is the retarded Green's function. This comes with
the properties 
\begin{description} 
\item[\qquad {\sf R1}:] $G_{\rm R}(x,x')$ satisfies the homogeneous wave
equation, 
\begin{equation}
(\Box - \xi R) G_{\rm R}(x,x') = 0;  
\label{13.5.5}
\end{equation} 
\item[\qquad {\sf R2}:] $G_{\rm R}(x,x')$ agrees with the retarded
Green's function if $x$ is in the chronological future of $x'$, 
\begin{equation}
G_{\rm R}(x,x') = G_+(x,x') \qquad \mbox{when $x \in I^+(x')$}; 
\label{13.5.6}
\end{equation} 
\item[\qquad {\sf R3}:] $G_{\rm R}(x,x')$ vanishes if $x$ is in the 
chronological past of $x'$,  
\begin{equation}
G_{\rm R}(x,x') = 0 \qquad \mbox{when $x \in I^-(x')$}.  
\label{13.5.7}
\end{equation} 
\end{description}
Property {\sf R1} follows directly from Eq.~(\ref{13.5.4}) and
property {\sf S1} of the singular Green's function. Properties 
{\sf R2} and {\sf R3} follow from {\sf S3} and the fact that the
retarded Green's function vanishes if $x$ is in past of $x'$. The
properties of the radiative Green's function ensure that the
corresponding radiative field will be smooth at the world line, and 
will depend only on the past history of the scalar charge.  

We must still show that such singular and radiative Green's functions 
can be constructed. This relies on the existence of a two-point
function $H(x,x')$ that would possess the properties  
\begin{description} 
\item[\qquad {\sf H1}:] $H(x,x')$ satisfies the homogeneous wave
equation, 
\begin{equation}
(\Box - \xi R) H(x,x') = 0; 
\label{13.5.8}
\end{equation} 
\item[\qquad {\sf H2}:] $H(x,x')$ is symmetric in its arguments, 
\begin{equation}
H(x',x) = H(x,x'); 
\label{13.5.9}
\end{equation} 
\item[\qquad {\sf H3}:] $H(x,x')$ agrees with the retarded 
Green's function if $x$ is in the chronological future of $x'$,  
\begin{equation}
H(x,x') = G_+(x,x') \qquad \mbox{when $x \in I^+(x')$};  
\label{13.5.10}
\end{equation} 
\item[\qquad {\sf H4}:] $H(x,x')$ agrees with the advanced  
Green's function if $x$ is in the chronological past of $x'$,  
\begin{equation}
H(x,x') = G_-(x,x') \qquad \mbox{when $x \in I^-(x')$}.   
\label{13.5.11}
\end{equation} 
\end{description}
With a biscalar $H(x,x')$ satisfying these relations, a singular
Green's function defined by    
\begin{equation}
G_{\rm S}(x,x') = \frac{1}{2} \Bigl[ G_+(x,x') + G_-(x,x') - H(x,x')
\Bigr]
\label{13.5.12}
\end{equation}  
will satisfy all the properties listed previously: {\sf S1} comes as a
consequence of {\sf H1} and the fact that both the advanced and the
retarded Green's functions are solutions to the inhomogeneous wave
equation, {\sf S2} follows directly from {\sf H2} and the definition
of Eq.~(\ref{13.5.12}), and {\sf S3} comes as a consequence of 
{\sf H3}, {\sf H4} and the properties of the retarded and advanced
Green's functions. 

The question is now: does such a function $H(x,x')$ exist? I will
present a plausibility argument for an affirmative answer. Later in
this section we will see that $H(x,x')$ is guaranteed to exist in the
local convex neighbourhood of $x'$, where it is equal to
$V(x,x')$. And in Sec.~\ref{13.6} we will see that there exist
particular spacetimes for which $H(x,x')$ can be defined globally.   

To satisfy all of {\sf H1}--{\sf H4} might seem a tall order, but it
should be possible. We first note that property {\sf H4} is not
independent from the rest: it follows from {\sf H2}, {\sf H3},
and the reciprocity relation (\ref{13.3.1}) satisfied by the retarded
and advanced Green's functions. Let $x \in I^-(x')$, so that $x' \in
I^+(x)$. Then $H(x,x') = H(x',x)$ by {\sf H2}, and by {\sf H3} this is
equal to $G_+(x',x)$. But by the reciprocity relation this is also
equal to $G_-(x,x')$, and we have obtained {\sf H4}. Alternatively,
and this shall be our point of view in the next paragraph, we can
think of {\sf H3} as following from {\sf H2} and {\sf H4}.  

Because $H(x,x')$ satisfies the homogeneous wave equation (property 
{\sf H1}), it can be given the Kirkhoff representation of
Eq.~(\ref{13.4.2}): if $\Sigma$ is a spacelike hypersurface in the
past of both $x$ and $x'$, then    
\[
H(x,x') = -\frac{1}{4\pi} \int_\Sigma \Bigl( G_+(x,x'')
\nabla^{\alpha''} H(x'',x') - H(x'',x') \nabla^{\alpha''} G_+(x,x'')  
\Bigr)\, d\Sigma_{\alpha''},
\]
where $d\Sigma_{\alpha''}$ is a surface element on $\Sigma$. The 
hypersurface can be partitioned into two segments, $\Sigma^-(x')$ and
$\Sigma - \Sigma^-(x')$, with $\Sigma^-(x')$ denoting the intersection
of $\Sigma$ with $I^-(x')$. To enforce {\sf H4} it suffices to choose
for $H(x,x')$ initial data on $\Sigma^-(x')$ that agree with the
initial data for the advanced Green's function; because both functions
satisfy the homogeneous wave equation in $I^-(x')$, the agreement will
be preserved in all of the domain of dependence of $\Sigma^-(x')$. The
data on $\Sigma - \Sigma^-(x')$ is still free, and it should be
possible to choose it so as to make $H(x,x')$ symmetric. Assuming that
this can be done, we see that {\sf H2} is enforced and we conclude 
that the properties {\sf H1}, {\sf H2}, {\sf H3}, and {\sf H4} can all
be satisfied. 

When $x$ is restricted to the normal convex neighbourhood of $x'$,
properties {\sf H1}--{\sf H4} imply that  
\begin{equation} 
H(x,x') = V(x,x'); 
\label{13.5.13}
\end{equation} 
it should be stressed here that while $H(x,x')$ is assumed to be
defined globally in the entire spacetime, the existence of $V(x,x')$
is limited to ${\cal N}(x')$. With Eqs.~(\ref{13.2.1}) and
(\ref{13.5.12}) we find that the singular Green's function is given
explicitly by   
\begin{equation} 
G_{\rm S}(x,x') = \frac{1}{2} U(x,x') \delta(\sigma) - \frac{1}{2}
V(x,x') \theta(\sigma)
\label{13.5.14}
\end{equation}
in the normal convex neighbourhood. Equation (\ref{13.5.14}) shows
very clearly that the singular Green's function does not distinguish 
between past and future (property {\sf S2}), and that its support
excludes $I^\pm(x')$, in which $\theta(\sigma) = 0$ (property 
{\sf S3}). From Eq.~(\ref{13.5.4}) we get an analogous expression for
the radiative Green's function: 
\begin{equation} 
G_{\rm R}(x,x') = \frac{1}{2} U(x,x') \Bigl[ \delta_+(\sigma) -
\delta_-(\sigma) \Bigr] + V(x,x') \Bigl[ \theta_+(-\sigma) +
\frac{1}{2} \theta(\sigma) \Bigr]. 
\label{13.5.15}
\end{equation} 
This reveals directly that the radiative Green's function coincides
with $G_+(x,x')$ in $I^+(x')$, in which $\theta(\sigma) = 0$ and
$\theta_+(-\sigma) = 1$ (property {\sf R2}), and that its support does  
not include $I^-(x')$, in which $\theta(\sigma) = \theta_+(-\sigma) =
0$ (property {\sf R3}).     

\subsection{Example: Cosmological Green's functions}  
\label{13.6} 

To illustrate the general theory outlined in the previous subsections
we consider here the specific case of a minimally-coupled ($\xi=0$)
scalar field in a cosmological spacetime with metric  
\begin{equation}
ds^2 = a^2(\eta)(-d\eta^2 + dx^2 + dy^2 + dz^2), 
\label{13.6.1}
\end{equation}
where $a(\eta)$ is the scale factor expressed in terms of
conformal time. For concreteness we take the universe to be matter 
dominated, so that $a(\eta) = C \eta^2$, where $C$ is a constant. This
spacetime is one of the very few for which Green's functions can be
explicitly constructed. The calculation presented here was first
carried out by Burko, Harte, and Poisson \cite{BHP}; it can be
extended to other cosmologies. 

To solve Green's equation $\Box G(x,x') = -4\pi \delta_4(x,x')$ we
first introduce a reduced Green's function $g(x,x')$ defined by  
\begin{equation}
G(x,x') = \frac{g(x,x')}{a(\eta) a(\eta')}. 
\label{13.6.2}
\end{equation} 
Substitution yields 
\begin{equation}
\biggl( - \frac{\partial^2}{\partial \eta^2} + \nabla^2 +
\frac{2}{\eta^2} \biggr) g(x,x') = 
-4\pi \delta(\eta - \eta') \delta_3(\bm{x} - \bm{x'}),    
\label{13.6.3}
\end{equation} 
where $\bm{x} = (x,y,z)$ is a vector in three-dimensional flat space,
and $\nabla^2$ is the Laplacian operator in this space. We next expand 
$g(x,x')$ in terms of plane-wave solutions to Laplace's equation,  
\begin{equation}
g(x,x') = \frac{1}{(2\pi)^3}\, \int \tilde{g}(\eta,\eta';\bm{k})\,
e^{i \bm{k} \cdot (\bm{x} - \bm{x'})}\, d^3 k,
\label{13.6.4} 
\end{equation}
and we substitute this back into Eq.~(\ref{13.6.3}). The result, after
also Fourier transforming $\delta_3(\bm{x}-\bm{x'})$, is an ordinary
differential equation for $\tilde{g}(\eta,\eta';\bm{k})$: 
\begin{equation}
\biggl( \frac{d^2}{d\eta^2} + k^2 - \frac{2}{\eta^2} \biggr)
\tilde{g} = 4\pi \delta(\eta - \eta'), 
\label{13.6.5}
\end{equation}
where $k^2 = \bm{k} \cdot \bm{k}$. To generate the retarded
Green's function we set  
\begin{equation}
\tilde{g}_+(\eta,\eta';\bm{k}) = \theta(\eta-\eta')\, 
\hat{g}(\eta,\eta';k),  
\label{13.6.6}
\end{equation}
in which we indicate that $\hat{g}$ depends only on the modulus of the 
vector $\bm{k}$. To generate the advanced Green's function we would
set instead $\tilde{g}_-(\eta,\eta';\bm{k}) = \theta(\eta'-\eta)\, 
\hat{g}(\eta,\eta';k)$. The following manipulations will refer
specifically to the retarded Green's function; they are easily 
adapted to the case of the advanced Green's function.  

Substitution of Eq.~(\ref{13.6.6}) into Eq.~(\ref{13.6.5}) reveals
that $\hat{g}$ must satisfy the homogeneous equation  
\begin{equation}
\biggl( \frac{d^2}{d\eta^2} + k^2 - \frac{2}{\eta^2} \biggr)
\hat{g} = 0, 
\label{13.6.7}
\end{equation}
together with the boundary conditions 
\begin{equation}
\hat{g}(\eta=\eta';k) = 0, \qquad
\frac{d\hat{g}}{d\eta}(\eta=\eta';k) = 4\pi. 
\label{13.6.8}
\end{equation} 
Inserting Eq.~(\ref{13.6.6}) into Eq.~(\ref{13.6.4}) and integrating
over the angular variables associated with the vector $\bm{k}$ yields   
\begin{equation}
g_+(x,x') = \frac{\theta(\Delta \eta)}{2\pi^2 R} \int_0^\infty  
\hat{g}(\eta,\eta';k)\, k \sin(kR)\, dk, 
\label{13.6.9}
\end{equation}
where $\Delta \eta \equiv \eta - \eta'$ and 
$R \equiv |\bm{x}-\bm{x'}|$.  

Equation (\ref{13.6.7}) has $\cos (k \Delta \eta) - (k \eta)^{-1} 
\sin (k \Delta \eta)$ and $\sin (k \Delta \eta) + (k \eta)^{-1} 
\cos (k \Delta \eta)$ as linearly independent solutions, and
$\hat{g}(\eta,\eta';k)$ must be given by a linear superposition. The
coefficients can be functions of $\eta'$, and after imposing  
Eqs.~(\ref{13.6.8}) we find that the appropriate combination is       
\begin{equation}
\hat{g}(\eta,\eta';k) = \frac{4\pi}{k}\, \biggl[ 
\biggl( 1 + \frac{1}{k^2 \eta \eta'} \biggr) \sin(k \Delta \eta) 
- \frac{\Delta \eta}{k \eta \eta'}\, \cos(k \Delta \eta) 
\biggr]. 
\label{13.6.10}
\end{equation}
Substituting this into Eq.~(\ref{13.6.9}) and using the identity 
$(2/\pi) \int_0^\infty \sin(\omega x) \sin(\omega x')\, d\omega
= \delta(x-x') - \delta(x+x')$ yields 
\[
g_+(x,x') = \frac{\delta(\Delta \eta - R)}{R} 
+ \frac{\theta(\Delta \eta)}{\eta \eta'}\, 
\frac{2}{\pi} \int_0^\infty \frac{1}{k}\,  
\sin(k \Delta \eta) \cos(k R)\, dk
\]
after integration by parts. The integral evaluates to 
$\theta(\Delta \eta - R)$. 

We have arrived at 
\begin{equation}
g_+(x,x') = \frac{\delta(\eta - \eta' 
- |\bm{x}-\bm{x'}|)}{|\bm{x}-\bm{x'}|}  
+ \frac{\theta(\eta - \eta' - |\bm{x}-\bm{x'}|)}{\eta \eta'}  
\label{13.6.11} 
\end{equation}
for our final expression for the retarded Green's function. The
advanced Green's function is given instead by 
\begin{equation}
g_-(x,x') = \frac{\delta(\eta - \eta' 
+ |\bm{x}-\bm{x'}|)}{|\bm{x}-\bm{x'}|}  
+ \frac{\theta(-\eta + \eta' - |\bm{x}-\bm{x'}|)}{\eta \eta'}.   
\label{13.6.12} 
\end{equation}
The distributions $g_\pm(x,x')$ are solutions to the reduced Green's
equation of Eq.~(\ref{13.6.3}). The actual Green's functions are
obtained by substituting Eqs.~(\ref{13.6.11}) and (\ref{13.6.12}) into 
Eq.~(\ref{13.6.2}). We note that the support of the retarded Green's
function is given by $\eta - \eta' \geq |\bm{x}-\bm{x'}|$, while the
support of the advanced Green's function is given by $\eta - \eta'
\leq -|\bm{x}-\bm{x'}|$.  

It may be verified that the symmetric two-point function 
\begin{equation}
h(x,x') = \frac{1}{\eta \eta'} 
\label{13.6.13}
\end{equation}
satisfies all of the properties {\sf H1}--{\sf H4} listed in
Sec.~\ref{13.5}; it may thus be used to define singular and radiative 
Green's functions. According to
Eq.~(\ref{13.5.12}) the singular Green's function is given by 
\begin{eqnarray}
g_{\rm S}(x,x') &=& \frac{1}{2|\bm{x}-\bm{x'}|} \Bigl[ 
\delta(\eta - \eta' - |\bm{x}-\bm{x'}|)
+ \delta(\eta - \eta' + |\bm{x}-\bm{x'}|) \Bigr] 
\nonumber \\ & & \mbox{} 
+ \frac{1}{2\eta \eta'} \Bigr[ 
\theta(\eta - \eta' - |\bm{x}-\bm{x'}|)
- \theta(\eta - \eta' + |\bm{x}-\bm{x'}|) \Bigr]
\label{13.6.14}
\end{eqnarray} 
and its support is limited to the interval $-|\bm{x}-\bm{x'}| \leq
\eta-\eta' \leq |\bm{x}-\bm{x'}|$. According to Eq.~(\ref{13.5.4}) the
radiative Green's function is given by 
\begin{eqnarray}
g_{\rm R}(x,x') &=& \frac{1}{2|\bm{x}-\bm{x'}|} \Bigl[ 
\delta(\eta - \eta' - |\bm{x}-\bm{x'}|)
- \delta(\eta - \eta' + |\bm{x}-\bm{x'}|) \Bigr] 
\nonumber \\ & & \mbox{} 
+ \frac{1}{2\eta \eta'} \Bigr[ 
\theta(\eta - \eta' - |\bm{x}-\bm{x'}|)
+ \theta(\eta - \eta' + |\bm{x}-\bm{x'}|) \Bigr]; 
\label{13.6.15}
\end{eqnarray} 
its support is given by $\eta-\eta' \geq - |\bm{x}-\bm{x'}|$ and for 
$\eta-\eta' \geq |\bm{x}-\bm{x'}|$ the radiative Green's function
agrees with the retarded Green's function. 

As a final observation we note that for this cosmological spacetime,  
the normal convex neighbourhood of any point $x$ consists of the whole
spacetime manifold (which excludes the cosmological singularity at
$a = 0$). The Hadamard construction of the Green's functions is
therefore valid globally, a fact that is immediately revealed by
Eqs.~(\ref{13.6.11}) and (\ref{13.6.12}).   

\section{Electromagnetic Green's functions} 
\label{14} 

\subsection{Equations of electromagnetism} 
\label{14.1} 

The electromagnetic field tensor $F_{\alpha\beta} = \nabla_\alpha 
A_\beta - \nabla_\beta A_\alpha$ is expressed in terms of a vector
potential $A_{\alpha}$. In the Lorenz gauge $\nabla_\alpha A^\alpha =  
0$, the vector potential satisfies the wave equation  
\begin{equation} 
\Box A^\alpha - R^\alpha_{\ \beta} A^\beta = -4\pi j^\alpha, 
\label{14.1.1} 
\end{equation}
where $\Box = g^{\alpha\beta} \nabla_\alpha \nabla_\beta$ is the wave 
operator, $R^\alpha_{\ \beta}$ the Ricci tensor, and $j^\alpha$ a
prescribed current density. The wave equation
enforces the condition $\nabla_\alpha j^\alpha = 0$, which expresses
charge conservation.  

The solution to the wave equation is written as  
\begin{equation}
A^\alpha(x) = \int G^\alpha_{\ \beta'}(x,x') j^{\beta'}(x')
\sqrt{-g'}\, d^4 x', 
\label{14.1.2} 
\end{equation}
in terms of a Green's function $G^\alpha_{\ \beta'}(x,x')$ that
satisfies 
\begin{equation} 
\Box G^\alpha_{\ \beta'}(x,x') - R^\alpha_{\ \beta}(x) 
G^\beta_{\ \beta'}(x,x') = -4\pi g^\alpha_{\ \beta'}(x,x')
\delta_4(x,x'),  
\label{14.1.3} 
\end{equation}
where $g^\alpha_{\ \beta'}(x,x')$ is a parallel propagator and
$\delta_4(x,x')$ an invariant Dirac distribution. The parallel
propagator is inserted on the right-hand side of Eq.~(\ref{14.1.3}) to 
keep the index structure of the equation consistent from side to side;
because $g^\alpha_{\ \beta'}(x,x') \delta_4(x,x')$ is distributionally
equal to $[g^\alpha_{\ \beta'}] \delta_4(x,x') = 
\delta^{\alpha'}_{\ \beta'} \delta_4(x,x')$, it could have been
replaced by either $\delta^{\alpha'}_{\ \beta'}$ 
or $\delta^{\alpha}_{\ \beta}$. It is easy to check that by virtue of  
Eq.~(\ref{14.1.3}), the vector potential of Eq.~(\ref{14.1.2})
satisfies the wave equation of Eq.~(\ref{14.1.1}).   

We will assume that the retarded Green's function 
$G^{\ \alpha}_{+\beta'}(x,x')$, which is nonzero if $x$ is in the
causal future of $x'$, and the advanced Green's function 
$G^{\ \alpha}_{-\beta'}(x,x')$, which is nonzero if $x$ is in the
causal past of $x'$, exist as distributions and can be defined
globally in the entire spacetime.    

\subsection{Hadamard construction of the Green's functions} 
\label{14.2} 

Assuming throughout this subsection that $x$ is in the normal convex
neighbourhood of $x'$, we make the ansatz 
\begin{equation}
G^{\ \alpha}_{\pm\beta'}(x,x') = U^\alpha_{\ \beta'}(x,x')
\delta_\pm(\sigma) + V^\alpha_{\ \beta'}(x,x')
\theta_\pm(-\sigma), 
\label{14.2.1}
\end{equation}
where $\theta_\pm(-\sigma)$, $\delta_\pm(\sigma)$ are the
light-cone distributions introduced in Sec.~\ref{12.2}, and where   
$U^\alpha_{\ \beta'}(x,x')$, $V^\alpha_{\ \beta'}(x,x')$ are smooth 
bitensors. 

To conveniently manipulate the Green's functions we shift $\sigma$ by 
a small positive quantity $\epsilon$. The Green's functions are then 
recovered by the taking the limit of 
\[
G^{\epsilon\ \alpha}_{\pm \ \beta'}(x,x') \equiv 
U^\alpha_{\ \beta'}(x,x') \delta_\pm(\sigma+\epsilon) 
+ V^{\alpha}_{\ \beta'}(x,x') \theta_\pm(-\sigma-\epsilon)
\]
as $\epsilon \to 0^+$. When we substitute this into the left-hand side
of Eq.~(\ref{14.1.3}) and then take the limit, we obtain  
\begin{eqnarray*}
\Box G^{\ \alpha}_{\pm \beta'} - R^\alpha_{\ \beta} 
G^{\ \beta}_{\pm \beta'} &=&  
- 4\pi \delta_4(x,x') U^\alpha_{\ \beta'}  
+ \delta'_\pm(\sigma) \Bigl\{ 2 U^\alpha_{\ \beta';\gamma}
  \sigma^\gamma 
  + (\sigma^\gamma_{\ \gamma} - 4) U^\alpha_{\ \beta'} \Bigr\} 
\\ & & \mbox{} 
+ \delta_\pm(\sigma) \Bigl\{ -2 V^\alpha_{\ \beta';\gamma}
  \sigma^\gamma
  + (2 - \sigma^\gamma_{\ \gamma}) V^\alpha_{\ \beta'}
  + \Box U^\alpha_{\ \beta'} 
  - R^\alpha_{\ \beta} U^\beta_{\ \beta'} \Bigr\}
\\ & & \mbox{} 
+ \theta_\pm(-\sigma) \Bigl\{ \Box V^\alpha_{\ \beta'} 
  - R^\alpha_{\ \beta} V^\beta_{\ \beta'} \Bigr\} 
\end{eqnarray*}
after a routine computation similar to the one presented at the
beginning of Sec.~\ref{13.2}. Comparison with Eq.~(\ref{14.1.3})
returns: (i) the equations   
\begin{equation}
\bigl[ U^\alpha_{\ \beta'} \bigr] = \bigl[g^\alpha_{\ \beta'} \bigr]
= \delta^{\alpha'}_{\ \beta'} 
\label{14.2.2}
\end{equation}
and 
\begin{equation} 
2 U^\alpha_{\ \beta';\gamma} \sigma^\gamma 
  + (\sigma^\gamma_{\ \gamma} - 4) U^\alpha_{\ \beta'} = 0
\label{14.2.3}
\end{equation}
that determine $U^\alpha_{\ \beta'}(x,x')$; (ii) the equation 
\begin{equation} 
\check{V}^\alpha_{\ \beta';\gamma} \sigma^\gamma 
+ \frac{1}{2} (\sigma^\gamma_{\ \gamma} - 2) \check{V}^\alpha_{\ \beta'}  
= \frac{1}{2} \bigl( \Box U^\alpha_{\ \beta'} 
  - R^\alpha_{\ \beta} U^\beta_{\ \beta'} \bigr) \Bigr|_{\sigma=0} 
\label{14.2.4}
\end{equation} 
that determines $\check{V}^\alpha_{\ \beta'}(x,x')$, the restriction
of $V^\alpha_{\ \beta'}(x,x')$ on the light cone $\sigma(x,x') = 0$; 
and (iii) the wave equation  
\begin{equation} 
\Box V^\alpha_{\ \beta'} - R^\alpha_{\ \beta} V^\beta_{\ \beta'} = 0
\label{14.2.5} 
\end{equation}
that determines $V^\alpha_{\ \beta'}(x,x')$ inside the light cone.  

Equation (\ref{14.2.3}) can be integrated along the unique geodesic 
$\beta$ that links $x'$ to $x$. The initial conditions are provided 
by Eq.~(\ref{14.2.2}), and if we set $U^\alpha_{\ \beta'}(x,x') = 
g^\alpha_{\ \beta'}(x,x') U(x,x')$, we find that these equations
reduce to Eqs.~(\ref{13.2.4}) and (\ref{13.2.3}),
respectively. According to Eq.~(\ref{13.2.5}), then, we have   
\begin{equation}
U^\alpha_{\ \beta'}(x,x') = g^\alpha_{\ \beta'}(x,x')
\Delta^{1/2}(x,x'),  
\label{14.2.6}
\end{equation} 
which reduces to 
\begin{equation} 
U^\alpha_{\ \beta'} = g^\alpha_{\ \beta'} \Bigl( 1 + \frac{1}{12}
R_{\gamma'\delta'} \sigma^{\gamma'} \sigma^{\delta'} + O(\lambda^3) 
\Bigr) 
\label{14.2.7}
\end{equation}
near coincidence, with $\lambda$ denoting the affine-parameter
distance between $x'$ and $x$. Differentiation of this relation gives  
\begin{eqnarray}
U^\alpha_{\ \beta';\gamma} &=& \frac{1}{2} g^{\gamma'}_{\ \gamma} 
\Bigl( g^\alpha_{\ \alpha'} R^{\alpha'}_{\ \beta'\gamma'\delta'} 
- \frac{1}{3} g^\alpha_{\ \beta'} R_{\gamma'\delta'} \Bigr) 
\sigma^{\delta'} + O(\lambda^2),
\label{14.2.8} \\  
U^\alpha_{\ \beta';\gamma'} &=& \frac{1}{2} 
\Bigl( g^\alpha_{\ \alpha'} R^{\alpha'}_{\ \beta'\gamma'\delta'} 
+ \frac{1}{3} g^\alpha_{\ \beta'} R_{\gamma'\delta'} \Bigr) 
\sigma^{\delta'} + O(\lambda^2),
\label{14.2.9}
\end{eqnarray}
and eventually, 
\begin{equation} 
\bigl[ \Box U^\alpha_{\ \beta'} \bigr] = \frac{1}{6}
\delta^{\alpha'}_{\ \beta'} R(x'). 
\label{14.2.10} 
\end{equation} 
 
Similarly, Eq.~(\ref{14.2.4}) can be integrated along each null
geodesic that generates the null cone $\sigma(x,x')=0$. The initial
values are obtained by taking the coincidence limit of this equation,
using Eqs.~(\ref{14.2.2}), (\ref{14.2.10}), and the additional
relation $[\sigma^\gamma_{\ \gamma}] = 4$. We arrive at   
\begin{equation} 
\bigl[ V^\alpha_{\ \beta'} \bigr] = -\frac{1}{2} \Bigl(
R^{\alpha'}_{\ \beta'} - \frac{1}{6} \delta^{\alpha'}_{\ \beta'} R'
\Bigr). 
\label{14.2.11}
\end{equation}
With the characteristic data obtained by integrating
Eq.~(\ref{14.2.4}), the wave equation of Eq.~(\ref{14.2.5}) admits a
unique solution.  

To summarize, the retarded and advanced electromagnetic Green's
functions are given by Eq.~(\ref{14.2.1}) with 
$U^\alpha_{\ \beta'}(x,x')$ given by Eq.~(\ref{14.2.6}) and     
$V^\alpha_{\ \beta'}(x,x')$ determined by Eq.~(\ref{14.2.5}) and the 
characteristic data constructed with Eqs.~(\ref{14.2.4}) and
(\ref{14.2.11}). It should be emphasized that the construction
provided in this subsection is restricted to ${\cal N}(x')$, the
normal convex neighbourhood of the reference point $x'$.   

\subsection{Reciprocity and Kirchhoff representation} 
\label{14.3} 

Like their scalar counterparts, the (globally defined) electromagnetic
Green's functions satisfy a reciprocity relation, the statement of
which is  
\begin{equation} 
G^-_{\beta'\alpha}(x',x) = G^+_{\alpha\beta'}(x,x'). 
\label{14.3.1} 
\end{equation} 
The derivation of Eq.~(\ref{14.3.1}) is virtually identical to what
was presented in Sec.~\ref{13.3}, and we shall not present the
details. It suffices to mention that it is based on the identities    
\[
G^+_{\alpha\beta'}(x,x') \Bigl( \Box G^{\ \alpha}_{-\gamma''}(x,x'') -  
R^\alpha_{\ \gamma} G^{\ \gamma}_{-\gamma''}(x,x'') \Bigl) = -4\pi
G^+_{\alpha\beta'}(x,x') g^\alpha_{\ \gamma''}(x,x'') \delta_4(x,x'') 
\]
and 
\[
G^-_{\alpha\gamma''}(x,x'') \Bigl( \Box G^{\ \alpha}_{+\beta'}(x,x') - 
R^\alpha_{\ \gamma} G^{\ \gamma}_{+\beta'}(x,x') \Bigl) = -4\pi
G^-_{\alpha\gamma''}(x,x'') g^\alpha_{\ \beta'}(x,x') \delta_4(x,x').   
\]
A direct consequence of the reciprocity relation is 
\begin{equation}
V_{\beta'\alpha}(x',x) = V_{\alpha\beta'}(x,x'), 
\label{14.3.2} 
\end{equation} 
the statement that the bitensor $V_{\alpha\beta'}(x,x')$ is symmetric
in its indices and arguments.  

The Kirchhoff representation for the electromagnetic vector potential
is formulated as follows. Suppose that $A^\alpha(x)$ satisfies the   
{\it homogeneous} version of Eq.~(\ref{14.1.1}) and that initial
values $A^{\alpha'}(x')$, $n^{\beta'} \nabla_{\beta'} A^{\alpha'}(x')$
are specified on a spacelike hypersurface $\Sigma$. Then the value of
the potential at a point $x$ in the future of $\Sigma$ is given by   
\begin{equation}
A^{\alpha}(x) = -\frac{1}{4\pi} \int_{\Sigma} \biggl(
G^{\ \alpha}_{+\beta'}(x,x') \nabla^{\gamma'} A^{\beta'}(x') -
A^{\beta'}(x') \nabla^{\gamma'} G^{\ \alpha}_{+\beta'}(x,x') \biggr)\, 
d\Sigma_{\gamma'}, 
\label{14.3.3} 
\end{equation}
where $d\Sigma_{\gamma'} = -n_{\gamma'} dV$ is a surface element on
$\Sigma$; $n_{\gamma'}$ is the future-directed unit normal and $dV$
is the invariant volume element on the hypersurface. The derivation of 
Eq.~(\ref{14.3.3}) is virtually identical to what was presented in
Sec.~\ref{13.4}.    

\subsection{Singular and radiative Green's functions} 
\label{14.4} 

We shall now construct singular and radiative Green's functions for
the electromagnetic field. The treatment here parallels closely what
was presented in Sec.~\ref{13.5}, and the reader is referred to that
section for a more complete discussion.  

We begin by introducing the bitensor $H^\alpha_{\ \beta'}(x,x')$ with
properties  
\begin{description} 
\item[\qquad {\sf H1}:] $H^\alpha_{\ \beta'}(x,x')$ satisfies the
homogeneous wave equation, 
\begin{equation}
\Box H^\alpha_{\ \beta'}(x,x') 
- R^\alpha_{\ \beta}(x) H^\beta_{\ \beta'}(x,x') = 0;  
\label{14.4.1} 
\end{equation} 
\item[\qquad {\sf H2}:] $H^\alpha_{\ \beta'}(x,x')$ is symmetric in
its indices and arguments,   
\begin{equation}
H_{\beta'\alpha}(x',x) = H_{\alpha\beta'}(x,x'); 
\label{14.4.2}
\end{equation} 
\item[\qquad {\sf H3}:] $H^\alpha_{\ \beta'}(x,x')$ agrees with the
retarded Green's function if $x$ is in the chronological future of
$x'$,   
\begin{equation}
H^\alpha_{\ \beta'}(x,x') = G^{\ \alpha}_{+\beta'}(x,x') \qquad
\mbox{when $x \in I^+(x')$};   
\label{14.4.3} 
\end{equation} 
\item[\qquad {\sf H4}:] $H^\alpha_{\ \beta'}(x,x')$ agrees with the
advanced Green's function if $x$ is in the chronological past of $x'$,   
\begin{equation}
H^\alpha_{\ \beta'}(x,x') = G^{\ \alpha}_{-\beta'}(x,x') \qquad
\mbox{when $x \in I^-(x')$}.    
\label{14.4.4} 
\end{equation} 
\end{description}
It is easy to prove that property {\sf H4} follows from {\sf H2}, 
{\sf H3}, and the reciprocity relation (\ref{14.3.1}) satisfied by the 
retarded and advanced Green's functions. That such a bitensor exists
can be argued along the same lines as those presented in
Sec.~\ref{13.5}.  

Equipped with the bitensor $H^\alpha_{\ \beta'}(x,x')$ we define the
singular Green's function to be 
\begin{equation}
G^{\ \alpha}_{{\rm S}\,\beta'}(x,x') = \frac{1}{2} \Bigl[ 
G^{\ \alpha}_{+\beta'}(x,x') 
+ G^{\ \alpha}_{-\beta'}(x,x')
- H^\alpha_{\ \beta'}(x,x') \Bigr]. 
\label{14.4.5}
\end{equation} 
This comes with the properties  
\begin{description} 
\item[\qquad {\sf S1}:] $G^{\ \alpha}_{{\rm S}\,\beta'}(x,x')$
satisfies the inhomogeneous wave equation,     
\begin{equation} 
\Box G^{\ \alpha}_{{\rm S}\,\beta'}(x,x') - R^\alpha_{\ \beta}(x)  
G^{\ \beta}_{{\rm S}\,\beta'}(x,x') = -4\pi g^\alpha_{\ \beta'}(x,x') 
\delta_4(x,x'); 
\label{14.4.6}  
\end{equation} 
\item[\qquad {\sf S2}:] $G^{\ \alpha}_{{\rm S}\,\beta'}(x,x')$ is
symmetric in its indices and arguments,  
\begin{equation} 
G^{\rm S}_{\beta'\alpha}(x',x) = G^{\rm S}_{\alpha\beta'}(x,x'); 
\label{14.4.7}
\end{equation} 
\item[\qquad {\sf S3}:] $G^{\ \alpha}_{{\rm S}\,\beta'}(x,x')$
vanishes if $x$ is in the chronological past or future of $x'$,  
\begin{equation}
G^{\ \alpha}_{{\rm S}\,\beta'}(x,x') = 0 \qquad 
\mbox{when $x \in I^\pm(x')$}.  
\label{14.4.8} 
\end{equation} 
\end{description}  
These can be established as consequences of {\sf H1}--{\sf H4} and the
properties of the retarded and advanced Green's functions.  

The radiative Green's function is then defined by 
\begin{equation} 
G^{\ \,\alpha}_{{\rm R}\,\beta'}(x,x') = 
G^{\ \alpha}_{+\beta'}(x,x') 
- G^{\ \alpha}_{{\rm S}\,\beta'}(x,x'), 
\label{14.4.9}
\end{equation}
and it comes with the properties 
\begin{description} 
\item[\qquad {\sf R1}:] $G^{\ \,\alpha}_{{\rm R}\,\beta'}(x,x')$
satisfies the homogeneous wave equation, 
\begin{equation}
\Box G^{\ \,\alpha}_{{\rm R}\,\beta'}(x,x') - R^\alpha_{\ \beta}(x)  
G^{\ \,\beta}_{{\rm R}\,\beta'}(x,x') = 0;  
\label{14.4.10} 
\end{equation} 
\item[\qquad {\sf R2}:] $G^{\ \,\alpha}_{{\rm R}\,\beta'}(x,x')$
agrees with the retarded Green's function if $x$ is in the
chronological future of $x'$,  
\begin{equation}
G^{\ \,\alpha}_{{\rm R}\,\beta'}(x,x') = 
G^{\ \alpha}_{+\beta'}(x,x') 
\qquad \mbox{when $x \in I^+(x')$}; 
\label{14.4.11} 
\end{equation} 
\item[\qquad {\sf R3}:] $G^{\ \,\alpha}_{{\rm R}\,\beta'}(x,x')$
vanishes if $x$ is in the chronological past of $x'$,  
\begin{equation}
G^{\ \,\alpha}_{{\rm R}\,\beta'}(x,x') = 0 
\qquad \mbox{when $x \in I^-(x')$}.  
\label{14.4.12} 
\end{equation} 
\end{description}
Those follow immediately from {\sf S1}--{\sf S3} and the properties of
the retarded Green's function. 

When $x$ is restricted to the normal convex neighbourhood of $x'$, we
have the explicit relations 
\begin{eqnarray} 
H^\alpha_{\ \beta'}(x,x') &=& V^\alpha_{\ \beta'}(x,x'), 
\label{14.4.13} \\ 
G^{\ \alpha}_{{\rm S}\,\beta'}(x,x') &=& 
\frac{1}{2} U^\alpha_{\ \beta'}(x,x') \delta(\sigma) - \frac{1}{2} 
V^\alpha_{\ \beta'}(x,x') \theta(\sigma), 
\label{14.4.14} \\ 
G^{\ \,\alpha}_{{\rm R}\,\beta'}(x,x') &=& 
\frac{1}{2} U^\alpha_{\ \beta'}(x,x') \Bigl[ \delta_+(\sigma) -
\delta_-(\sigma) \Bigr] + V^\alpha_{\ \beta'}(x,x') \Bigl[
\theta_+(-\sigma) + \frac{1}{2} \theta(\sigma) \Bigr]. 
\label{14.4.15}
\end{eqnarray} 
From these we see clearly that the singular Green's function does not
distinguish between past and future (property {\sf S2}), and that its 
support excludes $I^\pm(x')$ (property {\sf S3}). We see also that
the radiative Green's function coincides with 
$G^{\ \alpha}_{+\beta'}(x,x')$ in $I^+(x')$ (property {\sf R2}), and  
that its support does not include $I^-(x')$ (property {\sf R3}). 
 
\section{Gravitational Green's functions}
\label{15} 

\subsection{Equations of linearized gravity}  
\label{15.1} 

We are given a background spacetime for which the metric
$g_{\alpha\beta}$ satisfies the Einstein field equations 
{\it in vacuum}. We then perturb the metric from $g_{\alpha\beta}$ to 
\begin{equation} 
{\sf g}_{\alpha\beta} = g_{\alpha\beta} + h_{\alpha\beta}.   
\label{15.1.1}
\end{equation} 
The metric perturbation $h_{\alpha\beta}$ is assumed to be small, and
when working out the Einstein field equations to be satisfied by the
new metric ${\sf g}_{\alpha\beta}$, we work consistently to
first order in $h_{\alpha\beta}$. To simplify the expressions we use
the trace-reversed potentials $\gamma_{\alpha\beta}$ defined by      
\begin{equation} 
\gamma_{\alpha\beta} = h_{\alpha\beta} - \frac{1}{2} \bigl(
g^{\gamma\delta} h_{\gamma\delta} \bigr) g_{\alpha\beta},  
\label{15.1.2}
\end{equation} 
and we impose the Lorenz gauge condition,  
\begin{equation}
\gamma^{\alpha\beta}_{\ \ \ ;\beta} = 0. 
\label{15.1.3} 
\end{equation} 
In this equation, and in all others below, indices are raised and
lowered with the background metric $g_{\alpha\beta}$. Similarly, the
connection involved in Eq.~(\ref{15.1.3}), and in all other equations
below, is the one that is compatible with the background metric. If 
$T^{\alpha\beta}$ is the perturbing stress-energy tensor, then by
virtue of the linearized Einstein field equations the perturbation
field obeys the wave equation   
\begin{equation} 
\Box \gamma^{\alpha\beta} + 2 R_{\gamma\ \delta}^{\ \alpha\ \beta}
\gamma^{\gamma\delta} = -16\pi T^{\alpha\beta}, 
\label{15.1.4}
\end{equation} 
in which $\Box = g^{\alpha\beta} \nabla_\alpha \nabla_\beta$ is the
wave operator and $R_{\gamma\alpha\delta\beta}$ the Riemann tensor. In   
first-order perturbation theory, the stress-energy tensor must be
conserved in the background spacetime: 
$T^{\alpha\beta}_{\ \ \ ;\beta} = 0$. 

The solution to the wave equation is written as 
\begin{equation}
\gamma^{\alpha\beta}(x) = 4\int 
G^{\alpha\beta}_{\ \ \gamma'\delta'}(x,x') 
T^{\gamma'\delta'}(x') \sqrt{-g'}\, d^4 x', 
\label{15.1.5} 
\end{equation}
in terms of a Green's function 
$G^{\alpha\beta}_{\ \ \gamma'\delta'}(x,x')$ that satisfies \cite{SWG}  
\begin{equation} 
\Box G^{\alpha\beta}_{\ \ \gamma'\delta'}(x,x') + 
2 R_{\gamma\ \delta}^{\ \alpha\ \beta} (x)  
G^{\gamma\delta}_{\ \ \gamma'\delta'}(x,x')
= -4\pi g^{(\alpha}_{\ \gamma'}(x,x')
g^{\beta)}_{\ \delta'}(x,x') \delta_4(x,x'),  
\label{15.1.6} 
\end{equation}
where $g^\alpha_{\ \gamma'}(x,x')$ is a parallel propagator and
$\delta_4(x,x')$ an invariant Dirac functional. The parallel
propagators are inserted on the right-hand side of Eq.~(\ref{15.1.6})
to keep the index structure of the equation consistent from side to
side; in particular, both sides of the equation are symmetric in
$\alpha$ and $\beta$, and in $\gamma'$ and $\delta'$. It is easy to
check that by virtue of Eq.~(\ref{15.1.6}), the perturbation field of  
Eq.~(\ref{15.1.5}) satisfies the wave equation of Eq.~(\ref{15.1.4}).      
Once $\gamma_{\alpha\beta}$ is known, the metric perturbation can  
be reconstructed from the relation $h_{\alpha\beta} =
\gamma_{\alpha\beta} - \frac{1}{2} (g^{\gamma\delta}
\gamma_{\gamma\delta}) g_{\alpha\beta}$. 

We will assume that the retarded Green's function 
$G^{\ \alpha\beta}_{+\ \gamma'\delta'}(x,x')$, which is nonzero if $x$
is in the causal future of $x'$, and the advanced Green's function 
$G^{\ \alpha\beta}_{-\ \gamma'\delta'}(x,x')$, which is nonzero if $x$
is in the causal past of $x'$, exist as distributions and can be
defined globally in the entire background spacetime.     

\subsection{Hadamard construction of the Green's functions} 
\label{15.2}  

Assuming throughout this subsection that $x$ is in the normal convex
neighbourhood of $x'$, we make the ansatz 
\begin{equation}
G^{\ \alpha\beta}_{\pm\ \gamma'\delta'}(x,x') = 
U^{\alpha\beta}_{\ \ \gamma'\delta'}(x,x') 
\delta_\pm(\sigma) 
+ V^{\alpha\beta}_{\ \ \gamma'\delta'}(x,x') \theta_\pm(-\sigma),  
\label{15.2.1}  
\end{equation}
where $\theta_\pm(-\sigma)$, $\delta_\pm(\sigma)$ are the
light-cone distributions introduced in Sec.~\ref{12.2}, and where   
$U^{\alpha\beta}_{\ \ \gamma'\delta'}(x,x')$, 
$V^{\alpha\beta}_{\ \ \gamma'\delta'}(x,x')$ are smooth bitensors. 

To conveniently manipulate the Green's functions we shift $\sigma$ by  
a small positive quantity $\epsilon$. The Green's functions are then 
recovered by the taking the limit of 
\[
G^{\epsilon\ \alpha\beta}_{\pm\ \ \gamma'\delta'}(x,x') = 
U^{\alpha\beta}_{\ \ \gamma'\delta'}(x,x') 
\delta_\pm(\sigma+\epsilon) 
+ V^{\alpha\beta}_{\ \ \gamma'\delta'}(x,x')
\theta_\pm(-\sigma-\epsilon)
\]
as $\epsilon \to 0^+$. When we substitute this into the left-hand side 
of Eq.~(\ref{15.1.6}) and then take the limit, we obtain  
\begin{eqnarray*}
\Box G^{\ \alpha\beta}_{\pm \ \gamma'\delta'}  
+ 2 R_{\gamma\ \delta}^{\ \alpha\ \beta} 
G^{\ \gamma\delta}_{\pm \ \gamma'\delta'}   
&=& 
- 4\pi \delta_4(x,x') U^{\alpha\beta}_{\ \ \gamma'\delta'}   
+ \delta'_\pm(\sigma) \Bigl\{ 
   2 U^{\alpha\beta}_{\ \ \gamma'\delta';\gamma} \sigma^\gamma  
  + (\sigma^\gamma_{\ \gamma} - 4) U^{\alpha\beta}_{\ \ \gamma'\delta'}
\Bigr\}  
\\ & & \mbox{} 
+ \delta_\pm(\sigma) \Bigl\{ 
  -2 V^{\alpha\beta}_{\ \ \gamma'\delta';\gamma} \sigma^\gamma 
  + (2 - \sigma^\gamma_{\ \gamma}) 
    V^{\alpha\beta}_{\ \ \gamma'\delta'}  
  + \Box U^{\alpha\beta}_{\ \ \gamma'\delta'} 
  + 2 R_{\gamma\ \delta}^{\ \alpha\ \beta} 
    U^{\gamma\delta}_{\ \ \gamma'\delta'} 
  \Bigr\}
\\ & & \mbox{} 
+ \theta_\pm(-\sigma) \Bigl\{ 
\Box V^{\alpha\beta}_{\ \ \gamma'\delta'} 
  + 2 R_{\gamma\ \delta}^{\ \alpha\ \beta} 
    V^{\gamma\delta}_{\ \ \gamma'\delta'}
\Bigr\}
\end{eqnarray*}
after a routine computation similar to the one presented at the
beginning of Sec.~\ref{13.2}. Comparison with Eq.~(\ref{15.1.6})
returns: (i) the equations  
\begin{equation}
\bigl[ U^{\alpha\beta}_{\ \ \gamma'\delta'} \bigr] = 
\Bigl[ g^{(\alpha}_{\ \gamma'} g^{\beta)}_{\ \delta'} \Bigr]
= \delta^{(\alpha'}_{\ \ \gamma'} \delta^{\beta')}_{\ \delta'}
\label{15.2.2} 
\end{equation}
and 
\begin{equation} 
2 U^{\alpha\beta}_{\ \ \gamma'\delta';\gamma} \sigma^\gamma 
  + (\sigma^\gamma_{\ \gamma} - 4) 
  U^{\alpha\beta}_{\ \ \gamma'\delta'} = 0 
\label{15.2.3} 
\end{equation}
that determine $U^{\alpha\beta}_{\ \ \gamma'\delta'}(x,x')$; (ii) the 
equation  
\begin{equation} 
\check{V}^{\alpha\beta}_{\ \ \gamma'\delta';\gamma} \sigma^\gamma   
+ \frac{1}{2} (\sigma^\gamma_{\ \gamma} - 2) 
\check{V}^{\alpha\beta}_{\ \ \gamma'\delta'}  
= \frac{1}{2} \bigl( \Box U^{\alpha\beta}_{\ \ \gamma'\delta'}  
+ 2 R_{\gamma\ \delta}^{\ \alpha\ \beta} 
    U^{\gamma\delta}_{\ \ \gamma'\delta'} \bigr) \Bigr|_{\sigma=0}  
\label{15.2.4}  
\end{equation} 
that determine $\check{V}^{\alpha\beta}_{\ \ \gamma'\delta'}(x,x')$,
the restriction of $V^{\alpha\beta}_{\ \ \gamma'\delta'}(x,x')$ on the
light cone $\sigma(x,x') = 0$; and (iii) the wave equation 
\begin{equation} 
\Box V^{\alpha\beta}_{\ \ \gamma'\delta'} 
  + 2 R_{\gamma\ \delta}^{\ \alpha\ \beta} 
    V^{\gamma\delta}_{\ \ \gamma'\delta'} = 0 
\label{15.2.5}
\end{equation}
that determines $V^{\alpha\beta}_{\ \ \gamma'\delta'}(x,x')$ inside
the light cone.  

Equation (\ref{15.2.3}) can be integrated along the unique geodesic  
$\beta$ that links $x'$ to $x$. The initial conditions are provided 
by Eq.~(\ref{15.2.2}), and if we set 
$U^{\alpha\beta}_{\ \ \gamma'\delta'}(x,x') = g^{(\alpha}_{\ \gamma'} 
g^{\beta)}_{\ \delta'} U(x,x')$, we find that these equations reduce
to Eqs.~(\ref{13.2.4}) and (\ref{13.2.3}), respectively. According to 
Eq.~(\ref{13.2.5}), then, we have   
\begin{equation}
U^{\alpha\beta}_{\ \ \gamma'\delta'}(x,x') = 
g^{(\alpha}_{\ \gamma'}(x,x') g^{\beta)}_{\ \delta'}(x,x') 
\Delta^{1/2}(x,x'), 
\label{15.2.6} 
\end{equation} 
which reduces to 
\begin{equation} 
U^{\alpha\beta}_{\ \ \gamma'\delta'} = 
g^{(\alpha}_{\ \gamma'} g^{\beta)}_{\ \delta'} 
\Bigl( 1 + O(\lambda^3) \Bigr) 
\label{15.2.7} 
\end{equation}
near coincidence, with $\lambda$ denoting the affine-parameter
distance between $x'$ and $x$; there is no term of order $\lambda^2$
because by assumption, the background Ricci tensor vanishes at $x'$
(as it does in the entire spacetime). Differentiation of this relation
gives     
\begin{eqnarray}
U^{\alpha\beta}_{\ \ \gamma'\delta';\epsilon} &=& \frac{1}{2}  
g^{(\alpha}_{\ \alpha'} g^{\beta)}_{\ \beta'} 
g^{\epsilon'}_{\ \epsilon}  
\Bigl( R^{\alpha'}_{\ \gamma'\epsilon'\iota'} 
\delta^{\beta'}_{\ \delta'} 
+ R^{\alpha'}_{\ \delta'\epsilon'\iota'} \delta^{\beta'}_{\ \gamma'}
\Bigr) \sigma^{\iota'} + O(\lambda^2),
\label{15.2.8} \\ 
U^{\alpha\beta}_{\ \ \gamma'\delta';\epsilon'} &=& \frac{1}{2}  
g^{(\alpha}_{\ \alpha'} g^{\beta)}_{\ \beta'} 
\Bigl( R^{\alpha'}_{\ \gamma'\epsilon'\iota'} 
\delta^{\beta'}_{\ \delta'} 
+ R^{\alpha'}_{\ \delta'\epsilon'\iota'} 
\delta^{\beta'}_{\ \gamma'} \Bigr)   
\sigma^{\iota'} + O(\lambda^2), 
\label{15.2.9} 
\end{eqnarray}
and eventually, 
\begin{equation} 
\bigl[ \Box U^{\alpha\beta}_{\ \ \gamma'\delta'} ] = 0;
\label{15.2.10}
\end{equation} 
this last result follows from the fact that 
$[U^{\alpha\beta}_{\ \ \gamma'\delta';\epsilon\iota}]$ is
antisymmetric in the last pair of indices.    

Similarly, Eq.~(\ref{15.2.4}) can be integrated along each null
geodesic that generates the null cone $\sigma(x,x')=0$. The initial
values are obtained by taking the coincidence limit of this equation,
using Eqs.~(\ref{15.2.2}), (\ref{15.2.10}), and the additional
relation $[\sigma^\gamma_{\ \gamma}] = 4$. We arrive at  
\begin{equation} 
\bigl[ V^{\alpha\beta}_{\ \ \gamma'\delta'} \bigr] = \frac{1}{2}
\Bigl( R^{\alpha'\ \beta'}_{\ \gamma'\ \delta'} + 
R^{\beta'\ \alpha'}_{\ \gamma'\ \delta'} \Bigr).    
\label{15.2.11} 
\end{equation}
With the characteristic data obtained by integrating
Eq.~(\ref{15.2.4}), the wave equation of Eq.~(\ref{15.2.5}) admits a
unique solution.   

To summarize, the retarded and advanced gravitational Green's
functions are given by Eq.~(\ref{15.2.1}) with  
$U^{\alpha\beta}_{\ \ \gamma'\delta'}(x,x')$ given by
Eq.~(\ref{15.2.6}) and $V^{\alpha\beta}_{\ \ \gamma'\delta'}(x,x')$
determined by Eq.~(\ref{15.2.5}) and the characteristic data
constructed with Eqs.~(\ref{15.2.4}) and (\ref{15.2.11}). It should be
emphasized that the construction provided in this subsection is
restricted to ${\cal N}(x')$, the normal convex neighbourhood of the
reference point $x'$.    

\subsection{Reciprocity and Kirchhoff representation}
\label{15.3}   

The (globally defined) gravitational Green's functions satisfy the
reciprocity relation   
\begin{equation} 
G^-_{\gamma'\delta'\alpha\beta}(x',x) =
G^+_{\alpha\beta\gamma'\delta'}(x,x').  
\label{15.3.1} 
\end{equation} 
The derivation of this result is virtually identical to what was
presented in Secs.~\ref{13.3} and \ref{14.3}. A direct consequence of
the reciprocity relation is the statement
\begin{equation}
V_{\gamma'\delta'\alpha\beta}(x',x) =
V_{\alpha\beta\gamma'\delta'}(x,x').  
\label{15.3.2}
\end{equation} 

The Kirchhoff representation for the trace-reversed gravitational
perturbation $\gamma_{\alpha\beta}$ is formulated as follows. Suppose
that $\gamma^{\alpha\beta}(x)$ satisfies the homogeneous version of
Eq.~(\ref{15.1.4}) and that initial values 
$\gamma^{\alpha'\beta'}(x')$, $n^{\gamma'} \nabla_{\gamma'}
\gamma^{\alpha'\beta'}(x')$ are specified on a spacelike hypersurface
$\Sigma$. Then the value of the perturbation field at a point $x$ in
the future of $\Sigma$ is given by   
\begin{equation}
\gamma^{\alpha\beta}(x) = -\frac{1}{4\pi} \int_{\Sigma} \biggl( 
G^{\ \alpha\beta}_{+\ \gamma'\delta'}(x,x') \nabla^{\epsilon'}
\gamma^{\gamma'\delta'}(x') - \gamma^{\gamma'\delta'}(x') 
\nabla^{\epsilon'} G^{\ \alpha\beta}_{+\ \gamma'\delta'}(x,x')
\biggr)\, d\Sigma_{\epsilon'},  
\label{15.3.3}  
\end{equation}
where $d\Sigma_{\epsilon'} = -n_{\epsilon'} dV$ is a surface element
on $\Sigma$; $n_{\epsilon'}$ is the future-directed unit normal and
$dV$ is the invariant volume element on the hypersurface. The
derivation of Eq.~(\ref{15.3.3}) is virtually identical to what was
presented in Secs.~\ref{13.4} and \ref{14.3}.     

\subsection{Singular and radiative Green's functions} 
\label{15.4} 

We shall now construct singular and radiative Green's functions for
the linearized gravitational field. The treatment here
parallels closely what was presented in Secs.~\ref{13.5} and
\ref{14.4}.  

We begin by introducing the bitensor 
$H^{\alpha\beta}_{\ \ \gamma'\delta'}(x,x')$ with properties  
\begin{description} 
\item[\qquad {\sf H1}:] $H^{\alpha\beta}_{\ \ \gamma'\delta'}(x,x')$
satisfies the homogeneous wave equation, 
\begin{equation}
\Box H^{\alpha\beta}_{\ \ \gamma'\delta'}(x,x') 
  + 2 R_{\gamma\ \delta}^{\ \alpha\ \beta}(x) 
    H^{\gamma\delta}_{\ \ \gamma'\delta'}(x,x')  
= 0; 
\label{15.4.1} 
\end{equation} 
\item[\qquad {\sf H2}:] $H^{\alpha\beta}_{\ \ \gamma'\delta'}(x,x')$
is symmetric in its indices and arguments,   
\begin{equation}
H_{\gamma'\delta'\alpha\beta}(x',x) 
= H_{\alpha\beta\gamma'\delta'}(x,x'); 
\label{15.4.2}
\end{equation} 
\item[\qquad {\sf H3}:] $H^{\alpha\beta}_{\ \ \gamma'\delta'}(x,x')$
agrees with the retarded Green's function if $x$ is in the
chronological future of $x'$,   
\begin{equation}
H^{\alpha\beta}_{\ \ \gamma'\delta'}(x,x') 
= G^{\ \alpha\beta}_{+\ \gamma'\delta'}(x,x') \qquad
\mbox{when $x \in I^+(x')$};   
\label{15.4.3} 
\end{equation} 
\item[\qquad {\sf H4}:] $H^{\alpha\beta}_{\ \ \gamma'\delta'}(x,x')$
agrees with the advanced Green's function if $x$ is in the
chronological past of $x'$,    
\begin{equation}
H^{\alpha\beta}_{\ \ \gamma'\delta'}(x,x') 
= G^{\ \alpha\beta}_{-\ \gamma'\delta'}(x,x') \qquad
\mbox{when $x \in I^-(x')$}.    
\label{15.4.4} 
\end{equation} 
\end{description}
It is easy to prove that property {\sf H4} follows from {\sf H2}, 
{\sf H3}, and the reciprocity relation (\ref{15.3.1}) satisfied by the  
retarded and advanced Green's functions. That such a bitensor exists
can be argued along the same lines as those presented in
Sec.~\ref{13.5}.  

Equipped with $H^{\alpha\beta}_{\ \ \gamma'\delta'}(x,x')$ we define
the singular Green's function to be  
\begin{equation}
G^{\ \alpha\beta}_{{\rm S}\ \,\gamma'\delta'}(x,x') = \frac{1}{2}
\Bigl[ G^{\ \alpha\beta}_{+\ \gamma'\delta'}(x,x') 
+ G^{\ \alpha\beta}_{-\ \gamma'\delta'}(x,x') 
- H^{\alpha\beta}_{\ \ \gamma'\delta'}(x,x') \Bigr]. 
\label{15.4.5}
\end{equation} 
This comes with the properties  
\begin{description} 
\item[\qquad {\sf S1}:] 
$G^{\ \alpha\beta}_{{\rm S}\ \,\gamma'\delta'}(x,x')$ satisfies the
inhomogeneous wave equation,      
\begin{equation} 
\Box G^{\ \alpha\beta}_{{\rm S}\ \,\gamma'\delta'}(x,x') +  
2 R_{\gamma\ \delta}^{\ \alpha\ \beta} (x)  
G^{\ \gamma\delta}_{{\rm S}\ \,\gamma'\delta'}(x,x') 
= -4\pi g^{(\alpha}_{\ \gamma'}(x,x')
g^{\beta)}_{\ \delta'}(x,x') \delta_4(x,x');   
\label{15.4.6}  
\end{equation} 
\item[\qquad {\sf S2}:] 
$G^{\ \alpha\beta}_{{\rm S}\ \,\gamma'\delta'}(x,x')$ is symmetric in
its indices and arguments,   
\begin{equation} 
G^{\rm S}_{\gamma'\delta'\alpha\beta}(x',x) 
= G^{\rm S}_{\alpha\beta\gamma'\delta'}(x,x');  
\label{15.4.7}
\end{equation} 
\item[\qquad {\sf S3}:] 
$G^{\ \alpha\beta}_{{\rm S}\ \,\gamma'\delta'}(x,x')$ vanishes if $x$ is 
in the chronological past or future of $x'$,  
\begin{equation}
G^{\ \alpha\beta}_{{\rm S}\ \,\gamma'\delta'}(x,x') = 0 \qquad  
\mbox{when $x \in I^\pm(x')$}.  
\label{15.4.8} 
\end{equation} 
\end{description}  
These can be established as consequences of {\sf H1}--{\sf H4} and the 
properties of the retarded and advanced Green's functions.  

The radiative Green's function is then defined by 
\begin{equation}  
G^{\ \,\alpha\beta}_{{\rm R}\ \,\gamma'\delta'}(x,x') = 
G^{\ \alpha\beta}_{+\ \gamma'\delta'}(x,x') 
- G^{\ \alpha\beta}_{{\rm S}\ \,\gamma'\delta'}(x,x'), 
\label{15.4.9}
\end{equation}
and it comes with the properties 
\begin{description} 
\item[\qquad {\sf R1}:] 
$G^{\ \,\alpha\beta}_{{\rm R}\ \,\gamma'\delta'}(x,x')$ satisfies the 
homogeneous wave equation,  
\begin{equation}
\Box G^{\ \,\alpha\beta}_{{\rm R}\ \,\gamma'\delta'}(x,x') +  
2 R_{\gamma\ \delta}^{\ \alpha\ \beta} (x)  
G^{\ \,\gamma\delta}_{{\rm R}\ \,\gamma'\delta'}(x,x') = 0; 
\label{15.4.10} 
\end{equation} 
\item[\qquad {\sf R2}:] 
$G^{\ \,\alpha\beta}_{{\rm R}\ \,\gamma'\delta'}(x,x')$ agrees with
the retarded Green's function if $x$ is in the chronological future of 
$x'$,   
\begin{equation}
G^{\ \,\alpha\beta}_{{\rm R}\ \,\gamma'\delta'}(x,x') = 
G^{\ \alpha\beta}_{+\ \gamma'\delta'}(x,x') 
\qquad \mbox{when $x \in I^+(x')$}; 
\label{15.4.11} 
\end{equation} 
\item[\qquad {\sf R3}:] 
$G^{\ \,\alpha\beta}_{{\rm R}\ \,\gamma'\delta'}(x,x')$ vanishes if
$x$ is in the chronological past of $x'$,  
\begin{equation}
G^{\ \,\alpha\beta}_{{\rm R}\ \,\gamma'\delta'}(x,x') = 0 
\qquad \mbox{when $x \in I^-(x')$}.  
\label{15.4.12} 
\end{equation} 
\end{description}
Those follow immediately from {\sf S1}--{\sf S3} and the properties of 
the retarded Green's function. 

When $x$ is restricted to the normal convex neighbourhood of $x'$, we 
have the explicit relations 
\begin{eqnarray} 
H^{\alpha\beta}_{\ \ \gamma'\delta'}(x,x') &=& 
V^{\alpha\beta}_{\ \ \gamma'\delta'}(x,x'), 
\label{15.4.13} \\ 
G^{\ \alpha\beta}_{{\rm S}\ \,\gamma'\delta'}(x,x') &=& 
\frac{1}{2} U^{\alpha\beta}_{\ \ \gamma'\delta'}(x,x') \delta(\sigma)
- \frac{1}{2} V^{\alpha\beta}_{\ \ \gamma'\delta'}(x,x')
  \theta(\sigma),  
\label{15.4.14} \\ 
G^{\ \,\alpha\beta}_{{\rm R}\ \,\gamma'\delta'}(x,x') &=&
\frac{1}{2} U^{\alpha\beta}_{\ \ \gamma'\delta'}(x,x')
\Bigl[ \delta_+(\sigma) - \delta_-(\sigma) \Bigr] 
+ V^{\alpha\beta}_{\ \ \gamma'\delta'}(x,x') \Bigl[ 
\theta_+(-\sigma) + \frac{1}{2} \theta(\sigma) \Bigr]. 
\label{15.4.15}
\end{eqnarray} 
From these we see clearly that the singular Green's function does not
distinguish between past and future (property {\sf S2}), and that its 
support excludes $I^\pm(x')$ (property {\sf S3}). We see also that 
the radiative Green's function coincides with 
$G^{\ \alpha\beta}_{+\ \gamma'\delta'}(x,x')$ in $I^+(x')$ (property
{\sf R2}), and that its support does not include $I^-(x')$ (property 
{\sf R3}).

\newpage
\hrule
\hrule
\part{Motion of point particles} 
\label{part4}
\hrule
\hrule
\vspace*{.25in} 
%
\section{Motion of a scalar charge} 
\label{16} 

\subsection{Dynamics of a point scalar charge}   
\label{16.1} 

A point particle carries a scalar charge $q$ and moves on a world line 
$\gamma$ described by relations $z^\mu(\lambda)$, in which $\lambda$
is an arbitrary parameter. The particle generates a scalar potential 
$\Phi(x)$ and a field $\Phi_\alpha(x) \equiv \nabla_\alpha
\Phi(x)$. The dynamics of the entire system is governed by the action 
\begin{equation}
S = S_{\rm field} + S_{\rm particle} + S_{\rm interaction}, 
\label{16.1.1}
\end{equation} 
where $S_{\rm field}$ is an action functional for a free scalar field 
in a spacetime with metric $g_{\alpha\beta}$, $S_{\rm particle}$ is
the action of a free particle moving on a world line $\gamma$ in this  
spacetime, and $S_{\rm interaction}$ is an interaction term that
couples the field to the particle. 

The field action is given by  
\begin{equation}
S_{\rm field} = -\frac{1}{8\pi} \int \bigl( g^{\alpha\beta}
\Phi_{\alpha} \Phi_{\beta} + \xi R \Phi^2 \bigr) \sqrt{-g}\, d^4 x,  
\label{16.1.2}
\end{equation} 
where the integration is over all of spacetime; the field is coupled
to the Ricci scalar $R$ by an arbitrary constant $\xi$. The particle
action is   
\begin{equation} 
S_{\rm particle} = -m_0 \int_\gamma d\tau, 
\label{16.1.3}
\end{equation}
where $m_0$ is the bare mass of the particle and $d\tau =
\sqrt{-g_{\mu\nu}(z) \dot{z}^\mu \dot{z}^\nu}\, d\lambda$ is the
differential of proper time along the world line; we use an overdot on 
$z^\mu(\lambda)$ to indicate differentiation with respect to the
parameter $\lambda$. Finally, the interaction term is given by  
\begin{equation}  
S_{\rm interaction} = q \int_\gamma \Phi(z)\, d\tau = q \int
\Phi(x) \delta_4(x,z)\sqrt{-g}\, d^4x d\tau. 
\label{16.1.4}
\end{equation}
Notice that both $S_{\rm particle}$ and $S_{\rm interaction}$ are
invariant under a reparameterization $\lambda \to \lambda'(\lambda)$
of the world line.  

Demanding that the total action be stationary under a variation
$\delta\Phi(x)$ of the field configuration yields the wave equation  
\begin{equation} 
\bigl( \Box - \xi R \bigr) \Phi(x) = -4\pi \mu(x) 
\label{16.1.5}
\end{equation} 
for the scalar potential, with a charge density $\mu(x)$ defined by  
\begin{equation}
\mu(x) = q \int_\gamma \delta_4(x,z)\, d\tau. 
\label{16.1.6}
\end{equation} 
These equations determine the field $\Phi_\alpha(x)$ once the motion
of the scalar charge is specified. On the other hand, demanding that
the total action be stationary under a variation 
$\delta z^\mu(\lambda)$ of the world line yields the equations of
motion  
\begin{equation} 
m(\tau) \frac{D u^\mu}{d\tau} = q \bigl( g^{\mu\nu} + u^\mu u^\nu
\bigr) \Phi_{\nu}(z) 
\label{16.1.7}
\end{equation} 
for the scalar charge. We have here adopted $\tau$ as the parameter on
the world line, and introduced the four-velocity $u^\mu(\tau) \equiv 
dz^\mu/d\tau$. The dynamical mass that appears in Eq.~(\ref{16.1.7})
is defined by $m(\tau) = m_0 - q \Phi(z)$, which can also be written
in differential form as 
\begin{equation} 
\frac{d m}{d\tau} = -q \Phi_{\mu}(z) u^\mu. 
\label{16.1.8}
\end{equation} 
It should be clear that Eqs.~(\ref{16.1.7}) and (\ref{16.1.8}) are
valid only in a formal sense, because the scalar potential obtained
from Eqs.~(\ref{16.1.5}) and (\ref{16.1.6}) diverges on the world
line. Before we can make sense of these equations we have to analyze
the field's singularity structure near the world line.   

\subsection{Retarded potential near the world line} 
\label{16.2}

The retarded solution to Eq.~(\ref{16.1.5}) is $\Phi(x) = \int
G_+(x,x') \mu(x') \sqrt{g'}\, d^4x'$, where $G_+(x,x')$ is the
retarded Green's function introduced in Sec.~\ref{13}. After
substitution of Eq.~(\ref{16.1.6}) we obtain
\begin{equation}
\Phi(x) = q \int_\gamma G_+(x,z)\, d\tau, 
\label{16.2.1}
\end{equation}
in which $z(\tau)$ gives the description of the world line
$\gamma$. Because the retarded Green's function is defined globally in
the entire spacetime, Eq.~(\ref{16.2.1}) applies to any field point
$x$. 

\begin{figure}[b]
\vspace*{2.2in}
\special{hscale=35 vscale=35 hoffset=115.0 voffset=-55.0
         psfile=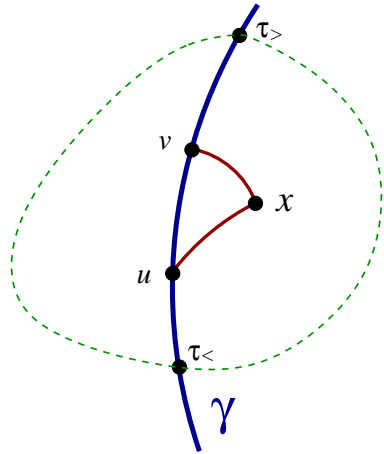}
\caption{The region within the dashed boundary represents the normal 
convex neighbourhood of the point $x$. The world line $\gamma$ enters 
the neighbourhood at proper time $\tau_<$ and exits at proper time
$\tau_>$. Also shown are the retarded point $z(u)$ and the advanced
point $z(v)$.}   
\end{figure} 

We now specialize Eq.~(\ref{16.2.1}) to a point $x$ near the world
line; see Fig.~9. We let ${\cal N}(x)$ be the normal convex
neighbourhood of this point, and we assume that the world line
traverses ${\cal N}(x)$. Let $\tau_<$ be the value of the proper-time
parameter at which $\gamma$ enters ${\cal N}(x)$ from the past, and
let $\tau_>$ be its value when the world line leaves 
${\cal N}(x)$. Then Eq.~(\ref{16.2.1}) can be broken down into the
three integrals    
\[
\Phi(x) = q \int_{-\infty}^{\tau_<} G_+(x,z)\, d\tau 
+ q \int_{\tau_<}^{\tau_>} G_+(x,z)\, d\tau 
+ q \int_{\tau_>}^{\infty} G_+(x,z)\, d\tau. 
\]
The third integration vanishes because $x$ is then in the past of
$z(\tau)$, and $G_+(x,z) = 0$. For the second integration, $x$ is the 
normal convex neighbourhood of $z(\tau)$, and the retarded Green's 
function can be expressed in the Hadamard form produced in 
Sec.~\ref{13.2}. This gives   
\[ 
\int_{\tau_<}^{\tau_>} G_+(x,z)\, d\tau = \int_{\tau_<}^{\tau_>}
U(x,z) \delta_+(\sigma)\, d\tau + \int_{\tau_<}^{\tau_>} V(x,z)
\theta_+(-\sigma)\, d\tau, 
\] 
and to evaluate this we refer back to Sec.~\ref{9} and let $x' 
\equiv z(u)$ be the retarded point associated with $x$; these points
are related by $\sigma(x,x') = 0$ and $r \equiv \sigma_{\alpha'}
u^{\alpha'}$ is the retarded distance between $x$ and the world
line. We resume the index convention of Sec.~\ref{9}: to tensors at
$x$ we assign indices $\alpha$, $\beta$, etc.; to tensors at $x'$
we assign indices $\alpha'$, $\beta'$, etc.; and to tensors at a
generic point $z(\tau)$ on the world line we assign indices $\mu$,
$\nu$, etc.    

To perform the first integration we change variables from $\tau$
to $\sigma$, noticing that $\sigma$ increases as $z(\tau)$ passes
through $x'$. The change of $\sigma$ on the world line is given
by $d\sigma \equiv \sigma(x,z+dz) - \sigma(x,z) = \sigma_\mu u^\mu\,
d\tau$, and we find that the first integral evaluates to
$U(x,z)/(\sigma_\mu u^\mu)$ with $z$ identified with $x'$. The second
integration is cut off at $\tau = u$ by the step function, and we
obtain our final expression for the retarded potential of a point
scalar charge:  
\begin{equation}
\Phi(x) = \frac{q}{r} U(x,x') + q \int_{\tau_<}^u V(x,z)\, d\tau + q
\int_{-\infty}^{\tau_<} G_+(x,z)\, d\tau.   
\label{16.2.2}
\end{equation} 
This expression applies to a point $x$ sufficiently close to the world
line that there exists a nonempty intersection between 
${\cal N}(x)$ and $\gamma$.  

\subsection{Field of a scalar charge in retarded coordinates}  
\label{16.3} 

When we differentiate the potential of Eq.~(\ref{16.2.2}) we must keep
in mind that a variation in $x$ induces a variation in $x'$ because
the new points $x + \delta x$ and $x' + \delta x'$ must also be linked
by a null geodesic --- you may refer back to Sec.~\ref{9.2} for a
detailed discussion. This means, for example, that the total variation
of $U(x,x')$ is $\delta U = U(x+\delta x,x'+\delta x') - U(x,x') =
U_{;\alpha} \delta x^\alpha + U_{;\alpha'} u^{\alpha'}\, 
\delta u$. The gradient of the scalar potential is therefore given by    
\begin{equation}
\Phi_{\alpha}(x) = -\frac{q}{r^2} U(x,x') \partial_\alpha r + 
\frac{q}{r} U_{;\alpha}(x,x') + \frac{q}{r} U_{;\alpha'}(x,x')
u^{\alpha'} \partial_\alpha u + q V(x,x') \partial_\alpha u +  
\Phi^{\rm tail}_{\alpha}(x),   
\label{16.3.1}
\end{equation}
where the ``tail integral'' is defined by    
\begin{eqnarray}
\Phi^{\rm tail}_{\alpha}(x) &=& q \int_{\tau_<}^u \nabla_\alpha 
V(x,z)\, d\tau + q \int_{-\infty}^{\tau_<} \nabla_\alpha G_+(x,z)\, 
d\tau \nonumber \\  
&=& q \int_{-\infty}^{u^-} \nabla_\alpha G_+(x,z)\, d\tau.   
\label{16.3.2}
\end{eqnarray} 
In the second form of the definition we integrate $\nabla_\alpha 
G_+(x,z)$ from $\tau = -\infty$ to almost $\tau = u$, but we cut the  
integration short at $\tau = u^- \equiv u - 0^+$ to avoid the singular  
behaviour of the retarded Green's function at $\sigma = 0$. This
limiting procedure gives rise to the first form of the definition,
with the advantage that the integral need not be broken down into
contributions that refer to ${\cal N}(x)$ and its complement,
respectively.  

We shall now expand $\Phi_{\alpha}(x)$ in powers of $r$, and express 
the results in terms of the retarded coordinates $(u,r,\Omega^a)$
introduced in Sec.~\ref{9}. It will be convenient to decompose
$\Phi_{\alpha}(x)$ in the tetrad
$(\base{\alpha}{0},\base{\alpha}{a})$ that is obtained by parallel
transport of $(u^{\alpha'},\base{\alpha'}{a})$ on the null geodesic
that links $x$ to $x' \equiv z(u)$; this construction is detailed in 
Sec.~\ref{9}. Note that throughout this section we set
$\omega_{ab} = 0$, where $\omega_{ab}$ is the rotation tensor defined
by Eq.~(\ref{9.1.1}): the tetrad vectors $\base{\alpha'}{a}$ are taken
to be Fermi-Walker transported on $\gamma$. The expansion relies on
Eq.~(\ref{9.5.3}) for $\partial_\alpha u$, Eq.~(\ref{9.5.5}) for
$\partial_\alpha r$, and we shall need    
\begin{equation}
U(x,x') = 1 + \frac{1}{12} r^2 \bigl( R_{00} + 2 R_{0a} \Omega^a + 
R_{ab} \Omega^a \Omega^b \bigr) + O(r^3), 
\label{16.3.3}
\end{equation} 
which follows from Eq.~(\ref{13.2.7}) and the relation 
$\sigma^{\alpha'} = -r (u^{\alpha'} + \Omega^a \base{\alpha'}{a})$
first encountered in Eq.~(\ref{9.2.3}); recall that
\[
R_{00}(u) = R_{\alpha'\beta'} u^{\alpha'} u^{\beta'}, 
\qquad 
R_{0a}(u) = R_{\alpha'\beta'} u^{\alpha'} \base{\beta'}{a},
\qquad 
R_{ab}(u) = R_{\alpha'\beta'} \base{\alpha'}{a} \base{\beta'}{b}
\]
are frame components of the Ricci tensor evaluated at $x'$. We shall
also need the expansions    
\begin{equation} 
U_{;\alpha}(x,x') = \frac{1}{6} r g^{\alpha'}_{\ \alpha} 
\bigl( R_{\alpha'0} + R_{\alpha' b} \Omega^b \bigr) + O(r^2)   
\label{16.3.4} 
\end{equation}
and 
\begin{equation} 
U_{;\alpha'}(x,x') u^{\alpha'} = -\frac{1}{6} r \bigl( R_{00} 
+ R_{0a} \Omega^a \bigr) + O(r^2) 
\label{16.3.5}
\end{equation}
which follow from Eqs.~(\ref{13.2.8}); recall from Eq.~(\ref{9.1.4}) 
that the parallel propagator can be expressed as 
$g^{\alpha'}_{\ \alpha} = u^{\alpha'} \base{0}{\alpha} 
+ \base{\alpha'}{a} \base{a}{\alpha}$. And finally, we shall need  
\begin{equation}
V(x,x') = \frac{1}{12} \bigl( 1 - 6\xi \bigr) R + O(r), 
\label{16.3.6}
\end{equation} 
a relation that was first established in Eq.~(\ref{13.2.10}); here
$R \equiv R(u)$ is the Ricci scalar evaluated at $x'$.  

Collecting all these results gives    
\begin{eqnarray} 
\Phi_0(u,r,\Omega^a) &\equiv& \Phi_{\alpha}(x) \base{\alpha}{0}(x)
\nonumber \\  
&=& \frac{q}{r} a_a \Omega^a + \frac{1}{2} q R_{a0b0}\Omega^a
\Omega^b + \frac{1}{12} \bigl( 1 - 6\xi \bigr) q R 
+ \Phi_0^{\rm tail} + O(r), 
\label{16.3.7} \\ 
\Phi_a(u,r,\Omega^a) &\equiv& \Phi_{\alpha}(x) \base{\alpha}{a}(x)
\nonumber \\  
&=& -\frac{q}{r^2} \Omega_a - \frac{q}{r} a_b \Omega^b \Omega_a -
\frac{1}{3} q R_{b0c0} \Omega^b \Omega^c \Omega_a - \frac{1}{6} q
\bigl( R_{a0b0} \Omega^b - R_{ab0c} \Omega^b \Omega^c \bigr) 
\nonumber \\ & & \mbox{}
+ \frac{1}{12} q \bigl[ R_{00} - R_{bc} \Omega^b\Omega^c -
(1-6\xi) R \bigr] \Omega_a + \frac{1}{6} q \bigl( R_{a0} + R_{ab} 
\Omega^b \bigr) + \Phi_{a}^{\rm tail} + O(r),  \qquad
\label{16.3.8}
\end{eqnarray} 
where $a_a = a_{\alpha'} \base{\alpha'}{a}$ are the frame  
components of the acceleration vector, 
\[
R_{a0b0}(u) = R_{\alpha'\gamma'\beta'\delta'} \base{\alpha'}{a}
u^{\gamma'}  \base{\beta'}{b} u^{\delta'}, \qquad 
R_{ab0c}(u) = R_{\alpha'\gamma'\beta'\delta'} \base{\alpha'}{a}
\base{\gamma'}{b} u^{\beta'} \base{\delta'}{c}
\]
are frame components of the Riemann tensor evaluated at $x'$, and
\begin{equation} 
\Phi_0^{\rm tail}(u) = \Phi_{\alpha'}^{\rm tail}(x')
u^{\alpha'},  
\qquad  
\Phi_a^{\rm tail}(u) = \Phi_{\alpha'}^{\rm tail}(x') 
\base{\alpha'}{a}  
\label{16.3.9}
\end{equation}
are the frame components of the tail integral evaluated at 
$x'$. Equations (\ref{16.3.7}) and (\ref{16.3.8}) show clearly 
that $\Phi_{\alpha}(x)$ is singular on the world line: the field
diverges as $r^{-2}$ when $r\to 0$, and many of the terms that stay 
bounded in the limit depend on $\Omega^a$ and therefore possess a
directional ambiguity at $r=0$.  

\subsection{Field of a scalar charge in Fermi normal coordinates}   
\label{16.4} 

The gradient of the scalar potential can also be expressed in the
Fermi normal coordinates of Sec.~\ref{8}. To effect this translation
we make $\bar{x} \equiv z(t)$ the new
reference point on the world line. We resume here the notation of 
Sec.~\ref{10} and assign indices $\bar{\alpha}$, $\bar{\beta}$,
\ldots to tensors at $\bar{x}$. The Fermi normal coordinates are
denoted $(t,s,\omega^a)$, and we let $(\bar{e}^\alpha_0,
\bar{e}^\alpha_a)$ be the tetrad at $x$ that is obtained by parallel
transport of $(u^{\bar{\alpha}}, \base{\bar{\alpha}}{a})$ on the
spacelike geodesic that links $x$ to $\bar{x}$.  

Our first task is to decompose $\Phi_{\alpha}(x)$ in the tetrad
$(\bar{e}^\alpha_0, \bar{e}^\alpha_a)$, thereby defining $\bar{\Phi}_0 
\equiv \Phi_{\alpha} \bar{e}^\alpha_0$ and $\bar{\Phi}_a \equiv
\Phi_{\alpha} \bar{e}^\alpha_a$. For this purpose we use 
Eqs.~(\ref{10.3.1}), (\ref{10.3.2}), (\ref{16.3.7}), and
(\ref{16.3.8}) to obtain  
\begin{eqnarray*} 
\bar{\Phi}_0 &=& 
\Bigl[ 1 + O(r^2) \Bigr] \Phi_0 + \Bigl[ r \bigl( 1 - a_b \Omega^b 
\bigr) a^a + \frac{1}{2} r^2 \dot{a}^a + \frac{1}{2} r^2 
R^a_{\ 0b0} \Omega^b + O(r^3) \Bigr] \Phi_a \\ 
&=& -\frac{1}{2} q \dot{a}_a \Omega^a + \frac{1}{12} (1-6\xi) q R +
\bar{\Phi}_0^{\rm tail} + O(r) 
\end{eqnarray*} 
and 
\begin{eqnarray*} 
\bar{\Phi}_a &=& 
\Bigl[ \delta^b_{\ a} + \frac{1}{2} r^2 a^b a_a - \frac{1}{2} r^2
R^b_{\ a0c} \Omega^c + O(r^3) \Bigr] \Phi_b + \Bigl[ r a_a + O(r^2)
\Bigr] \Phi_0 \\ 
&=& -\frac{q}{r^2} \Omega_a - \frac{q}{r} a_b \Omega^b \Omega_a 
+ \frac{1}{2} q a_b \Omega^b a_a 
- \frac{1}{3} q R_{b0c0} \Omega^b \Omega^c \Omega_a 
- \frac{1}{6} q R_{a0b0} \Omega^b 
- \frac{1}{3} q R_{ab0c} \Omega^b \Omega^c  
\nonumber \\ & & \mbox{}
+ \frac{1}{12} q \bigl[ R_{00} - R_{bc} \Omega^b\Omega^c 
- (1-6\xi) R \bigr] \Omega_a 
+ \frac{1}{6} q \bigl( R_{a0} + R_{ab} \Omega^b \bigr) 
+ \bar{\Phi}_{a}^{\rm tail} + O(r), 
\end{eqnarray*} 
where all frame components are still evaluated at $x'$, except for 
$\bar{\Phi}_{0}^{\rm tail}$ and $\bar{\Phi}_{a}^{\rm tail}$ which are
evaluated at $\bar{x}$.   

We must still translate these results into the Fermi normal
coordinates $(t,s,\omega^a)$. For this we involve Eqs.~(\ref{10.2.1}),
(\ref{10.2.2}), and (\ref{10.2.3}), from which we deduce, for example,   
\begin{eqnarray*} 
\frac{1}{r^2}\Omega_a &=& \frac{1}{s^2}\omega_a + \frac{1}{2s} a_a  
- \frac{3}{2s} a_b \omega^b \omega_a - \frac{3}{4} a_b \omega^b a_a 
+ \frac{15}{8} \bigl( a_b \omega^b \bigr)^2 \omega_a 
+ \frac{3}{8} \dot{a}_0 \omega_a - \frac{1}{3} \dot{a}_a 
\\ & & \mbox{} 
+ \dot{a}_b \omega^b \omega_a 
+ \frac{1}{6} R_{a0b0} \omega^b    
- \frac{1}{2} R_{b0c0} \omega^b \omega^c \omega_a 
- \frac{1}{3} R_{ab0c} \omega^b \omega^c + O(s) 
\end{eqnarray*}
and 
\[
\frac{1}{r} a_b \Omega^b \Omega_a = \frac{1}{s} a_b \omega^b \omega_a 
+ \frac{1}{2} a_b \omega^b a_a 
- \frac{3}{2} \bigl( a_b \omega^b \bigr)^2 \omega_a   
- \frac{1}{2} \dot{a}_0 \omega_a - \dot{a}_b \omega^b \omega_a 
+ O(s), 
\]
in which all frame components (on the right-hand side of these
relations) are now evaluated at $\bar{x}$; to obtain the second
relation we expressed $a_a(u)$ as $a_a(t) - s \dot{a}_a(t) + O(s^2)$
since according to Eq.~(\ref{10.2.1}), $u = t - s + O(s^2)$.  

Collecting these results yields 
\begin{eqnarray} 
\bar{\Phi}_0(t,s,\omega^a) &\equiv& \Phi_{\alpha}(x)
\bar{e}^\alpha_0(x) 
\nonumber \\  
&=& -\frac{1}{2} q \dot{a}_a \omega^a + \frac{1}{12} (1-6\xi) q R + 
\bar{\Phi}_0^{\rm tail} + O(s), 
\label{16.4.1} \\  
\bar{\Phi}_a(t,s,\omega^a) &\equiv& \Phi_{\alpha}(x)
\bar{e}^\alpha_a(x) 
\nonumber \\  
&=& -\frac{q}{s^2} \omega_a 
- \frac{q}{2s} \bigl( a_a - a_b \omega^b \omega_a \bigr)  
+ \frac{3}{4} q a_b \omega^b a_a 
- \frac{3}{8} q \bigl( a_b \omega^b \bigr)^2 \omega_a 
+ \frac{1}{8} q \dot{a}_0 \omega_a 
+ \frac{1}{3} q \dot{a}_a 
\nonumber \\ & & \mbox{} 
- \frac{1}{3} q R_{a0b0} \omega^b 
+ \frac{1}{6} q R_{b0c0} \omega^b \omega^c \omega_a 
+ \frac{1}{12} q \bigl[ R_{00} - R_{bc} \omega^b\omega^c 
- (1-6\xi) R \bigr] \omega_a 
\nonumber \\ & & \mbox{} 
+ \frac{1}{6} q \bigl( R_{a0} + R_{ab} \omega^b \bigr) 
+ \bar{\Phi}_{a}^{\rm tail} + O(s). 
\label{16.4.2}
\end{eqnarray} 
In these expressions, $a_a(t) = a_{\bar{\alpha}}
\base{\bar{\alpha}}{a}$ are the frame components of the acceleration
vector evaluated at $\bar{x}$, $\dot{a}_0(t) =
\dot{a}_{\bar{\alpha}} u^{\bar{\alpha}}$ and $\dot{a}_a(t) =  
\dot{a}_{\bar{\alpha}} \base{\bar{\alpha}}{a}$ are frame components of
its covariant derivative, $R_{a0b0}(t) =
R_{\bar{\alpha}\bar{\gamma}\bar{\beta}\bar{\delta}}
\base{\bar{\alpha}}{a} u^{\bar{\gamma}} \base{\bar{\beta}}{b}
u^{\bar{\delta}}$ are frame components of the Riemann tensor evaluated
at $\bar{x}$, 
\[
R_{00}(t) = R_{\bar{\alpha}\bar{\beta}} u^{\bar{\alpha}}
u^{\bar{\beta}},  
\qquad 
R_{0a}(t) = R_{\bar{\alpha}\bar{\beta}} u^{\bar{\alpha}}
\base{\bar{\beta}}{a}, 
\qquad 
R_{ab}(t) = R_{\bar{\alpha}\bar{\beta}} \base{\bar{\alpha}}{a}
\base{\bar{\beta}}{b} 
\]
are frame components of the Ricci tensor, and $R(t)$ is the Ricci 
scalar evaluated at $\bar{x}$. Finally, we have that 
\begin{equation} 
\bar{\Phi}_0^{\rm tail}(t) = 
\Phi_{\bar{\alpha}}^{\rm tail}(\bar{x}) u^{\bar{\alpha}},   
\qquad  
\bar{\Phi}_a^{\rm tail}(t) =
\Phi_{\bar{\alpha}}^{\rm tail}(\bar{x}) \base{\bar{\alpha}}{a}   
\label{16.4.3}
\end{equation}
are the frame components of the tail integral --- see
Eq.~(\ref{16.3.2}) --- evaluated at $\bar{x} \equiv z(t)$. 

We shall now compute the averages of $\bar{\Phi}_0$ and $\bar{\Phi}_a$ 
over $S(t,s)$, a two-surface of constant $t$ and $s$; these will
represent the mean value of the field at a fixed proper distance away
from the world line, as measured in a reference frame that is
momentarily comoving with the particle. The two-surface is
charted by angles $\theta^A$ ($A = 1, 2$) and it is described,
in the Fermi normal coordinates, by the parametric relations
$\hat{x}^a = s \omega^a(\theta^A)$; a canonical choice of
parameterization is $\omega^a = (\sin\theta\cos\phi,
\sin\theta\sin\phi, \cos\theta)$. Introducing the transformation
matrices $\omega^a_A \equiv \partial \omega^a/\partial \theta^A$, we
find from Eq.~(\ref{8.5.5}) that the induced metric on $S(t,s)$ is
given by    
\begin{equation} 
ds^2 = s^2 \Bigl[ \omega_{AB} - \frac{1}{3} s^2 R_{AB} + O(s^3) 
\Bigr]\, d\theta^A d\theta^B, 
\label{16.4.4}
\end{equation} 
where $\omega_{AB} \equiv \delta_{ab} \omega^a_A \omega^b_B$ is the 
metric of the unit two-sphere, and where $R_{AB} \equiv R_{acbd}
\omega^a_A \omega^c \omega^b_B \omega^d$ depends on $t$ and the angles
$\theta^A$. From this we infer that the element of surface area is
given by 
\begin{equation} 
d{\cal A} = s^2 \Bigl[ 1 - \frac{1}{6} s^2 R^c_{\ acb}(t) \omega^a 
\omega^b + O(s^3) \Bigr]\, d\Omega, 
\label{16.4.5}
\end{equation}
where $d\Omega = \sqrt{\mbox{det}[\omega_{AB}]}\, d^2\theta$ is an
element of solid angle --- in the canonical parameterization, $d\Omega
= \sin\theta\, d\theta d\phi$. Integration of Eq.~(\ref{16.4.5})
produces the total surface area of $S(t,s)$, and ${\cal A} 
= 4\pi s^2 [1 - \frac{1}{18} s^2 R^{ab}_{\ \ ab} + O(s^3)]$.  

The averaged fields are defined by 
\begin{equation} 
\bigl\langle \bar{\Phi}_0 \bigr\rangle(t,s) = \frac{1}{\cal A} 
\oint_{S(t,s)} \bar{\Phi}_0(t,s,\theta^A)\, d{\cal A}, \qquad  
\bigl\langle \bar{\Phi}_a \bigr\rangle(t,s) = \frac{1}{\cal A} 
\oint_{S(t,s)} \bar{\Phi}_a(t,s,\theta^A)\, d{\cal A}, 
\label{16.4.6}
\end{equation} 
where the quantities to be integrated are scalar functions of the
Fermi normal coordinates. The results 
\begin{equation}
\frac{1}{4\pi} \oint \omega^a\, d\Omega = 0, \qquad 
\frac{1}{4\pi} \oint \omega^a \omega^b\, d\Omega = 
\frac{1}{3} \delta^{ab}, \qquad  
\frac{1}{4\pi} \oint \omega^a \omega^b \omega^c\, d\Omega = 0, 
\label{16.4.7}
\end{equation} 
are easy to establish, and we obtain 
\begin{eqnarray} 
\bigl\langle \bar{\Phi}_0 \bigr\rangle &=& \frac{1}{12} (1-6\xi)
q R + \bar{\Phi}_0^{\rm tail} + O(s), 
\label{16.4.8} \\ 
\bigl\langle \bar{\Phi}_a \bigr\rangle &=& -\frac{q}{3s} a_a +
\frac{1}{3} q \dot{a}_a + \frac{1}{6} q R_{a0} + 
\bar{\Phi}_a^{\rm tail} + O(s).  
\label{16.4.9}
\end{eqnarray} 
The averaged field is still singular on the world line. Regardless, we
shall take the formal limit $s \to 0$ of the expressions displayed in 
Eqs.~(\ref{16.4.8}) and (\ref{16.4.9}). In the limit the tetrad
$(\bar{e}^\alpha_0, \bar{e}^\alpha_a)$ reduces to 
$(u^{\bar{\alpha}}, \base{\bar{\alpha}}{a})$, and we can
reconstruct the field at $\bar{x}$ by invoking the completeness
relations $\delta^{\bar{\alpha}}_{\ \bar{\beta}} =
-u^{\bar{\alpha}} u_{\bar{\beta}} + \base{\bar{\alpha}}{a}
\base{a}{\bar{\beta}}$. We thus obtain 
\begin{equation} 
\bigl\langle \Phi_{\bar{\alpha}} \bigr\rangle = \lim_{s\to 0} \biggl(
- \frac{q}{3s} \biggr) a_{\bar{\alpha}}  
- \frac{1}{12} (1-6\xi) q R u_{\bar{\alpha}} 
+ q \bigl( g_{\bar{\alpha}\bar{\beta}} 
+ u_{\bar{\alpha}} u_{\bar{\beta}} \bigr) 
\biggl( \frac{1}{3} \dot{a}^{\bar{\beta}} 
+ \frac{1}{6} R^{\bar{\beta}}_{\ \bar{\gamma}} u^{\bar{\gamma}}
\biggr) + \Phi_{\bar{\alpha}}^{\rm tail}, 
\label{16.4.10}
\end{equation}
where the tail integral can be copied from Eq.~(\ref{16.3.2}), 
\begin{equation} 
\Phi_{\bar{\alpha}}^{\rm tail}(\bar{x}) = q \int_{-\infty}^{t^-} 
\nabla_{\bar{\alpha}} G_+(\bar{x},z)\, d\tau. 
\label{16.4.11}
\end{equation} 
The tensors appearing in Eq.~(\ref{16.4.10}) all refer to $\bar{x}
\equiv z(t)$, which now stands for an arbitrary point on the world
line $\gamma$.   

\subsection{Singular and radiative fields} 
\label{16.5} 

The singular potential 
\begin{equation} 
\Phi^{\rm S}(x) = q \int_\gamma G_{\rm S}(x,z)\, d\tau
\label{16.5.1}
\end{equation} 
is the (unphysical) solution to Eqs.~(\ref{16.1.5}) and (\ref{16.1.6})
that is obtained by adopting the singular Green's function of
Eq.~(\ref{13.5.12}) instead of the retarded Green's function. As we
shall see, the resulting singular field $\Phi^{\rm S}_\alpha(x)$
reproduces the singular behaviour of the retarded solution; the
difference, $\Phi^{\rm R}_\alpha(x) = \Phi_\alpha(x) 
- \Phi^{\rm S}_\alpha(x)$, is smooth on the world line.     

To evaluate the integral of Eq.~(\ref{16.5.1}) we assume once more
that $x$ is sufficiently close to $\gamma$ that the world line
traverses ${\cal N}(x)$; refer back to Fig.~9. As before we let
$\tau_<$ and $\tau_>$ be the values of the proper-time parameter at
which $\gamma$ enters and leaves ${\cal N}(x)$, respectively. Then
Eq.~(\ref{16.5.1}) can be broken down into the three integrals   
\[
\Phi^{\rm S}(x) = q \int_{-\infty}^{\tau_<} G_{\rm S}(x,z)\, d\tau  
+ q \int_{\tau_<}^{\tau_>} G_{\rm S}(x,z)\, d\tau 
+ q \int_{\tau_>}^{\infty} G_{\rm S}(x,z)\, d\tau.  
\]
The first integration vanishes because $x$ is then in the
chronological future of $z(\tau)$, and $G_{\rm S}(x,z) = 0$ by
Eq.~(\ref{13.5.3}). Similarly, the third integration vanishes because
$x$ is then in the chronological past of $z(\tau)$. For the second
integration, $x$ is the normal convex neighbourhood of $z(\tau)$, the
singular Green's function can be expressed in the Hadamard form of
Eq.~(\ref{13.5.14}), and we have      
\[ 
\int_{\tau_<}^{\tau_>} G_{\rm S}(x,z)\, d\tau = 
\frac{1}{2} \int_{\tau_<}^{\tau_>} U(x,z) \delta_+(\sigma)\, d\tau 
+ \frac{1}{2} \int_{\tau_<}^{\tau_>} U(x,z) \delta_-(\sigma)\, d\tau 
- \frac{1}{2} \int_{\tau_<}^{\tau_>} V(x,z) \theta(\sigma)\, d\tau.  
\] 
To evaluate these we re-introduce the retarded point $x' \equiv z(u)$
and let $x'' \equiv z(v)$ be the {\it advanced point} associated with
$x$; we recall from Sec.~\ref{10.4} that these points are related by
$\sigma(x,x'') = 0$ and that $r_{\rm adv} \equiv - \sigma_{\alpha''}
u^{\alpha''}$ is the advanced distance between $x$ and the world
line. 

To perform the first integration we change variables from $\tau$
to $\sigma$, noticing that $\sigma$ increases as $z(\tau)$ passes
through $x'$; the integral evaluates to $U(x,x')/r$. We do the same
for the second integration, but we notice now that $\sigma$ decreases
as $z(\tau)$ passes through $x''$; the integral evaluates to
$U(x,x'')/r_{\rm adv}$. The third integration is restricted to the
interval $u \leq \tau \leq v$ by the step function, and we obtain our
final expression for the singular potential of a point scalar charge:   
\begin{equation}
\Phi^{\rm S}(x) = \frac{q}{2r} U(x,x') 
+ \frac{q}{2r_{\rm adv}} U(x,x'') 
- \frac{1}{2} q \int_u^v V(x,z)\, d\tau.   
\label{16.5.2} 
\end{equation} 
We observe that $\Phi^{\rm S}(x)$ depends on the state of motion of 
the scalar charge between the retarded time $u$ and the advanced time
$v$; contrary to what was found in Sec.~\ref{16.2} for the retarded
potential, there is no dependence on the particle's remote past.          

We use the techniques of Sec.~\ref{16.3} to differentiate the
potential of Eq.~(\ref{16.5.2}). We find 
\begin{eqnarray}
\Phi^{\rm S}_{\alpha}(x) &=& 
-\frac{q}{2r^2} U(x,x') \partial_\alpha r  
- \frac{q}{2 {r_{\rm adv}}^2} U(x,x'') \partial_\alpha r_{\rm adv} 
+ \frac{q}{2r} U_{;\alpha}(x,x') 
+ \frac{q}{2r} U_{;\alpha'}(x,x') u^{\alpha'} \partial_\alpha u 
\nonumber \\  & & \mbox{} 
+ \frac{q}{2r_{\rm adv}} U_{;\alpha}(x,x'') 
+ \frac{q}{2r_{\rm adv}} U_{;\alpha''}(x,x'') u^{\alpha''}
\partial_\alpha v + \frac{1}{2} q V(x,x') \partial_\alpha u 
- \frac{1}{2} q V(x,x'') \partial_\alpha v 
\nonumber \\  & & \mbox{} 
- \frac{1}{2} q \int_u^v \nabla_\alpha V(x,z)\, d\tau, 
\label{16.5.3}
\end{eqnarray}
and we would like to express this as an expansion in powers of
$r$. For this we shall rely on results already established in 
Sec.~\ref{16.3}, as well as additional expansions that will involve
the advanced point $x''$. Those we develop now.  

We recall first that a relation between retarded and advanced times
was worked out in Eq.~(\ref{10.4.2}), that an expression for the
advanced distance was displayed in Eq.~(\ref{10.4.3}), and that
Eqs.~(\ref{10.4.4}) and (\ref{10.4.5}) give expansions for
$\partial_\alpha v$ and $\partial_\alpha r_{\rm adv}$,
respectively. 

To derive an expansion for $U(x,x'')$ we follow the general method of 
Sec.~\ref{10.4} and define a function $U(\tau) \equiv U(x,z(\tau))$
of the proper-time parameter on $\gamma$. We have that 
\[
U(x,x'') \equiv U(v) = U(u+\Delta^{\!\prime}) = U(u) + \dot{U}(u) 
\Delta^{\!\prime} + \frac{1}{2} \ddot{U}(u) \Delta^{\!\prime 2} 
+ O\bigl( \Delta^{\!\prime 3} \bigr), 
\]
where overdots indicate differentiation with respect to $\tau$, and
where $\Delta^{\!\prime} \equiv v-u$. The leading term $U(u) \equiv
U(x,x')$ was worked out in Eq.~(\ref{16.3.3}), and the derivatives of 
$U(\tau)$ are given by 
\[
\dot{U}(u) = U_{;\alpha'} u^{\alpha'} = -\frac{1}{6} r \bigl( R_{00} 
+ R_{0a} \Omega^a \bigr) + O(r^2)
\]
and 
\[
\ddot{U}(u) = U_{;\alpha'\beta'} u^{\alpha'} u^{\beta'} + U_{;\alpha'}
a^{\alpha'} = \frac{1}{6} R_{00} + O(r),
\]
according to Eqs.~(\ref{16.3.5}) and (\ref{13.2.8}). Combining these
results together with Eq.~(\ref{10.4.2}) for $\Delta^{\!\prime}$ gives   
\begin{equation} 
U(x,x'') = 1 + \frac{1}{12} r^2 \bigl( R_{00} - 2 R_{0a} \Omega^a +
R_{ab} \Omega^a \Omega^b \bigr) + O(r^3), 
\label{16.5.4}
\end{equation}
which should be compared with Eq.~(\ref{16.3.3}). It should be
emphasized that in Eq.~(\ref{16.5.4}) and all equations below, the
frame components of the Ricci tensor are evaluated at the retarded
point $x' \equiv z(u)$, and not at the advanced point. The preceding
computation gives us also an expansion for 
$U_{;\alpha''} u^{\alpha''} \equiv \dot{U}(v) =
\dot{U}(u) + \ddot{U}(u) \Delta^{\!\prime} + O(\Delta^{\!\prime 2})$. 
This becomes     
\begin{equation}
U_{;\alpha''}(x,x'') u^{\alpha''} = \frac{1}{6} r \bigl( R_{00} 
- R_{0a} 
\Omega^a \bigr) + O(r^2), 
\label{16.5.5}
\end{equation}
which should be compared with Eq.~(\ref{16.3.5}).  

We proceed similarly to derive an expansion for
$U_{;\alpha}(x,x'')$. Here we introduce the functions
$U_{\alpha}(\tau) \equiv U_{;\alpha}(x,z(\tau))$ and express
$U_{;\alpha}(x,x'')$ as $U_\alpha(v) = U_\alpha(u) + \dot{U}_\alpha(u)
\Delta^{\!\prime} + O(\Delta^{\!\prime 2})$. The leading term
$U_\alpha(u) \equiv U_{;\alpha}(x,x')$ was computed in
Eq.~(\ref{16.3.4}), and  
\[
\dot{U}_\alpha(u) = U_{;\alpha\beta'} u^{\beta'} = -\frac{1}{6}
g^{\alpha'}_{\ \alpha} R_{\alpha'0} + O(r) 
\]
follows from Eq.~(\ref{13.2.8}). Combining these results together with
Eq.~(\ref{10.4.2}) for $\Delta^{\!\prime}$ gives
\begin{equation}
U_{;\alpha}(x,x'') = -\frac{1}{6} r g^{\alpha'}_{\ \alpha} 
\bigr( R_{\alpha'0} - R_{\alpha'b} \Omega^b \Bigr) + O(r^2), 
\label{16.5.6}
\end{equation}
and this should be compared with Eq.~(\ref{16.3.4}).  

The last expansion we shall need is 
\begin{equation}
V(x,x'') = \frac{1}{12} \bigl( 1 - 6\xi \bigr) R + O(r), 
\label{16.5.7} 
\end{equation} 
which follows at once from Eq.~(\ref{16.3.6}) and the fact that
$V(x,x'') - V(x,x') = O(r)$; the Ricci scalar is evaluated at the
retarded point $x'$. 

It is now a straightforward (but tedious) matter to substitute these
expansions (all of them!)\ into Eq.~(\ref{16.5.3}) and obtain the
projections of the singular field $\Phi^{\rm S}_\alpha(x)$ in the
same tetrad $(\base{\alpha}{0}, \base{\alpha}{a})$ that was employed
in Sec.~\ref{16.3}. This gives 
\begin{eqnarray} 
\Phi^{\rm S}_0(u,r,\Omega^a) &\equiv& \Phi^{\rm S}_{\alpha}(x)
\base{\alpha}{0}(x) 
\nonumber \\  
&=& \frac{q}{r} a_a \Omega^a 
+ \frac{1}{2} q R_{a0b0}\Omega^a \Omega^b + O(r), 
\label{16.5.8} \\ 
\Phi^{\rm S}_a(u,r,\Omega^a) &\equiv& \Phi^{\rm S}_{\alpha}(x)
\base{\alpha}{a}(x) 
\nonumber \\  
&=& -\frac{q}{r^2} \Omega_a - \frac{q}{r} a_b \Omega^b \Omega_a 
- \frac{1}{3} q \dot{a}_a 
- \frac{1}{3} q R_{b0c0} \Omega^b \Omega^c \Omega_a 
- \frac{1}{6} q \bigl( R_{a0b0} \Omega^b 
- R_{ab0c} \Omega^b \Omega^c \bigr)  
\nonumber \\ & & \mbox{}
+ \frac{1}{12} q \bigl[ R_{00} - R_{bc} \Omega^b\Omega^c 
- (1-6\xi) R \bigr] \Omega_a 
+ \frac{1}{6} q R_{ab} \Omega^b, 
\label{16.5.9}
\end{eqnarray} 
in which all frame components are evaluated at the retarded point 
$x' \equiv z(u)$. Comparison of these expressions with 
Eqs.~(\ref{16.3.7}) and (\ref{16.3.8}) reveals that the retarded and
singular fields share the same singularity structure.    

The difference between the retarded field of Eqs.~(\ref{16.3.7}),
(\ref{16.3.8}) and the singular field of Eqs.~(\ref{16.5.8}),
(\ref{16.5.9}) defines the radiative field 
$\Phi^{\rm R}_\alpha(x)$. Its tetrad components are 
\begin{eqnarray} 
\Phi^{\rm R}_0 &=& \frac{1}{12} (1-6\xi) q R + \Phi_0^{\rm tail} 
+ O(r),  
\label{16.5.10} \\ 
\Phi^{\rm R}_a &=& \frac{1}{3} q \dot{a}_a + \frac{1}{6} q R_{a0} 
+ \Phi_a^{\rm tail} + O(r), 
\label{16.5.11}
\end{eqnarray} 
and we see that $\Phi^{\rm R}_\alpha(x)$ is a smooth vector field on
the world line. There is therefore no obstacle in evaluating the
radiative field directly at $x=x'$, where the tetrad
$(\base{\alpha}{0},\base{\alpha}{a})$ becomes $(u^{\alpha'}, 
\base{\alpha'}{a})$. Reconstructing the field at $x'$ from its
frame components, we obtain 
\begin{equation} 
\Phi^{\rm R}_{\alpha'}(x') = 
- \frac{1}{12} (1-6\xi) q R u_{\alpha'} 
+ q \bigl( g_{\alpha'\beta'} + u_{\alpha'} u_{\beta'} \bigr) 
\biggl( \frac{1}{3} \dot{a}^{\beta'} 
+ \frac{1}{6} R^{\beta'}_{\ \gamma'} u^{\gamma'}
\biggr) + \Phi_{\alpha'}^{\rm tail}, 
\label{16.5.12}
\end{equation}
where the tail term can be copied from Eq.~(\ref{16.3.2}), 
\begin{equation} 
\Phi_{\alpha'}^{\rm tail}(x') = q \int_{-\infty}^{u^-} 
\nabla_{\alpha'} G_+(x',z)\, d\tau. 
\label{16.5.13} 
\end{equation} 
The tensors appearing in Eq.~(\ref{16.5.12}) all refer to the 
retarded point $x' \equiv z(u)$, which now stands for an  
arbitrary point on the world line $\gamma$.  
  
\subsection{Equations of motion} 
\label{16.6}

The retarded field $\Phi_\alpha(x)$ of a point scalar charge is
singular on the world line, and this behaviour makes it difficult to
understand how the field is supposed to act on the particle and affect
its motion. The field's singularity structure was analyzed in
Secs.~\ref{16.3} and \ref{16.4}, and in Sec.~\ref{16.5} it was shown
to originate from the singular field $\Phi^{\rm S}_\alpha(x)$; the
radiative field $\Phi^{\rm R}_\alpha(x) = \Phi_\alpha(x) - \Phi^{\rm
S}_\alpha(x)$ was then shown to be smooth on the world line. 

To make sense of the retarded field's action on the particle we
temporarily model the scalar charge not as a point particle, but as a  
small hollow shell that appears spherical when observed in a reference 
frame that is momentarily comoving with the particle; the shell's
radius is $s_0$ in Fermi normal coordinates, and it is independent of
the angles contained in the unit vector $\omega^a$. The {\it net
force} acting at proper time $\tau$ on this hollow shell is the
average of $q\Phi_\alpha(\tau,s_0,\omega^a)$ over the surface of the 
shell. This was worked out at the end of Sec.~\ref{16.4}, and ignoring
terms that disappear in the limit $s_0 \to 0$, we obtain 
\begin{equation} 
q\bigl\langle \Phi_\mu \bigr\rangle = -(\delta m) a_\mu       
- \frac{1}{12} (1-6\xi) q^2 R u_{\mu} 
+ q^2 \bigl( g_{\mu\nu} + u_{\mu} u_{\nu} \bigr) 
\biggl( \frac{1}{3} \dot{a}^{\nu} 
+ \frac{1}{6} R^{\nu}_{\ \lambda} u^{\lambda}
\biggr) + q\Phi_{\mu}^{\rm tail},
\label{16.6.1}
\end{equation}
where 
\begin{equation} 
\delta m \equiv \lim_{s_0 \to 0} \frac{q^2}{3 s_0} 
\label{16.6.2}
\end{equation}
is formally a divergent quantity and 
\begin{equation} 
q \Phi_\mu^{\rm tail} = q^2 \int_{-\infty}^{\tau^-}  
\nabla_{\mu} G_+\bigl(z(\tau),z(\tau')\bigr)\, d\tau' 
\label{16.6.3}
\end{equation}
is the tail part of the force; all tensors in Eq.~(\ref{16.6.1}) are
evaluated at an arbitrary point $z(\tau)$ on the world line.  

Substituting Eqs.~(\ref{16.6.1}) and (\ref{16.6.3}) into
Eq.~(\ref{16.1.7}) gives rise to the equations of motion  
\begin{equation} 
\bigl( m + \delta m) a^\mu = q^2 \bigl( \delta^\mu_{\ \nu} 
+ u^\mu u_\nu \bigr) \Biggl[ \frac{1}{3} \dot{a}^{\nu} 
+ \frac{1}{6} R^{\nu}_{\ \lambda} u^{\lambda} 
+ \int_{-\infty}^{\tau^-}   
\nabla^{\nu} G_+\bigl(z(\tau),z(\tau')\bigr)\, d\tau' \Biggr] 
\label{16.6.4} 
\end{equation} 
for the scalar charge, with $m \equiv m_0 -q \Phi(z)$ denoting
the (also formally divergent) dynamical mass of the particle. We see
that $m$ and $\delta m$ combine in Eq.~(\ref{16.6.4}) to form the 
particle's observed mass $m_{\rm obs}$, which is taken to be finite
and to give a true measure of the particle's inertia. All diverging
quantities have thus disappeared into the process of mass
renormalization. Substituting Eqs.~(\ref{16.6.1}) and (\ref{16.6.3})
into Eq.~(\ref{16.1.8}), in which we replace $m$ by $m_{\rm obs} = m +
\delta m$, returns an expression for the rate of change of the
observed mass,  
\begin{equation}
\frac{d m_{\rm obs}}{d\tau} = - \frac{1}{12} (1-6\xi) q^2 R 
- q^2 u^\mu \int_{-\infty}^{\tau^-} \nabla_{\mu}
G_+\bigl(z(\tau),z(\tau')\bigr)\, d\tau'. 
\label{16.6.5}
\end{equation} 
That the observed mass is {\it not} conserved is a remarkable property
of the dynamics of a scalar charge in a curved spacetime. Physically, 
this corresponds to the fact that in a spacetime with a time-dependent  
metric, a scalar charge radiates monopole waves and the radiated
energy comes at the expense of the particle's inertial mass.  

Apart from the term proportional to $\delta m$, the averaged field of
Eq.~(\ref{16.6.1}) has exactly the same form as the radiative field of
Eq.~(\ref{16.5.12}), which we re-express as 
\begin{equation} 
q \Phi^{\rm R}_{\mu} = - \frac{1}{12} (1-6\xi) q^2 R u_{\mu}  
+ q^2 \bigl( g_{\mu\nu} + u_{\mu} u_{\nu} \bigr) 
\biggl( \frac{1}{3} \dot{a}^{\nu} 
+ \frac{1}{6} R^{\nu}_{\ \lambda} u^{\lambda}
\biggr) + q\Phi_{\mu}^{\rm tail}. 
\label{16.6.6}
\end{equation}
The force acting on the point particle can therefore be thought of as 
originating from the (smooth) radiative field, while the singular
field simply contributes to the particle's inertia. After mass
renormalization, Eqs.~(\ref{16.6.4}) and (\ref{16.6.5}) are equivalent 
to the statements 
\begin{equation} 
m a^\mu = q \bigl( g^{\mu\nu} + u^\mu u^\nu \bigr) 
\Phi^{\rm R}_{\nu}(z), \qquad 
\frac{d m}{d\tau} = -q u^\mu \Phi^{\rm R}_{\mu}(z), 
\label{16.6.7}
\end{equation} 
where we have dropped the superfluous label ``obs'' on the
particle's observed mass. Another argument in support of the claim
that the motion of the particle should be affected by the radiative
field only was presented in Sec.~\ref{13.5}. 

The equations of motion displayed in Eqs.~(\ref{16.6.4}) and
(\ref{16.6.5}) are third-order differential equations for the
functions $z^\mu(\tau)$. It is well known that such a system of
equations admits many unphysical solutions, such as runaway situations
in which the particle's acceleration increases exponentially with
$\tau$, even in the absence of any external force \cite{dirac,
jackson, poisson}. And indeed, our equations of motion do not yet
incorporate an external force which presumably is mostly responsible
for the particle's acceleration. Both defects can be cured in one
stroke. We shall take the point of view,  
the only admissible one in a classical treatment, that a point
particle is merely an idealization for an extended object whose
internal structure --- the details of its charge distribution --- can
be considered to be irrelevant. This view automatically implies that
our equations are meant to provide only an {\it approximate}
description of the object's motion. It can then be shown
\cite{poisson, flanaganwald} that within the context of this
approximation, it is consistent to replace, on the 
right-hand side of the equations of motion, any occurrence of the
acceleration vector by $f_{\rm ext}^\mu/m$, where $f_{\rm ext}^\mu$ is
the external force acting on the particle. Because $f_{\rm ext}^\mu$
is a prescribed quantity, differentiation of the external force does
not produce higher derivatives of the functions $z^\mu(\tau)$, and the 
equations of motion are properly of the second order.  

We shall therefore write, in the final analysis, the equations of
motion in the form 
\begin{equation} 
m \frac{D u^\mu}{d\tau} = f_{\rm ext}^\mu 
+ q^2 \bigl( \delta^\mu_{\ \nu} + u^\mu u_\nu \bigr) 
\Biggl[ \frac{1}{3m} \frac{D f_{\rm ext}^\nu}{d \tau}   
+ \frac{1}{6} R^{\nu}_{\ \lambda} u^{\lambda} 
+ \int_{-\infty}^{\tau^-}   
\nabla^{\nu} G_+\bigl(z(\tau),z(\tau')\bigr)\, d\tau' \Biggr] 
\label{16.6.8} 
\end{equation}   
and 
\begin{equation}
\frac{d m}{d\tau} = - \frac{1}{12} (1-6\xi) q^2 R 
- q^2 u^\mu \int_{-\infty}^{\tau^-} \nabla_{\mu}
G_+\bigl(z(\tau),z(\tau')\bigr)\, d\tau',  
\label{16.6.9}
\end{equation} 
where $m$ denotes the observed inertial mass of the scalar charge, and
where all tensors are evaluated at $z(\tau)$. We recall that the tail
integration must be cut short at $\tau' = \tau^- \equiv \tau - 0^+$ to
avoid the singular behaviour of the retarded Green's function at
coincidence; this procedure was justified at the beginning of
Sec.~\ref{16.3}. Equations (\ref{16.6.8}) and (\ref{16.6.9}) were
first derived by Theodore C.\ Quinn in 2000 \cite{quinn}. In his paper
Quinn also establishes that the total work done by the scalar
self-force matches the amount of energy radiated away by the particle.      

\section{Motion of an electric charge} 
\label{17} 

\subsection{Dynamics of a point electric charge}   
\label{17.1} 

A point particle carries an electric charge $e$ and moves on a world
line $\gamma$ described by relations $z^\mu(\lambda)$, in which
$\lambda$ is an arbitrary parameter. The particle generates a vector
potential $A^\alpha(x)$ and an electromagnetic field
$F_{\alpha\beta}(x) = \nabla_\alpha A_\beta - \nabla_\beta
A_\alpha$. The dynamics of the entire system is governed by the action  
\begin{equation}
S = S_{\rm field} + S_{\rm particle} + S_{\rm interaction}, 
\label{17.1.1}
\end{equation} 
where $S_{\rm field}$ is an action functional for a free
electromagnetic field in a spacetime with metric $g_{\alpha\beta}$,
$S_{\rm particle}$ is the action of a free particle moving on a world
line $\gamma$ in this spacetime, and $S_{\rm interaction}$ is an
interaction term that couples the field to the particle. 

The field action is given by  
\begin{equation}
S_{\rm field} = -\frac{1}{16\pi} \int F_{\alpha\beta} F^{\alpha\beta} 
\sqrt{-g}\, d^4 x,  
\label{17.1.2}
\end{equation} 
where the integration is over all of spacetime. The particle action is    
\begin{equation} 
S_{\rm particle} = -m \int_\gamma d\tau, 
\label{17.1.3}
\end{equation}
where $m$ is the bare mass of the particle and $d\tau =
\sqrt{-g_{\mu\nu}(z) \dot{z}^\mu \dot{z}^\nu}\, d\lambda$ is the
differential of proper time along the world line; we use an overdot to
indicate differentiation with respect to the parameter
$\lambda$. Finally, the interaction term is given by  
\begin{equation}  
S_{\rm interaction} = e \int_\gamma A_\mu(z) \dot{z}^\mu\, d\lambda =
e \int A_\alpha(x) g^\alpha_{\ \mu}(x,z) \dot{z}^\mu \delta_4(x,z)
\sqrt{-g}\, d^4x d\lambda.   
\label{17.1.4}
\end{equation}
Notice that both $S_{\rm particle}$ and $S_{\rm interaction}$ are
invariant under a reparameterization $\lambda \to \lambda'(\lambda)$
of the world line.  

Demanding that the total action be stationary under a variation
$\delta A^\alpha(x)$ of the vector potential yields Maxwell's
equations 
\begin{equation} 
F^{\alpha\beta}_{\ \ \ ;\beta} = 4\pi j^\alpha 
\label{17.1.5}
\end{equation} 
with a current density $j^\alpha(x)$ defined by  
\begin{equation}
j^\alpha(x) = e \int_\gamma g^{\alpha}_{\ \mu}(x,z) \dot{z}^\mu 
\delta_4(x,z)\, d\lambda.  
\label{17.1.6}
\end{equation} 
These equations determine the electromagnetic field $F_{\alpha\beta}$
once the motion of the electric charge is specified. On the other
hand, demanding that the total action be stationary under a variation  
$\delta z^\mu(\lambda)$ of the world line yields the equations of
motion  
\begin{equation} 
m \frac{D u^\mu}{d\tau} = e F^\mu_{\ \nu}(z) u^\nu  
\label{17.1.7}
\end{equation} 
for the electric charge. We have adopted $\tau$ as the parameter on
the world line, and introduced the four-velocity $u^\mu(\tau)
\equiv dz^\mu/d\tau$. 

The electromagnetic field $F_{\alpha\beta}$ is invariant under a gauge
transformation of the form $A_\alpha \to A_\alpha + \nabla_\alpha
\Lambda$, in which $\Lambda(x)$ is an arbitrary scalar function. This 
function can always be chosen so that the vector potential satisfies
the Lorenz gauge condition,  
\begin{equation}
\nabla_\alpha A^\alpha = 0. 
\label{17.1.8}
\end{equation} 
Under this condition the Maxwell equations of Eq.~(\ref{17.1.5})
reduce to a wave equation for the vector potential, 
\begin{equation} 
\Box A^\alpha - R^\alpha_{\ \beta} A^\beta = -4\pi j^\alpha, 
\label{17.1.9}
\end{equation}
where $\Box = g^{\alpha\beta} \nabla_\alpha \nabla_\beta$ is the wave 
operator and $R^\alpha_{\ \beta}$ is the Ricci tensor. Having adopted
$\tau$ as the parameter on the world line, we can re-express the
current density of Eq.~(\ref{17.1.6}) as 
\begin{equation}
j^\alpha(x) = e \int_\gamma g^{\alpha}_{\ \mu}(x,z) u^\mu 
\delta_4(x,z)\, d\tau, 
\label{17.1.10}
\end{equation} 
and we shall use Eqs.~(\ref{17.1.9}) and (\ref{17.1.10}) to determine 
the electromagnetic field of a point electric charge. The motion of
the particle is in principle determined by Eq.~(\ref{17.1.7}), but 
because the vector potential obtained from Eq.~(\ref{17.1.9}) is
singular on the world line, these equations have only formal validity.   
Before we can make sense of them we will have to analyze the field's
singularity structure near the world line. The calculations to be
carried out parallel closely those presented in Sec.~\ref{16} for the
case of a scalar charge; the details will therefore be kept to a 
minimum and the reader is referred to Sec.~\ref{16} for
additional information.     

\subsection{Retarded potential near the world line} 
\label{17.2}

The retarded solution to Eq.~(\ref{17.1.9}) is $A^\alpha(x) = \int 
G^{\ \alpha}_{+\beta'}(x,x') j^{\beta'}(x') \sqrt{g'}\, d^4x'$, where
$G^{\ \alpha}_{+\beta'}(x,x')$ is the retarded Green's function
introduced in Sec.~\ref{14}. After substitution of Eq.~(\ref{17.1.10})
we obtain 
\begin{equation}
A^\alpha(x) = e \int_\gamma G^{\ \alpha}_{+\mu}(x,z) u^\mu\, d\tau, 
\label{17.2.1}
\end{equation}
in which $z^\mu(\tau)$ gives the description of the world line
$\gamma$ and $u^\mu(\tau) = dz^\mu/d\tau$. Because the retarded
Green's function is defined globally in the entire spacetime,
Eq.~(\ref{17.2.1}) applies to any field point $x$. 

We now specialize Eq.~(\ref{17.2.1}) to a point $x$ close to the world 
line. We let ${\cal N}(x)$ be the normal convex neighbourhood
of this point, and we assume that the world line traverses 
${\cal N}(x)$; refer back to Fig.~9. As in Sec.~\ref{16.2} we let
$\tau_<$ and $\tau_>$ be the values of the proper-time parameter at
which $\gamma$ enters and leaves ${\cal N}(x)$, respectively. Then
Eq.~(\ref{17.2.1}) can be expressed as 
\[
A^\alpha(x) = e \int_{-\infty}^{\tau_<} G^{\ \alpha}_{+\mu}(x,z)
u^\mu\, d\tau + e \int_{\tau_<}^{\tau_>} G^{\ \alpha}_{+\mu}(x,z)
u^\mu\, d\tau + e \int_{\tau_>}^{\infty} G^{\ \alpha}_{+\mu}(x,z)
u^\mu\, d\tau.  
\]
The third integration vanishes because $x$ is then in the past of
$z(\tau)$, and $G^{\ \alpha}_{+\mu}(x,z) = 0$. For the second
integration, $x$ is the normal convex neighbourhood of $z(\tau)$, and
the retarded Green's function can be expressed in the Hadamard form
produced in Sec.~\ref{14.2}. This gives   
\[ 
\int_{\tau_<}^{\tau_>} G^{\ \alpha}_{+\mu}(x,z) u^\mu\, d\tau 
= \int_{\tau_<}^{\tau_>} U^\alpha_{\ \mu}(x,z) u^\mu
\delta_+(\sigma)\, d\tau 
+ \int_{\tau_<}^{\tau_>} V^\alpha_{\ \mu}(x,z) u^\mu 
\theta_+(-\sigma)\, d\tau, 
\] 
and to evaluate this we let $x' \equiv z(u)$ be the retarded point
associated with $x$; these points are related by $\sigma(x,x') = 0$
and $r \equiv \sigma_{\alpha'} u^{\alpha'}$ is the retarded distance
between $x$ and the world line. To perform the first integration we
change variables from $\tau$ to $\sigma$, noticing that $\sigma$
increases as $z(\tau)$ passes through $x'$; the integral evaluates to  
$U^\alpha_{\ \beta'} u^{\beta'}/r$. The second integration is cut off
at $\tau = u$ by the step function, and we obtain our final expression
for the vector potential of a point electric charge:  
\begin{equation}
A^\alpha(x) = \frac{e}{r} U^\alpha_{\ \beta'}(x,x') u^{\beta'} 
+ e \int_{\tau_<}^u V^\alpha_{\ \mu}(x,z) u^\mu\, d\tau 
+ e \int_{-\infty}^{\tau_<} G^{\ \alpha}_{+\mu}(x,z) u^\mu\, d\tau.    
\label{17.2.2}
\end{equation} 
This expression applies to a point $x$ sufficiently close to the world 
line that there exists a nonempty intersection between 
${\cal N}(x)$ and $\gamma$.  

\subsection{Electromagnetic field in retarded coordinates}  
\label{17.3} 

When we differentiate the vector potential of Eq.~(\ref{17.2.2}) we
must keep in mind that a variation in $x$ induces a variation in $x'$,
because the new points $x + \delta x$ and
$x' + \delta x'$ must also be linked by a null geodesic. Taking this
into account, we find that the gradient of the vector potential is
given by     
\begin{equation}
\nabla_\beta A_\alpha(x) = 
-\frac{e}{r^2} U_{\alpha \beta'} u^{\beta'} \partial_\beta r  
+ \frac{e}{r} U_{\alpha \beta';\beta} u^{\beta'} 
+ \frac{e}{r} \Bigl( U_{\alpha \beta';\gamma'} u^{\beta'}
   u^{\gamma'} 
+ U_{\alpha \beta'} a^{\beta'} \Bigr) \partial_\beta u   
+ e V_{\alpha \beta'} u^{\beta'} \partial_\beta u 
+ A^{\rm tail}_{\alpha\beta}(x),  
\label{17.3.1}
\end{equation}
where the ``tail integral'' is defined by    
\begin{eqnarray}
A^{\rm tail}_{\alpha\beta}(x) &=& e \int_{\tau_<}^u 
\nabla_\beta V_{\alpha\mu}(x,z) u^\mu\, d\tau 
+ e \int_{-\infty}^{\tau_<} 
\nabla_\beta G_{+\alpha\mu}(x,z) u^\mu\, d\tau \nonumber \\    
&=& e \int_{-\infty}^{u^-} 
\nabla_\beta G_{+\alpha\mu}(x,z) u^\mu\, d\tau.  
\label{17.3.2}
\end{eqnarray} 
The second form of the definition, in which we integrate the gradient
of the retarded Green's function from $\tau = -\infty$ to $\tau = u^-
\equiv u - 0^+$ to avoid the singular behaviour of the retarded
Green's function at $\sigma = 0$, is equivalent to the first form.      

We shall now expand $F_{\alpha\beta} = \nabla_\alpha A_\beta 
- \nabla_\beta A_\alpha$ in powers of $r$, and express the result in
terms of the retarded coordinates $(u,r,\Omega^a)$ introduced in
Sec.~\ref{9}. It will be convenient to decompose the electromagnetic
field in the tetrad $(\base{\alpha}{0},\base{\alpha}{a})$ that is
obtained by parallel transport of $(u^{\alpha'},\base{\alpha'}{a})$ on
the null geodesic that links $x$ to $x' \equiv z(u)$; this
construction is detailed in Sec.~\ref{9}. Note that throughout this
section we set $\omega_{ab} = 0$, where $\omega_{ab}$ is the
rotation tensor defined by Eq.~(\ref{9.1.1}): the tetrad vectors
$\base{\alpha'}{a}$ are taken to be Fermi-Walker transported on
$\gamma$. We recall from Eq.~(\ref{9.1.4}) that the parallel
propagator can be expressed as 
$g^{\alpha'}_{\ \alpha} = u^{\alpha'} \base{0}{\alpha} 
+ \base{\alpha'}{a} \base{a}{\alpha}$. The expansion relies on
Eq.~(\ref{9.5.3}) for $\partial_\alpha u$, Eq.~(\ref{9.5.5}) for
$\partial_\alpha r$, and we shall need    
\begin{equation}
U_{\alpha \beta'} u^{\beta'} = g^{\alpha'}_{\ \alpha} \biggl[
u_{\alpha'} + \frac{1}{12} r^2 \bigl( R_{00} + 2 R_{0a} \Omega^a 
+ R_{ab} \Omega^a \Omega^b \bigr)u_{\alpha'} + O(r^3) \biggr],  
\label{17.3.3}
\end{equation} 
which follows from Eq.~(\ref{14.2.7}) and the relation 
$\sigma^{\alpha'} = -r (u^{\alpha'} + \Omega^a \base{\alpha'}{a})$ 
first encountered in Eq.~(\ref{9.2.3}). We shall also need the
expansions     
\begin{equation} 
U_{\alpha \beta';\beta} u^{\beta'} = -\frac{1}{2} r  
g^{\alpha'}_{\ \alpha} g^{\beta'}_{\ \beta} \biggl[ 
R_{\alpha'0\beta'0} + R_{\alpha' 0\beta'c} \Omega^c 
- \frac{1}{3} \bigl( R_{\beta'0} + R_{\beta' c} \Omega^c 
\bigr) u_{\alpha'} + O(r) \biggr]    
\label{17.3.4}  
\end{equation}
and 
\begin{equation} 
U_{\alpha\beta';\gamma'} u^{\beta'}u^{\gamma'} +
U_{\alpha\beta'} a^{\beta'} = g^{\alpha'}_{\ \alpha} \biggl[ 
a_{\alpha'} + \frac{1}{2} r R_{\alpha'0b0}\Omega^b 
-\frac{1}{6} r \bigl( R_{00} + R_{0b} \Omega^b \bigr) u_{\alpha'} 
+ O(r^2) \biggr]   
\label{17.3.5}
\end{equation}
that follow from Eqs.~(\ref{14.2.7})--(\ref{14.2.9}). And finally, 
we shall need    
\begin{equation}
V_{\alpha\beta'} u^{\beta'} = 
-\frac{1}{2} g^{\alpha'}_{\ \alpha} \biggl[
R_{\alpha'0} - \frac{1}{6} R u_{\alpha'} + O(r) \biggr], 
\label{17.3.6}
\end{equation} 
a relation that was first established in Eq.~(\ref{14.2.11}).  

Collecting all these results gives    
\begin{eqnarray} 
F_{a0}(u,r,\Omega^a) &\equiv& F_{\alpha\beta}(x) 
\base{\alpha}{a}(x) \base{\beta}{0}(x) 
\nonumber \\  
&=& \frac{e}{r^2} \Omega_a 
- \frac{e}{r} \bigl( a_a - a_b \Omega^b \Omega_a \bigr) 
+ \frac{1}{3} e R_{b0c0} \Omega^b \Omega^c \Omega_a 
- \frac{1}{6} e \bigl( 5R_{a0b0} \Omega^b + R_{ab0c} \Omega^b \Omega^c
\bigr) 
\nonumber \\ & & \mbox{}
+ \frac{1}{12} e \bigl( 5 R_{00} + R_{bc} \Omega^b\Omega^c + R \bigr)
\Omega_a 
+ \frac{1}{3} e R_{a0} - \frac{1}{6} e R_{ab} \Omega^b 
+ F_{a0}^{\rm tail} + O(r), 
\label{17.3.7} \\ 
F_{ab}(u,r,\Omega^a) &\equiv& F_{\alpha\beta}(x) 
\base{\alpha}{a}(x) \base{\beta}{b}(x) 
\nonumber \\  
&=& \frac{e}{r} \bigl( a_a \Omega_b - \Omega_a a_b \bigr) 
+ \frac{1}{2} e \bigl( R_{a0bc} - R_{b0ac} + R_{a0c0} \Omega_b 
- \Omega_a R_{b0c0} \bigr) \Omega^c 
\nonumber \\ & & \mbox{}
- \frac{1}{2} e \bigl( R_{a0} \Omega_b - \Omega_a R_{b0} \bigr)
+ F_{ab}^{\rm tail} + O(r),  
\label{17.3.8}
\end{eqnarray} 
where 
\begin{equation} 
F_{a0}^{\rm tail} = F_{\alpha'\beta'}^{\rm tail}(x') \base{\alpha'}{a}
u^{\beta'}, \qquad 
F_{ab}^{\rm tail} = F_{\alpha'\beta'}^{\rm tail}(x') \base{\alpha'}{a}
\base{\beta'}{b} 
\label{17.3.9}
\end{equation}
are the frame components of the tail integral; this is
obtained from Eq.~(\ref{17.3.2}) evaluated at $x'$:   
\begin{equation}
F_{\alpha'\beta'}^{\rm tail}(x') = 2 e \int_{-\infty}^{u^-}
\nabla_{[\alpha'} G_{+\beta']\mu}(x',z) u^\mu\, d\tau.  
\label{17.3.10}
\end{equation} 
It should be emphasized that in Eqs.~(\ref{17.3.7}) and
(\ref{17.3.8}), all frame components are evaluated at the retarded
point $x' \equiv z(u)$ associated with $x$; for example, $a_a \equiv
a_a(u) \equiv a_{\alpha'} \base{\alpha'}{a}$. It is clear from these
equations that the electromagnetic field $F_{\alpha\beta}(x)$ is
singular on the world line. 

\subsection{Electromagnetic field in Fermi normal coordinates}   
\label{17.4} 

We now wish to express the electromagnetic field in the Fermi normal
coordinates of Sec.~\ref{8}; as before those will be denoted
$(t,s,\omega^a)$. The translation will be carried out as in
Sec.~\ref{16.4}, and we will decompose the field in the tetrad 
$(\bar{e}^\alpha_0, \bar{e}^\alpha_a)$ that is obtained by parallel
transport of $(u^{\bar{\alpha}}, \base{\bar{\alpha}}{a})$ on the
spacelike geodesic that links $x$ to the simultaneous point
$\bar{x} \equiv z(t)$.    

Our first task is to decompose $F_{\alpha\beta}(x)$ in the tetrad 
$(\bar{e}^\alpha_0, \bar{e}^\alpha_a)$, thereby defining $\bar{F}_{a0}  
\equiv F_{\alpha\beta} \bar{e}^\alpha_a \bar{e}^\beta_0$ and
$\bar{F}_{ab} \equiv F_{\alpha\beta} \bar{e}^\alpha_a
\bar{e}^\beta_b$. For this purpose we use Eqs.~(\ref{10.3.1}),
(\ref{10.3.2}), (\ref{17.3.7}), and (\ref{17.3.8}) to obtain  
\begin{eqnarray*} 
\bar{F}_{a0} &=& 
\frac{e}{r^2} \Omega_a 
- \frac{e}{r} \bigl( a_a - a_b \Omega^b \Omega_a \bigr) 
+ \frac{1}{2} e a_b \Omega^b a_a 
+ \frac{1}{2} e \dot{a}_0 \Omega_a 
- \frac{5}{6} e R_{a0b0} \Omega^b 
+ \frac{1}{3} e R_{b0c0} \Omega^b \Omega^c \Omega_a 
\\ & & \mbox{}
+ \frac{1}{3} e R_{ab0c} \Omega^b \Omega^c 
+ \frac{1}{12} e \bigl( 5 R_{00} + R_{bc} \Omega^b\Omega^c + R \bigr)
\Omega_a 
+ \frac{1}{3} e R_{a0} - \frac{1}{6} e R_{ab} \Omega^b 
+ \bar{F}_{a0}^{\rm tail} + O(r) 
\end{eqnarray*} 
and 
\[
\bar{F}_{ab} = 
\frac{1}{2} e \bigl( \Omega_a \dot{a}_b - \dot{a}_a \Omega_b \bigr) 
+ \frac{1}{2} e \bigl( R_{a0bc} - R_{b0ac} \bigr) \Omega^c 
- \frac{1}{2} e \bigl( R_{a0} \Omega_b - \Omega_a R_{b0} \bigr)
+ \bar{F}_{ab}^{\rm tail} + O(r), 
\]
where all frame components are still evaluated at $x'$, except for 
\[
\bar{F}_{a0}^{\rm tail} \equiv 
F^{\rm tail}_{\bar{\alpha}\bar{\beta}}(\bar{x})
\base{\bar{\alpha}}{a} u^{\bar{\beta}}, \qquad 
\bar{F}_{ab}^{\rm tail} \equiv 
F^{\rm tail}_{\bar{\alpha}\bar{\beta}}(\bar{x})
\base{\bar{\alpha}}{a} \base{\bar{\beta}}{b}, 
\] 
which are evaluated at $\bar{x}$.   

We must still translate these results into the Fermi normal
coordinates $(t,s,\omega^a)$. For this we involve Eqs.~(\ref{10.2.1}),
(\ref{10.2.2}), and (\ref{10.2.3}), and we recycle some computations
that were first carried out in Sec.~\ref{16.4}. After some algebra, we
arrive at       
\begin{eqnarray} 
\bar{F}_{a0}(t,s,\omega^a) &\equiv& F_{\alpha\beta}(x) 
\bar{e}^\alpha_a(x) \bar{e}^\beta_0(x) 
\nonumber \\ 
&=& \frac{e}{s^2} \omega_a 
- \frac{e}{2s} \bigl( a_a + a_b \omega^b \omega_a \bigr) 
+ \frac{3}{4} e a_b \omega^b a_a 
+ \frac{3}{8} e \bigl( a_b \omega^b \bigr)^2 \omega_a 
+ \frac{3}{8} e \dot{a}_0 \omega_a
+ \frac{2}{3} e \dot{a}_a  
\nonumber \\ & & \mbox{}
- \frac{2}{3} e R_{a0b0} \omega^b 
- \frac{1}{6} e R_{b0c0} \omega^b \omega^c \omega_a 
+ \frac{1}{12} e \bigl( 5 R_{00} + R_{bc} \omega^b\omega^c + R \bigr) 
\omega_a 
\nonumber \\ & & \mbox{}
+ \frac{1}{3} e R_{a0} - \frac{1}{6} e R_{ab} \omega^b 
+ \bar{F}_{a0}^{\rm tail} + O(s), 
\label{17.4.1} \\ 
\bar{F}_{ab}(t,s,\omega^a) &\equiv& F_{\alpha\beta}(x) 
\bar{e}^\alpha_a(x) \bar{e}^\beta_b(x) 
\nonumber \\
&=& \frac{1}{2} e \bigl( \omega_a \dot{a}_b - \dot{a}_a \omega_b
\bigr)   
+ \frac{1}{2} e \bigl( R_{a0bc} - R_{b0ac} \bigr) \omega^c 
- \frac{1}{2} e \bigl( R_{a0} \omega_b - \omega_a R_{b0} \bigr)
\nonumber \\ & & \mbox{}
+ \bar{F}_{ab}^{\rm tail} + O(s),
\label{17.4.2}
\end{eqnarray} 
where all frame components are now evaluated at $\bar{x} \equiv z(t)$;
for example, $a_a \equiv a_a(t) \equiv
a_{\bar{\alpha}} \base{\bar{\alpha}}{a}$. 

Our next task is to compute the averages of $\bar{F}_{a0}$ and
$\bar{F}_{ab}$ over $S(t,s)$, a two-surface of constant $t$ and $s$.  
These are defined by 
\begin{equation} 
\bigl\langle \bar{F}_{a0} \bigr\rangle(t,s) = \frac{1}{\cal A}  
\oint_{S(t,s)} \bar{F}_{a0}(t,s,\omega^a)\, d{\cal A}, \qquad   
\bigl\langle \bar{F}_{ab} \bigr\rangle(t,s) = \frac{1}{\cal A} 
\oint_{S(t,s)} \bar{F}_{ab}(t,s,\omega^a)\, d{\cal A},  
\label{17.4.3}
\end{equation} 
where $d{\cal A}$ is the element of surface area on $S(t,s)$, and 
${\cal A} = \oint d {\cal A}$. Using the methods developed in
Sec.~\ref{16.4}, we find 
\begin{eqnarray} 
\bigl\langle \bar{F}_{a0} \bigr\rangle &=& -\frac{2e}{3s} a_a 
+ \frac{2}{3} e \dot{a}_a + \frac{1}{3} e R_{a0} 
+ \bar{F}_{a0}^{\rm tail} + O(s), 
\label{17.4.4} \\ 
\bigl\langle \bar{F}_{ab} \bigr\rangle &=& 
\bar{F}_{ab}^{\rm tail} + O(s). 
\label{17.4.5}
\end{eqnarray} 
The averaged field is singular on the world line, but we nevertheless
take the formal limit $s \to 0$ of the expressions displayed in 
Eqs.~(\ref{17.4.4}) and (\ref{17.4.5}). In the limit the tetrad 
$(\bar{e}^\alpha_0, \bar{e}^\alpha_a)$ becomes 
$(u^{\bar{\alpha}}, \base{\bar{\alpha}}{a})$, and we can easily 
reconstruct the field at $\bar{x}$ from its frame components. 
We thus obtain  
\begin{equation} 
\bigl\langle F_{\bar{\alpha}\bar{\beta}} \bigr\rangle = 
\lim_{s\to 0} \biggl( - \frac{4e}{3s} \biggr) 
u_{[\bar{\alpha}} a_{\bar{\beta}]}   
+ 2e u_{[\bar{\alpha}} \bigl( g_{\bar{\beta}]\bar{\gamma}}  
+ u_{\bar{\beta}]} u_{\bar{\gamma}} \bigr) 
\biggl( \frac{2}{3} \dot{a}^{\bar{\gamma}} 
+ \frac{1}{3} R^{\bar{\gamma}}_{\ \bar{\delta}} u^{\bar{\delta}}
\biggr) + F_{\bar{\alpha}\bar{\beta}}^{\rm tail}, 
\label{17.4.6}
\end{equation}
where the tail term can be copied from Eq.~(\ref{17.3.10}), 
\begin{equation} 
F_{\bar{\alpha}\bar{\beta}}^{\rm tail}(\bar{x}) = 
2 e \int_{-\infty}^{t^-}
\nabla_{[\bar{\alpha}} G_{+\bar{\beta}]\mu}(\bar{x},z) u^\mu\, d\tau. 
\label{17.4.7}
\end{equation} 
The tensors appearing in Eq.~(\ref{17.4.6}) all refer to $\bar{x}
\equiv z(t)$, which now stands for an    
arbitrary point on the world line $\gamma$.  

\subsection{Singular and radiative fields} 
\label{17.5} 

The singular vector potential 
\begin{equation} 
A^\alpha_{\rm S}(x) = e \int_\gamma 
G^{\ \alpha}_{{\rm S}\,\mu}(x,z) u^\mu\, d\tau 
\label{17.5.1}
\end{equation} 
is the (unphysical) solution to Eqs.~(\ref{17.1.9}) and
(\ref{17.1.10}) that is obtained by adopting the singular Green's
function of Eq.~(\ref{14.4.5}) instead of the retarded Green's
function. We will see that the singular field
$F^{\rm S}_{\alpha\beta}$ reproduces the singular behaviour of the
retarded solution, and that the difference, $F^{\rm R}_{\alpha\beta} 
= F_{\alpha\beta} - F^{\rm S}_{\alpha\beta}$, is smooth on the world
line.      

To evaluate the integral of Eq.~(\ref{17.5.1}) we assume once more 
that $x$ is sufficiently close to $\gamma$ that the world line
traverses ${\cal N}(x)$; refer back to Fig.~9. As before we let
$\tau_<$ and $\tau_>$ be the values of the proper-time parameter at
which $\gamma$ enters and leaves ${\cal N}(x)$, respectively. Then
Eq.~(\ref{17.5.1}) becomes   
\[
A^\alpha_{\rm S}(x) = e \int_{-\infty}^{\tau_<} 
G^{\ \alpha}_{{\rm S}\,\mu}(x,z) u^\mu \, d\tau  
+ e \int_{\tau_<}^{\tau_>} G^{\ \alpha}_{{\rm S}\,\mu}(x,z) u^\mu\,
d\tau + e \int_{\tau_>}^{\infty} G^{\ \alpha}_{{\rm S}\,\mu}(x,z)
u^\mu\, d\tau.   
\]
The first integration vanishes because $x$ is then in the
chronological future of $z(\tau)$, and 
$G^{\ \alpha}_{{\rm S}\,\mu}(x,z) = 0$ by
Eq.~(\ref{14.4.8}). Similarly, the third integration vanishes because 
$x$ is then in the chronological past of $z(\tau)$. For the second
integration, $x$ is the normal convex neighbourhood of $z(\tau)$, the
singular Green's function can be expressed in the Hadamard form of
Eq.~(\ref{14.4.14}), and we have      
\begin{eqnarray*} 
\int_{\tau_<}^{\tau_>} G^{\ \alpha}_{{\rm S}\,\mu}(x,z) u^\mu\, 
d\tau &=& \frac{1}{2} \int_{\tau_<}^{\tau_>} U^\alpha_{\ \mu}(x,z)
u^\mu \delta_+(\sigma)\, d\tau + \frac{1}{2} \int_{\tau_<}^{\tau_>} 
U^\alpha_{\ \mu}(x,z) u^\mu \delta_-(\sigma)\, d\tau \\ 
& & \mbox{} 
- \frac{1}{2} \int_{\tau_<}^{\tau_>} V^\alpha_{\ \mu}(x,z) u^\mu 
\theta(\sigma)\, d\tau.   
\end{eqnarray*} 
To evaluate these we let $x' \equiv z(u)$ and $x'' \equiv z(v)$ be the
retarded and advanced points associated with $x$, respectively. To
perform the first integration we change variables from $\tau$ to
$\sigma$, noticing that $\sigma$ increases as $z(\tau)$ passes
through $x'$; the integral evaluates to $U^\alpha_{\ \beta'}
u^{\beta'}/r$. We do the same for the second integration, but we
notice now that $\sigma$ decreases as $z(\tau)$ passes through $x''$;
the integral evaluates to 
$U^\alpha_{\ \beta''} u^{\beta''}/r_{\rm adv}$, where $r_{\rm adv}
\equiv - \sigma_{\alpha''} u^{\alpha''}$ is the advanced distance
between $x$ and the world line. The third integration is restricted to
the interval $u \leq \tau \leq v$ by the step function, and we obtain
the expression
\begin{equation}
A^\alpha_{\rm S}(x) = \frac{e}{2r} U^\alpha_{\ \beta'} u^{\beta'}  
+ \frac{e}{2r_{\rm adv}} U^\alpha_{\ \beta''} u^{\beta''} 
- \frac{1}{2} e \int_u^v V^\alpha_{\ \mu}(x,z) u^\mu\, d\tau   
\label{17.5.2} 
\end{equation} 
for the singular vector potential. 

Differentiation of Eq.~(\ref{17.5.2}) yields 
\begin{eqnarray}
\nabla_\beta A_\alpha^{\rm S}(x) &=& 
-\frac{e}{2r^2} U_{\alpha\beta'} u^{\beta'} \partial_\beta r   
- \frac{e}{2 {r_{\rm adv}}^2} U_{\alpha\beta''} u^{\beta''}
  \partial_\beta r_{\rm adv}  
+ \frac{e}{2r} U_{\alpha\beta';\beta} u^{\beta'}  
\nonumber \\  & & \mbox{} 
+ \frac{e}{2r} \Bigl( U_{\alpha \beta';\gamma'} u^{\beta'}
   u^{\gamma'} 
+ U_{\alpha\beta'} a^{\beta'} \Bigr) \partial_\beta u   
+ \frac{e}{2r_{\rm adv}} U_{\alpha\beta'';\beta} u^{\beta''}   
\nonumber \\  & & \mbox{} 
+ \frac{e}{2r_{\rm adv}} \Bigl( U_{\alpha \beta'';\gamma''}
   u^{\beta''} u^{\gamma''} 
+ U_{\alpha\beta''} a^{\beta''} \Bigr) \partial_\beta v   
+ \frac{1}{2} e V_{\alpha \beta'} u^{\beta'} \partial_\beta u   
\nonumber \\  & & \mbox{} 
- \frac{1}{2} e V_{\alpha \beta''} u^{\beta''} \partial_\beta v    
- \frac{1}{2} e \int_u^v \nabla_\beta V_{\alpha\mu}(x,z) u^\mu\,
d\tau, 
\label{17.5.3}
\end{eqnarray}
and we would like to express this as an expansion in powers of
$r$. For this we will rely on results already established in 
Sec.~\ref{17.3}, as well as additional expansions that will involve
the advanced point $x''$. We recall that a relation between retarded
and advanced times was worked out in Eq.~(\ref{10.4.2}), that an
expression for the advanced distance was displayed in
Eq.~(\ref{10.4.3}), and that Eqs.~(\ref{10.4.4}) and (\ref{10.4.5})
give expansions for $\partial_\alpha v$ and 
$\partial_\alpha r_{\rm adv}$, respectively.  

To derive an expansion for $U_{\alpha\beta''} u^{\beta''}$ we follow
the general method of Sec.~\ref{10.4} and introduce the functions
$U_\alpha(\tau) \equiv U_{\alpha\mu}(x,z) u^\mu$. We have 
that  
\[
U_{\alpha\beta''} u^{\beta''} \equiv U_\alpha(v) = U_\alpha(u) 
+ \dot{U}_\alpha(u) \Delta^{\!\prime} + \frac{1}{2} \ddot{U}_\alpha(u)
\Delta^{\!\prime 2} + O\bigl( \Delta^{\!\prime 3} \bigr),  
\]
where overdots indicate differentiation with respect to $\tau$, and
$\Delta^{\!\prime} \equiv v-u$. The leading term $U_\alpha(u)
\equiv U_{\alpha\beta'} u^{\beta'}$ was worked out in
Eq.~(\ref{17.3.3}), and the derivatives of $U_\alpha(\tau)$ are given
by  
\[
\dot{U}_\alpha(u) = U_{\alpha\beta';\gamma'} u^{\beta'} u^{\gamma'} +
U_{\alpha\beta'} a^{\beta'} = g^{\alpha'}_{\ \alpha} \biggl[ 
a_{\alpha'} + \frac{1}{2} r R_{\alpha'0b0}\Omega^b 
-\frac{1}{6} r \bigl( R_{00} + R_{0b} \Omega^b \bigr) u_{\alpha'} 
+ O(r^2) \biggr]   
\] 
and 
\[
\ddot{U}_\alpha(u) = U_{\alpha\beta';\gamma'\delta'} u^{\beta'}
u^{\gamma'} u^{\delta'} + U_{\alpha\beta';\gamma'} \bigl( 2 a^{\beta'}
u^{\gamma'} + u^{\beta'} a^{\gamma'} \bigr) + U_{\alpha\beta'}
\dot{a}^{\beta'} = g^{\alpha'}_{\ \alpha} \biggl[ \dot{a}_{\alpha'} 
+ \frac{1}{6} R_{00} u_{\alpha'} + O(r) \biggr], 
\]
according to Eqs.~(\ref{17.3.5}) and (\ref{14.2.9}). Combining these
results together with Eq.~(\ref{10.4.2}) for $\Delta^{\!\prime}$ gives    
\begin{eqnarray} 
U_{\alpha\beta''} u^{\beta''} &=& g^{\alpha'}_{\ \alpha} \biggl[ 
u_{\alpha'} + 2r \bigl(1 - r a_b \Omega^b \bigr) a_{\alpha'} 
+ 2r^2 \dot{a}_{\alpha'} + r^2 R_{\alpha'0b0}\Omega^b \qquad \qquad
\nonumber \\ & & \qquad \mbox{} 
+ \frac{1}{12} r^2 \bigl( R_{00} - 2 R_{0a} \Omega^a 
+ R_{ab} \Omega^a \Omega^b \bigr)u_{\alpha'} 
+ O(r^3) \biggr],  
\label{17.5.4}
\end{eqnarray} 
which should be compared with Eq.~(\ref{17.3.3}). It should be
emphasized that in Eq.~(\ref{17.5.4}) and all equations below, all
frame components are evaluated at the retarded point $x'$, and not at
the advanced point. The preceding computation gives us also
an expansion for  
\[
U_{\alpha\beta'';\gamma''} u^{\beta''} u^{\gamma''} +
U_{\alpha\beta''} a^{\beta''} \equiv \dot{U}_\alpha(v) =
\dot{U}_\alpha(u) + \ddot{U}_\alpha(u) \Delta^{\!\prime} +
O(\Delta^{\!\prime 2}), 
\]
which becomes     
\begin{equation}
U_{\alpha\beta'';\gamma''} u^{\beta''} u^{\gamma''} +
U_{\alpha\beta''} a^{\beta''} = g^{\alpha'}_{\ \alpha} \biggl[  
a_{\alpha'} + 2 r \dot{a}_{\alpha'} 
+ \frac{1}{2} r R_{\alpha'0b0}\Omega^b  
+ \frac{1}{6} r \bigl( R_{00} - R_{0b} \Omega^b \bigr) u_{\alpha'} 
+ O(r^2) \biggr],   
\label{17.5.5}
\end{equation}
and which should be compared with Eq.~(\ref{17.3.5}).   

We proceed similarly to derive an expansion for
$U_{\alpha\beta'';\beta} u^{\beta''}$. Here we introduce the functions 
$U_{\alpha\beta}(\tau) \equiv U_{\alpha\mu;\beta}u^\mu$ and express
$U_{\alpha\beta'';\beta} u^{\beta''}$ as $U_{\alpha\beta}(v) =
U_{\alpha\beta}(u) + \dot{U}_{\alpha\beta}(u) \Delta^{\!\prime} 
+ O(\Delta^{\!\prime 2})$. The leading term $U_{\alpha\beta}(u) \equiv
U_{\alpha\beta';\beta}u^{\beta'}$ was computed in Eq.~(\ref{17.3.4}),
and   
\[
\dot{U}_{\alpha\beta}(u) = U_{\alpha\beta';\beta\gamma'} u^{\beta'}
u^{\gamma'} + U_{\alpha\beta';\beta} a^{\beta'} = \frac{1}{2}
g^{\alpha'}_{\ \alpha} g^{\beta'}_{\ \beta} \biggl[
R_{\alpha'0\beta'0} - \frac{1}{3} u_{\alpha'} R_{\beta'0} + O(r)
\biggr] 
\]
follows from Eq.~(\ref{14.2.8}). Combining these results together with 
Eq.~(\ref{10.4.2}) for $\Delta^{\!\prime}$ gives
\begin{equation}
U_{\alpha \beta'';\beta} u^{\beta''} = \frac{1}{2} r  
g^{\alpha'}_{\ \alpha} g^{\beta'}_{\ \beta} \biggl[ 
R_{\alpha'0\beta'0} - R_{\alpha'0\beta'c} \Omega^c 
- \frac{1}{3} \bigl( R_{\beta'0} - R_{\beta' c} \Omega^c 
\bigr) u_{\alpha'} + O(r) \biggr],     
\label{17.5.6}
\end{equation}
and this should be compared with Eq.~(\ref{17.3.4}). The last
expansion we shall need is  
\begin{equation}
V_{\alpha\beta''} u^{\beta''} = 
-\frac{1}{2} g^{\alpha'}_{\ \alpha} \biggl[
R_{\alpha'0} - \frac{1}{6} R u_{\alpha'} + O(r) \biggr],
\label{17.5.7} 
\end{equation} 
which follows at once from Eq.~(\ref{17.3.6}).

It is now a straightforward (but still tedious) matter to substitute
these expansions into Eq.~(\ref{17.5.3}) to obtain the projections of
the singular electromagnetic field $F^{\rm S}_{\alpha\beta} =
\nabla_\alpha A^{\rm S}_\beta - \nabla_\beta A^{\rm S}_\alpha$ in the  
same tetrad $(\base{\alpha}{0}, \base{\alpha}{a})$ that was employed
in Sec.~\ref{17.3}. This gives 
\begin{eqnarray} 
F^{\rm S}_{a0}(u,r,\Omega^a) &\equiv& F^{\rm S}_{\alpha\beta}(x)  
\base{\alpha}{a}(x) \base{\beta}{0}(x) 
\nonumber \\  
&=& \frac{e}{r^2} \Omega_a 
- \frac{e}{r} \bigl( a_a - a_b \Omega^b \Omega_a \bigr) 
- \frac{2}{3} e \dot{a}_a 
+ \frac{1}{3} e R_{b0c0} \Omega^b \Omega^c \Omega_a 
- \frac{1}{6} e \bigl( 5R_{a0b0} \Omega^b + R_{ab0c} \Omega^b \Omega^c
\bigr) 
\nonumber \\ & & \mbox{}
+ \frac{1}{12} e \bigl( 5 R_{00} + R_{bc} \Omega^b\Omega^c + R \bigr)
\Omega_a 
- \frac{1}{6} e R_{ab} \Omega^b 
+ O(r), 
\label{17.5.8} \\ 
F^{\rm S}_{ab}(u,r,\Omega^a) &\equiv& F^{\rm S}_{\alpha\beta}(x)  
\base{\alpha}{a}(x) \base{\beta}{b}(x) 
\nonumber \\  
&=& \frac{e}{r} \bigl( a_a \Omega_b - \Omega_a a_b \bigr) 
+ \frac{1}{2} e \bigl( R_{a0bc} - R_{b0ac} + R_{a0c0} \Omega_b 
- \Omega_a R_{b0c0} \bigr) \Omega^c 
\nonumber \\ & & \mbox{}
- \frac{1}{2} e \bigl( R_{a0} \Omega_b - \Omega_a R_{b0} \bigr)
+ O(r),  
\label{17.5.9}
\end{eqnarray} 
in which all frame components are evaluated at the retarded point 
$x'$. Comparison of these expressions with Eqs.~(\ref{17.3.7}) and
(\ref{17.3.8}) reveals that the retarded and singular fields share the
same singularity structure.    

The difference between the retarded field of Eqs.~(\ref{17.3.7}),
(\ref{17.3.8}) and the singular field of Eqs.~(\ref{17.5.8}),
(\ref{17.5.9}) defines the radiative field 
$F^{\rm R}_{\alpha\beta}(x)$. Its tetrad components are 
\begin{eqnarray} 
F^{\rm R}_{a0} &=& \frac{2}{3} e \dot{a}_a + \frac{1}{3} e R_{a0} 
+ F_{a0}^{\rm tail} + O(r),  
\label{17.5.10} \\ 
F^{\rm R}_{ab} &=& F_{ab}^{\rm tail} + O(r), 
\label{17.5.11}
\end{eqnarray} 
and we see that $F^{\rm R}_{\alpha\beta}$ is a smooth tensor field
on the world line. There is therefore no obstacle in evaluating the
radiative field directly at $x=x'$, where the tetrad
$(\base{\alpha}{0},\base{\alpha}{a})$ becomes $(u^{\alpha'}, 
\base{\alpha'}{a})$. Reconstructing the field at $x'$ from its 
frame components, we obtain 
\begin{equation} 
F^{\rm R}_{\alpha'\beta'}(x') = 
2e u_{[\alpha'} \bigl( g_{\beta']\gamma'}  
+ u_{\beta']} u_{\gamma'} \bigr) 
\biggl( \frac{2}{3} \dot{a}^{\gamma'} 
+ \frac{1}{3} R^{\gamma'}_{\ \delta'} u^{\delta'}
\biggr) + F_{\alpha'\beta'}^{\rm tail}, 
\label{17.5.12}
\end{equation}
where the tail term can be copied from Eq.~(\ref{17.3.10}), 
\begin{equation} 
F_{\alpha'\beta'}^{\rm tail}(x') = 2 e \int_{-\infty}^{u^-}
\nabla_{[\alpha'} G_{+\beta']\mu}(x',z) u^\mu\, d\tau.  
\label{17.5.13} 
\end{equation} 
The tensors appearing in Eq.~(\ref{17.5.12}) all refer to the  
retarded point $x' \equiv z(u)$, which now stands for an  
arbitrary point on the world line $\gamma$.  
  
\subsection{Equations of motion} 
\label{17.6}

The retarded field $F_{\alpha\beta}$ of a point electric charge is 
singular on the world line, and this behaviour makes it difficult to
understand how the field is supposed to act on the particle and exert
a force. The field's singularity structure was analyzed in
Secs.~\ref{17.3} and \ref{17.4}, and in Sec.~\ref{17.5} it was shown 
to originate from the singular field $F^{\rm S}_{\alpha\beta}$; the 
radiative field $F^{\rm R}_{\alpha\beta} = F_{\alpha\beta} - 
F^{\rm S}_{\alpha\beta}$ was then shown to be smooth on the world
line.  

To make sense of the retarded field's action on the particle we follow
the discussion of Sec.~\ref{16.6} and temporarily picture the electric
charge as a spherical hollow shell; the shell's radius is $s_0$ in
Fermi normal coordinates, and it is independent of the angles
contained in the unit vector $\omega^a$. The {\it net force} acting at
proper time $\tau$ on this shell is proportional to the average of 
$F_{\alpha\beta}(\tau,s_0,\omega^a)$ over the shell's surface. This
was worked out at the end of Sec.~\ref{17.4}, and ignoring terms that
disappear in the limit $s_0 \to 0$, we obtain  
\begin{equation} 
e\bigl\langle F_{\mu\nu} \bigr\rangle u^\nu = -(\delta m) a_\mu        
+ e^2 \bigl( g_{\mu\nu} + u_{\mu} u_{\nu} \bigr) 
\biggl( \frac{2}{3} \dot{a}^{\nu} 
+ \frac{1}{3} R^{\nu}_{\ \lambda} u^{\lambda}
\biggr) + eF_{\mu\nu}^{\rm tail} u^\nu, 
\label{17.6.1}
\end{equation}
where 
\begin{equation} 
\delta m \equiv \lim_{s_0 \to 0} \frac{2 e^2}{3 s_0} 
\label{17.6.2}
\end{equation}
is formally a divergent quantity and 
\begin{equation} 
eF_{\mu\nu}^{\rm tail} u^\nu = 2 e^2 u^\nu \int_{-\infty}^{\tau^-}     
\nabla_{[\mu} G_{+\nu]\lambda'}\bigl(z(\tau),z(\tau') \bigr)
u^{\lambda'}\, d\tau' 
\label{17.6.3}
\end{equation}
is the tail part of the force; all tensors in Eq.~(\ref{17.6.1}) are 
evaluated at an arbitrary point $z(\tau)$ on the world line.  

Substituting Eqs.~(\ref{17.6.1}) and (\ref{17.6.3}) into
Eq.~(\ref{17.1.7}) gives rise to the equations of motion  
\begin{equation} 
\bigl( m + \delta m) a^\mu = e^2 \bigl( \delta^\mu_{\ \nu} 
+ u^\mu u_\nu \bigr) \biggl( \frac{2}{3} \dot{a}^{\nu} 
+ \frac{1}{3} R^{\nu}_{\ \lambda} u^{\lambda} \biggr)  
+ 2 e^2 u_\nu \int_{-\infty}^{\tau^-}     
\nabla^{[\mu} G^{\ \nu]}_{+\,\lambda'}\bigl(z(\tau),z(\tau')\bigr)   
u^{\lambda'}\, d\tau' 
\label{17.6.4} 
\end{equation} 
for the electric charge, with $m$ denoting the (also formally
divergent) bare mass of the particle. We see that $m$ and $\delta m$ 
combine in Eq.~(\ref{17.6.4}) to form the particle's observed mass
$m_{\rm obs}$, which is finite and gives a true measure of the
particle's inertia. All diverging quantities have thus disappeared
into the procedure of mass renormalization.  

Apart from the term proportional to $\delta m$, the averaged force of 
Eq.~(\ref{17.6.1}) has exactly the same form as the force that arises
from the radiative field of Eq.~(\ref{17.5.12}), which we express as 
\begin{equation} 
e F^{\rm R}_{\mu\nu} u^\nu = e^2 \bigl( g_{\mu\nu} 
+ u_{\mu} u_{\nu} \bigr) \biggl( \frac{2}{3} \dot{a}^{\nu} 
+ \frac{1}{3} R^{\nu}_{\ \lambda} u^{\lambda} \biggr) 
+ eF_{\mu\nu}^{\rm tail} u^\nu. 
\label{17.6.5}
\end{equation}
The force acting on the point particle can therefore be thought of as 
originating from the (smooth) radiative field, while the singular
field simply contributes to the particle's inertia. After mass
renormalization, Eq.~(\ref{17.6.4}) is equivalent to the statement  
\begin{equation} 
m a_\mu = e F^{\rm R}_{\mu\nu}(z) u^\nu,
\label{17.6.6}
\end{equation} 
where we have dropped the superfluous label ``obs'' on the 
particle's observed mass. 

For the final expression of the equations of motion we follow the 
discussion of Sec.~\ref{16.6} and allow an external force 
$f^\mu_{\rm ext}$ to act on the particle, and we replace, on the 
right-hand side of the equations, the acceleration vector by
$f^\mu_{\rm ext}/m$. This produces  
\begin{equation} 
m \frac{D u^\mu}{d\tau} = f_{\rm ext}^\mu 
+ e^2 \bigl( \delta^\mu_{\ \nu} + u^\mu u_\nu \bigr) 
\biggl( \frac{2}{3m} \frac{D f_{\rm ext}^\nu}{d \tau}   
+ \frac{1}{3} R^{\nu}_{\ \lambda} u^{\lambda} \biggr)  
+ 2 e^2 u_\nu \int_{-\infty}^{\tau^-}     
\nabla^{[\mu} G^{\ \nu]}_{+\,\lambda'}\bigl(z(\tau),z(\tau')\bigr)   
u^{\lambda'}\, d\tau',  
\label{17.6.7} 
\end{equation}   
in which $m$ denotes the observed inertial mass of the electric charge 
and all tensors are evaluated at $z(\tau)$, the current position of
the particle on the world line; the primed indices in the tail
integral refer to the point $z(\tau')$, which represents a prior
position. We recall that the integration must be cut short at $\tau' =
\tau^- \equiv \tau - 0^+$ to avoid the singular behaviour of the
retarded Green's function at coincidence; this procedure was justified
at the beginning of Sec.~\ref{17.3}. Equation (\ref{17.6.7}) was first
derived (without the Ricci-tensor term) by Bryce S.\ DeWitt and Robert
W.\ Brehme in 1960 \cite{dewittbrehme}, and then corrected by J.M.\
Hobbs in 1968 \cite{hobbs}. An alternative derivation was produced by
Theodore C.\ Quinn and Robert M.\ Wald in 1997 \cite{QW1}. In a
subsequent publication \cite{QW2}, Quinn and Wald proved that the
total work done by the electromagnetic self-force matches the energy
radiated away by the particle.

%
\section{Motion of a point mass} 
\label{18} 

\subsection{Dynamics of a point mass} 
\label{18.1}

In this section we consider the motion of a point particle of mass $m$
subjected to its own gravitational field. The particle moves on a
world line $\gamma$ in a curved spacetime whose background metric 
$g_{\alpha\beta}$ is assumed to be a {\it vacuum} solution to the
Einstein field equations. We shall suppose that $m$ is small, so that
the perturbation $h_{\alpha\beta}$ created by the particle can also
be considered to be small; it will obey a linear wave equation in the  
background spacetime. This linearization of the field equations will
allow us to fit the problem of determining the motion of a point mass
within the framework developed in Secs.~\ref{16} and \ref{17}, and we 
shall obtain the equations of motion by following the same general
line of reasoning. We shall find that $\gamma$ is not a geodesic of
the background spacetime because $h_{\alpha\beta}$ acts on the
particle and induces an acceleration of order $m$; the
motion is geodesic in the test-mass limit only.  

Our discussion in this first subsection is largely formal: as in 
Secs.~\ref{16.1} and \ref{17.1} we insert the point particle in the
background spacetime and ignore the fact that the field it produces is
singular on the world line. To make sense of the formal equations of
motion will be our goal in the following subsections. The problem of
determining the motion of a small mass in a background spacetime will
be reconsidered in Sec.~\ref{19} from a different and more satisfying
premise: there the small body will be modeled as a black hole instead
of as a point particle, and the singular behaviour of the perturbation 
will automatically be eliminated.       

Let a point particle of mass $m$ move on a world line $\gamma$ in a
curved spacetime with metric ${\sf g}_{\alpha\beta}$. This is the
{\it total} metric of the {\it perturbed} spacetime, and it depends on
$m$ as well as all other relevant parameters. At a later stage of the
discussion the total metric will be broken down into a ``background''
part $g_{\alpha\beta}$ that is independent of $m$, and a
``perturbation'' part $h_{\alpha\beta}$ that is proportional to
$m$. The world line is described by relations $z^\mu(\lambda)$ in
which $\lambda$ is an arbitrary parameter --- this will later be
identified with proper time $\tau$ in the {\it background}
spacetime. In this and the following sections we will use 
{\sf sans-serif symbols} to denote tensors that refer to the perturbed
spacetime; tensors in the background spacetime will be denoted, as
usual, by italic symbols.        

The particle's action functional is 
\begin{equation} 
S_{\rm particle} = - m \int_\gamma \sqrt{ -{\sf g}_{\mu\nu}
\dot{z}^\mu \dot{z}^\nu }\, d\lambda 
\label{18.1.1}
\end{equation}
where $\dot{z}^\mu = dz^\mu/d\lambda$ is tangent to the world line and
the metric is evaluated at $z$. We assume that the particle provides
the {\it only} source of matter in the spacetime --- an explanation
will be provided at the end of this subsection --- so that the
Einstein field equations take the form of   
\begin{equation}
{\sf G}^{\alpha\beta} = 8\pi {\sf T}^{\alpha\beta}, 
\label{18.1.2}
\end{equation} 
where ${\sf G}^{\alpha\beta}$ is the Einstein tensor constructed from 
${\sf g}_{\alpha\beta}$ and 
\begin{equation} 
{\sf T}^{\alpha\beta}(x) = m \int_\gamma 
\frac{ {\sf g}^\alpha_{\ \mu}(x,z) {\sf g}^\beta_{\ \nu}(x,z)  
\dot{z}^\mu \dot{z}^\nu}{ \sqrt{ -{\sf g}_{\mu\nu}
\dot{z}^\mu \dot{z}^\nu }}\, \delta_4(x,z)\, d\lambda 
\label{18.1.3}
\end{equation} 
is the particle's stress-energy tensor, obtained by functional 
differentiation of $S_{\rm particle}$ with respect to 
${\sf g}_{\alpha\beta}(x)$; the parallel propagators appear naturally 
by expressing ${\sf g}_{\mu\nu}$ as ${\sf g}^\alpha_{\ \mu}   
{\sf g}^\beta_{\ \nu} {\sf g}_{\alpha\beta}$. 

On a formal level the metric ${\sf g}_{\alpha\beta}$ is obtained by 
solving the Einstein field equations, and the world line is determined
by solving the equations of energy-momentum conservation, which follow
from the field equations. From Eqs.~(\ref{4.3.2}),
(\ref{12.1.3}), and (\ref{18.1.3}) we obtain  
\[
\nabla_\beta {\sf T}^{\alpha\beta} = m \int_\gamma \frac{d}{d\lambda} 
\biggl( \frac{ {\sf g}^\alpha_{\ \mu} \dot{z}^\mu }{     
\sqrt{ -{\sf g}_{\mu\nu} \dot{z}^\mu \dot{z}^\nu }} \biggr)
\delta_4(x,z)\, d\lambda,    
\]
and additional manipulations reduce this to 
\[
\nabla_\beta {\sf T}^{\alpha\beta} = m \int_\gamma  
\frac{ {\sf g}^\alpha_{\ \mu} }{     
\sqrt{ -{\sf g}_{\mu\nu} \dot{z}^\mu \dot{z}^\nu }}
\biggl( \frac{ {\sf D} \dot{z}^\mu}{d\lambda} - {\sf k}
\dot{z}^\mu \biggr) \delta_4(x,z)\, d\lambda,  
\]
where ${\sf D} \dot{z}^\mu/d\lambda$ is the covariant acceleration and
${\sf k}$ is a scalar field on the world line. Energy-momentum 
conservation therefore produces the geodesic equation   
\begin{equation} 
\frac{ {\sf D} \dot{z}^\mu}{d\lambda} = {\sf k} \dot{z}^\mu, 
\label{18.1.4} 
\end{equation} 
and 
\begin{equation} 
{\sf k} \equiv \frac{1}{\sqrt{ -{\sf g}_{\mu\nu}
\dot{z}^\mu \dot{z}^\nu }} \frac{d}{d\lambda} 
\sqrt{ -{\sf g}_{\mu\nu} \dot{z}^\mu \dot{z}^\nu}  
\label{18.1.5}
\end{equation} 
measures the failure of $\lambda$ to be an affine parameter on the 
geodesic $\gamma$.  

At this stage we begin treating $m$ as a formal expansion parameter, 
and we write  
\begin{equation}
{\sf g}_{\alpha\beta} = g_{\alpha\beta} + h_{\alpha\beta} + O(m^2), 
\label{18.1.6}
\end{equation}
with $g_{\alpha\beta}$ denoting the $m \to 0$ limit of the total
metric ${\sf g}_{\alpha\beta}$, and $h_{\alpha\beta} = O(m)$ the
first-order correction. We shall refer to $g_{\alpha\beta}$ as the
``metric of the background spacetime'' and to $h_{\alpha\beta}$ as the
``perturbation'' produced by the particle. We similarly write 
\begin{equation}
{\sf G}^{\alpha\beta}[{\sf g}] = G^{\alpha\beta}[g] +
H^{\alpha\beta}[g;h] + O(m^2) 
\label{18.1.7}
\end{equation} 
for the Einstein tensor, and 
\begin{equation}
{\sf T}^{\alpha\beta} = T^{\alpha\beta} + O(m^2) 
\label{18.1.8}
\end{equation}
for the particle's stress-energy tensor. The leading term
$T^{\alpha\beta}(x)$ describes the stress-energy tensor of a test 
particle of mass $m$ that moves on a world line $\gamma$ in a
background spacetime with metric $g_{\alpha\beta}$. If we choose
$\lambda$ to be proper time $\tau$ as measured in this
spacetime, then Eq.~(\ref{18.1.3}) implies 
\begin{equation} 
T^{\alpha\beta}(x) = m \int_\gamma g^\alpha_{\ \mu}(x,z) 
g^\beta_{\ \nu}(x,z) u^\mu u^\nu \delta_4(x,z)\, d\tau, 
\label{18.1.9}
\end{equation} 
where $u^\mu(\tau) = d z^\mu/d\tau$ is the particle's four-velocity. 

We have already stated that the particle is the only source of matter
in the spacetime, and the metric $g_{\alpha\beta}$ must therefore be 
a solution to the vacuum field equations: $G^{\alpha\beta}[g] =
0$. Equations (\ref{18.1.2}), (\ref{18.1.7}), and (\ref{18.1.8}) then
imply $H^{\alpha\beta}[g;h] = 8\pi T^{\alpha\beta}$, in which both
sides of the equation are of order $m$. To simplify the expression of
the first-order correction to the Einstein tensor we introduce the
trace-reversed gravitational potentials 
\begin{equation}
\gamma_{\alpha\beta} = h_{\alpha\beta} - \frac{1}{2} \bigl(
g^{\gamma\delta} h_{\gamma\delta} \bigr) g_{\alpha\beta}, 
\label{18.1.10}
\end{equation} 
and we impose the Lorenz gauge condition 
\begin{equation} 
\gamma^{\alpha\beta}_{\ \ \ ;\beta} = 0. 
\label{18.1.11}
\end{equation} 
Here and below it is understood that indices are lowered and raised
with the background metric and its inverse, respectively, and that
covariant differentiation refers to a connection that is compatible
with $g_{\alpha\beta}$. We then have $H^{\alpha\beta} = -\frac{1}{2}(
\Box \gamma^{\alpha\beta} + 2 R_{\gamma\ \delta}^{\ \alpha\ \beta}
\gamma^{\gamma\delta} )$ and Eq.~(\ref{18.1.2}) reduces to 
\begin{equation} 
\Box \gamma^{\alpha\beta} + 2 R_{\gamma\ \delta}^{\ \alpha\ \beta} 
\gamma^{\gamma\delta} = -16\pi T^{\alpha\beta}, 
\label{18.1.12}
\end{equation}
where $\Box = g^{\alpha\beta} \nabla_\alpha \nabla_\beta$ is the wave
operator and $T^{\alpha\beta}$ is defined by Eq.~(\ref{18.1.9}). We
have here a linear wave equation for the potentials
$\gamma_{\alpha\beta}$, and this equation can be placed on an equal
footing with Eq.~(\ref{16.1.5}) for the potential $\Phi$ associated
with a point scalar charge, and Eq.~(\ref{17.1.9}) for the vector
potential $A^\alpha$ associated with a point electric charge.   

The equations of motion for the point mass are obtained by
substituting the expansion of Eq.~(\ref{18.1.6}) into
Eqs.~(\ref{18.1.4}) and (\ref{18.1.5}). The perturbed connection is
easily computed to be $\Gamma^\alpha_{\ \beta\gamma} + \frac{1}{2}(
h^\alpha_{\ \beta;\gamma} + h^\alpha_{\ \gamma;\beta} -
h_{\beta\gamma}^{\ \ \ ;\alpha})$, and this leads to 
\[
\frac{{\sf D} \dot{z}^\mu}{d\tau} = \frac{D u^\mu}{d\tau} +
\frac{1}{2} \bigl( h^\mu_{\ \nu;\lambda} + h^\mu_{\ \lambda;\nu} -
h_{\nu\lambda}^{\ \ ;\mu} \bigr) u^\nu u^\lambda + O(m^2),  
\]
having once more selected proper time $\tau$ (as measured in the
background spacetime) as the parameter on the world line. On the other
hand, Eq.~(\ref{18.1.5}) gives 
\[
{\sf k} = -\frac{1}{2} h_{\nu\lambda;\rho} u^\nu u^\lambda u^\rho -
h_{\nu\lambda} u^\nu a^\lambda + O(m^2), 
\]
where $a^\lambda = D u^\lambda/d\tau$ is the particle's acceleration
vector. Since it is clear that the acceleration will be of order $m$,
the second term can be discarded and we obtain 
\[
\frac{D u^\mu}{d\tau} = -\frac{1}{2} \Bigl( h^\mu_{\ \nu;\lambda} +
h^\mu_{\ \lambda;\nu} - h_{\nu\lambda}^{\ \ ;\mu} + u^\mu
h_{\nu\lambda;\rho} u^\rho \Bigr) u^\nu u^\lambda + O(m^2).
\]
Keeping the error term implicit, we shall express this in the
equivalent form 
\begin{equation}
\frac{D u^\mu}{d\tau} = -\frac{1}{2} \bigl( g^{\mu\nu} + u^\mu
u^\nu \bigr) \bigl( 2 h_{\nu\lambda;\rho} - h_{\lambda\rho;\nu} \bigr)
u^\lambda u^\rho, 
\label{18.1.13} 
\end{equation}   
which emphasizes the fact that the acceleration is orthogonal to
the four-velocity. 

It should be clear that Eq.~(\ref{18.1.13}) is valid only in a formal
sense, because the potentials obtained from Eqs.~(\ref{18.1.12})
diverge on the world line. The nonlinearity of the Einstein field
equations makes this problem even worse here than for the scalar and
electromagnetic cases, because the
singular behaviour of the perturbation might render meaningless a
formal expansion of ${\sf g}_{\alpha\beta}$ in powers of $m$. Ignoring
this issue for the time being (we shall return to it in
Sec.~\ref{19}), we will proceed as in Secs.~\ref{16} and \ref{17} 
and attempt, with a careful analysis of the field's singularity
structure, to make sense of these equations.  

To conclude this subsection I should explain why it is desirable to 
restrict our discussion to spacetimes that contain no matter except
for the point particle. Suppose, in contradiction with this
assumption, that the background spacetime contains a distribution of
matter around which the particle is moving. (The corresponding vacuum
situation has the particle moving around a black hole. Notice that we
are still assuming that the particle moves in a region of spacetime in
which there is no matter; the issue is whether we can allow for a
distribution of matter {\it somewhere else}.) Suppose also that the
matter distribution is described by a collection of matter fields
$\Psi$. Then the field equations satisfied by the matter have the
schematic form $E[\Psi;g] = 0$, and the metric is determined by the
Einstein field equations $G[g] = 8\pi M[\Psi;g]$, in which $M[\Psi;g]$
stands for the matter's stress-energy tensor. We now insert the
point particle in the spacetime, and recognize that this displaces
the background solution $(\Psi,g)$ to a new solution ($\Psi
+ \delta \Psi,g + \delta g)$. The perturbations are determined by the
coupled set of equations $E[\Psi+\delta \Psi;g+\delta g] = 0$ and $G[g
+ \delta g] = 8\pi M[\Psi+\delta \Psi;g + \delta g] + 8\pi
T[g]$. After linearization these take the form of 
\[
E_\Psi \cdot \delta \Psi + E_g \cdot \delta g = 0, \qquad 
G_g \cdot \delta g = 8\pi \bigl( M_\Psi \cdot \delta \Psi + M_g \cdot
\delta g + T \bigr), 
\]
where $E_\Psi$, $E_g$, $M_\Phi$, and $M_g$ are suitable differential
operators acting on the perturbations. This is a {\it coupled set} of
partial differential equations for the perturbations $\delta \Psi$ and
$\delta g$. These equations are linear, but they are much more
difficult to deal with than the single equation for $\delta g$ that
was obtained in the vacuum case. And although it is still possible to
solve the coupled set of equations via a Green's function technique,
the degree of difficulty is such that we will not attempt this
here. We shall, therefore, continue to restrict our attention to the
case of a point particle moving in a vacuum (globally Ricci-flat)
background spacetime. 

\subsection{Retarded potentials near the world line} 
\label{18.2}

The retarded solution to Eq.~(\ref{18.1.12}) is
$\gamma^{\alpha\beta}(x) = 4 \int G^{\ \alpha\beta}_{+\
\gamma'\delta'}(x,x') T^{\gamma'\delta'}(x') \sqrt{-g'}\, d^4 x'$, 
where $G^{\ \alpha\beta}_{+\ \gamma'\delta'}(x,x')$ is the retarded
Green's function introduced in Sec.~\ref{15}. After substitution of
the stress-energy tensor of Eq.~(\ref{18.1.9}) we obtain 
\begin{equation} 
\gamma^{\alpha\beta}(x) = 4 m \int_\gamma 
G^{\ \alpha\beta}_{+\ \mu\nu}(x,z) u^\mu u^\nu\, d\tau, 
\label{18.2.1}
\end{equation}
in which $z^\mu(\tau)$ gives the description of the world line
$\gamma$ and $u^\mu = d z^\mu/d\tau$. Because the retarded Green's
function is defined globally in the entire background spacetime,
Eq.~(\ref{18.2.1}) describes the gravitational perturbation created by
the particle at any point $x$ in that spacetime. 

For a more concrete expression we must take $x$ to be in a
neighbourhood of the world line. The following manipulations follow
closely those performed in Sec.~\ref{16.2} for the case of a scalar 
charge, and in Sec.~\ref{17.2} for the case of an electric
charge. Because these manipulations are by now familiar, it will be
sufficient here to present only the main steps. There are two 
important simplifications that occur in the case of a massive
particle. First, for the purposes of computing
$\gamma^{\alpha\beta}(x)$ to first order in $m$, it is sufficient to 
take the world line to be a {\it geodesic} of the background
spacetime: the deviations from geodesic motion that we are in the
process of calculating are themselves of order $m$ and would
affect $\gamma^{\alpha\beta}(x)$ at order $m^2$ only. We shall
therefore be allowed to set  
\begin{equation}
a^\mu = 0 = \dot{a}^\mu 
\label{18.2.2} 
\end{equation}
in our computations. Second, because we take $g_{\alpha\beta}$ to be a 
solution to the vacuum field equations, we are also allowed to set  
\begin{equation}
R_{\mu\nu}(z) = 0 
\label{18.2.3}
\end{equation} 
in our computations. 

With the understanding that $x$ is close to the world line (refer back
to Fig.~9), we substitute the Hadamard construction of
Eq.~(\ref{15.2.1}) into Eq.~(\ref{18.2.1}) and integrate over the
portion of $\gamma$ that is contained in ${\cal N}(x)$. The result is  
\begin{equation}
\gamma^{\alpha\beta}(x) = \frac{4m}{r} 
U^{\alpha\beta}_{\ \ \gamma'\delta'}(x,x') u^{\gamma'} u^{\delta'} 
+ 4 m \int_{\tau_<}^u V^{\alpha\beta}_{\ \ \mu\nu}(x,z) u^\mu u^\nu\,
d\tau + 4 m \int_{-\infty}^{\tau_<} 
G^{\ \alpha\beta}_{+\ \mu\nu}(x,z) u^\mu u^\nu\, d\tau, 
\label{18.2.4}
\end{equation}
in which primed indices refer to the retarded point $x' \equiv z(u)$ 
associated with $x$, $r \equiv \sigma_{\alpha'} u^{\alpha'}$ is the
retarded distance from $x'$ to $x$, and $\tau_<$ is the proper time at
which $\gamma$ enters ${\cal N}(x)$ from the past. 

In the following subsections we shall refer to
$\gamma_{\alpha\beta}(x)$ as the {\it gravitational potentials} at $x$
produced by a particle of mass $m$ moving on the world line $\gamma$, 
and to $\gamma_{\alpha\beta;\gamma}(x)$ as the {\it gravitational
field} at $x$. To compute this is our next task.   

\subsection{Gravitational field in retarded coordinates} 
\label{18.3}
    
Keeping in mind that $x'$ and $x$ are related by $\sigma(x,x') = 0$, a
straightforward computation reveals that the covariant derivatives of
the gravitational potentials are given by  
\begin{eqnarray}
\gamma_{\alpha\beta;\gamma}(x) &=& -\frac{4m}{r^2}
U_{\alpha\beta\alpha'\beta'} u^{\alpha'} u^{\beta'} \partial_\gamma r 
+ \frac{4m}{r} U_{\alpha\beta\alpha'\beta';\gamma} u^{\alpha'}
u^{\beta'} 
+ \frac{4m}{r} U_{\alpha\beta\alpha'\beta';\gamma'} u^{\alpha'}
u^{\beta'} u^{\gamma'} \partial_\gamma u
\nonumber \\ & & \mbox{} 
+ 4m V_{\alpha\beta\alpha'\beta'} u^{\alpha'} u^{\beta'}
\partial_\gamma u 
+ \gamma_{\alpha\beta\gamma}^{\rm tail}(x), 
\label{18.3.1}
\end{eqnarray} 
where the ``tail integral'' is defined by 
\begin{eqnarray} 
\gamma_{\alpha\beta\gamma}^{\rm tail}(x) &=& 4 m \int_{\tau_<}^u
\nabla_\gamma V_{\alpha\beta\mu\nu}(x,z) u^\mu u^\nu\, d\tau
+ 4 m \int_{-\infty}^{\tau_<} \nabla_\gamma
G_{+\alpha\beta\mu\nu}(x,z) u^\mu u^\nu\, d\tau 
\nonumber \\ 
&=& 4m \int_{-\infty}^{u^-} \nabla_\gamma
G_{+\alpha\beta\mu\nu}(x,z) u^\mu u^\nu\, d\tau.  
\label{18.3.2}
\end{eqnarray} 
The second form of the definition, in which the integration is cut 
short at $\tau = u^- \equiv u - 0^+$ to avoid the singular behaviour
of the retarded Green's function at $\sigma = 0$, is equivalent to the
first form. 

We wish to express $\gamma_{\alpha\beta;\gamma}(x)$ in the retarded
coordinates of Sec.~\ref{9}, as an expansion in powers of $r$. For
this purpose we decompose the field in the tetrad
$(\base{\alpha}{0},\base{\alpha}{a})$ that is obtained by parallel
transport of $(u^{\alpha'},\base{\alpha'}{a})$ on the null geodesic
that links $x$ to $x'$; this construction is detailed in
Sec.~\ref{9}. Note that throughout this section we set
$\omega_{ab} = 0$, where $\omega_{ab}$ is the rotation tensor defined
by Eq.~(\ref{9.1.1}): the tetrad vectors $\base{\alpha'}{a}$ are taken
to be parallel transported on $\gamma$. We recall from
Eq.~(\ref{9.1.4}) that the parallel propagator can be expressed as  
$g^{\alpha'}_{\ \alpha} = u^{\alpha'} \base{0}{\alpha}  
+ \base{\alpha'}{a} \base{a}{\alpha}$. The expansion relies on
Eq.~(\ref{9.5.3}) for $\partial_\gamma u$ and Eq.~(\ref{9.5.5}) for 
$\partial_\gamma r$, both specialized to the case of geodesic motion
--- $a_a = 0$. We shall also need 
\begin{equation} 
U_{\alpha\beta\alpha'\beta'} u^{\alpha'} u^{\beta'} = 
g^{\alpha'}_{\ (\alpha} g^{\beta'}_{\ \beta)} \Bigl[ u_{\alpha'}
u_{\beta'} + O(r^3) \Bigr], 
\label{18.3.3}
\end{equation} 
which follows from Eq.~(\ref{15.2.7}), 
\begin{eqnarray} 
U_{\alpha\beta\alpha'\beta';\gamma} u^{\alpha'} u^{\beta'} &=&  
g^{\alpha'}_{\ (\alpha} g^{\beta'}_{\ \beta)} g^{\gamma'}_{\ \gamma}
\Bigl[ -r\bigl( R_{\alpha'0\gamma'0} + R_{\alpha'0\gamma'd} \Omega^d
\bigr) u_{\beta'} + O(r^2) \Bigr], 
\label{18.3.4} \\ 
U_{\alpha\beta\alpha'\beta';\gamma'} u^{\alpha'} u^{\beta'}
u^{\gamma'} &=& g^{\alpha'}_{\ (\alpha} g^{\beta'}_{\ \beta)} \Bigl[ 
r R_{\alpha'0d0} \Omega^d u_{\beta'} + O(r^2) \Bigr], 
\label{18.3.5}
\end{eqnarray} 
which follow from Eqs.~(\ref{15.2.8}) and (\ref{15.2.9}),
respectively, as well as the relation $\sigma^{\alpha'} = -r
(u^{\alpha'} + \Omega^a \base{\alpha'}{a})$ first encountered
in Eq.~(\ref{9.2.3}). And finally, we shall need  
\begin{equation}
V_{\alpha\beta\alpha'\beta'} u^{\alpha'} u^{\beta'} = 
g^{\alpha'}_{\ (\alpha} g^{\beta'}_{\ \beta)} \Bigl[
R_{\alpha'0\beta'0} + O(r) \Bigr], 
\label{18.3.6}
\end{equation} 
which follows from Eq.~(\ref{15.2.11}).  

Making these substitutions in Eq.~(\ref{18.1.3}) and projecting
against various members of the tetrad gives 
\begin{eqnarray} 
\gamma_{000}(u,r,\Omega^a) &\equiv& \gamma_{\alpha\beta;\gamma}(x) 
\base{\alpha}{0}(x) \base{\beta}{0}(x) \base{\gamma}{0}(x) 
= 2m R_{a0b0} \Omega^a \Omega^b + \gamma^{\rm tail}_{000} + O(r), 
\label{18.3.7} \\ 
\gamma_{0b0}(u,r,\Omega^a) &\equiv& \gamma_{\alpha\beta;\gamma}(x) 
\base{\alpha}{0}(x) \base{\beta}{b}(x) \base{\gamma}{0}(x) 
= -4m R_{b0c0} \Omega^c + \gamma^{\rm tail}_{0b0} + O(r),
\label{18.3.8} \\ 
\gamma_{ab0}(u,r,\Omega^a) &\equiv& \gamma_{\alpha\beta;\gamma}(x) 
\base{\alpha}{a}(x) \base{\beta}{b}(x) \base{\gamma}{0}(x) 
= 4m R_{a0b0} + \gamma^{\rm tail}_{ab0} + O(r), 
\label{18.3.9} \\ 
\gamma_{00c}(u,r,\Omega^a) &\equiv& \gamma_{\alpha\beta;\gamma}(x) 
\base{\alpha}{0}(x) \base{\beta}{0}(x) \base{\gamma}{c}(x) 
\nonumber \\ 
&=& -4m \biggl[ \Bigl(\frac{1}{r^2} + \frac{1}{3} R_{a0b0} \Omega^a
\Omega^b \Bigr) \Omega_c + \frac{1}{6} R_{c0b0} \Omega^b - \frac{1}{6}
R_{ca0b} \Omega^a \Omega^b \biggr] + \gamma^{\rm tail}_{00c} + O(r),\qquad 
\label{18.3.10} \\ 
\gamma_{0bc}(u,r,\Omega^a) &\equiv& \gamma_{\alpha\beta;\gamma}(x) 
\base{\alpha}{0}(x) \base{\beta}{b}(x) \base{\gamma}{c}(x) 
\nonumber \\ 
&=& 2m \bigl( R_{b0c0} + R_{b0cd} \Omega^d + R_{b0d0} \Omega^d
\Omega_c \bigr) + \gamma^{\rm tail}_{0bc} + O(r), 
\label{18.3.11} \\ 
\gamma_{abc}(u,r,\Omega^a) &\equiv& \gamma_{\alpha\beta;\gamma}(x) 
\base{\alpha}{a}(x) \base{\beta}{b}(x) \base{\gamma}{c}(x) 
= -4m R_{a0b0} \Omega_c + \gamma^{\rm tail}_{abc} + O(r), 
\label{18.3.12}
\end{eqnarray} 
where, for example, $R_{a0b0}(u) \equiv
R_{\alpha'\gamma'\beta'\delta'} \base{\alpha'}{a} u^{\gamma'}
\base{\beta'}{b} u^{\delta'}$ are frame components of the Riemann
tensor evaluated at $x' \equiv z(u)$. We have also introduced the
frame components of the tail part of the gravitational field, which
are obtained from Eq.~(\ref{18.3.2}) evaluated at $x'$ instead of $x$;
for example, $\gamma^{\rm tail}_{000} = u^{\alpha'} u^{\beta'}
u^{\gamma'} \gamma^{\rm tail}_{\alpha'\beta'\gamma'}(x')$. We 
may note here that while $\gamma_{00c}$ is the only component of the
gravitational field that diverges when $r \to 0$, the other components
are nevertheless singular because of their dependence on
the unit vector $\Omega^a$; the only exception is $\gamma_{ab0}$,
which is smooth.  

\subsection{Gravitational field in Fermi normal coordinates}   
\label{18.4} 

The translation of the results contained in
Eqs.~(\ref{18.3.7})--(\ref{18.3.12}) into the Fermi normal coordinates
of Sec.~\ref{8} proceeds as in Secs.~\ref{16.4} and \ref{17.4}, but 
is simplified by the fact that here, the world line can be taken to
be a geodesic. We may thus set $a_a = \dot{a}_0 = \dot{a}_a = 0$ in
Eqs.~(\ref{10.3.1}) and (\ref{10.3.2}) that relate the tetrad
$(\bar{e}^\alpha_0,\bar{e}^\alpha_a)$ to
$(\base{\alpha}{0},\base{\alpha}{a})$, as well as in
Eqs.~(\ref{10.2.1})--(\ref{10.2.3}) that relate the Fermi normal
coordinates $(t,s,\omega^a)$ to the retarded coordinates. We recall
that the Fermi normal coordinates refer to a point $\bar{x} \equiv
z(t)$ on the world line that is linked to $x$ by a spacelike geodesic
that intersects $\gamma$ orthogonally. 

The translated results are     
\begin{eqnarray} 
\bar{\gamma}_{000}(t,s,\omega^a) &\equiv&
\gamma_{\alpha\beta;\gamma}(x)  
\bar{e}^\alpha_0(x) \bar{e}^\beta_0(x) \bar{e}^\gamma_0(x) 
= \bar{\gamma}^{\rm tail}_{000} + O(s),  
\label{18.4.1} \\ 
\bar{\gamma}_{0b0}(t,s,\omega^a) &\equiv&
\gamma_{\alpha\beta;\gamma}(x)  
\bar{e}^\alpha_0(x) \bar{e}^\beta_b(x) \bar{e}^\gamma_0(x) 
= -4m R_{b0c0} \omega^c + \bar{\gamma}^{\rm tail}_{0b0} + O(s),
\label{18.4.2} \\ 
\bar{\gamma}_{ab0}(t,s,\omega^a) &\equiv&
\gamma_{\alpha\beta;\gamma}(x)  
\bar{e}^\alpha_a(x) \bar{e}^\beta_b(x) \bar{e}^\gamma_0(x) 
= 4m R_{a0b0} + \bar{\gamma}^{\rm tail}_{ab0} + O(s), 
\label{18.4.3} \\ 
\bar{\gamma}_{00c}(t,s,\omega^a) &\equiv&
\gamma_{\alpha\beta;\gamma}(x)  
\bar{e}^\alpha_0(x) \bar{e}^\beta_0(x) \bar{e}^\gamma_c(x) 
\nonumber \\ 
&=& -4m \biggl[ \Bigl(\frac{1}{s^2} - \frac{1}{6} R_{a0b0} \omega^a
\omega^b \Bigr) \omega_c + \frac{1}{3} R_{c0b0} \omega^b \biggr] 
+ \bar{\gamma}^{\rm tail}_{00c} + O(s), 
\label{18.4.4} \\ 
\bar{\gamma}_{0bc}(t,s,\omega^a) &\equiv&
\gamma_{\alpha\beta;\gamma}(x)  
\bar{e}^\alpha_0(x) \bar{e}^\beta_b(x) \bar{e}^\gamma_c(x) 
= 2m \bigl( R_{b0c0} + R_{b0cd} \omega^d \bigr) 
+ \bar{\gamma}^{\rm tail}_{0bc} + O(s), 
\label{18.4.5} \\ 
\bar{\gamma}_{abc}(t,s,\omega^a) &\equiv&
\gamma_{\alpha\beta;\gamma}(x)  
\bar{e}^\alpha_a(x) \bar{e}^\beta_b(x) \bar{e}^\gamma_c(x) 
= -4m R_{a0b0} \omega_c + \bar{\gamma}^{\rm tail}_{abc} + O(s), 
\label{18.4.6}
\end{eqnarray} 
where all frame components are now evaluated at $\bar{x}$ instead of
$x'$.  

It is then a simple matter to average these results over a two-surface
of constant $t$ and $s$. Using the area element of Eq.~(\ref{16.4.5})
and definitions analogous to those of Eq.~(\ref{16.4.6}), we obtain 
\begin{eqnarray}
\langle \bar{\gamma}_{000} \rangle &=& \bar{\gamma}^{\rm tail}_{000}  
+ O(s), 
\label{18.4.7} \\ 
\langle \bar{\gamma}_{0b0} \rangle &=& \bar{\gamma}^{\rm tail}_{0b0}  
+ O(s), 
\label{18.4.8} \\ 
\langle \bar{\gamma}_{ab0} \rangle &=& 4 m R_{a0b0} 
+ \bar{\gamma}^{\rm tail}_{ab0} + O(s), 
\label{18.4.9} \\ 
\langle \bar{\gamma}_{00c} \rangle &=& \bar{\gamma}^{\rm tail}_{00c}  
+ O(s), 
\label{18.4.10} \\ 
\langle \bar{\gamma}_{0bc} \rangle &=& 2 m R_{b0c0} 
+ \bar{\gamma}^{\rm tail}_{0bc} + O(s), 
\label{18.4.11} \\ 
\langle \bar{\gamma}_{abc} \rangle &=& \bar{\gamma}^{\rm tail}_{abc}  
+ O(s).   
\label{18.4.12} 
\end{eqnarray} 
The averaged gravitational field is smooth in the limit $s \to 0$, in
which the tetrad $(\bar{e}^\alpha_0,\bar{e}^\alpha_a)$ coincides with 
$(u^{\bar{\alpha}},\base{\bar{\alpha}}{a})$. Reconstructing the field 
at $\bar{x}$ from its frame components gives 
\begin{equation}
\langle \gamma_{\bar{\alpha}\bar{\beta};\bar{\gamma}} \rangle = 
-4 m \Bigl( u_{(\bar{\alpha}}
R_{\bar{\beta})\bar{\delta}\bar{\gamma}\bar{\epsilon}}
+ R_{\bar{\alpha}\bar{\delta}\bar{\beta}\bar{\epsilon}}
u_{\bar{\gamma}} \Bigr) u^{\bar{\delta}} u^{\bar{\epsilon}} 
+ \gamma^{\rm tail}_{\bar{\alpha}\bar{\beta}\bar{\gamma}}, 
\label{18.4.13}
\end{equation}
where the tail term can be copied from Eq.~(\ref{18.3.2}), 
\begin{equation} 
\gamma^{\rm tail}_{\bar{\alpha}\bar{\beta}\bar{\gamma}}(\bar{x}) 
= 4m \int_{-\infty}^{t^-} \nabla_{\bar{\gamma}}
G_{+\bar{\alpha}\bar{\beta}\mu\nu}(\bar{x},z) u^\mu u^\nu\, d\tau. 
\label{18.4.14}
\end{equation}
The tensors that appear in Eq.~(\ref{18.4.13}) all refer to the  
simultaneous point $\bar{x} \equiv z(t)$, which can now be treated as
an arbitrary point on the world line $\gamma$. 

\subsection{Singular and radiative fields}
\label{18.5} 

The singular gravitational potentials
\begin{equation} 
\gamma^{\alpha\beta}_{\rm S}(x) = 4 m \int_\gamma 
G^{\ \alpha\beta}_{{\rm S}\ \mu\nu}(x,z) u^\mu u^\nu\, d\tau 
\label{18.5.1}
\end{equation}
are solutions to the wave equation of
Eq.~(\ref{18.1.12}); the singular Green's function was introduced in
Sec.~\ref{15.4}. We will see that the singular field 
$\gamma^{\rm S}_{\alpha\beta;\gamma}$ reproduces the singular
behaviour of the retarded solution near the world line, and that the
difference, $\gamma^{\rm R}_{\alpha\beta;\gamma} = 
\gamma_{\alpha\beta;\gamma} - \gamma^{\rm S}_{\alpha\beta;\gamma}$, is
smooth on the world line. 

To evaluate the integral of Eq.~(\ref{18.5.1}) we take $x$ to be close
to the world line (see Fig.~9), and we invoke Eq.~(\ref{15.4.8}) as
well as the Hadamard construction of Eq.~(\ref{15.4.14}). This gives 
\begin{equation} 
\gamma^{\alpha\beta}_{\rm S}(x) = \frac{2m}{r} 
U^{\alpha\beta}_{\ \ \gamma'\delta'} u^{\gamma'} u^{\delta'} 
+ \frac{2m}{r_{\rm adv}} U^{\alpha\beta}_{\ \ \gamma''\delta''}
u^{\gamma''} u^{\delta''} 
- 2 m \int_u^v V^{\alpha\beta}_{\ \ \mu\nu}(x,z) u^\mu u^\nu\, d\tau, 
\label{18.5.2}
\end{equation} 
where primed indices refer to the retarded point $x' \equiv z(u)$,
double-primed indices refer to the advanced point $x'' \equiv z(v)$,
and where $r_{\rm adv} \equiv -\sigma_{\alpha''} u^{\alpha''}$ is the
advanced distance between $x$ and the world line. 

Differentiation of Eq.~(\ref{18.5.2}) yields 
\begin{eqnarray} 
\gamma^{\rm S}_{\alpha\beta;\gamma}(x) &=&    
-\frac{2m}{r^2}
U_{\alpha\beta\alpha'\beta'} u^{\alpha'} u^{\beta'} \partial_\gamma r 
-\frac{2m}{{r_{\rm adv}}^2}
U_{\alpha\beta\alpha''\beta''} u^{\alpha''} u^{\beta''}
\partial_\gamma r_{\rm adv} 
+ \frac{2m}{r} U_{\alpha\beta\alpha'\beta';\gamma} u^{\alpha'}
u^{\beta'} 
\nonumber \\ & & \mbox{} 
+ \frac{2m}{r} U_{\alpha\beta\alpha'\beta';\gamma'} u^{\alpha'}
u^{\beta'} u^{\gamma'} \partial_\gamma u
+ \frac{2m}{r_{\rm adv}} U_{\alpha\beta\alpha''\beta'';\gamma}
u^{\alpha''} u^{\beta''} 
+ \frac{2m}{r_{\rm adv}} U_{\alpha\beta\alpha''\beta'';\gamma''}
u^{\alpha''} u^{\beta''} u^{\gamma''} \partial_\gamma v
\nonumber \\ & & \mbox{} 
+ 2m V_{\alpha\beta\alpha'\beta'} u^{\alpha'} u^{\beta'}
\partial_\gamma u    
- 2m V_{\alpha\beta\alpha''\beta''} u^{\alpha''} u^{\beta''}
\partial_\gamma v
- 2m \int_u^v \nabla_\gamma V_{\alpha\beta\mu\nu}(x,z) 
u^\mu u^\nu\, d\tau, 
\label{18.5.3}
\end{eqnarray}
and we would like to express this as an expansion in powers of
$r$. For this we will rely on results already established in 
Sec.~\ref{18.3}, as well as additional expansions that will involve 
the advanced point $x''$. We recall that a relation between retarded
and advanced times was worked out in Eq.~(\ref{10.4.2}), that an
expression for the advanced distance was displayed in
Eq.~(\ref{10.4.3}), and that Eqs.~(\ref{10.4.4}) and (\ref{10.4.5})
give expansions for $\partial_\gamma v$ and 
$\partial_\gamma r_{\rm adv}$, respectively; these results can be
simplified by setting $a_a = \dot{a}_0 = \dot{a}_a = 0$, which is
appropriate in this computation.  

To derive an expansion for $U_{\alpha\beta\alpha''\beta''}
u^{\alpha''} u^{\beta''}$ we follow the general method of
Sec.~\ref{10.4} and introduce the functions $U_{\alpha\beta}(\tau)
\equiv U_{\alpha\beta\mu\nu}(x,z) u^\mu u^\nu$. We have that  
\[
U_{\alpha\beta\alpha''\beta''} u^{\alpha''} u^{\beta''} \equiv
U_{\alpha\beta}(v) = U_{\alpha\beta}(u)  
+ \dot{U}_{\alpha\beta}(u) \Delta^{\!\prime} 
+ \frac{1}{2} \ddot{U}_{\alpha\beta}(u) \Delta^{\!\prime 2}  
+ O\bigl( \Delta^{\!\prime 3} \bigr),  
\]
where overdots indicate differentiation with respect to $\tau$ and
$\Delta^{\!\prime} \equiv v-u$. The leading term $U_{\alpha\beta}(u) 
\equiv U_{\alpha\beta\alpha'\beta'} u^{\alpha'} u^{\beta'}$ was worked
out in Eq.~(\ref{18.3.3}), and the derivatives of
$U_{\alpha\beta}(\tau)$ are given by  
\[
\dot{U}_{\alpha\beta}(u) = U_{\alpha\beta\alpha'\beta';\gamma'}
u^{\alpha'} u^{\beta'} u^{\gamma'} = g^{\alpha'}_{\ (\alpha} 
g^{\beta'}_{\ \beta)} \Bigl[ r R_{\alpha'0d0}\Omega^d u_{\beta'}  
+ O(r^2) \Bigr]   
\] 
and 
\[
\ddot{U}_{\alpha\beta}(u) =
U_{\alpha\beta\alpha'\beta';\gamma'\delta'} u^{\alpha'} u^{\beta'}
u^{\gamma'} u^{\delta'} = O(r),  
\] 
according to Eqs.~(\ref{18.3.5}) and (\ref{15.2.9}). Combining these
results together with Eq.~(\ref{10.4.2}) for $\Delta^{\!\prime}$ gives    
\begin{equation} 
U_{\alpha\beta\alpha''\beta''} u^{\alpha''} u^{\beta''} = 
g^{\alpha'}_{\ (\alpha} g^{\beta'}_{\ \beta)} \Bigl[ u_{\alpha'}
u_{\beta'} + 2 r^2 R_{\alpha'0d0} \Omega^d u_{\beta'} + O(r^3) \Bigr], 
\label{18.5.4}
\end{equation} 
which should be compared with Eq.~(\ref{18.3.3}). It should be
emphasized that in Eq.~(\ref{18.5.4}) and all equations below, all
frame components are evaluated at the retarded point $x'$,
and not at the advanced point. The preceding computation gives us also
an expansion for  
\[
U_{\alpha\beta\alpha''\beta'';\gamma''} u^{\alpha'} u^{\beta''}
u^{\gamma''} = \dot{U}_{\alpha\beta}(u) + \ddot{U}_{\alpha\beta}(u)
\Delta^{\!\prime} + O(\Delta^{\!\prime 2}), 
\]
which becomes     
\begin{equation}
U_{\alpha\beta\alpha''\beta'';\gamma''} u^{\alpha''} u^{\beta''}
u^{\gamma''} = g^{\alpha'}_{\ (\alpha} g^{\beta'}_{\ \beta)} \Bigl[ 
r R_{\alpha'0d0} \Omega^d u_{\beta'} + O(r^2) \Bigr], 
\label{18.5.5}
\end{equation}
and which is identical to Eq.~(\ref{18.3.5}).    

We proceed similarly to obtain an expansion for
$U_{\alpha\beta\alpha''\beta'';\gamma} u^{\alpha''} u^{\beta''}$. Here
we introduce the functions $U_{\alpha\beta\gamma}(\tau) \equiv
U_{\alpha\beta\mu\nu;\gamma} u^\mu u^\nu$ and express
$U_{\alpha\beta\alpha''\beta'';\gamma} u^{\alpha''} u^{\beta''}$ as
$U_{\alpha\beta\gamma}(v) = U_{\alpha\beta\gamma}(u) +
\dot{U}_{\alpha\beta\gamma}(u) \Delta^{\!\prime} 
+ O(\Delta^{\!\prime 2})$. The leading term $U_{\alpha\beta\gamma}(u)
\equiv U_{\alpha\beta\alpha'\beta';\gamma} u^{\alpha'} u^{\beta'}$ was
computed in Eq.~(\ref{18.3.4}), and   
\[
\dot{U}_{\alpha\beta\gamma}(u) =
U_{\alpha\beta\alpha'\beta';\gamma\gamma'} u^{\alpha'} u^{\beta'}
u^{\gamma'} = g^{\alpha'}_{\ (\alpha} g^{\beta'}_{\ \beta)}
g^{\gamma'}_{\ \gamma} \Bigl[ R_{\alpha'0\gamma'0} u_{\beta'} + O(r)
\Bigr] 
\]
follows from Eq.~(\ref{15.2.8}). Combining these results together with  
Eq.~(\ref{10.4.2}) for $\Delta^{\!\prime}$ gives
\begin{equation}
U_{\alpha\beta\alpha''\beta'';\gamma} u^{\alpha''} u^{\beta''} =    
g^{\alpha'}_{\ (\alpha} g^{\beta'}_{\ \beta)} g^{\gamma'}_{\ \gamma} 
\Bigl[ r\bigl( R_{\alpha'0\gamma'0} 
- R_{\alpha'0\gamma'd} \Omega^d \bigr) u_{\beta'} + O(r^2) \Bigr],     
\label{18.5.6}
\end{equation}
and this should be compared with Eq.~(\ref{18.3.4}). The last
expansion we shall need is  
\begin{equation}
V_{\alpha\beta\alpha''\beta''} u^{\alpha''} u^{\beta''} = 
g^{\alpha'}_{\ (\alpha} g^{\beta'}_{\ \beta)} \Bigl[
R_{\alpha'0\beta'0} + O(r) \Bigr], 
\label{18.5.7}
\end{equation} 
which is identical to Eq.~(\ref{18.3.6}). 

We obtain the frame components of the singular gravitational field by
substituting these expansions into Eq.~(\ref{18.5.3}) and projecting
against the tetrad  $(\base{\alpha}{0},\base{\alpha}{a})$. 
After some algebra we arrive at    
\begin{eqnarray} 
\gamma^{\rm S}_{000}(u,r,\Omega^a) &\equiv& 
\gamma^{\rm S}_{\alpha\beta;\gamma}(x) 
\base{\alpha}{0}(x) \base{\beta}{0}(x) \base{\gamma}{0}(x) 
= 2m R_{a0b0} \Omega^a \Omega^b + O(r), 
\label{18.5.8} \\ 
\gamma^{\rm S}_{0b0}(u,r,\Omega^a) &\equiv& 
\gamma^{\rm S}_{\alpha\beta;\gamma}(x) 
\base{\alpha}{0}(x) \base{\beta}{b}(x) \base{\gamma}{0}(x) 
= -4m R_{b0c0} \Omega^c + O(r),
\label{18.5.9} \\ 
\gamma^{\rm S}_{ab0}(u,r,\Omega^a) &\equiv& 
\gamma^{\rm S}_{\alpha\beta;\gamma}(x) 
\base{\alpha}{a}(x) \base{\beta}{b}(x) \base{\gamma}{0}(x) 
= O(r), 
\label{18.5.10} \\ 
\gamma^{\rm S}_{00c}(u,r,\Omega^a) &\equiv& 
\gamma^{\rm S}_{\alpha\beta;\gamma}(x) 
\base{\alpha}{0}(x) \base{\beta}{0}(x) \base{\gamma}{c}(x) 
\nonumber \\ 
&=& -4m \biggl[ \Bigl(\frac{1}{r^2} + \frac{1}{3} R_{a0b0} \Omega^a
\Omega^b \Bigr) \Omega_c + \frac{1}{6} R_{c0b0} \Omega^b - \frac{1}{6}
R_{ca0b} \Omega^a \Omega^b \biggr] + O(r),
\label{18.5.11} \\ 
\gamma^{\rm S}_{0bc}(u,r,\Omega^a) &\equiv& 
\gamma^{\rm S}_{\alpha\beta;\gamma}(x) 
\base{\alpha}{0}(x) \base{\beta}{b}(x) \base{\gamma}{c}(x) 
= 2m \bigl( R_{b0cd} \Omega^d + R_{b0d0} \Omega^d
\Omega_c \bigr) + O(r), 
\label{18.5.12} \\ 
\gamma^{\rm S}_{abc}(u,r,\Omega^a) &\equiv& 
\gamma^{\rm S}_{\alpha\beta;\gamma}(x) 
\base{\alpha}{a}(x) \base{\beta}{b}(x) \base{\gamma}{c}(x) 
= -4m R_{a0b0} \Omega_c + O(r), 
\label{18.5.13}
\end{eqnarray} 
in which all frame components are evaluated at the retarded point
$x'$. Comparison of these expressions with 
Eqs.~(\ref{18.3.7})--(\ref{18.3.12}) reveals identical singularity 
structures for the retarded and singular gravitational fields.  

The difference between the retarded field of
Eqs.~(\ref{18.3.7})--(\ref{18.3.12}) and the singular field of
Eqs.~(\ref{18.5.8})--(\ref{18.5.13}) defines the radiative
gravitational field $\gamma^{\rm R}_{\alpha\beta;\gamma}$. Its tetrad
components are 
\begin{eqnarray} 
\gamma^{\rm R}_{000} &=& \gamma^{\rm tail}_{000} + O(r), 
\label{18.5.14} \\ 
\gamma^{\rm R}_{0b0} &=& \gamma^{\rm tail}_{0b0} + O(r),
\label{18.5.15} \\ 
\gamma^{\rm R}_{ab0} &=& 4m R_{a0b0} + \gamma^{\rm tail}_{ab0} + O(r), 
\label{18.5.16} \\ 
\gamma^{\rm R}_{00c} &=& \gamma^{\rm tail}_{00c} + O(r),
\label{18.5.17} \\ 
\gamma^{\rm R}_{0bc} &=& 2m R_{b0c0} + \gamma^{\rm tail}_{0bc} + O(r),  
\label{18.5.18} \\ 
\gamma^{\rm R}_{abc} &=& \gamma^{\rm tail}_{abc} + O(r), 
\label{18.5.19}
\end{eqnarray} 
and we see that $\gamma^{\rm R}_{\alpha\beta;\gamma}$ is smooth in the
limit $r \to 0$. We may therefore evaluate the radiative field
directly at $x=x'$, where the tetrad $(\base{\alpha}{0},
\base{\alpha}{a})$ coincides with $(u^{\alpha'},
\base{\alpha'}{a})$. After reconstructing the field at $x'$ from its
frame components, we obtain    
\begin{equation}
\gamma^{\rm R}_{\alpha'\beta';\gamma'}(x') =  
-4 m \Bigl( u_{(\alpha'} R_{\beta')\delta'\gamma'\epsilon'} 
+ R_{\alpha'\delta'\beta'\epsilon'} u_{\gamma'} \Bigr) u^{\delta'}
u^{\epsilon'} + \gamma^{\rm tail}_{\alpha'\beta'\gamma'},  
\label{18.5.20}
\end{equation}
where the tail term can be copied from Eq.~(\ref{18.3.2}), 
\begin{equation} 
\gamma^{\rm tail}_{\alpha'\beta'\gamma'}(x') 
= 4m \int_{-\infty}^{u^-} \nabla_{\gamma'}
G_{+\alpha'\beta'\mu\nu}(x',z) u^\mu u^\nu\, d\tau. 
\label{18.5.21}
\end{equation}
The tensors that appear in Eq.~(\ref{18.5.21}) all refer to the  
retarded point $x' \equiv z(u)$, which can now be treated as an
arbitrary point on the world line $\gamma$.  

\subsection{Equations of motion}
\label{18.6}

The retarded gravitational field $\gamma_{\alpha\beta;\gamma}$ of a
point particle is singular on the world line, and this behaviour makes
it difficult to understand how the field is supposed to act on the
particle and influence its motion. The field's singularity structure
was analyzed in Secs.~\ref{18.3} and \ref{18.4}, and in
Sec.~\ref{18.5} it was shown to originate from the singular field
$\gamma^{\rm S}_{\alpha\beta;\gamma}$; the radiative field 
$\gamma^{\rm R}_{\alpha\beta;\gamma}$ was then shown to be smooth on
the world line. 

To make sense of the retarded field's action on the particle we can
follow the discussions of Sec.~\ref{16.6} and \ref{17.6} and postulate
that the self gravitational field of the point particle is either
$\langle \gamma_{\mu\nu;\lambda} \rangle$, as worked out in
Eq.~(\ref{18.4.13}), or $\gamma^{\rm R}_{\mu\nu;\lambda}$, as worked
out in Eq.~(\ref{18.5.20}). These regularized fields are both given by  
\begin{equation}
\gamma^{\rm reg}_{\mu\nu;\lambda} 
= -4 m \Bigl( u_{(\mu} R_{\nu)\rho\lambda\xi}
+ R_{\mu\rho\nu\xi} u_\lambda \Bigr) u^\rho u^\xi 
+ \gamma^{\rm tail}_{\mu\nu\lambda}
\label{18.6.1}
\end{equation}
and 
\begin{equation}
\gamma^{\rm tail}_{\mu\nu\lambda} = 4 m \int_{-\infty}^{\tau^-}
\nabla_\lambda G_{+\mu\nu\mu'\nu'}\bigl( z(\tau), z(\tau') \bigr)
u^{\mu'} u^{\nu'}\, d\tau', 
\label{18.6.2} 
\end{equation}     
in which all tensors are now evaluated at an arbitrary point $z(\tau)$
on the world line $\gamma$.  

The actual gravitational perturbation $h_{\alpha\beta}$ is obtained by
inverting Eq.~(\ref{18.1.10}), which leads to $h_{\mu\nu;\lambda} =
\gamma_{\mu\nu;\gamma} - \frac{1}{2} g_{\mu\nu} 
\gamma^\rho_{\ \rho;\lambda}$. Substituting Eq.~(\ref{18.6.1}) yields  
\begin{equation} 
h^{\rm reg}_{\mu\nu;\lambda} 
= -4 m \Bigl( u_{(\mu} R_{\nu)\rho\lambda\xi}
+ R_{\mu\rho\nu\xi} u_\lambda \Bigr) u^\rho u^\xi 
+ h^{\rm tail}_{\mu\nu\lambda}, 
\label{18.6.3}
\end{equation}
where the tail term is given by the trace-reversed counterpart to 
Eq.~(\ref{18.6.2}): 
\begin{equation}
h^{\rm tail}_{\mu\nu\lambda} = 4 m \int_{-\infty}^{\tau^-}
\nabla_\lambda \biggl( G_{+\mu\nu\mu'\nu'}
- \frac{1}{2} g_{\mu\nu} G^{\ \ \rho}_{+\ \rho\mu'\nu'}
\biggr) \bigl( z(\tau), z(\tau')\bigr) u^{\mu'} u^{\nu'}\, d\tau'. 
\label{18.6.4} 
\end{equation}     
When this regularized field is substituted into Eq.~(\ref{18.1.13}),
we find that the terms that depend on the Riemann tensor cancel out,
and we are left with 
\begin{equation}
\frac{D u^\mu}{d\tau} = -\frac{1}{2} \bigl( g^{\mu\nu} + u^\mu
u^\nu \bigr) \bigl( 2 h^{\rm tail}_{\nu\lambda\rho} 
- h^{\rm tail}_{\lambda\rho\nu} \bigr) u^\lambda u^\rho. 
\label{18.6.5} 
\end{equation}   
We see that only the tail term is involved
in the final form of the equations of motion. The tail integral of
Eq.~(\ref{18.6.4}) involves the current position $z(\tau)$ of the
particle, at which all tensors with unprimed indices are evaluated,
as well as all prior positions $z(\tau')$, at which all tensors with
primed indices are evaluated. The tail integral is cut short at
$\tau' = \tau^- \equiv \tau - 0^+$ to avoid the singular behaviour of
the retarded Green's function at coincidence; this limiting procedure
was justified at the beginning of Sec.~\ref{18.3}.    
 
Equation (\ref{18.6.5}) was first derived by Yasushi Mino, Misao
Sasaki, and Takahiro Tanaka in 1997 \cite{MST}. (An incomplete
treatment had been given previously by Morette-DeWitt and Ging
\cite{moretteging}.) An alternative derivation was then produced, also
in 1997, by Theodore C.\ Quinn and Robert M.\ Wald \cite{QW1}. These
equations are now known as the MiSaTaQuWa equations of motion. It
should be noted that Eq.~(\ref{18.6.5}) is formally equivalent to the
statement that the point particle moves on a geodesic in a spacetime
with metric $g_{\alpha\beta} + h^{\rm R}_{\alpha\beta}$, where 
$h^{\rm R}_{\alpha\beta}$ is the radiative metric perturbation
obtained by trace-reversal of the potentials 
$\gamma^{\rm R}_{\alpha\beta} \equiv \gamma_{\alpha\beta} 
- \gamma^{\rm S}_{\alpha\beta}$; this perturbed metric is smooth on
the world line, and it is a solution to the vacuum field
equations. This elegant interpretation of the MiSaTaQuWa equations was  
proposed in 2002 by Steven Detweiler and Bernard F.\ Whiting
\cite{detweilerwhiting}. Quinn and Wald \cite{QW2} have shown that
under some conditions, the total work done by the gravitational
self-force is equal to the energy radiated (in gravitational waves) by
the particle.          

\subsection{Gauge dependence of the equations of motion} 
\label{18.7}

The equations of motion derived in the preceding subsection refer to a  
specific choice of gauge for the metric perturbation $h_{\alpha\beta}$
produced by a point particle of mass $m$. We indeed recall that back
at Eq.~(\ref{18.1.11}) we imposed the Lorenz gauge condition 
$\gamma^{\alpha\beta}_{\ \ \ ;\beta} = 0$ on the gravitational
potentials $\gamma_{\alpha\beta} \equiv h_{\alpha\beta} - \frac{1}{2}  
(g^{\gamma\delta} h_{\gamma\delta}) g_{\alpha\beta}$. By
virtue of this condition we found that the potentials satisfy the
wave equation of Eq.~(\ref{18.1.12}) in a background spacetime with
metric $g_{\alpha\beta}$. The hyperbolic nature of this equation
allowed us to identify the retarded solution as the physically
relevant solution, and the equations of motion were obtained by
removing the singular part of the retarded field. It seems clear that
the Lorenz condition is a most appropriate choice of gauge.      

Once the equations of motion have been formulated, however, the
freedom of performing a gauge transformation (either away from the 
Lorenz gauge, or within the class of Lorenz gauges) should be 
explored. A gauge transformation will affect the form of the equations
of motion: these must depend on the choice of coordinates, and there
is no reason to expect Eq.~(\ref{18.6.5}) to be invariant
under a gauge transformation. Our purpose in this subsection is 
to work out how the equations of motion change under such a
transformation. This issue was first examined by Barack and Ori
\cite{BO1}. 

We introduce a coordinate transformation of the form 
\begin{equation}
x^\alpha \to x^\alpha + \xi^\alpha, 
\label{18.7.1}
\end{equation} 
where $x^\alpha$ are the coordinates of the background spacetime, and 
$\xi^\alpha$ is a vector field that we take to be of order $m$. We
assume that $\xi^\alpha$ is smooth in a neighbourhood of the world
line $\gamma$. The coordinate transformation changes the background
metric according to 
\[
g_{\alpha\beta} \to g_{\alpha\beta} - \xi_{\alpha;\beta} -
\xi_{\beta;\alpha} + O(m^2), 
\]
and this change can be interpreted as a gauge transformation of the
metric perturbation created by the moving particle:  
\begin{equation}
h_{\alpha\beta} \to h_{\alpha\beta} - \xi_{\alpha;\beta} 
- \xi_{\beta;\alpha}. 
\label{18.7.2}
\end{equation} 
This, in turn, produces a change in the particle's acceleration:  
\begin{equation} 
a^\mu \to a^\mu + a[\xi]^\mu, 
\label{18.7.3}
\end{equation} 
where $a^\mu$ is the acceleration of Eq.~(\ref{18.6.5}) and
$a[\xi]^\mu$ is the ``gauge acceleration'' generated by the vector  
field $\xi^\alpha$.

To compute the gauge acceleration we substitute Eq.~(\ref{18.7.2})
into Eq.~(\ref{18.1.13}), and we simplify the result by invoking
Ricci's identity, $\xi_{\lambda;\nu\rho} - \xi_{\lambda;\rho\nu} 
= R_{\nu\rho\omega\lambda} \xi^\omega$, and the fact that $a^\mu =
O(m)$. The final expression is 
\begin{equation}
a[\xi]^\mu = \bigl( \delta^\mu_{\ \nu} + u^\mu u_\nu \bigr) \biggl( 
\frac{D^2 \xi^\nu}{d\tau^2} + R^\nu_{\ \rho\omega\lambda} u^\rho
\xi^\omega u^\lambda \biggr), 
\label{18.7.4}
\end{equation} 
where $D^2 \xi^\nu/d\tau^2 = (\xi^\nu_{\ ;\mu} u^\mu)_{;\rho} u^\rho$
is the second covariant derivative of $\xi^\nu$ in the direction of
the world line. The expression within the large brackets is familiar
from the equation of geodesic deviation, which states that this
quantity vanishes if $\xi^\mu$ is a deviation vector between two
neighbouring geodesics. Equation (\ref{18.7.3}), with
$a[\xi]^\mu$ given by Eq.~(\ref{18.7.4}), is therefore a generalized
version of this statement.

\section{Motion of a small black hole}
\label{19} 
    
\subsection{Matched asymptotic expansions} 
\label{19.1}

The derivation of the MiSaTaQuWa equations of motion presented in
Sec.~\ref{18} was framed within the paradigm introduced in
Secs.~\ref{16} and \ref{17} to describe the motion of a point scalar
charge, and a point electric charge, respectively. While this paradigm 
is well suited to fields that satisfy linear wave equations, it is not
the best conceptual starting point in the nonlinear context of general 
relativity. The linearization of the Einstein field equations with
respect to the small parameter $m$ did allow us to use the same
mathematical techniques as in Secs.~\ref{16} and \ref{17}, but the
validity of the perturbative method must be critically examined when
the gravitational potentials are allowed to be singular. So while 
Eq.~(\ref{18.6.5}) does indeed give the correct equations of motion
when $m$ is small, its previous derivation leaves much to be
desired. In this section I provide another derivation that is
entirely free of conceptual and technical pitfalls. Here the point
mass will be replaced by a nonrotating black hole, and the
perturbation's singular behaviour on the world line will be replaced 
by a well-behaved metric at the event horizon. We will use the
powerful technique of {\it matched asymptotic expansions}
\cite{manasse, kates, thornehartle, death, alvi, detweiler}.   

The problem presents itself with a clean separation of length scales,
and the method relies entirely on this. On the one hand we have the
length scale associated with the small black hole, which is set by its
mass $m$. On the other hand we have the length scale associated with
the background spacetime in which the black hole moves, which is set
by the radius of curvature $\cal R$; formally this is defined so that
a typical component of the background spacetime's Riemann tensor is
equal to $1/{\cal R}^2$ up to a numerical factor of order unity. We
demand that $m/{\cal R} \ll 1$. As before we assume that the
background spacetime contains no matter, so that its metric is a
solution to the Einstein field equations in vacuum. 

For example, suppose that our small black hole of mass $m$ is on an
orbit of radius $b$ around another black hole of mass $M$. Then 
${\cal R} \sim b \sqrt{b/M} > b$ and we take $m$ to be much smaller
than the orbital separation. Notice that the time scale over which the
background geometry changes is of the order of the orbital period 
$b \sqrt{b/M} \sim {\cal R}$, so that this does not constitute a
separate scale. Similarly, the inhomogeneity scale --- the length
scale over which the Riemann tensor of the background spacetime
changes --- is of order $b \sim {\cal R} \sqrt{M/b} < {\cal R}$ and
also does not constitute an independent scale. (In this discussion we
have considered $b/M$ to be of order unity, so as to represent
a strong-field, fast-motion situation.)  

\begin{figure}[b]
\vspace*{2.8in}
\special{hscale=35 vscale=35 hoffset=100.0 voffset=210.0
         angle = -90.0 psfile=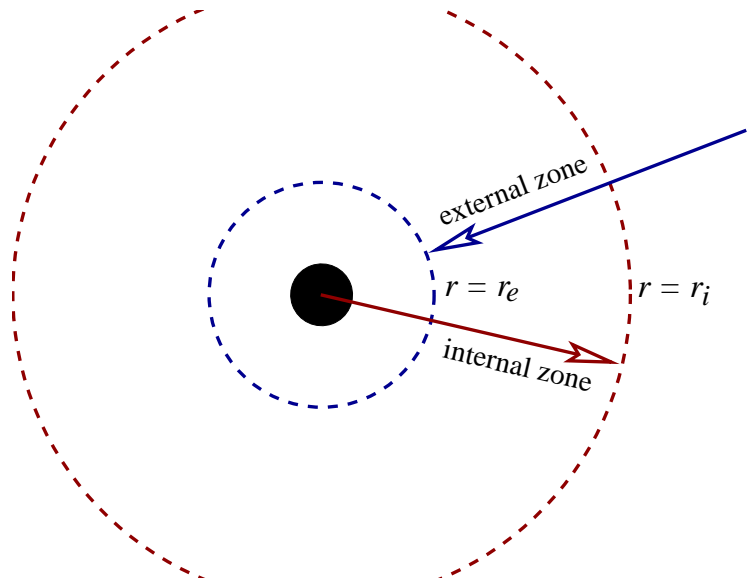}
\caption{A black hole, represented by the black disk, is immersed in
a background spacetime. The internal zone extends from $r=0$ to $r=r_i  
\ll {\cal R}$, while the external zone extends from $r=r_e \gg m$ to
$r=\infty$. When $m \ll {\cal R}$ there exists a buffer zone that
extends from $r=r_e$ to $r=r_i$. In the buffer zone $m/r$ and
$r/{\cal R}$ are both small.}   
\end{figure} 

Let $r$ be a meaningful measure of distance from the small black hole, 
and let us consider a region of spacetime defined by $r < r_i$, where
$r_i$ is a constant that is much smaller than ${\cal R}$. This
inequality defines a narrow world tube that surrounds the small black 
hole, and we shall call this region the {\it internal zone}; see
Fig.~10. In the internal zone the gravitational field is dominated by
the black hole, and the metric can be expressed as     
\begin{equation}
{\sf g}(\mbox{internal zone}) = g(\mbox{black hole}) 
+ H_1/{\cal R} + H_2/{\cal R}^2 + \cdots, 
\label{19.1.1}
\end{equation}
where $g(\mbox{black hole})$ is the metric of a nonrotating black hole
in isolation (as given by the unperturbed Schwarzschild solution),
while $H_1$ and $H_2$ are corrections associated with the
conditions in the external universe. The metric of Eq.~(\ref{19.1.1})
represents a black hole that is distorted by the tidal gravitational
field of the external universe, and $H_1$, $H_2$ are functions of $m$
and the spacetime coordinates that can be obtained by solving the
Einstein field equations. They must be such that the spacetime
possesses a regular event horizon near $r=2m$, and such that
${\sf g}(\mbox{internal zone})$ agrees with the metric of the external
universe --- the metric of the background spacetime in the absence of
the black hole --- when $r \gg m$. As we shall see in Sec.~\ref{19.2},
$H_1$ actually vanishes and the small correction $H_2/{\cal R}^2$ can
be obtained by employing the well-developed tools of black-hole
perturbation theory \cite{reggewheeler, vishveshwara, zerilli}.      

Consider now a region of spacetime defined by $r > r_e$, where $r_e$
is a constant that is much larger than $m$; this region will be called 
the {\it external zone} (see Fig.~10). In the external zone the
gravitational field is dominated by the conditions in the external
universe, and the metric can be expressed as 
\begin{equation}
{\sf g}(\mbox{external zone}) = g(\mbox{background spacetime}) 
+ m h_1 + m^2 h_2 + \cdots, 
\label{19.1.2}
\end{equation}
where $g(\mbox{background spacetime})$ is the unperturbed metric of
the background spacetime in which the black hole is moving, while
$h_1$ and $h_2$ are corrections associated with the hole's
presence; these are functions of ${\cal R}$ and the spacetime
coordinates that can be obtained by solving the Einstein
field equations. We shall truncate Eq.~(\ref{19.1.2}) to first order
in $m$, and $m h_1$ will be calculated in Sec.~\ref{19.3} by
linearizing the field equations about the metric of the background
spacetime. In the external zone the perturbation associated with the
presence of a black hole cannot be distinguished from the perturbation
produced by a point particle of the same mass, and $m h_1$ will
therefore be obtained by solving Eq.~(\ref{18.1.12}) in the background 
spacetime. 

The metric ${\sf g}(\mbox{external zone})$ returned by the procedure
described in the preceding paragraph is a functional of a world line
$\gamma$ that represents the motion of the small black hole in the
background spacetime. Our goal is to obtain a description of this
world line, in the form of equations of motion to be satisfied by the
black hole; these equations will be formulated in the background
spacetime.  It is important to understand that fundamentally, $\gamma$
exists only as an external-zone construct: It is only in the external
zone that the black hole can be thought of as moving on a world line;
in the internal zone the black hole is revealed as an extended object
and the notion of a world line describing its motion is no longer 
meaningful.  

Equations (\ref{19.1.1}) and (\ref{19.1.2}) give two different
expressions for the metric of the same spacetime; the first is valid
in the internal zone $r < r_i \ll {\cal R}$, while the second is valid
in the external zone $r > r_e \gg m$. The fact that ${\cal R} \gg m$
allows us to define a {\it buffer zone} in which $r$ is restricted to
the interval $r_e < r < r_i$. In the buffer zone $r$ is simultaneously
much larger than $m$ and much smaller than ${\cal R}$ --- a typical
value might be $\sqrt{m {\cal R}}$ --- and Eqs.~(\ref{19.1.1}),
(\ref{19.1.2}) are simultaneously valid. Since the two metrics are the
same up to a diffeomorphism, these expressions must agree. And since 
${\sf g}(\mbox{external zone})$ is a functional of a world line
$\gamma$ while ${\sf g}(\mbox{internal zone})$ contains no such
information, {\it matching the metrics necessarily determines the
motion of the small black hole in the background spacetime}. What we
have here is a beautiful implementation of the general observation
that the motion of self-gravitating bodies is determined by the 
Einstein field equations.      

It is not difficult to recognize that the metrics of
Eqs.~(\ref{19.1.1}), (\ref{19.1.2}) can be matched in the buffer
zone. When $r \gg m$ in the internal zone the metric of the
unperturbed black hole can be expanded as $g(\mbox{black hole}) = \eta
\oplus m/r \oplus m^2/r^2 \oplus \cdots$, where $\eta$ is the metric
of flat spacetime (in asymptotically inertial coordinates) and the
symbol $\oplus$ means ``and a term of the form\ldots''. On the other
hand, dimensional analysis dictates that $H_1/{\cal R}$ be of the form
$r/{\cal R} \oplus m/{\cal R} \oplus m^2 /(r {\cal R}) \oplus \cdots$
while $H_2/{\cal R}^2$ should be expressed as $r^2/{\cal R}^2 \oplus 
m r/{\cal R}^2 \oplus m^2/{\cal R}^2 \oplus \cdots$. Altogether we
obtain  
\begin{eqnarray} 
{\sf g}(\mbox{buffer zone}) &=& \eta \oplus m/r \oplus m^2/r^2 \oplus
\cdots  
\nonumber \\ & & \mbox{} \oplus
r/{\cal R} \oplus m/{\cal R} \oplus m^2 /(r {\cal R}) \oplus \cdots
\nonumber \\ & & \mbox{} \oplus
r^2/{\cal R}^2 \oplus m r/{\cal R}^2 \oplus m^2/{\cal R}^2 \oplus
\cdots  
\nonumber \\ & & \mbox{} \oplus
\cdots 
\label{19.1.3} 
\end{eqnarray} 
for the buffer-zone metric. If instead we approach the buffer zone
from the opposite side, letting $r$ be much smaller than ${\cal R}$ in
the external zone, we have that the metric of the background spacetime
can be expressed as $g(\mbox{background spacetime}) = \eta \oplus 
r/{\cal R} \oplus r^2/{\cal R}^2 \oplus \cdots$, where the expansion 
now uses world-line based coordinates such as the Fermi normal
coordinates of Sec.~\ref{8} or the retarded coordinates of
Sec.~\ref{9}. On dimensional grounds we also have $m h_1 = m/r \oplus
m/{\cal R} \oplus m r/{\cal R}^2 \oplus \cdots$ and $m^2 h_2 = m^2/r^2
\oplus m^2/(r{\cal R}) \oplus m^2/{\cal R}^2 \oplus
\cdots$. Altogether this gives   
\begin{eqnarray} 
{\sf g}(\mbox{buffer zone}) &=&    
\eta \oplus r/{\cal R} \oplus r^2/{\cal R}^2 \oplus \cdots
\nonumber \\ & & \mbox{} \oplus 
m/r \oplus m/{\cal R} \oplus m r/{\cal R}^2 \oplus \cdots
\nonumber \\ & & \mbox{} \oplus 
m^2/r^2 \oplus m^2/(r {\cal R}) \oplus m^2/{\cal R}^2 \oplus \cdots 
\nonumber \\ & & \mbox{} \oplus
\cdots 
\label{19.1.4} 
\end{eqnarray} 
for the buffer-zone metric. Apart from a different ordering of terms,
the metrics of Eqs.~(\ref{19.1.3}) and (\ref{19.1.4}) have identical
forms.  

Matching the metrics of Eqs.~(\ref{19.1.1}) and (\ref{19.1.2}) in the
buffer zone can be carried out in practice only after performing a 
transformation from the external coordinates used to express 
${\sf g}(\mbox{external zone})$ to the internal coordinates employed
for ${\sf g}(\mbox{internal zone})$. The details of this coordinate
transformation will be described in Sec.~\ref{19.4}, and the end
result of matching --- the MiSaTaQuWa equations of motion --- will be
revealed in Sec.~\ref{19.5}. 

\subsection{Metric in the internal zone} 
\label{19.2}   

To flesh out the ideas contained in the previous subsection we 
first calculate the internal-zone metric and replace
Eq.~(\ref{19.1.1}) by a more concrete expression. We recall that the
internal zone is defined by $r < r_i \ll {\cal R}$, where $r$ is a
suitable measure of distance from the black hole.  

We begin by expressing $g(\mbox{black hole})$, the Schwarzschild
metric of an isolated black hole of mass $m$, in terms of retarded
Eddington-Finkelstein coordinates $(\bar{u},\bar{r},\bar{\theta}^A)$,
where $\bar{u}$ is retarded time, $\bar{r}$ the usual areal radius,
and $\bar{\theta}^A = (\bar{\theta},\bar{\phi})$ are two angles
spanning the two-spheres of constant $\bar{u}$ and $\bar{r}$. The
metric is given by 
\begin{equation}
ds^2 = - f\, d\bar{u}^2 - 2\, d\bar{u} d\bar{r} + \bar{r}^2\,
d\bar{\Omega}^2, 
\qquad 
f = 1 - \frac{2m}{\bar{r}}, 
\label{19.2.1}
\end{equation}
where $d\bar{\Omega}^2 = \bar{\Omega}_{AB}\, d\bar{\theta}^A
d\bar{\theta}^B = d\bar{\theta}^2 + \sin^2\bar{\theta}\,
d\bar{\phi}^2$ is the line element on the unit two-sphere. In the
limit $r \gg m$ this metric achieves the asymptotic values  
\[
g_{\bar{u}\bar{u}} \to -1, \qquad 
g_{\bar{u}\bar{r}} = -1, \qquad
g_{\bar{u}\bar{A}} = 0, \qquad  
g_{\bar{A}\bar{B}} = \bar{r}^2\, \bar{\Omega}_{AB};
\]
these are appropriate for a black hole immersed in a flat
spacetime charted by retarded coordinates. 

The corrections $H_1$ and $H_2$ in Eq.~(\ref{19.1.1}) 
encode the information that our black hole is not isolated but
in fact immersed in an external universe whose metric becomes 
$g(\mbox{background spacetime})$ asymptotically. In the internal zone
the metric of the background spacetime can be expanded in powers of
$\bar{r}/{\cal R}$ and expressed in a form that can be directly
imported from Sec.~\ref{9}. If we assume for the moment that the
``world line'' $\bar{r} = 0$ has no acceleration in the background 
spacetime (a statement that will be justified shortly) then the   
asymptotic values of ${\sf g}(\mbox{internal zone})$
must be given by Eqs.~(\ref{9.8.15})--(\ref{9.8.18}):  
\[ 
{\sf g}_{\bar{u}\bar{u}} \to - 1 - \bar{r}^2 \bar{{\cal E}}^* 
+ O(\bar{r}^3/{\cal R}^3), \qquad 
{\sf g}_{\bar{u}\bar{r}} = - 1,
\]
\[
{\sf g}_{\bar{u}\bar{A}} \to 
\frac{2}{3} \bar{r}^3 \bigl( \bar{{\cal E}}^*_A   
+ \bar{{\cal B}}^*_A \bigr) + O(\bar{r}^4/{\cal R}^3), \qquad 
{\sf g}_{\bar{A}\bar{B}} \to \bar{r}^2 \bar{\Omega}_{AB} 
- \frac{1}{3} \bar{r}^4 \bigl( \bar{{\cal E}}^*_{AB} 
+ \bar{{\cal B}}^*_{AB} \bigr) + O(\bar{r}^5/{\cal R}^3),    
\]
where 
\begin{equation} 
\bar{{\cal E}}^* = {\cal E}_{ab} \bar{\Omega}^a \bar{\Omega}^b, 
\qquad 
\bar{{\cal E}}^*_A = {\cal E}_{ab} \bar{\Omega}^a_A \bar{\Omega}^b, 
\qquad
\bar{{\cal E}}^*_{AB} = 2 {\cal E}_{ab} \bar{\Omega}^a_A
\bar{\Omega}^b_B + \bar{{\cal E}}^* \bar{\Omega}_{AB} 
\label{19.2.2}
\end{equation}
and
\begin{equation}  
\bar{{\cal B}}^*_A = \varepsilon_{abc} \bar{\Omega}^a_A
\bar{\Omega}^b {\cal B}^c_{\ d} \bar{\Omega}^d,
\qquad
\bar{{\cal B}}^*_{AB} = 2 \varepsilon_{acd} \bar{\Omega}^c 
{\cal B}^d_{\ b} \bar{\Omega}^{(a}_A \bar{\Omega}^{b)}_B  
\label{19.2.3} 
\end{equation}  
are the tidal gravitational fields that were first introduced in
Sec.~\ref{9.8}. Recall that $\bar{\Omega}^a =
(\sin\bar{\theta}\cos\bar{\phi}, \sin\bar{\theta}\sin\bar{\phi}, 
\linebreak \cos\bar{\theta})$ and $\bar{\Omega}^a_A = 
\partial \bar{\Omega}^a/\partial \bar{\theta}^A$. Apart from an
angular dependence made explicit by these relations, the tidal fields
depend on $\bar{u}$ through the frame components ${\cal E}_{ab} \equiv
R_{a0b0} = O(1/{\cal R}^2)$ and ${\cal B}^a_{\ b} \equiv \frac{1}{2}
\varepsilon^{acd} R_{0bcd} = O(1/{\cal R}^2)$ of the Riemann tensor. 
(This is the Riemann tensor of the background
spacetime evaluated at $\bar{r}=0$.) Notice that we have incorporated
the fact that the Ricci tensor vanishes at $\bar{r}=0$: the black hole
moves in a vacuum spacetime.     

The modified asymptotic values lead us to the following ansatz for the
internal-zone metric: 
\begin{eqnarray} 
{\sf g}_{\bar{u}\bar{u}} &=& - f \bigr[ 1 + \bar{r}^2 e_1(\bar{r})  
\bar{\cal E}^* \bigr] + O(\bar{r}^3/{\cal R}^3), 
\label{19.2.4} \\ 
{\sf g}_{\bar{u}\bar{r}} &=& - 1,
\label{19.2.5} \\  
{\sf g}_{\bar{u}\bar{A}} &=& \frac{2}{3} \bar{r}^3 \bigl[ e_2(\bar{r})   
\bar{\cal E}^*_A + b_2(\bar{r}) \bar{\cal B}^*_A \bigr] 
+ O(\bar{r}^4/{\cal R}^3),
\label{19.2.6} \\   
{\sf g}_{\bar{A}\bar{B}} &=& \bar{r}^2 \bar{\Omega}_{AB} 
- \frac{1}{3} \bar{r}^4    
\big[ e_3(\bar{r}) \bar{\cal E}^*_{AB} + b_3(\bar{r}) 
\bar{\cal B}^*_{AB} \bigr] + O(\bar{r}^5/{\cal R}^3).
\label{19.2.7}
\end{eqnarray}
The five unknown functions $e_1$, $e_2$, $e_3$, $b_2$, and $b_3$ can
all be determined by solving the Einstein field equations; they must
all approach unity when $r \gg m$ and be well-behaved at $r=2m$
(so that the tidally distorted black hole will have a nonsingular
event horizon). It is clear from Eqs.~(\ref{19.2.4})--(\ref{19.2.7}) 
that the assumed deviation of ${\sf g}(\mbox{internal zone})$ with
respect to $g(\mbox{black hole})$ scales as $1/{\cal R}^2$. It is
therefore of the form of Eq.~(\ref{19.1.1}) with $H_1 = 0$. The fact
that $H_1$ vanishes comes as a consequence of our previous assumption
that the ``world line'' $\bar{r} = 0$ has a zero acceleration in the
background spacetime; a nonzero acceleration of order $1/{\cal R}$
would bring terms of order $1/{\cal R}$ to the metric, and $H_1$ would
then be nonzero.    

Why is the assumption of no acceleration justified? As I shall explain
in the next paragraph (and you might also refer back to the discussion
of Sec.~\ref{18.7}), the reason is simply that it reflects a choice of
coordinate system: setting the acceleration to zero amounts to
adopting a specific --- and convenient --- gauge condition. This gauge
differs from the Lorenz gauge adopted in Sec.~\ref{18}, and it will be
our choice in this subsection only; in the following subsections we
will return to the Lorenz gauge, and the acceleration will be seen to
return to its standard MiSaTaQuWa expression.        

Inspection of Eqs.~(\ref{19.2.2}) and (\ref{19.2.3}) reveals that the 
angular dependence of the metric perturbation is generated entirely by 
scalar, vectorial, and tensorial spherical harmonics of degree
$\ell=2$. In particular, $H_2$ contains no $\ell=0$ and $\ell=1$
modes, and this statement reflects a choice of gauge
condition. Zerilli has shown \cite{zerilli} that a perturbation of the
Schwarzschild spacetime with $\ell = 0$ corresponds to a shift in the
mass parameter. As Thorne and Hartle have shown \cite{thornehartle}, a
black hole interacting with its environment will undergo a change of
mass, but this effect is of order $m^3/{\cal R}^2$ and thus beyond the
level of accuracy of our calculations. There is therefore no need to
include $\ell = 0$ terms in $H_2$. Similarly, it was shown by Zerilli
that odd-parity perturbations of degree $\ell=1$ correspond to a shift
in the black hole's angular-momentum parameters. As Thorne and Hartle
have shown, a change of angular momentum is quadratic in the hole's
angular momentum, and we can ignore this effect when dealing with a
nonrotating black hole. There is therefore no need to include
odd-parity, $\ell=1$ terms in $H_2$. Finally, Zerilli has shown
that in a vacuum spacetime, even-parity perturbations of degree
$\ell=1$ correspond to a change of coordinate system --- these
modes are pure gauge. Since we have the freedom to adopt any gauge
condition, we can exclude even-parity, $\ell=1$ terms
from the perturbed metric. This leads us to
Eqs.~(\ref{19.2.4})--(\ref{19.2.7}), which contain only $\ell=2$
perturbation modes; the even-parity modes are contained in those terms
that involve ${\cal E}_{ab}$, while the odd-parity modes are
associated with ${\cal B}_{ab}$. The perturbed metric contains also  
higher multipoles, but those come at a higher order in $1/{\cal R}$; 
for example, the terms of order $1/{\cal R}^3$ include $\ell=3$ 
modes. We conclude that Eqs.~(\ref{19.2.4})--(\ref{19.2.7}) is a 
sufficiently general ansatz for the perturbed metric in the internal
zone.  

There remains the task of finding the functions $e_1$, $e_2$, $e_3$,
$b_2$, and $b_3$. For this it is sufficient to take, say, 
${\cal E}_{12} = {\cal E}_{21}$ and ${\cal B}_{12} = {\cal B}_{21}$ as
the only nonvanishing components of the tidal fields 
${\cal E}_{ab}$ and ${\cal B}_{ab}$. And since the equations for
even-parity and odd-parity perturbations decouple, each case can be
considered separately. Including only even-parity perturbations,  
Eqs.~(\ref{19.2.4})--(\ref{19.2.7}) become  
\[
{\sf g}_{\bar{u}\bar{u}} = - f \bigr( 1 + \bar{r}^2 e_1   
{\cal E}_{12} \sin^2\bar{\theta} \sin2\bar\phi \bigr), 
\qquad 
{\sf g}_{\bar{u}\bar{r}} = - 1, 
\qquad 
{\sf g}_{\bar{u}\bar{\theta}} = \frac{2}{3} \bar{r}^3 e_2   
{\cal E}_{12} \sin\bar{\theta} \cos\bar{\theta} \sin2\bar{\phi}, 
\]
\[ 
{\sf g}_{\bar{u}\bar{\phi}} = \frac{2}{3} \bar{r}^3 e_2   
{\cal E}_{12} \sin^2\bar{\theta} \cos2\bar{\phi}, 
\qquad 
{\sf g}_{\bar{\theta}\bar{\theta}} = \bar{r}^2 - \frac{1}{3} \bar{r}^4      
e_3 {\cal E}_{12} (1 + \cos^2\bar{\theta}) \sin2\bar{\phi}, 
\]
\[
{\sf g}_{\bar{\theta}\bar{\phi}} = - \frac{2}{3} \bar{r}^4      
e_3 {\cal E}_{12} \sin\bar{\theta}\cos\bar{\theta} \cos2\bar{\phi},  
\qquad 
{\sf g}_{\bar{\phi}\bar{\phi}} = \bar{r}^2\sin^2\bar{\theta} +
\frac{1}{3} \bar{r}^4 e_3 {\cal E}_{12} \sin^2\bar{\theta} (1 +
\cos^2\bar{\theta}) \sin2\bar{\phi}. 
\] 
This metric is then substituted into the vacuum Einstein field
equations, ${\sf G}_{\alpha\beta} = 0$. Calculating the Einstein
tensor is simplified by linearizing with respect to ${\cal E}_{12}$
and discarding its derivatives with respect to $\bar{u}$: Since the
time scale over which ${\cal E}_{ab}$ changes is of order ${\cal R}$,
the ratio between temporal and spatial derivatives is of order
$\bar{r}/{\cal R}$ and therefore small in the internal zone; the
temporal derivatives can be consistently neglected. The field
equations produce ordinary differential equations to be satisfied by 
the functions $e_1$, $e_2$, and $e_3$. Those are easily decoupled, and
demanding that the functions all approach unity as $r \to \infty$ and
be well-behaved at $r=2m$ yields the unique solutions 
\begin{equation} 
e_1(\bar{r}) = e_2(\bar{r}) = f, \qquad  
e_3(\bar{r}) = 1 - \frac{2m^2}{\bar{r}^2}. 
\label{19.2.8}
\end{equation} 
Switching now to odd-parity perturbations,  
Eqs.~(\ref{19.2.4})--(\ref{19.2.7}) become  
\[
{\sf g}_{\bar{u}\bar{u}} = - f 
\qquad 
{\sf g}_{\bar{u}\bar{r}} = - 1, 
\qquad 
{\sf g}_{\bar{u}\bar{\theta}} = -\frac{2}{3} \bar{r}^3 b_2   
{\cal B}_{12} \sin\bar{\theta} \cos2\bar{\phi}, 
\qquad 
{\sf g}_{\bar{u}\bar{\phi}} = \frac{2}{3} \bar{r}^3 b_2   
{\cal B}_{12} \sin^2\bar{\theta} \cos\bar{\theta} \sin2\bar{\phi},  
\]
\[ 
{\sf g}_{\bar{\theta}\bar{\theta}} = \bar{r}^2 + \frac{2}{3} \bar{r}^4      
b_3 {\cal B}_{12} \cos\bar{\theta} \cos2\bar{\phi}, 
\qquad
{\sf g}_{\bar{\theta}\bar{\phi}} = - \frac{1}{3} \bar{r}^4      
b_3 {\cal B}_{12} \sin\bar{\theta} (1 + \cos^2\bar{\theta})
\sin2\bar{\phi},   
\]
\[ 
{\sf g}_{\bar{\phi}\bar{\phi}} = \bar{r}^2\sin^2\bar{\theta} -
\frac{2}{3} \bar{r}^4 b_3 {\cal B}_{12} \sin^2\bar{\theta}
\cos\bar{\theta} \cos2\bar{\phi}.  
\] 
Following the same procedure, we arrive at 
\begin{equation} 
b_2(\bar{r}) = f, \qquad  
b_3(\bar{r}) = 1. 
\label{19.2.9}
\end{equation} 
Substituting Eqs.~(\ref{19.2.8}) and (\ref{19.2.9}) into
Eqs.~(\ref{19.2.4})--(\ref{19.2.7}) returns our final expression for  
the metric in the internal zone. 

It shall prove convenient to transform 
${\sf g}(\mbox{internal zone})$ from the quasi-spherical coordinates
$(\bar{r},\bar{\theta}^A)$ to a set of quasi-Cartesian coordinates
$\bar{x}^a = \bar{r} \bar{\Omega}^a(\bar{\theta}^A)$. The
transformation rules are worked out in Sec.~\ref{9.7} and further
illustrated in Sec.~\ref{9.8}. This gives 
\begin{eqnarray} 
{\sf g}_{\bar{u}\bar{u}} &=& -f \bigl(1 + \bar{r}^2 f \bar{\cal E}^*  
\bigr) + O(\bar{r}^3/{\cal R}^3), 
\label{19.2.10} \\ 
{\sf g}_{\bar{u}\bar{a}} &=& -\bar{\Omega}_a + \frac{2}{3} \bar{r}^2 
f \bigl( \bar{\cal E}^*_a + \bar{\cal B}^*_a \bigr) 
+ O(\bar{r}^3/{\cal R}^3),  
\label{19.2.11} \\ 
{\sf g}_{\bar{a}\bar{b}} &=& \delta_{ab} 
- \bar{\Omega}_a \bar{\Omega}_b  
- \frac{1}{3} \bar{r}^2 \biggl( 1 - 2\frac{m^2}{\bar{r}^2} \biggr)  
\bar{\cal E}^*_{ab} - \frac{1}{3} \bar{r}^2 \bar{\cal B}^*_{ab} 
+ O(\bar{r}^3/{\cal R}^3), 
\label{19.2.12}
\end{eqnarray} 
where $f = 1 -2m/\bar{r}$ and where the tidal fields 
\begin{eqnarray}  
\bar{\cal E}^* &=& {\cal E}_{ab} \bar{\Omega}^a \bar{\Omega}^b, 
\label{19.2.13} \\ 
\bar{\cal E}^*_a &=& \bigl(\delta_a^{\ b} - \bar{\Omega}_a
\bar{\Omega}^b \bigr) {\cal E}_{bc} \bar{\Omega}^c, 
\label{19.2.14} \\ 
\bar{\cal E}^*_{ab} &=& 2 {\cal E}_{ab} 
- 2\bar{\Omega}_a {\cal E}_{bc} \bar{\Omega}^c  
- 2\bar{\Omega}_b {\cal E}_{ac} \bar{\Omega}^c
+ (\delta_{ab} + \bar{\Omega}_a \bar{\Omega}_b) \bar{\cal E}^*, 
\label{19.2.15} \\ 
\bar{\cal B}^*_a &=& \varepsilon_{abc} \bar{\Omega}^b {\cal B}^c_{\ d}
\bar{\Omega}^d,  
\label{19.2.16} \\ 
\bar{\cal B}^*_{ab} &=& \varepsilon_{acd} \bar{\Omega}^c 
{\cal B}^d_{\ e} \bigl(\delta^e_{\ b} - \bar{\Omega}^e \bar{\Omega}_b
\bigr) + \varepsilon_{bcd} \bar{\Omega}^c {\cal B}^d_{\ e} 
\bigl(\delta^e_{\ a} - \bar{\Omega}^e \bar{\Omega}_a \bigr) 
\label{19.2.17}
\end{eqnarray}    
were first introduced in Sec.~\ref{9.8}. The metric of
Eqs.~(\ref{19.2.10})--(\ref{19.2.12}) represents the spacetime
geometry of a black hole immersed in an external universe and 
distorted by its tidal gravitational fields.  

\subsection{Metric in the external zone} 
\label{19.3}

We next move on to the external zone and seek to replace
Eq.~(\ref{19.1.2}) by a more concrete expression; recall that the  
external zone is defined by $m \ll r_e < r$. As was pointed out in
Sec.~\ref{19.1}, in the external zone the gravitational perturbation
associated with the presence of a black hole cannot be distinguished
from the perturbation produced by a point particle of the same mass;
it can therefore be obtained by solving Eq.~(\ref{18.1.12}) in a
background spacetime with metric 
$g(\mbox{background spacetime})$. The external-zone metric is
decomposed as 
\begin{equation}
{\sf g}_{\alpha\beta} = g_{\alpha\beta} + h_{\alpha\beta}, 
\label{19.3.1}
\end{equation}
where $g_{\alpha\beta}$ is the metric of the background spacetime and
$h_{\alpha\beta} = O(m)$ is the perturbation; we shall work
consistently to first order in $m$ and systematically discard all
terms of higher order. We relate $h_{\alpha\beta}$ to trace-reversed
potentials $\gamma_{\alpha\beta}$, 
\begin{equation}
h_{\alpha\beta} = \gamma_{\alpha\beta} - \frac{1}{2} \bigl(
g^{\gamma\delta} \gamma_{\gamma\delta} \bigr) g_{\alpha\beta}, 
\label{19.3.2}
\end{equation}
and solving the linearized field equations produces 
\begin{equation} 
\gamma_{\alpha\beta}(x) = 4 m \int_\gamma 
G_{+\alpha\beta\mu\nu}(x,z) u^\mu u^\nu\, d\tau,    
\label{19.3.3}
\end{equation}
where $z^\mu(\tau)$ gives the description of the world line
$\gamma$, $\tau$ is proper time in the background spacetime, 
$u^\mu = d z^\mu/d\tau$ is the four-velocity, and 
$G^{\ \alpha\beta}_{+\ \mu\nu}(x,z)$ is the retarded Green's function
associated with Eq.~(\ref{18.1.12}); the potentials of
Eq.~(\ref{19.3.3}) satisfy the Lorenz-gauge condition 
$\gamma^{\alpha\beta}_{\ \ \ ;\beta} = 0$. As was pointed out in 
Sec.~\ref{19.1}, $\gamma_{\alpha\beta}$ (and therefore
$h_{\alpha\beta}$) are functionals of a world line $\gamma$ that will
be determined by matching ${\sf g}(\mbox{external zone})$ to 
${\sf g}(\mbox{internal zone})$.   

We now place ourselves in the buffer zone (where $m \ll r \ll 
{\cal R}$ and where the matching will take place) and work toward
expressing ${\sf g}(\mbox{external zone})$ as an expansion in powers
of $r/{\cal R}$. For this purpose we will adopt the retarded
coordinates $(u,r\Omega^a)$ of Sec.~\ref{9} and rely on the machinery
developed there.   

We begin with $g_{\alpha\beta}$, the metric of the background
spacetime. We have seen in Sec.~\ref{9.8} that if the world line
$\gamma$ is a geodesic, if the vectors $\base{\mu}{a}$ are
parallel transported on the world line, and if the Ricci tensor
vanishes on $\gamma$, then the metric takes the form given by 
Eqs.~(\ref{9.8.12})--(\ref{9.8.14}). This form, however, is too
restrictive for our purposes: We must allow $\gamma$ to have an
acceleration, and allow the basis vectors to be transported in the
most general way compatible with their orthonormality property; this 
transport law is given by Eq.~(\ref{9.1.1}),   
\begin{equation}
\frac{D \base{\mu}{a}}{d \tau} = a_a u^{\mu} 
+ \omega_a^{\ b} \base{\mu}{b},  
\label{19.3.4} 
\end{equation}
where $a_a(\tau) = a_{\mu} \base{\mu}{a}$ are the frame components of
the acceleration vector $a^\mu = D u^\mu/d\tau$, and
$\omega_{ab}(\tau) = -\omega_{ba}(\tau)$ is a rotation tensor to be 
determined. Anticipating that $a_a$ and $\omega_{ab}$ will both be
proportional to $m$, we express the metric of the background spacetime 
as  
\begin{eqnarray}  
g_{uu} &=& - 1 - 2 r a_a \Omega^a - r^2 {\cal E}^* 
+ O(r^3/{\cal R}^3), 
\label{19.3.5} \\ 
g_{ua} &=& -\Omega_a + r \bigr(\delta_a^{\ b} - \Omega_a
\Omega^b \bigr) a_b - r \omega_{ab} \Omega^b + \frac{2}{3} r^2 \bigl(
{\cal E}^*_a  + {\cal B}^*_a \bigr) + O(r^3/{\cal R}^3),  
\label{19.3.6} \\ 
g_{ab} &=& \delta_{ab} - \Omega_a \Omega_b - \frac{1}{3} r^2  
\bigl( {\cal E}^*_{ab} + {\cal B}^*_{ab} \bigr) + O(r^3/{\cal R}^3),   
\label{19.3.7}
\end{eqnarray}   
where ${\cal E}^*$, ${\cal E}^*_a$, ${\cal E}^*_{ab}$, ${\cal B}^*_a$,
and ${\cal B}^*_{ab}$ are the tidal gravitational fields first
introduced in Sec.~\ref{9.8}. The metric of
Eqs.~(\ref{19.3.5})--(\ref{19.3.7}) is obtained from the general form
of Eqs.~(\ref{9.6.3})--(\ref{9.6.5}) by neglecting quadratic terms in 
$a_a$ and $\omega_{ab}$ and specializing to a zero Ricci tensor. 

To express the perturbation $h_{\alpha\beta}$ as an expansion in
powers of $r/{\cal R}$ we first go back to Eq.~(\ref{18.2.4}) and
rewrite it in the form   
\begin{equation}
\gamma_{\alpha\beta}(x) = \frac{4m}{r} 
U_{\alpha\beta\gamma'\delta'}(x,x') u^{\gamma'} u^{\delta'} 
+ \gamma^{\rm tail}_{\alpha\beta}(x), 
\label{19.3.8}
\end{equation}
in which primed indices refer to the retarded point $x' \equiv z(u)$   
associated with $x$, and 
\begin{eqnarray} 
\gamma^{\rm tail}_{\alpha\beta}(x) &=& 
4 m \int_{\tau_<}^u V_{\alpha\beta\mu\nu}(x,z) u^\mu u^\nu\,
d\tau + 4 m \int_{-\infty}^{\tau_<} 
G_{+\alpha\beta\mu\nu}(x,z) u^\mu u^\nu\, d\tau 
\nonumber \\ 
&\equiv& 4 m \int_{-\infty}^{u^-} 
G_{+\alpha\beta\mu\nu}(x,z) u^\mu u^\nu\, d\tau 
\label{19.3.9}
\end{eqnarray}
is the ``tail part'' of the gravitational potentials (recall that
$\tau_<$ is the proper time at which $\gamma$ enters $x$'s normal
convex neighbourhood from the past). We next expand the first term on
the right-hand side of Eq.~(\ref{19.3.8}) with the help of
Eq.~(\ref{18.3.3}), and the tail term is expanded using
Eq.~(\ref{5.3.1}) in which we substitute Eq.~(\ref{18.3.6}) and the
familiar relation $\sigma^{\alpha'} = -r 
(u^{\alpha'} + \Omega^a \base{\alpha'}{a})$. This gives  
\begin{equation} 
\gamma_{\alpha\beta}(x) = g^{\alpha'}_{\ \alpha} g^{\beta'}_{\ \beta} 
\biggl[ \frac{4m}{r} u_{\alpha'} u_{\beta'} 
+ \gamma^{\rm tail}_{\alpha'\beta'}  
+ r \gamma^{\rm tail}_{\alpha'\beta'\gamma'} \bigl( u^{\gamma'} +
\Omega^c \base{\gamma'}{c} \bigr) + O(mr^2/{\cal R}^3) \biggr], 
\label{19.3.10}
\end{equation}
where $\gamma^{\rm tail}_{\alpha'\beta'}$ is the tensor of
Eq.~(\ref{19.3.9}) evaluated at $x'$, and where
\begin{equation} 
\gamma^{\rm tail}_{\alpha'\beta'\gamma'}(x')  
= 4m \int_{-\infty}^{u^-} \nabla_{\gamma'}
G_{+\alpha'\beta'\mu\nu}(x',z) u^\mu u^\nu\, d\tau 
\label{19.3.11}
\end{equation}
was first defined by Eq.~(\ref{18.5.21}).   

At this stage we introduce the trace-reversed fields 
\begin{eqnarray}
h^{\rm tail}_{\alpha'\beta'}(x') &=& 4m \int_{-\infty}^{u^-} 
\biggl( G_{+\alpha'\beta'\mu\nu} - \frac{1}{2} g_{\alpha'\beta'} 
G^{\ \ \delta'}_{+\ \delta'\mu\nu} \biggr)(x',z) u^\mu u^\nu\, d\tau,  
\label{19.3.12} \\ 
h^{\rm tail}_{\alpha'\beta'\gamma'}(x') &=& 4m \int_{-\infty}^{u^-}  
\nabla_{\gamma'} \biggl( G_{+\alpha'\beta'\mu\nu} - \frac{1}{2}
g_{\alpha'\beta'} G^{\ \ \delta'}_{+\ \delta'\mu\nu} \biggr)(x',z) 
u^\mu u^\nu\, d\tau  
\label{19.3.13}
\end{eqnarray}
and recognize that the metric perturbation obtained from
Eqs.~(\ref{19.3.2}) and (\ref{19.3.10}) is 
\begin{equation} 
h_{\alpha\beta}(x) = g^{\alpha'}_{\ \alpha} g^{\beta'}_{\ \beta} 
\biggl[ \frac{2m}{r} \Bigl( 2u_{\alpha'} u_{\beta'} 
+ g_{\alpha'\beta'} \Bigr) + h^{\rm tail}_{\alpha'\beta'}  
+ r h^{\rm tail}_{\alpha'\beta'\gamma'} \bigl( u^{\gamma'} +
\Omega^c \base{\gamma'}{c} \bigr) + O(mr^2/{\cal R}^3) \biggr].
\label{19.3.14}
\end{equation}
This is the desired expansion of the metric perturbation in powers of
$r/{\cal R}$. Our next task will be to calculate the components of
this tensor in the retarded coordinates $(u,r\Omega^a)$. 

The first step of this computation is to decompose $h_{\alpha\beta}$
in the tetrad $(\base{\alpha}{0},\base{\alpha}{a})$ that is obtained
by parallel transport of $(u^{\alpha'},\base{\alpha'}{a})$ on the null
geodesic that links $x$ to its corresponding retarded point $x' \equiv
z(u)$ on the world line. (The vectors are parallel transported in the
background spacetime.) The projections are 
\begin{eqnarray}
\hspace*{-15pt} h_{00}(u,r,\Omega^a) &\equiv& 
h_{\alpha\beta} \base{\alpha}{0} \base{\beta}{0} 
= \frac{2m}{r} + h^{\rm tail}_{00}(u) + r \bigl[ h^{\rm tail}_{000}(u)
+ h^{\rm tail}_{00c}(u) \Omega^c \bigr] + O(mr^2/{\cal R}^3), 
\label{19.3.15} \\ 
\hspace*{-15pt} h_{0b}(u,r,\Omega^a) &\equiv& 
h_{\alpha\beta} \base{\alpha}{0} \base{\beta}{b} 
= h^{\rm tail}_{0b}(u) + r \bigl[ h^{\rm tail}_{0b0}(u)
+ h^{\rm tail}_{0bc}(u) \Omega^c \bigr] + O(mr^2/{\cal R}^3), 
\label{19.3.16} \\ 
\hspace*{-15pt} h_{ab}(u,r,\Omega^a) &\equiv& 
h_{\alpha\beta} \base{\alpha}{a} \base{\beta}{b} 
= \frac{2m}{r}\delta_{ab} + h^{\rm tail}_{ab}(u) 
+ r \bigl[ h^{\rm tail}_{ab0}(u) 
+ h^{\rm tail}_{abc}(u) \Omega^c \bigr] + O(mr^2/{\cal R}^3); 
\label{19.3.17} 
\end{eqnarray} 
on the right-hand side we have the frame components of 
$h^{\rm tail}_{\alpha'\beta'}$ and 
$h^{\rm tail}_{\alpha'\beta'\gamma'}$ taken with respect
to the tetrad $(u^{\alpha'},\base{\alpha'}{a})$; these 
are functions of retarded time $u$ only. 

The perturbation is now expressed as  
\[
h_{\alpha\beta} = h_{00} \base{0}{\alpha} \base{0}{\beta} 
+ h_{0b} \bigl( \base{0}{\alpha} \base{b}{\beta} 
+ \base{b}{\alpha} \base{0}{\beta} \bigr)   
+ h_{ab} \base{a}{\alpha} \base{b}{\beta} 
\]
and its components are obtained by involving Eqs.~(\ref{9.6.1}) and 
(\ref{9.6.2}), which list the components of the tetrad vectors in the 
retarded coordinates; this is the second (and longest) step of the 
computation. Noting that $a_a$ and $\omega_{ab}$ can both be set equal
to zero in these equations (because they would produce negligible
terms of order $m^2$ in $h_{\alpha\beta}$), and that $S_{ab}$, $S_a$,
and $S$ can all be expressed in terms of the tidal fields 
${\cal E}^*$, ${\cal E}^*_a$, ${\cal E}^*_{ab}$, ${\cal B}^*_a$, and  
${\cal B}^*_{ab}$ using Eqs.~(\ref{9.8.9})--(\ref{9.8.11}), we arrive
at   
\begin{eqnarray} 
h_{uu} &=& \frac{2m}{r} + h^{\rm tail}_{00} + r \bigl( 2m {\cal E}^* 
+ h^{\rm tail}_{000} + h^{\rm tail}_{00a} \Omega^a \bigr) 
+ O(mr^2/{\cal R}^3), 
\label{19.3.18} \\ 
h_{ua} &=& \frac{2m}{r}\Omega_a + h^{\rm tail}_{0a} 
+ \Omega_a h^{\rm tail}_{00} + r \biggl[ 2m {\cal E}^* \Omega_a 
+ \frac{2m}{3} \bigl( {\cal E}^*_a + {\cal B}^*_a \bigr)    
\nonumber \\ & & \mbox{} 
+ h^{\rm tail}_{0a0} + \Omega_a h^{\rm tail}_{000} 
+ h^{\rm tail}_{0ab} \Omega^b + \Omega_a h^{\rm tail}_{00b} \Omega^b
\biggr] 
+ O(mr^2/{\cal R}^3),
\label{19.3.19} \\ 
h_{ab} &=& \frac{2m}{r} \bigl(\delta_{ab} + \Omega_a\Omega_b \bigr) 
+ \Omega_a \Omega_b h^{\rm tail}_{00} + \Omega_a h^{\rm tail}_{0b} 
+ \Omega_b h^{\rm tail}_{0a} + h^{\rm tail}_{ab} 
\nonumber \\ & & \mbox{} 
+ r \biggl[ -\frac{2m}{3} \Bigl( {\cal E}^*_{ab} 
+ \Omega_a {\cal E}^*_b + {\cal E}^*_a \Omega_b + {\cal B}^*_{ab} 
+ \Omega_a {\cal B}^*_b + \Omega_b {\cal B}^*_a \Bigr)    
+ \Omega_a \Omega_b \bigl( h^{\rm tail}_{000} + h^{\rm tail}_{00c}
\Omega^c \bigr) 
\nonumber \\ & & \mbox{} 
+ \Omega_a \bigl( h^{\rm tail}_{0b0} 
+ h^{\rm tail}_{0bc} \Omega^c \bigr) + \Omega_b 
\bigl( h^{\rm tail}_{0a0} + h^{\rm tail}_{0ac} \Omega^c \bigr) 
+ \bigl( h^{\rm tail}_{ab0} + h^{\rm tail}_{abc} \Omega^c \bigr)
\biggr] + O(mr^2/{\cal R}^3).
\label{19.3.20}
\end{eqnarray} 
These are the {\it coordinate components} of the metric perturbation 
$h_{\alpha\beta}$ in the retarded coordinates $(u,r\Omega^a)$,
expressed in terms of {\it frame components} of the tail fields 
$h^{\rm tails}_{\alpha'\beta'}$ and 
$h^{\rm tails}_{\alpha'\beta'\gamma'}$. The perturbation is expanded 
in powers of $r/{\cal R}$ and it also involves the tidal gravitational
fields of the background spacetime. 

The external-zone metric is obtained by adding $g_{\alpha\beta}$ as
given by Eqs.~(\ref{19.3.5})--(\ref{19.3.7}) to $h_{\alpha\beta}$ as
given by Eqs.~(\ref{19.3.18})--(\ref{19.3.20}). The final result is 
\begin{eqnarray} 
{\sf g}_{uu} &=& -1 - r^2 {\cal E}^* + O(r^3/{\cal R}^3) 
\nonumber \\ & & \mbox{} 
+ \frac{2m}{r} + h^{\rm tail}_{00} + r \bigl( 2m {\cal E}^* 
- 2 a_a \Omega^a + h^{\rm tail}_{000} + h^{\rm tail}_{00a} \Omega^a
\bigr) + O(mr^2/{\cal R}^3), 
\label{19.3.21} \\ 
{\sf g}_{ua} &=& -\Omega_a + \frac{2}{3} r^2 \bigl(
{\cal E}^*_a  + {\cal B}^*_a \bigr) + O(r^3/{\cal R}^3)
\nonumber \\ & & \mbox{} 
+ \frac{2m}{r}\Omega_a + h^{\rm tail}_{0a} 
+ \Omega_a h^{\rm tail}_{00} + r \biggl[ 2m {\cal E}^* \Omega_a 
+ \frac{2m}{3} \bigl( {\cal E}^*_a + {\cal B}^*_a \bigr)    
+ \bigr(\delta_a^{\ b} - \Omega_a \Omega^b \bigr) a_b 
\nonumber \\ & & \mbox{} 
- \omega_{ab} \Omega^b + h^{\rm tail}_{0a0} 
+ \Omega_a h^{\rm tail}_{000} + h^{\rm tail}_{0ab} \Omega^b 
+ \Omega_a h^{\rm tail}_{00b} \Omega^b \biggr] 
+ O(mr^2/{\cal R}^3),
\label{19.3.22} \\ 
{\sf g}_{ab} &=& \delta_{ab} - \Omega_a \Omega_b - \frac{1}{3} r^2   
\bigl( {\cal E}^*_{ab} + {\cal B}^*_{ab} \bigr) + O(r^3/{\cal R}^3) 
\nonumber \\ & & \mbox{}
+ \frac{2m}{r} \bigl(\delta_{ab} + \Omega_a\Omega_b \bigr) 
+ \Omega_a \Omega_b h^{\rm tail}_{00} + \Omega_a h^{\rm tail}_{0b} 
+ \Omega_b h^{\rm tail}_{0a} + h^{\rm tail}_{ab} 
\nonumber \\ & & \mbox{} 
+ r \biggl[ -\frac{2m}{3} \Bigl( {\cal E}^*_{ab} 
+ \Omega_a {\cal E}^*_b + {\cal E}^*_a \Omega_b + {\cal B}^*_{ab} 
+ \Omega_a {\cal B}^*_b + \Omega_b {\cal B}^*_a \Bigr)    
+ \Omega_a \Omega_b \bigl( h^{\rm tail}_{000} + h^{\rm tail}_{00c}
\Omega^c \bigr) 
\nonumber \\ & & \mbox{} 
+ \Omega_a \bigl( h^{\rm tail}_{0b0} 
+ h^{\rm tail}_{0bc} \Omega^c \bigr) + \Omega_b 
\bigl( h^{\rm tail}_{0a0} + h^{\rm tail}_{0ac} \Omega^c \bigr) 
+ \bigl( h^{\rm tail}_{ab0} + h^{\rm tail}_{abc} \Omega^c \bigr)
\biggr] + O(mr^2/{\cal R}^3).
\label{19.3.23}  
\end{eqnarray} 
Because $h^{\rm tails}_{\alpha'\beta'}$ is of order $m/{\cal R}$ and
$h^{\rm tails}_{\alpha'\beta'\gamma'}$ of order $m/{\cal R}^2$, we
see that the metric possesses the buffer-zone form ${\sf g} = \eta 
\oplus r^2/{\cal R}^2 \oplus m/r \oplus m/{\cal R} 
\oplus mr/{\cal R}^2$ that was anticipated in
Eq.~(\ref{19.1.4}). Notice that the expansion adopted here does not
contain a term at order $r/{\cal R}$ and presumes that $a_a$ and
$\omega_{ab}$ are both of order $m/{\cal R}^2$; this will be confirmed
in Sec.~\ref{19.5}.    

\subsection{Transformation from external to internal coordinates} 
\label{19.4} 

Comparison of Eqs.~(\ref{19.2.10})--(\ref{19.2.12}) and
Eqs.~(\ref{19.3.21})--(\ref{19.3.23}) reveals that the internal-zone
and external-zone metrics do no match in the buffer zone. But as the 
metrics are expressed in two different coordinate systems, this
mismatch is hardly surprising. A meaningful comparison of the two
metrics must therefore come after a transformation from the external 
coordinates $(u,r\Omega^a)$ to the internal coordinates
$(\bar{u},\bar{r}\bar{\Omega}^a)$. Our task in this subsection is to
construct this coordinate transformation. We shall proceed in three
stages. The first stage of the transformation,
$(u,r\Omega^a) \to (u',r'\Omega^{\prime a})$, will be seen to remove
unwanted terms of order $m/r$ in ${\sf g}_{\alpha\beta}$. The second
stage, $(u',r'\Omega^{\prime a}) \to 
(u'',r''\Omega^{\prime\prime a})$, will remove all terms of order
$m/{\cal R}$ in ${\sf g}_{\alpha'\beta'}$. Finally, the third stage
$(u'',r''\Omega^{\prime\prime a}) \to (\bar{u},\bar{r}\bar{\Omega}^a)$
will produce the desired internal coordinates. 

The first stage of the coordinate transformation is 
\begin{eqnarray} 
u' &=& u - 2m \ln r, 
\label{19.4.1} \\ 
x^{\prime a} &=& \Bigl( 1 + \frac{m}{r} \Bigr) x^a,   
\label{19.4.2}
\end{eqnarray} 
and it affects the metric at orders $m/r$ and $mr/{\cal R}^2$. This 
transformation redefines the radial coordinate --- $r \to r' = r + m$
--- and incorporates in $u'$ the gravitational time delay contributed
by the small mass $m$. After performing the coordinate transformation
the metric becomes    
\begin{eqnarray} 
{\sf g}_{u'u'} &=& -1 - r^{\prime 2} {\cal E}^{\prime *} 
+ O(r^{\prime 3}/{\cal R}^3)  
\nonumber \\ & & \mbox{} 
+ \frac{2m}{r'} + h^{\rm tail}_{00} + r' \bigl( 4m {\cal E}^{\prime *}  
- 2 a_a \Omega^{\prime a} + h^{\rm tail}_{000} + h^{\rm tail}_{00a}
\Omega^{\prime a} \bigr) + O(mr^{\prime 2}/{\cal R}^3),  
\label{19.4.3} \\ 
{\sf g}_{u'a'} &=& -\Omega'_a + \frac{2}{3} r^{\prime 2} \bigl(
{\cal E}^{\prime *}_a  + {\cal B}^{\prime *}_a \bigr) 
+ O(r^{\prime 3}/{\cal R}^3) 
\nonumber \\ & & \mbox{} 
+ h^{\rm tail}_{0a} 
+ \Omega'_a h^{\rm tail}_{00} + r' \biggl[ - \frac{4m}{3} 
\bigl( {\cal E}^{\prime *}_a + {\cal B}^{\prime *}_a \bigr)     
+ \bigr(\delta_a^{\ b} - \Omega'_a \Omega^{\prime b} \bigr) a_b  
\nonumber \\ & & \mbox{} 
- \omega_{ab} \Omega^{\prime b} + h^{\rm tail}_{0a0}  
+ \Omega'_a h^{\rm tail}_{000} + h^{\rm tail}_{0ab} \Omega^{\prime b}  
+ \Omega'_a h^{\rm tail}_{00b} \Omega^{\prime b} \biggr]  
+ O(mr^{\prime 2}/{\cal R}^3), 
\label{19.4.4} \\ 
{\sf g}_{a'b'} &=& \delta_{ab} - \Omega'_a \Omega'_b 
- \frac{1}{3} r^{\prime 2} \bigl( {\cal E}^{\prime *}_{ab} 
+ {\cal B}^{\prime *}_{ab} \bigr) + O(r^{\prime 3}/{\cal R}^3)  
\nonumber \\ & & \mbox{}
+ \Omega'_a \Omega'_b h^{\rm tail}_{00} + \Omega'_a h^{\rm tail}_{0b}  
+ \Omega'_b h^{\rm tail}_{0a} + h^{\rm tail}_{ab} 
+ r' \biggl[ \frac{2m}{3} \Bigl( {\cal E}^{\prime *}_{ab} 
+ \Omega'_a {\cal E}^{\prime *}_b 
+ {\cal E}^{\prime *}_a \Omega'_b 
\nonumber \\ & & \mbox{} 
+ {\cal B}^{\prime *}_{ab} 
+ \Omega'_a {\cal B}^{\prime *}_b 
+ \Omega'_b {\cal B}^{\prime *}_a \Bigr)     
+ \Omega'_a \Omega'_b \bigl( h^{\rm tail}_{000} + h^{\rm tail}_{00c}
\Omega^{\prime c} \bigr)  
+ \Omega'_a \bigl( h^{\rm tail}_{0b0} 
+ h^{\rm tail}_{0bc} \Omega^{\prime c} \bigr) 
\nonumber \\ & & \mbox{} 
+ \Omega'_b \bigl( h^{\rm tail}_{0a0} 
+ h^{\rm tail}_{0ac} \Omega^{\prime c} \bigr) 
+ \bigl( h^{\rm tail}_{ab0} 
+ h^{\rm tail}_{abc} \Omega^{\prime c} \bigr) \biggr] 
+ O(mr^{\prime 2}/{\cal R}^3).
\label{19.4.5}  
\end{eqnarray} 
This metric matches ${\sf g}(\mbox{internal zone})$ at orders $1$,
$r^{\prime 2}/{\cal R}^2$, and $m/r'$, but there is still a mismatch
at orders $m/{\cal R}$ and $mr'/{\cal R}^2$. 

The second stage of the coordinate transformation is  
\begin{eqnarray}
u'' &=& u' - \frac{1}{2} \int^{u'} h^{\rm tail}_{00}(u')\, du' 
- \frac{1}{2} r' \Bigl[ h^{\rm tail}_{00}(u') 
+ 2 h^{\rm tail}_{0a}(u') \Omega^{\prime a} 
+ h^{\rm tail}_{ab}(u') \Omega^{\prime a} \Omega^{\prime b} 
\Bigr], 
\label{19.4.6} \\ 
x''_a &=& x'_a 
+ \frac{1}{2} h^{\rm tail}_{ab}(u') x^{\prime b}, 
\label{19.4.7} 
\end{eqnarray}  
and it affects the metric at orders $m/{\cal R}$ and $mr/{\cal R}^2$. 
After performing this transformation the metric becomes   
\begin{eqnarray} 
\hspace*{-20pt} {\sf g}_{u''u''} &=& -1 
- r^{\prime\prime 2} {\cal E}^{\prime\prime *}  
+ O(r^{\prime\prime 3}/{\cal R}^3)  
\nonumber \\ & & \mbox{} 
+ \frac{2m}{r''} + r'' \biggl[ 4m {\cal E}^{\prime\prime *}  
- 2 \biggl(a_a -\frac{1}{2} h^{\rm tail}_{00a} + h^{\rm tail}_{0a0}
\biggr) \Omega^{\prime\prime a} \biggr] 
+ O(mr^{\prime\prime 2}/{\cal R}^3),  
\label{19.4.8} \\ 
\hspace*{-20pt} {\sf g}_{u''a''} &=& -\Omega''_a 
+ \frac{2}{3} r^{\prime\prime 2} \bigl(
{\cal E}^{\prime\prime *}_a  + {\cal B}^{\prime\prime *}_a \bigr)  
+ O(r^{\prime\prime 3}/{\cal R}^3) 
\nonumber \\ & & \mbox{} 
+ r'' \biggl[ - \frac{4m}{3} 
\bigl( {\cal E}^{\prime\prime *}_a + {\cal B}^{\prime\prime *}_a
\bigr) - 2 m {\cal E}_{ab} \Omega^{\prime\prime b} 
+ \bigr(\delta_a^{\ b} - \Omega''_a \Omega^{\prime\prime b}
\bigr) \biggl( a_b -\frac{1}{2} h^{\rm tail}_{00b} 
+ h^{\rm tail}_{0b0} \biggr) - \omega_{ab} \Omega^{\prime\prime b} 
\nonumber \\ & & \mbox{} 
+ \frac{1}{2} \Omega''_a h^{\rm tail}_{000} 
- \frac{1}{2} h^{\rm tail}_{ab0} \Omega^{\prime\prime b} 
+ h^{\rm tail}_{0ab} \Omega^{\prime\prime b}  
+ \frac{1}{2} \bigr(\delta_a^{\ b} 
+ \Omega''_a \Omega^{\prime\prime b} \bigr) h^{\rm tail}_{00b} \biggr]   
+ O(mr^{\prime\prime 2}/{\cal R}^3), 
\label{19.4.9} \\ 
\hspace*{-20pt} {\sf g}_{a''b''} &=& \delta_{ab} 
- \Omega''_a \Omega''_b  
- \frac{1}{3} r^{\prime\prime 2} \bigr( {\cal E}^{\prime\prime *}_{ab}  
+ {\cal B}^{\prime\prime *}_{ab} \bigr) 
+ O(r^{\prime\prime 3}/{\cal R}^3)   
\nonumber \\ & & \mbox{}
+ r'' \biggl[ \frac{2m}{3} \Bigl( {\cal E}^{\prime\prime *}_{ab} 
+ \Omega''_a {\cal E}^{\prime\prime *}_b 
+ {\cal E}^{\prime\prime *}_a \Omega''_b 
+ {\cal B}^{\prime\prime *}_{ab} 
+ \Omega''_a {\cal B}^{\prime\prime *}_b 
+ \Omega''_b {\cal B}^{\prime\prime *}_a \Bigr)     
+ \Omega''_a \Omega''_b \bigl( h^{\rm tail}_{000} + h^{\rm tail}_{00c}
\Omega^{\prime\prime c} \bigr)  
\nonumber \\ & & \mbox{} 
+ \Omega''_a \bigl( h^{\rm tail}_{0b0} 
+ h^{\rm tail}_{0bc} \Omega^{\prime\prime c} \bigr) 
+ \Omega''_b \bigl( h^{\rm tail}_{0a0} 
+ h^{\rm tail}_{0ac} \Omega^{\prime\prime c} \bigr) 
+ \bigl( h^{\rm tail}_{ab0} 
+ h^{\rm tail}_{abc} \Omega^{\prime\prime c} \bigr) \biggr] 
+ O(mr^{\prime\prime 2}/{\cal R}^3).
\label{19.4.10}  
\end{eqnarray} 
To arrive at these expressions we had to involve the relations  
\begin{equation} 
\frac{d}{du''} h^{\rm tail}_{00} = h^{\rm tail}_{000}, 
\qquad 
\frac{d}{du''} h^{\rm tail}_{0a} = h^{\rm tail}_{0a0}, 
\qquad
\frac{d}{du''} h^{\rm tail}_{ab} = 4m {\cal E}_{ab} 
+ h^{\rm tail}_{ab0} 
\label{19.4.11} 
\end{equation}
which are obtained by covariant differentiation of  
Eq.~(\ref{19.3.12}) with respect to $u$. The metric now matches 
${\sf g}(\mbox{internal zone})$ at orders $1$, 
$r^{\prime\prime 2}/{\cal R}^2$, $m/r''$, and $m/{\cal R}$, but there 
is still a mismatch at order $mr''/{\cal R}^2$. 

The third and final stage of the coordinate transformation is 
\begin{eqnarray} 
\bar{u} &=& u'' - \frac{1}{4} r^{\prime\prime 2} \Bigl[ 
h^{\rm tail}_{000} + \bigl(h^{\rm tail}_{00a} 
+ 2 h^{\rm tail}_{0a0} \bigr) \Omega^{\prime\prime a} 
+ \bigl( h^{\rm tail}_{ab0} + 2 h^{\rm tail}_{0ab} \bigr)    
\Omega^{\prime\prime a} \Omega^{\prime\prime b} 
+ h^{\rm tail}_{abc} \Omega^{\prime\prime a} \Omega^{\prime\prime b}
\Omega^{\prime\prime c} \Bigr], 
\label{19.4.12} \\ 
\bar{x}_a &=& \biggl( 1 + \frac{m}{3} r'' {\cal E}_{bc}   
\Omega^{\prime\prime b} \Omega^{\prime\prime c} \biggr)
x''_a + \frac{1}{2} r^{\prime\prime 2} \biggl[ 
-\frac{1}{2} h^{\rm tail}_{00a} + h^{\rm tail}_{0a0} 
+ \biggl( h^{\rm tail}_{0ab} - h^{\rm tail}_{0ba} 
+ h^{\rm tail}_{ab0} + \frac{4m}{3} {\cal E}_{ab} \biggr)
\Omega^{\prime\prime b} 
\nonumber \\ & & \mbox{} 
+ \bigl(Q_{abc} - Q_{bca} + Q_{cab} \bigr) 
\Omega^{\prime\prime b} \Omega^{\prime\prime c} \biggr], 
\label{19.4.13}
\end{eqnarray} 
where 
\begin{equation} 
Q_{abc} = \frac{1}{2} h^{\rm tail}_{abc} + \frac{m}{3}
\Bigl( \varepsilon_{acd} {\cal B}^d_{\ b} +
\varepsilon_{bcd} {\cal B}^d_{\ a} \Bigr). 
\label{19.4.14} 
\end{equation} 
This transformation puts the metric in its final form 
\begin{eqnarray} 
{\sf g}_{\bar{u}\bar{u}} &=& - 1 - \bar{r}^2 \bar{\cal E}^* 
+ O(\bar{r}^3/{\cal R}^3) 
\nonumber \\ & & \mbox{} 
+ \frac{2m}{\bar{r}} + \bar{r} \biggl[ 4m \bar{\cal E}^* 
- 2 \biggl( a_a -\frac{1}{2} h^{\rm tail}_{00a} 
+ h^{\rm tail}_{0a0} \biggr) \bar{\Omega}^a \biggr] 
+ O(m \bar{r}^2/{\cal R}^3), 
\label{19.4.15} \\ 
{\sf g}_{\bar{u}\bar{a}} &=& -\bar{\Omega}_a + \frac{2}{3} \bar{r}^2
\bigl( \bar{\cal E}^*_a + \bar{\cal B}^*_a \bigr) 
+ O(\bar{r}^3/{\cal R}^3) 
\nonumber \\ & & \mbox{} 
+ \bar{r} \biggl[ -\frac{4m}{3} \bigl( \bar{\cal E}^*_a 
+ \bar{\cal B}^*_a \bigr) + (\delta_a^{\ b} - \bar{\Omega}_a
\bar{\Omega}^b \bigr) \biggl( a_b -\frac{1}{2} h^{\rm tail}_{00b} 
+ h^{\rm tail}_{0b0} \biggr) 
\nonumber \\ & & \mbox{} 
- \bigl( \omega_{ab} 
- h^{\rm tail}_{0[ab]} \bigr) \bar{\Omega}^b \biggr] 
+ O(m \bar{r}^2/{\cal R}^3), 
\label{19.4.16} \\ 
{\sf g}_{\bar{a}\bar{b}} &=& \delta_{ab} - \bar{\Omega}_a
\bar{\Omega}_b - \frac{1}{3} \bar{r}^2 \bigl( \bar{\cal E}^*_{ab}  
+ \bar{\cal B}^*_{ab} \bigr) +  O(\bar{r}^3/{\cal R}^3) 
+ O(m \bar{r}^2/{\cal R}^3). 
\label{19.4.17} 
\end{eqnarray} 
Except for the terms involving $a_a$ and $\omega_{ab}$, this metric
is equal to ${\sf g}(\mbox{internal zone})$ as given by
Eqs.~(\ref{19.2.10})--(\ref{19.2.12}) linearized with respect to
$m$.    

\subsection{Motion of the black hole in the background spacetime} 
\label{19.5}  

A precise match between ${\sf g}(\mbox{external zone})$ and 
${\sf g}(\mbox{internal zone})$ is produced when we impose the
relations 
\begin{equation} 
a_a = \frac{1}{2} h^{\rm tail}_{00a} - h^{\rm tail}_{0a0}
\label{19.5.1}
\end{equation}
and 
\begin{equation}
\omega_{ab} = h^{\rm tail}_{0[ab]}. 
\label{19.5.2}
\end{equation} 
While Eq.~(\ref{19.5.1}) tells us how the black hole moves in the
background spacetime, Eq.~(\ref{19.5.2}) indicates that the vectors
$\base{\mu}{a}$ are not Fermi-Walker transported on the world line.  

The black hole's acceleration vector $a^\mu = a^a \base{\mu}{a}$ can
be constructed from the frame components listed in
Eq.~(\ref{19.5.1}). A straightforward computation gives 
\begin{equation}
a^\mu = -\frac{1}{2} \bigl( g^{\mu\nu} + u^\mu
u^\nu \bigr) \bigl( 2 h^{\rm tail}_{\nu\lambda\rho} 
- h^{\rm tail}_{\lambda\rho\nu} \bigr) u^\lambda u^\rho,  
\label{19.5.3} 
\end{equation}   
where the tail integral 
\begin{equation}
h^{\rm tail}_{\mu\nu\lambda} = 4 m \int_{-\infty}^{\tau^-}
\nabla_\lambda \biggl( G_{+\mu\nu\mu'\nu'}
- \frac{1}{2} g_{\mu\nu} G^{\ \ \rho}_{+\ \rho\mu'\nu'}
\biggr) \bigl( z(\tau), z(\tau')\bigr) u^{\mu'} u^{\nu'}\, d\tau' 
\label{19.5.4} 
\end{equation}     
was previously defined by Eq.~(\ref{19.3.13}). These are the
MiSaTaQuWa equations of motion, exactly as they were written down in 
Eq.~(\ref{18.6.5}). While the initial derivation of this result was
based upon formal manipulations of singular quantities, the present
derivation involves only well-behaved fields and is free of any
questionable aspect. Such a derivation, based on matched asymptotic
expansions, was first provided by Yasushi Mino, Misao Sasaki, and
Takahiro Tanaka in 1997 \cite{MST}.  

Substituting Eqs.~(\ref{19.5.1}) and (\ref{19.5.2}) into
Eq.~(\ref{19.3.4}) gives the following transport equation for the
tetrad vectors: 
\begin{equation} 
\frac{D \base{\mu}{a}}{d\tau} = -\frac{1}{2} u^\mu      
\bigl( 2 h^{\rm tail}_{\nu\lambda\rho} 
- h^{\rm tail}_{\nu\rho\lambda} \bigr) u^\nu \base{\lambda}{a} u^\rho 
+ \bigl(g^{\mu\rho} + u^\mu u^\rho \bigr) 
h^{\rm tail}_{\nu[\lambda\rho]} u^\nu \base{\lambda}{a}. 
\label{19.5.5}
\end{equation} 
This can also be written in the alternative form 
\begin{equation} 
\frac{D \base{\mu}{a}}{d\tau} = -\frac{1}{2} \Bigl( 
u^\mu \base{\lambda}{a} u^\rho + g^{\mu\lambda} \base{\rho}{a} 
- g^{\mu\rho} \base{\lambda}{a} \Bigr) u^\nu 
h^{\rm tail}_{\nu\lambda\rho} 
\label{19.5.6}
\end{equation} 
that was first proposed by Mino, Sasaki, and Tanaka. Both equations
state that in the background spacetime, the tetrad vectors are not
Fermi-Walker transported on $\gamma$; the rotation tensor is nonzero
and given by Eq.~(\ref{19.5.2}).   
  
%
\section{Concluding remarks} 
\label{20} 

I have presented a number of derivations of 
the equations that determine the motion of a point scalar charge $q$,
a point electric charge $e$, and a point mass $m$ in a specified
background spacetime. In this concluding section I summarize these 
derivations, and identify their strengths and weaknesses. I also
describe the challenges that lie ahead in the concrete evaluation of
the self-forces, most especially in the gravitational case.  

\subsection{Conservation of energy-momentum} 
\label{20.1} 

For each of the three cases (scalar, electromagnetic, and
gravitational) I have presented two different derivations of the
equations of motion. The first derivation is based on a spatial  
averaging of the retarded field, and the second is based on a
decomposition of the retarded field into singular and radiative
fields. In the gravitational case, a third derivation, based on
matched asymptotic expansions, was also presented. These derivations
will be reviewed below, but I want first to explain why I have omitted
to present a fourth derivation, based on energy-momentum conservation,
in spite of the fact that historically, it is one of the most
important.    

Conservation of energy-momentum was used by Dirac \cite{dirac} to
derive the equations of motion of a point electric charge in flat
spacetime, and the same method was adopted by DeWitt and Brehme
\cite{dewittbrehme} in their generalization of Dirac's work to curved
spacetimes. This method was also one of the starting points of Mino,
Sasaki, and Tanaka \cite{MST} in their calculation of the
gravitational self-force. I have not discussed this method for two 
reasons. First, it is technically more difficult to implement than the 
methods presented in this review (considerably longer computations are
involved). Second, it is difficult to endow this method with an
adequate level of rigour, to the point that it is perhaps less
convincing than the methods presented in this review. While the level
of rigour achieved in flat spacetime is now quite satisfactory
\cite{TVW}, I do not believe the same can be said of the
generalization to curved spacetimes. (But it should be possible to
improve on this matter.)   

The method is based on the conservation equation 
$T^{\alpha\beta}_{\ \ \ ;\beta} = 0$, where the stress-energy tensor
$T^{\alpha\beta}$ includes a contribution from the particle and a
contribution from the field; the particle's contribution is a Dirac
functional on the world line, and the field's contribution diverges as
$1/r^4$ near the world line. (I am using retarded coordinates in this
discussion.) While in flat spacetime the differential statement of
energy-momentum conservation can immediately be turned into an
integral statement, the same is not true in a curved spacetime (unless
the spacetime possesses at least one Killing vector). To proceed it is
necessary to rewrite the conservation equation as  
\[
0 = g^\mu_{\ \alpha} T^{\alpha\beta}_{\ \ \ ;\beta}  
= \bigl( g^\mu_{\ \alpha} T^{\alpha\beta} \bigr)_{;\beta} 
- g^\mu_{\ \alpha;\beta} T^{\alpha\beta}, 
\]
where $g^\mu_{\ \alpha}(z,x)$ is a parallel propagator from $x$ to an 
arbitrary point $z$ on the world line. Integrating this equation over
the interior of a world-tube segment that consists of a ``wall'' of
constant $r$ and two ``caps'' of constant $u$, we obtain 
\[
0 = 
\int_{\rm wall} g^\mu_{\ \alpha} T^{\alpha\beta} d\Sigma_\beta 
+ \int_{\rm caps} g^\mu_{\ \alpha} T^{\alpha\beta} d\Sigma_\beta 
+ \int_{\rm interior} g^\mu_{\ \alpha;\beta} T^{\alpha\beta}\, dV, 
\]
where $d\Sigma_\beta$ is a three-dimensional surface element and $dV$
an invariant, four-dimensional volume element. 

There is no obstacle in evaluating the wall integral, for which
$T^{\alpha\beta}$ reduces to the field's stress-energy tensor; for a
wall of radius $r$ the integral scales as $1/r^2$. The integrations
over the caps, however, are problematic: while the particle's
contribution to the stress-energy tensor is integrable, the
integration over the field's contribution goes as $\int_0^r
(r')^{-2}\, dr'$ and diverges. To properly regularize this integral 
requires great care, and the removal of all singular terms can be
achieved by mass renormalization \cite{dewittbrehme}. This issue
arises also in flat spacetime \cite{dirac}, and while it is plausible
that the rigourous distributional methods presented in Ref.~\cite{TVW}
could be generalized to curved spacetimes, this remains to be
done. More troublesome, however, is the interior integral, which does
not appear in flat spacetime. Because $g^\mu_{\ \alpha;\beta}$ scales
as $r$, this integral goes as $\int_0^r (r')^{-1}\, dr'$ and it also
diverges, albeit less strongly than the caps integration. While simply  
discarding this integral produces the correct equations of motion, it
would be desirable to go through a careful regularization of the
interior integration, and provide a convincing reason to discard
it altogether. To the best of my knowledge, this has not been done.            
 
\subsection{Averaging method} 
\label{20.2} 

To identify the strengths and weaknesses of the averaging method it  
is convenient to adopt the Detweiler-Whiting decomposition of the
retarded field into singular and radiative pieces. For concreteness I
shall focus my attention on the electromagnetic case, and write
\[
F_{\alpha\beta} = F^{\rm S}_{\alpha\beta} + F^{\rm R}_{\alpha\beta}.  
\]
Recall that this decomposition is unambiguous, and that the retarded 
and singular fields share the same singularity structure near the
world line. Recall also that the retarded and singular fields satisfy
the same field equations (with a distributional current density on the 
right-hand side), but that the radiative field is sourcefree. 

To formulate equations of motion for the point charge we temporarily 
model it as a spherical hollow shell, and we obtain the net force
acting on this object by averaging $F_{\alpha\beta}$ over the shell's
surface. (The averaging is performed in the shell's rest frame,
and the shell is spherical in the sense that its proper distance from
the world line is the same in all directions.) The averaged field is
next evaluated on the world line, in the limit of a zero-radius
shell. Because the radiative field is smooth on the world line, this
yields   
\[
e\langle F_{\mu\nu} \rangle u^\nu = e\langle F^{\rm S}_{\mu\nu}
\rangle u^\nu + e F^{\rm R}_{\mu\nu} u^\nu, 
\]
where 
\[
e\langle F^{\rm S}_{\mu\nu} \rangle u^\nu = -(\delta m) a_\mu, \qquad  
\delta m = \lim_{s\to 0} \biggl(\frac{2}{3} \frac{e^2}{s} \biggr) 
\]
and 
\[ 
e F^{\rm R}_{\mu\nu} u^\nu = e^2 \bigl( g_{\mu\nu} 
+ u_\mu u_\nu \bigr) \biggl( \frac{2}{3} \dot{a}^{\nu} 
+ \frac{1}{3} R^{\nu}_{\ \lambda} u^{\lambda} \biggr)  
+ 2 e^2 u^\nu \int_{-\infty}^{\tau^-}     
\nabla_{[\mu} G^+_{\ \nu]\lambda'}\bigl(z(\tau),z(\tau')\bigr)   
u^{\lambda'}\, d\tau'.
\]
The equations of motion are then postulated to be $m a_\mu = e\langle 
F_{\mu\nu} \rangle u^\nu$, where $m$ is the particle's bare
mass. With the preceding results we arrive at $m_{\rm obs} a_\mu = 
e F^{\rm R}_{\mu\nu} u^\nu$, where $m_{\rm obs} \equiv m + \delta m$
is the particle's observed (renormalized) inertial mass.  

The averaging method is sound, but it is not immune to criticism. A
first source of criticism concerns the specifics of the averaging
procedure, in particular, the choice of a spherical surface over any
other conceivable shape. Another source is a slight inconsistency of
the method that gives rise to the famous ``4/3 problem''
\cite{rohrlich}: the mass shift $\delta m$ is related to the shell's
electrostatic energy $E = e^2/(2s)$ by $\delta m = \frac{4}{3} E$
instead of the expected $\delta m = E$. This problem is
likely due \cite{ori-rosenthal:03} to the fact that the field that is
averaged over the surface of the shell is sourced by a point particle
and not by the shell itself. It is plausible that a more careful
treatment of the near-source field will eliminate both sources of
criticism: We can expect that the field produced by an extended
spherical object will give rise to a mass shift that equals the
object's electrostatic energy, and the object's spherical shape would
then fully justify a spherical averaging. (Considering other shapes
might also be possible, but one would prefer to keep the object's
structure simple and avoid introducing additional multipole moments.)
Further work is required to clean up these details. 

The averaging method is at the core of the approach followed by Quinn
and Wald \cite{QW1}, who also average the retarded field over a
spherical surface surrounding the particle. Their approach, however,
also incorporates a ``comparison axiom'' that allows them to avoid
renormalizing the mass.  

\subsection{Detweiler-Whiting axiom} 
\label{20.3} 

The Detweiler-Whiting decomposition of the retarded field becomes most 
powerful when it is combined with the Detweiler-Whiting axiom, which
asserts that  
\begin{verse} 
the singular field exerts no force on the particle (it merely
contributes to the particle's inertia); the entire self-force arises
from the action of the radiative field.
\end{verse} 
This axiom, which is motivated by the symmetric nature of the singular
field, and also its causal structure, gives rise to the equations of
motion $m a_\mu = e F^{\rm R}_{\mu\nu} u^\nu$, in agreement with the
averaging method (but with an implicit, instead of explicit, mass
shift). In this picture, the particle simply interacts with a free
radiative field (whose origin can be traced to the particle's past),
and the procedure of mass renormalization is sidestepped. In the
scalar and electromagnetic cases, the picture of a particle
interacting with a radiative field removes any tension between the
nongeodesic motion of the charge and the principle of equivalence. In
the gravitational case the Detweiler-Whiting axiom produces the
statement that the point mass $m$ moves on a geodesic in a spacetime
whose metric $g_{\alpha\beta} + h^{\rm R}_{\alpha\beta}$ is
nonsingular and a solution to the vacuum field equations. This is a  
conceptually powerful, and elegant, formulation of the MiSaTaQuWa
equations of motion. 

\subsection{Matched asymptotic expansions} 
\label{20.4} 

It is well known that in general relativity, the motion of gravitating
bodies is determined, along with the spacetime metric, by the Einstein
field equations; the equations of motion are not separately
imposed. This observation provides a means of deriving the MiSaTaQuWa
equations without having to rely on the fiction of a point mass. In
the method of matched asymptotic expansions, the small body is taken
to be a nonrotating black hole, and its metric perturbed by the tidal
gravitational field of the external universe is matched to the metric
of the external universe perturbed by the black hole. The equations of
motion are then recovered by demanding that the metric be a valid
solution to the vacuum field equations. This method, which was the 
second starting point of Mino, Sasaki, and Tanaka \cite{MST}, gives
what is by far the most compelling derivation of the MiSaTaQuWa
equations. Indeed, the method is entirely free of conceptual and
technical pitfalls --- there are no singularities (except deep inside
the black hole) and only retarded fields are employed.  

The introduction of a point mass in a nonlinear theory of gravitation 
would appear at first sight to be severely misguided. The lesson
learned here is that {\it one can in fact get away with it}. The
derivation of the MiSaTaQuWa equations of motion based on the method
of matched asymptotic expansions does indeed show that results
obtained on the basis of a point-particle description can be reliable,
in spite of all their questionable aspects. This is a remarkable
observation, and one that carries a lot of convenience: it is much
easier to implement the point-mass description than to perform the
matching of two metrics in two coordinate systems. 

\subsection{Evaluation of the gravitational self-force} 
\label{20.5} 
  
The concrete evaluation of the scalar, electromagnetic, and
gravitational self-forces is made challenging by the need to first
obtain the relevant retarded Green's function. Successes achieved in
the past were reviewed in Sec.~\ref{1.10}, and here I want to describe
the challenges that lie ahead. I will focus on the specific task of
computing the gravitational self-force acting on a point mass that
moves in a background Kerr spacetime. This case is especially
important because the motion of a small compact object around a
massive (galactic) black hole is a promising source of low-frequency
gravitational waves for the Laser Interferometer Space Antenna
\cite{LISA}; to calculate these waves requires an accurate description
of the motion, beyond the test-mass approximation which ignores the
object's radiation reaction.  

The gravitational self-acceleration is given by the MiSaTaQuWa
expression, which I write in the form 
\[
\frac{D u^\mu}{d\tau} = a^\mu \bigl[h^{\rm R} \bigr] \equiv  
-\frac{1}{2} \bigl( g^{\mu\nu} + u^\mu u^\nu \bigr) 
\bigl( 2 h^{\rm R}_{\nu\lambda;\rho} 
- h^{\rm R}_{\lambda\rho;\nu} \bigr) u^\lambda u^\rho, 
\]
where $h^{\rm R}_{\alpha\beta}$ is the radiative part of the metric
perturbation. Recall that this equation is equivalent to the statement
that the small body moves on a geodesic of a spacetime with metric
$g_{\alpha\beta} + h^{\rm R}_{\alpha\beta}$. Here $g_{\alpha\beta}$ is
the Kerr metric, and we wish to calculate $a^\mu[h^{\rm R}]$ for a
body moving in the Kerr spacetime. This calculation is challenging and
it involves a large number of steps.  

The first sequence of steps are concerned with the computation of the
(retarded) metric perturbation $h_{\alpha\beta}$ produced by a point 
particle moving on a specified geodesic of the Kerr spacetime. A
method for doing this was elaborated by Lousto and Whiting
\cite{loustowhiting} and Ori \cite{ori:03}, building on the
pioneering work of Teukolsky \cite{teukolsky:73}, Chrzanowski
\cite{chrzanowski:75}, and Wald \cite{wald:78}. The procedure consists
of (i) solving the Teukolsky equation for one of the Newman-Penrose
quantities $\psi_0$ and $\psi_4$ (which are complex components of the
Weyl tensor) produced by the point particle; (ii) obtaining from
$\psi_0$ or $\psi_4$ a related (Hertz) potential $\Psi$ by integrating
an ordinary differential equation; (iii) applying to $\Psi$ a number
of differential operators to obtain the metric perturbation in a 
radiation gauge that differs from the Lorenz gauge; and (iv)
performing a gauge transformation from the radiation gauge to
the Lorenz gauge.

It is well known that the Teukolsky equation separates when
$\psi_0$ or $\psi_4$ is expressed as a multipole expansion, summing
over modes with (spheroidal-harmonic) indices $l$ and $m$. In fact,
the procedure outlined above relies heavily on this mode
decomposition, and the metric perturbation returned at the end of the
procedure is also expressed as a sum over modes
$h^l_{\alpha\beta}$. (For each $l$, $m$ ranges from $-l$ to $l$, and
summation of $m$ over this range is henceforth understood.) From
these, mode contributions to the self-acceleration can be computed:
$a^\mu[h_l]$ is obtained from our preceding expression for the
self-acceleration by substituting $h^l_{\alpha\beta}$ in place of
$h^{\rm R}_{\alpha\beta}$. These mode contributions do not diverge on
the world line, but $a^\mu[h_l]$ is discontinuous at the radial
position of the particle. The sum over modes, on the other hand, does
not converge, because the ``bare'' acceleration (constructed from the 
retarded field $h_{\alpha\beta}$) is formally infinite.

The next sequence of steps is concerned with the regularization of 
each $a^\mu[h_l]$ by removing the contribution from 
$h^{\rm S}_{\alpha\beta}$ \cite{BMNOS, BO2, BO3, BO4, MNS, DMW}. The 
singular field can be constructed locally in a neighbourhood of the
particle, and then decomposed into modes of multipole order $l$. This
gives rise to modes $a^\mu[h^{\rm S}_l]$ for the singular part of the
self-acceleration; these are also finite and discontinuous, and their
sum over $l$ also diverges. But the true modes $a^\mu[h^{\rm R}_l] =
a^\mu[h_l] - a^\mu[h^{\rm S}_l]$ of the self-acceleration are
continuous at the radial position of the particle, and their sum does 
converge to the particle's acceleration. (It might be noted that
obtaining a mode decomposition of the singular field involves
providing an extension of $h^{\rm S}_{\alpha\beta}$ on a sphere of
constant radial coordinate, and then integrating over the angular
coordinates. The arbitrariness of the extension introduces
ambiguities in each $a^\mu[h^{\rm S}_l]$, but the ambiguity
disappears after summing over $l$.)

The self-acceleration is thus obtained by first
computing $a^\mu[h_l]$ from the metric perturbation derived
from $\psi_0$ or $\psi_4$, then computing the counterterms
$a^\mu[h^{\rm S}_l]$ by mode-decomposing the singular field, and
finally summing over all $a^\mu[h^{\rm R}_l] = a^\mu[h_l]
- a^\mu[h^{\rm S}_l]$. This procedure is lengthy and involved, and
thus far it has not been brought to completion, except for the special
case of a particle falling radially toward a nonrotating black hole
\cite{baracklousto}. In this regard it should be noted that the 
replacement of the central Kerr black hole by a Schwarzschild black
hole simplifies the task considerably. In particular, because there
exists a practical and well-developed formalism to describe the metric 
perturbations of a Schwarzschild spacetime \cite{reggewheeler,
vishveshwara, zerilli}, there is no necessity to rely on the Teukolsky
formalism and the complicated reconstruction of the metric variables. 

The procedure described above is lengthy and involved, but it is also
incomplete. The reason is that the metric perturbations
$h^l_{\alpha\beta}$ that can be recovered from $\psi_0$ or $\psi_4$ do
not by themselves sum up to the complete gravitational perturbation
produced by the moving particle. Missing are the perturbations derived
from the other Newman-Penrose quantities: $\psi_1$, $\psi_2$, and
$\psi_3$. While $\psi_1$ and $\psi_3$ can always be set to zero by an
appropriate choice of null tetrad, $\psi_2$ contains such important
physical information as the shifts in mass and angular-momentum
parameters produced by the particle \cite{wald:73}. Because the mode 
decompositions of $\psi_0$ and $\psi_4$ start at $l=2$, we might
colloquially say that what is missing from the above procedure are the
``$l=0$ and $l=1$'' modes of the metric perturbations. It is not
currently known how the procedure can be completed so as to
incorporate {\it all modes} of the metric perturbations. Specializing
to a Schwarzschild spacetime eliminates this difficulty, and in this
context the low multipole modes have been studied for the special case
of circular orbits \cite{nakano-etal:03, detweiler-poisson:03}. 

In view of these many difficulties (and I choose to stay silent on
others, for example, the issue of relating metric perturbations in
different gauges when the gauge transformation is singular on the
world line), it is perhaps not too surprising that such a small number
of concrete calculations have been presented to date. But progress in
dealing with these difficulties has been steady, and the situation
should change dramatically in the next few years.  

\subsection{Beyond the self-force} 
\label{20.6} 

The successful computation of the gravitational self-force is not the 
end of the road. After the difficulties reviewed in the preceding
subsection have all been removed and the motion of the small body
is finally calculated to order $m$, it will still be necessary to
obtain gauge-invariant information associated with the body's
corrected motion. Because the MiSaTaQuWa equations of motion are not
by themselves gauge-invariant, this step will necessitate going beyond 
the self-force. 

To see how this might be done, imagine that the small body is a 
pulsar, and that it emits light pulses at regular proper-time
intervals. The motion of the pulsar around the central black hole 
modulates the pulse frequencies as measured at infinity, and
information about the body's corrected motion is encoded in
the times-of-arrival of the pulses. Because these can be measured
directly by a distant observer, they clearly constitute
gauge-invariant information. But the times-of-arrival are determined
not only by the pulsar's motion, but also by the propagation of
radiation in the perturbed spacetime. This example shows that to
obtain gauge-invariant information, one must properly combine the
MiSaTaQuWa equations of motion with the metric perturbations.   

In the context of the Laser Interferometer Space Antenna, the relevant
observable is the instrument's response to a gravitational wave, which
is determined by gauge-invariant waveforms, $h_+$ and $h_\times$. To
calculate these is the ultimate goal of this research programme, and
the challenges that lie ahead go well beyond what I have described
thus far. To obtain the waveforms it will be necessary to solve the
Einstein field equations to {\it second order} in perturbation theory.  

To understand this, consider first the formulation of the first-order
problem. Schematically, one introduces a perturbation $h$ that
satisfies a wave equation $\Box h = T[z]$ in the background
spacetime, where $T[z]$ is the stress-energy tensor of the moving
body, which is a functional of the world line $z(\tau)$. In
first-order perturbation theory, the stress-energy tensor must be
conserved in the background spacetime, and $z(\tau)$ must describe a
geodesic. It follows that in first-order perturbation theory, the
waveforms constructed from the perturbation $h$ contain no information
about the body's corrected motion.

The first-order perturbation, however, can be used to correct the
motion, which is now described by the world line $z(\tau) + \delta
z(\tau)$. In a naive implementation of the self-force, one would now
re-solve the wave equation with a corrected stress-energy tensor,
$\Box h = T[z + \delta z]$, and the new waveforms constructed from
$h$ would then incorporate information about the corrected
motion. This implementation is naive because this information would
not be gauge-invariant. In fact, to be consistent one would have to
include {\it all} second-order terms in the wave equation, not 
just the ones that come from the corrected motion. Schematically, the
new wave equation would have the form of $\Box h = (1 + h) T[z +
\delta z] + (\nabla h)^2$, and this is much more difficult to solve
than the naive problem (if only because the source term is now
much more singular than the distributional singularity contained in
the stress-energy tensor). But provided one can find a way to make
this second-order problem well posed, and provided one can solve it
(or at least the relevant part of it), the waveforms constructed from
the second-order perturbation $h$ will be gauge invariant. In this
way, information about the body's corrected motion will have properly
been incorporated into the gravitational waveforms.       

The story is far from being over.   

\newpage
\bibliography{motion}

\end{document}